# The Spanish Square Kilometre Array White Book

**Editors**
Miguel Pérez-Torres
Lourdes Verdes-Montenegro
José Carlos Guirado
Antonio Alberdi
Jesús Martín-Pintado
Rafael Bachiller
Diego Herranz
Josep Miquel Girart
Simone Migliari
José Miguel Rodríguez-Espinosa

SKA
SQUARE KILOMETRE ARRAY
Spain

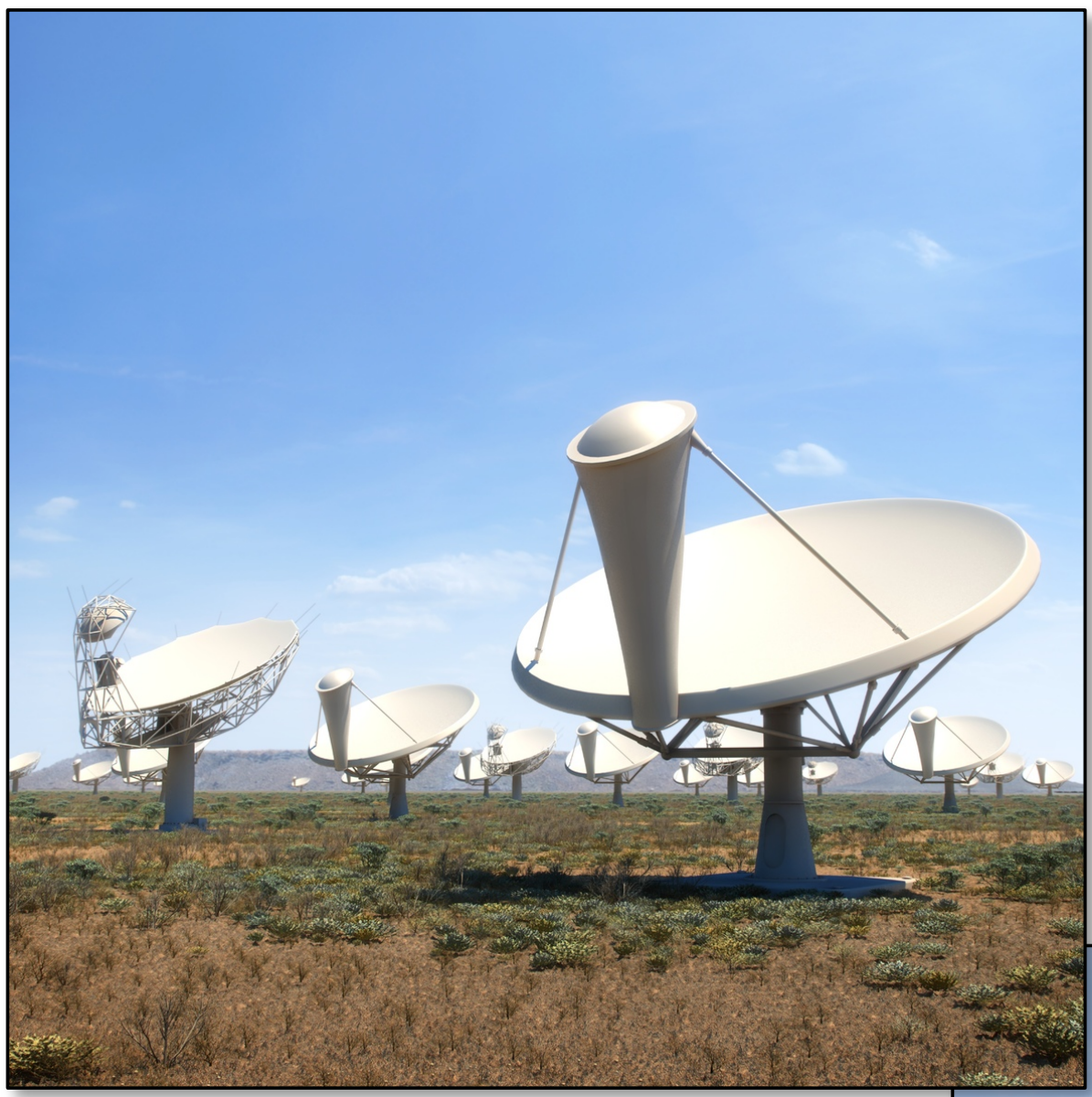

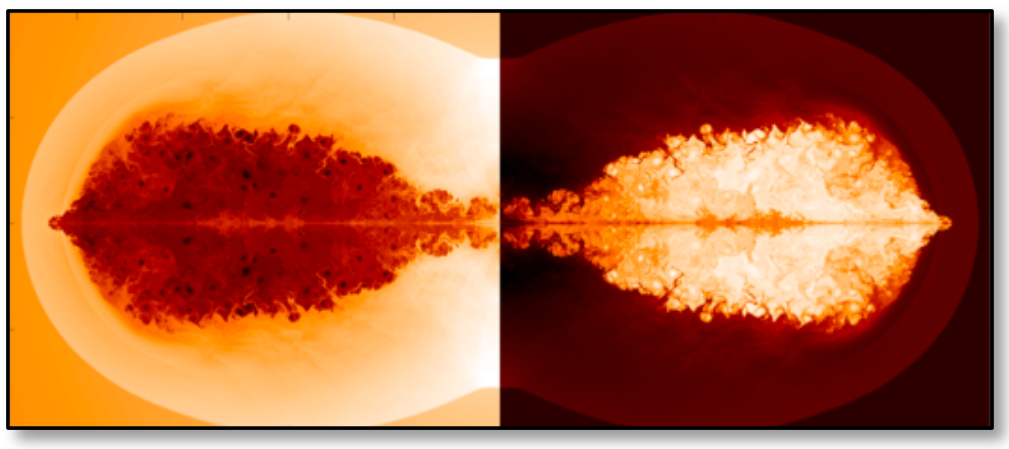

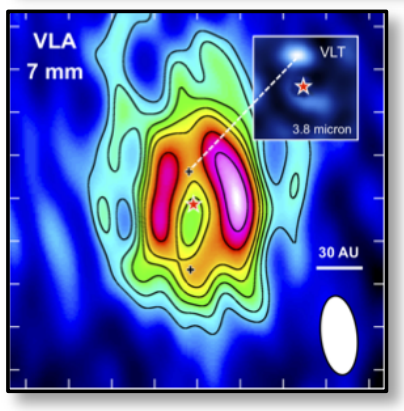

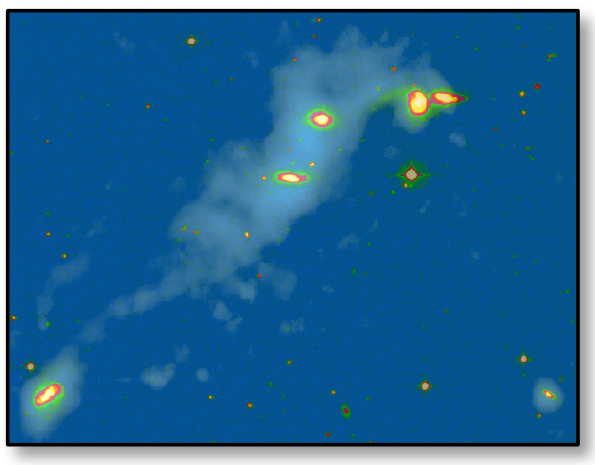

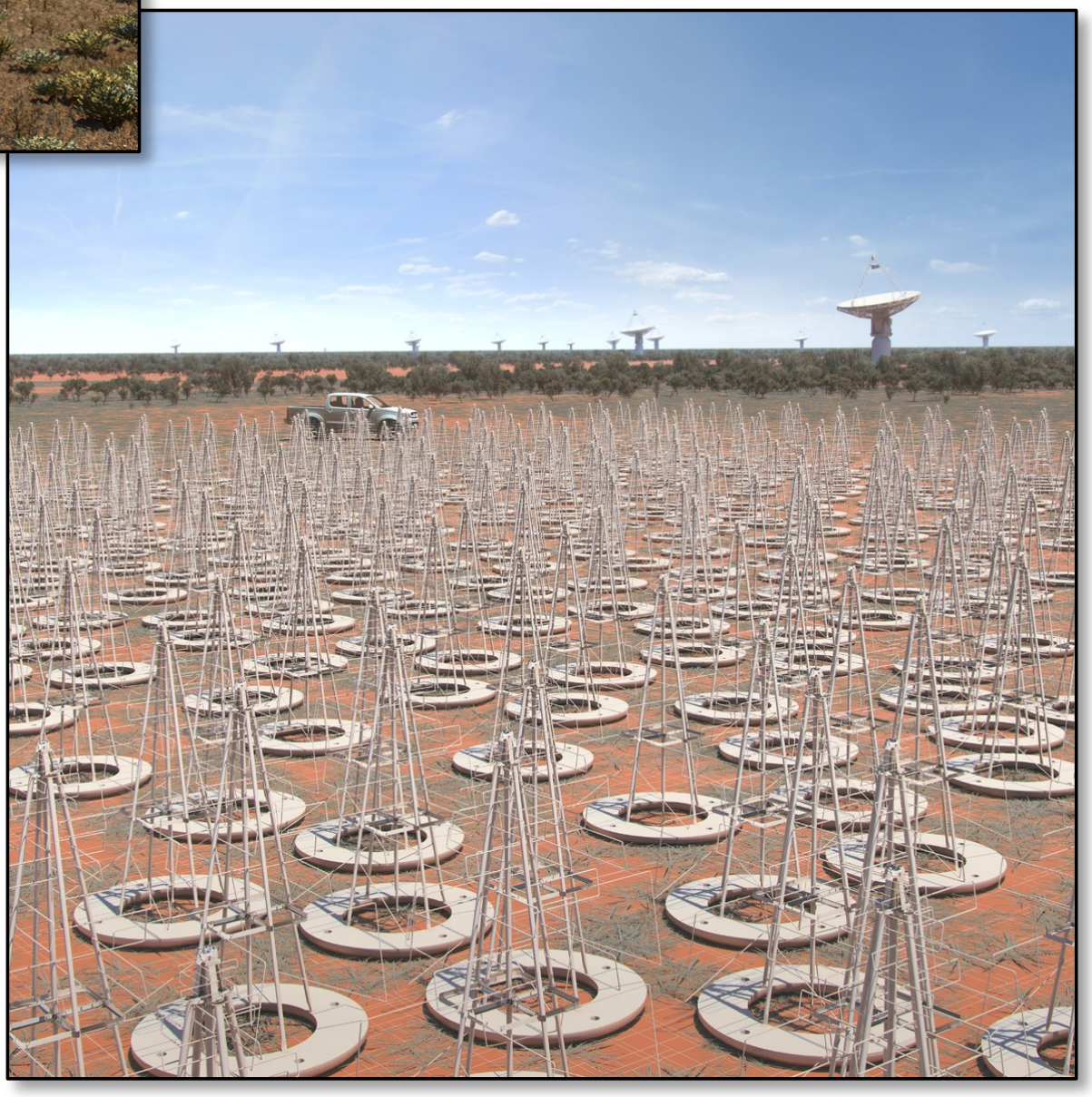

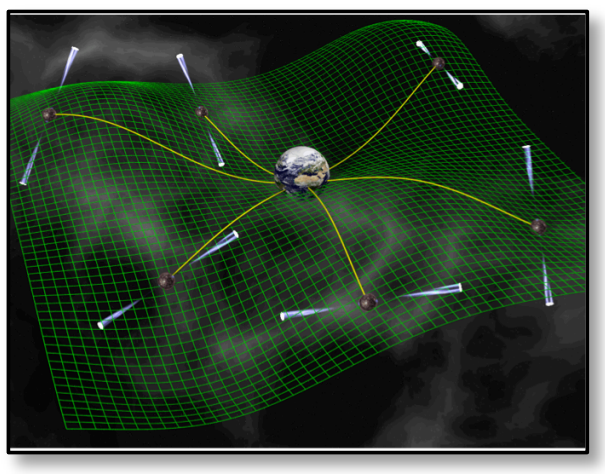

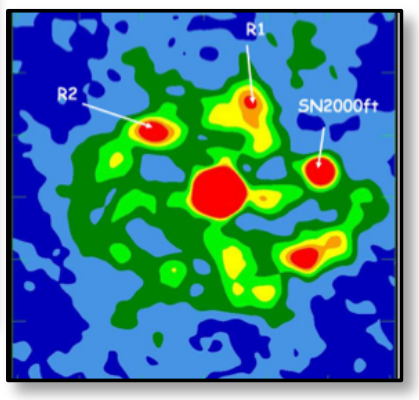

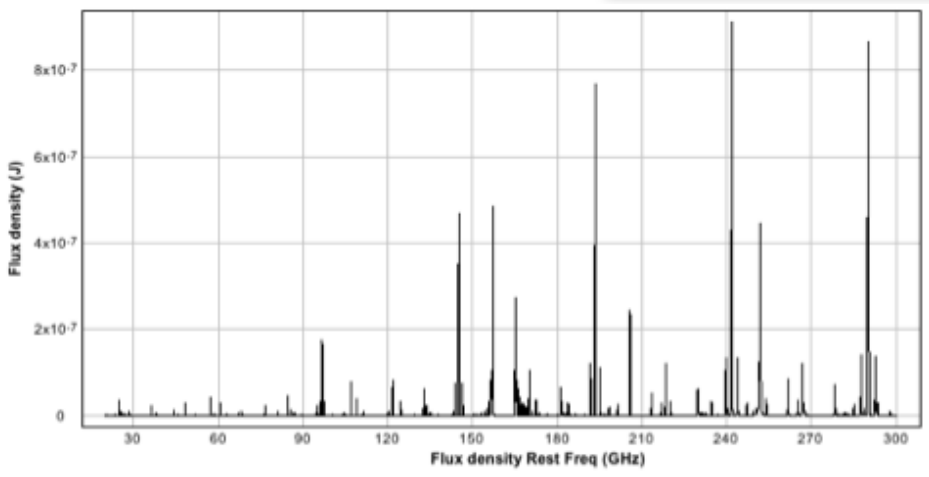

**Editors**
Miguel Pérez-Torres (Editor-in-chief; Instituto de Astrofísica de Andalucía, IAA-CSIC, Spain & Centro de Estudios de la Física del Cosmos de Aragón, CEFCA, Spain)
Lourdes Verdes-Montenegro (Instituto de Astrofísica de Andalucía, IAA-CSIC, Spain)
José Carlos Guirado (Observatori Astronòmic, Universitat de València, Spain; Departament d'Astronomia i Astrofísica, Universitat de València, Spain)
Antonio Alberdi (Instituto de Astrofísica de Andalucía, IAA-CSIC, Spain)
Jesús Martín-Pintado (Centro de Astrobiología, INTA-CSIC, Spain)
Rafael Bachiller (Observatorio Astronómico Nacional, IGN, Spain)
Diego Herranz (Instituto de Física de Cantabria, IFCA/CSIC-UC), Spain)
Josep Miquel Girart (Institut de Ciències de l'Espai, CSIC-IEEC, Spain)
Javier Gorgas (Departamento de Astrofísica y Física de la Atmósfera, Universidad Complutense de Madrid, Spain)
Carlos Hernández-Monteagudo (Centro de Estudios de la Física del Cosmos de Aragón, CEFCA, Spain)
Simone Migliari (European Space Astronomy Centre, European Space Agency, Spain; Departament d'Astronomia i Meteorologia, Universitat de Barcelona, Spain)
José Miguel Rodríguez-Espinosa (Instituto de Astrofísica de Canarias, IAC, Spain; Departamento de Astrofísica, Universidad de La Laguna, Spain)

# The Spanish Square Kilometre Array White Book

## Contributing authors

Acosta Pulido J. A. (Instituto de Astrofísica de Canarias), Spain), Agudo I. (Instituto de Astrofísica de Andalucía, CSIC, Spain), Alberdi A. (Instituto de Astrofí sica de Andalucía, CSIC, Spain), Alcolea J. (Observatorio Astronómico Nacional, IGN, Spain), Alfaro E.J. (Instituto de Astrofísica de Andalucía, CSIC, Spain), Alonso-Herrero A. (Instituto de Física de Cantabria, CSIC-UC, Spain), Anglada G. (Instituto de Astrofísica de Andalucía, CSIC, Spain), Arnalte-Mur P. (Observatori Astronòmic de la Universitat de València, Spain), Ascasibar Y. (Departamento de Física Teórica, Universidad Autónoma de Madrid, Spain), Ascaso B. (GEPI, Observatoire de Paris, CNRS, Université Paris Diderot, France), Azulay R. (Departamento de Astronomía y Astrofísica, Universidad de Valencia, Spain), Bachiller R. (Observatorio Astronómico Nacional, IGN, Spain), Báez-Rubio A. (Centro de Astrobiología, CSIC/INTA, Spain), Battaner E. (Departamento de Física Teórica y del Cosmos, Universidad de Granada, Spain; Instituto Carlos I de Física Teórica y Computacional, Universidad de Granada, Spain), Blasco J. (Instituto de Astrofísica de Andalucía, CSIC, Spain), Brook C.B. (Departamento de Física Teórica, Universidad Autónoma de Madrid, Spain), Bujarrabal V. (Observatorio Astronómico Nacional, IGN, Spain), Busquet G. (Instituto de Astrofísica de Andalucía, CSIC, Spain), Caballero-Garcia M. D. (Czech Technical University in Prague, Faculty of Electrical Engineering, Czech Republic), Carrasco-González C. (Centro de Radioastronomía y Astrofísica, UNAM, Mexico), Casares J. (Instituto de Astrofísica de Canarias, Spain; Departamento de Astrofísica, Universidad de La Laguna, Spain), Castro-Tirado A.J. (Instituto de Astrofísica de Andalucía, CSIC, Spain; Unidad Asociada Ingeniería de Sistemas y Automática, ISA-UMA, Universidad de Málaga, Spain), Colina L. (Astrophysics Department, Center for Astrobiology, CSIC-INTA, Spain), Colomer F. (Observatorio Astronómico Nacional, Alcalá de Henares, Spain), de Gregorio-Monsalvo I. (Joint ALMA Observatory, Chile), del Olmo A. (Instituto de Astrofísica de Andalucía, CSIC, Spain), Desmurs J-F (Observatorio Astronómico Nacional, IGN, Spain), Diego J.M. (Instituto de Física de Cantabria, CSIC-UC, Spain), Domínguez-Tenreiro R. (Departamento de Física Teórica, Universidad Autónoma de Madrid, Spain), Estalella R. (Departament d'Astronomia i Meteorologia, Univ. de Barcelona, Spain), Fernández-Soto A. (Instituto de Física de Cantabria, Universidad de Cantabria-CSIC, Spain; Unidad Asociada Observatorio Astronómico, Universitat de València-IFCA, Spain), Florido E. (Departamento de Física Teórica y del Cosmos, Universidad de Granada, Spain; Instituto Carlos I de Física Teórica y Computacional, Universidad de Granada, Spain), Font J. (Instituto de Astrofísica de Canarias, Spain), Font J.A. (Departamento de Astronomía y Astrofísica, Universitat de València, Spain; Observatori Astronòmic, Universitat de València, Spain), Fuente A. (Observatorio Astronómico Nacional, IGN, Spain), García-Burillo S. (Observatorio AstronæťŃico Nacional-Observatorio de Madrid, Spain), García-Benito R. (Instituto de Astrofísica de Andalucía, CSIC, Spain), García-Lorenzo B. (Instituto de Astrofísica de Canarias, Spain; Departamento de Astrofísica, Universidad de La Laguna, Spain), Gil de Paz A. (Dept. Astrofísica, U. Complutense, CC. Físicas, Spain), Girart J.M. (Institut de Ciències de l'Espai, CSIC-IEEC, Campus UAB, Spain), Goicoechea J.R. (Instituto de Ciencia de Materiales de Madrid, ICMM-CSIC, Spain), Gómez J.F. (Instituto de Astrofísica de Andalucía, CSIC, Spain), González-García M. (Group of Molecular Astrophysics, ICMM, CSIC, Spain), Gonzalez-Martin O. (Centro de Radioastronomía y Astrofísica, CRyA-UNAM, Mexico), González-Serrano J.I. (Instituto de Física de Cantabria (Universidad de Cantabria-CSIC),

Spain), Gorgas J. (Departamento de Astrofísica y Física de la Atmósfera, Universidad Complutense de Madrid, Madrid, Spain), Gorosabel[1] J. (Instituto de Astrofísica de Andalucía, CSIC, Spain; Unidad Asociada Grupo Ciencias Planetarias, UPV/EHU, IAA-CSIC, Departamento de Física Aplicada I, E.T.S. Ingeniería, Universidad del País Vasco, UPV/EHU, Spain; Ikerbasque, Basque Foundation for Science, Spain), Guijarro A. (Centro Astronómico Hispano Alemán, Calar Alto, Spain), Guirado J.C. (Observatori Astronòmic, Universitat de València, Spain; Departament d'Astronomia i Astrofísica, Universitat de València, Spain), Hernández-García L. (Instituto de Astrofísica de Andalucía, CSIC, Spain), Hernández-Monteagudo C. (Centro de Estudios de Física del Cosmos de Aragón, Spain), Herranz D. (Instituto de Física de Cantabria, CSIC-UC, Spain), Herrero-Illana R. (Instituto de Astrofísica de Andalucía, CSIC, Spain), Hu Y-D (Instituto de Astrofísica de Andalucía, CSIC, Spain), Huélamo N. (Centro de Astrobiología, INTA-CSIC, Spain), Huertas-Company M. (GEPI, Observatoire de Paris, CNRS, France), Iglesias-Páramo J. (Instituto de Astrofísica de Andalucía, CSIC, Spain), Jeong S. (Instituto de Astrofísica de Andalucía, CSIC, Spain), Jiménez-Serra I. (University College London, Department of Physics and Astronomy, UK), Knapen J.H. (Instituto de Astrofísica de Canarias, Spain; Departamento de Astrofísica, Universidad de La Laguna, Spain), Lineros R.A. (Instituto de Física Corpuscular, CSIC-UV, Spain), Lisenfeld U. (Departamento de Física Teórica y del Cosmos, Universidad de Granada, Spain; Instituto Universitario Carlos I de Física Teórica y Computacional, Facultad de Ciencias, Spain), Marcaide J.M. (Dept. Astronomia i Astrofisica, Universitat de València, Spain), Márquez I. (Instituto de Astrofísica de Andalucía, CSIC, Spain), Martí J. (Departamento de Física, Escuela Politécnica Superior de Jaén, Universidad de Jaén, Spain), Martí J.M. (Dept. d'Astronomia i Astrofísica. Universitat de València, Spain; Obs. Astronòmic. Universitat de València, Spain), Martínez-González E. (Instituto de Física de Cantabria, CSIC-UC, Spain), Martín-Pintado J. (Centro de Astrobiología, INTA-CSIC, Spain), Martí-Vidal I. (Onsala Space Obs., Chalmers University of Technology, Sweden), Masegosa J. (Instituto de Astrofísica de Andalucía, CSIC, Spain), Mayen-Gijon J.M. (Instituto de Astrofísica de Andalucía, CSIC, Spain), Mezcua M. (Harvard-Smithsonian Center for Astrophysics, USA), Migliari S. (European Space Astronomy Centre, European Space Agency, Spain; Departament d'Astronomia i Meteorologia, Universitat de Barcelona, Spain), Mimica P. (Dept. d'Astronomia i Astrofísica. Universitat de València, Spain), Moldón J. (Departament d'Astronomia i Meteorologia, Institut de Ciències del Cosmos, Universitat de Barcelona, IEEC-UB, Spain; ASTRON, the Netherlands Institute for Radio Astronomy, the Netherlands), Morata Ó. (nstitute of Astronomy and Astrophysics, Academia Sinica,Taiwan), Negueruela, I. (Departamento de Física, Ingeniería de Sistemas y Teoría de la Señal, Universidad de Alicante, Spain), Oates S.R. (Instituto de Astrofísica de Andalucía, CSIC, Spain), Osorio M. (Instituto de Astrofísica de Andalucía, CSIC, Spain), Palau A. (Centro de Radioastronomía y Astrofísica, Universidad Nacional Autónoma de México, México), Paredes J.M. (Departament d'Astronomia i Meteorologia, Institut de Ciències del Cosmos, Universitat de Barcelona, Spain), Perea J. (Instituto de Astrofísica de Andalucía, CSIC, Spain), Pérez-González P.G. (Departamento de Astrofísica, Facultad de CC. Físicas, Universidad Complutense de Madrid, Spain), Pérez-Montero E. (Instituto de Astrofísica de Andalucía, CSIC, Spain), Pérez-Torres M.A. (Instituto de Astrofísica de Andalucía, CSIC,

[1]Deceased

Spain; Centro de Estudios de la Física del Cosmos de Aragón, Spain; Departamento de Física Teórica, Universidad de Zaragoza, Spain), Perucho M. (Dept. d'Astronomia i Astrofísica, Universitat de València, Spain; Observatori Astronòmic, Universitat de València, Spain), Planelles S. (Astronomy Unit, Department of Physics, University of Trieste, Italy; INAF, Osservatorio Astronomico di Trieste, Italy), Pons J.A. (Departament de Fisica Aplicada, Universitat d'Alacant, Spain), Prieto A. (Instituto de Astrofísica de Canarias, Spain; Departamento de Astrofísica, Universidad de La Laguna, Spain), Quilis V. (Dept. d'Astronomia i Astrofísica. Universitat de València, Spain; Obs. Astronòmic. Universitat de València, Spain), Ramírez-Moreta P. (IAA-CSIC, Spain), Ramos Almeida C. (Instituto de Astrofísica de Canarias, Spain; Departamento de Astrofísica, Universidad de La Laguna, Spain), Rea N. (Anton Pannekoek Institute for Astronomy, University of Amsterdam, The Netherlands; Institute of Space Sciences, CSIC-IEEC, Campus UAB, Spain), Ribó M. (Departament d'Astronomia i Meteorologia, Institut de Ciències del Cosmos, Universitat de Barcelona, IEEC-UB, Spain; Serra Húnter Fellow), Rioja M.J. (Onsala Space Observatory, Chalmers Univ. of Technology, Sweden; Korea Astronomy and Space Science Institute, Korea; International Centre for Radio Astronomy Research, Univ. Western Australia, Australia; Observatorio Astronómico Nacional, Alcalá de Henares, Spain), Rodríguez Espinosa J.M. (Instituto de Astrofísica de Canarias, Spain; Departamento de Astrofísica, Universidad de La Laguna, Spain), Ros E. (Max-Planck-Institut für Radioastronomie, Germany; Dept. Astronomia i Astrofisica, Universitat de València, Spain; Observatori Astronòmic, Universitat de València, Spain), Rubiño-Martín J.A. (Instituto de Astrofísica de Canarias, Spain; Departamento de Astrofísica, Universidad de La Laguna, Spain; Instituto de Física Corpuscular, CSIC-UV, Spain), Ruiz-Granados B. (Instituto de Física de Cantabria, CSIC-Universidad de Cantabria, Spain), Sabater J. (Institute for Astronomy (IfA), University of Edinburgh, Royal Observatory, UK), Sánchez Contreras C. (Centro de Astrobiología, INTA-CSIC, Spain), Sánchez S. (Instituto de Astronomía, Universidad Nacional Autonóma de Mexico, México), Sánchez-Monge A. (I. Physikalisches Institut, Universität zu Köln, Germany), Sánchez-Ramírez R. (Instituto de Astrofísica de Andalucía, CSIC, Spain; Unidad Asociada Grupo Ciencias Planetarias, UPV/EHU, IAA-CSIC, Departamento de Física Aplicada I, E.T.S. Ingeniería, Universidad del País Vasco, UPV/EHU, Spain; Ikerbasque, Basque Foundation for Science, Spain), Sintes A.M. (Departament de Física, Universitat de les Illes Balears and Institut d'Estudis Espacials de Catalunya, Spain), Solanes J.M. (Departament d'Astronomia i Meteorologia and Institut de Ciències del Cosmos, Universitat de Barcelona, Spain), Sopuerta C.F. (Institut de Ciències de l'Espai, CSIC-IEEC, Campus UAB, Spain), Tafalla M. (Observatorio Astronómico Nacional-IGN, Spain), Tello J.C. (Instituto de Astrofísica de Andalucía, CSIC, Spain), Tercero B. (Instituto de Ciencia de Materiales de Madrid, CSIC), Spain), Toribio M.C. (Netherlands Institute for Radio Astronomy, ASTRON, The Netherlands), Torrelles J.M. (Instituto de Ciencias del Espacio, CSIC-UB/IEEC, Spain), Torres M. A. P. (European Southern Observatory, Chile; SRON Netherlands Institute for Space Research, The Netherlands), Usero A. (Obs. Astronómico Nacional-Obs. de Madrid, Spain), Verdes-Montenegro L. (Instituto de Astrofísica de Andalucía, CSIC, Spain), Vidal-García A. (UPMC-CNRS, UMR7095, Institut d'Astrophysique de Paris, France), Vielva P. (Instituto de Física de Cantabria, CSIC-UC, Spain), Vílchez J. (Instituto de Astrofísica de Andalucía, CSIC, Spain), Zhang B-B (Instituto de Astrofísica de Andalucía, CSIC, Spain)

# Foreword

Spain has a proud and strong tradition in radio astronomy. Spain has built and operates observatories at Yebes near Madrid, the IRAM 30-m dish on Pico Veleta in the Sierra Nevada near Granada, and is now constructing a small network of 12-m dishes for geodetic VLBI. Spanish astronomers make use of all major radio observatories around the world to conduct their science, as well as using facilities operating in wavebands other than radio. Spain is, of course, a member state of ESO. The Spanish astronomical community has maintained a close interest in the development of the Square Kilometre Array (SKA). Currently, approximately 2 M€ of staff time and other resources is focused on several of the SKA design consortia and, of course, Spanish astronomers are keen to see SKA built so they can conduct their science.

I am extremely impressed with the range of science in which Spanish astronomers are interested; it covers most of the key science areas in which SKA will deliver. As you, the reader, will see in the chapters in this White Book below, Spanish groups wish to use the SKA to participate in several of the areas identified as SKA key science, including the so-called Cosmic Dawn and the Epoch of Reionisation and fundamental physics through the detection and understanding of gravitational waves. They have strong interests in many other aspects of SKA-related science from the study of neutral hydrogen, the most common element in the Universe, through to how young stars form in our own Milky Way.

There are several chapters reminding us of the synergy that SKA will have with the other major instruments, such as ALMA, high-energy space observatories, Euclid, J-PAS and more. This is a demonstration of the modern multi-wavelength approach to astronomy and astrophysics; we do not truly understand an object, a surveyed region or a physical phenomenon until we have observed it across all of the available spectrum.

With this Spanish SKA White Book, the Spanish astronomy community has laid down a clear statement of their intent to be an integral part of the SKA project, to be engaged in the design and construction and to be leaders of SKA science. I congratulate all authors on developing an excellent set of chapters.

**Philip Diamond**
*Director General*
*SKA Organisation*

# Preface

During the last century, Radio astronomy has made fundamental contributions to our understanding of the Universe, as it allows the study of phenomena and components that cannot be observed with any other technique. Thanks to facilities like LOFAR, EVN, JVLA, NOEMA and ALMA[2], as well as other major worldwide ground-based radio telescopes and interferometers, and space and airborne observatories, radio astronomy covers now the range from submillimetre to metre wavelengths.

The Square Kilometre Array (SKA) will ensure that centimetre and metre wave Radio astronomy plays a central role in Astronomy for several decades. The SKA precursors (LOFAR, APERTIF, MWA, e-MERLIN, EVN, ATCA, JVLA, ASKAP, MeerKAT[3]) already show the potential of the new generation of centimetre wavelength radio interferometers with improvements in sensitivity, angular resolution, and field of view, which supersede, by at least one order of magnitude, the performances of present radio astronomical facilities.

The SKA, with a collecting area of one square kilometre, will have the potential to revolutionise essentially all areas of Astrophysics. In fact, the interest in SKA goes beyond radio astronomers, since it will be an extremely powerful tool for all astronomers, with relevance also in other fields of research in Physics and Astrobiology.

The Spanish scientific community officially showed its very strong interest in the SKA during the meeting "Science and Technological Opportunities in the SKA Era" of the Astronomy Infrastructures Network (RIA) in May 2011. Since then, several projects funded by the former MICINN led eventually to the creation of a "Spanish SKA network of academic institutions" and to the production of a feasibility study of the Spanish technological participation in the SKA construction.

At the beginning of 2014, the Steering Committee of the RIA issued a recommendation for Spain to explore the possibility of participate in the SKA project as a full member before the start of the construction phase in 2017. In October 2014, the "Spanish SKA day" meeting of the RIA brought together members of 18 academic institutions and 17 companies.

Spanish researchers and engineers are active members of several SKA working groups since 2012. Currently, 9 Spanish research centres and 11 companies contribute to the design of the SKA in 7 work packages, valued by the SKA Board at 2 M€. Since October 2013, a Spanish state representative is regularly invited to participate in the SKA Board meetings.

The Spanish participation as a full member in the SKA Project will multiply the economic value for the academic institutions and companies, as well as the scientific and technological return for Spain.

[2] European VLBI Network (EVN); LOw-Frequency ARray (LOFAR); Jansky Very Large Array (JVLA); NOrthern Extended Millimiter Array (NOEMA); Atacama Large Millimeter Array (ALMA)

[3] APERture Tile in Focus (APERTIF); Murchison Widefield Array (MWA); electronic Multi-Element Radio Linked Interferometre Network (e-MERLIN); Australian Telescope Compact Array (ATCA); Australian SKA Pathfinder (ASKAP); South African Karoo Array Telescope (MeerKAT)

This White Book is the result of the coordinated effort of 119 astronomers, and clearly shows the strong interest of the Spanish astrophysical community in the SKA. It also shows the scientific capability of the research groups and the synergies between them, as well as the synergies of the SKA with other astronomical observatories. All of this indicates a very promising potential for an optimum Scientific exploitation of the SKA in Spain.

**The Editors**

# Prólogo

La radioastronomía ha realizado a lo largo del último siglo aportaciones fundamentales a la comprensión del Universo, ya que nos permite estudiar componentes y fenómenos que no pueden observarse mediante ninguna otra técnica astronómica. Gracias a instrumentos como LOFAR, EVN, JVLA, NOEMA y ALMA[4], así como otros grandes radiotelescopios e interferómetros repartidos por todo el planeta y el espacio, la radioastronomía cubre un rango que va del métrico hasta el submilimétrico.

El Square Kilometre Array (SKA) garantizará que la radioastronomía a longitudes métricas y centimétricas continúe siendo gran protagonista de la astrofísica durante, al menos, varias décadas. De hecho, los instrumentos precursores del SKA (LOFAR, APERTIF, MWA, e-MERLIN, EVN, ATCA, JVLA, ASKAP, MeerKAT[5]) ya muestran el potencial de la nueva generación de radiotelescopios, con mejoras en sensibilidad, resolución angular, resolución espectral y campo de visión que superan en al menos un orden de magnitud las prestaciones de las infraestructuras radioastronómicas actuales.

Con un área colectora de un kilómetro cuadrado, el SKA será una herramienta con el potencial de revolucionar prácticamente todos los ámbitos de la astrofísica. Así, el SKA va mucho más allá del interés de los radioastrónomos, pues se revela como una herramienta poderosísima para todos los astrónomos, y que también será relevante en otros campos de la física y astrobiología.

La comunidad científica española mostró oficialmente su enorme interés en el proyecto SKA en mayo de 2011, durante una reunión de la Red de Infraestructuras de Astronomía (RIA) sobre "Ciencia y Oportunidades Tecnológicas en la era de SKA". Desde entonces, una serie de proyectos financiados por el extinto MICINN culminó en la creación de una red de instituciones académicas para el SKA y la elaboración del Estudio de viabilidad de la participación tecnológica española en el SKA. A principios de 2014, el Comité Directivo de la RIA emitió una recomendacion para que España explorara la posibilidad de participar en el proyecto SKA como miembro de pleno derecho antes del comienzo de la fase de construcion, en 2017. En octubre de 2014 tuvo lugar la reunión de la RIA "Spanish SKA day", que congregó a miembros de 18 instituciones académicas y 17 empresas.

Los investigadores españoles participan en varios grupos de trabajo del SKA desde 2012. Actualmente, 9 centros de investigación españoles y 11 empresas contribuyen a los esfuerzos de diseño del SKA en 7 paquetes de trabajo por un valor estimado de 2 M€, reconocido por el Board del SKA. Desde octubre de 2013 se invita regularmente a un representante del gobierno español a participar en las reuniones del Board del SKA. Una participación oficial de España en el proyecto multiplicaría el valor económico para instituciones y empresas, además del retorno científico y tecnológico para el país.

---

[4] European VLBI Network (EVN); LOw-Frequency ARray (LOFAR); Jansky Very Large Array (JVLA); NOrthern Extended Millimiter Array (NOEMA); Atacama Large Millimeter Array (ALMA)

[5] APERture Tile in Focus (APERTIF); Murchison Widefield Array (MWA); electronic Multi-Element Radio Linked Interferometre Network (e-MERLIN); Australian Telescope Compact Array (ATCA); Australian SKA Pathfinder (ASKAP); South African Karoo Array Telescope (MeerKAT)

Este Libro Blanco, fruto de un esfuerzo coordinado de 119 astrónomos, muestra el interés de la comunidad astrofísica española en el SKA, así como la capacidad científica de nuestros grupos de investigación y las sinergias existentes entre ellos, y del SKA con otros observatorios astronómicos. Todo ello representa un prometedor potencial para una óptima explotación científica del SKA en nuestro país.

**Los editores**

# Contents

**Magnetism in the Universe**

**The Cradle of Life and its Chemistry**

**Stellar Astrophysics**

## SKA Precursors

## Synergies with other Facilities

# Cosmology with the SKA: theoretical prospectives

**P. Arnalte-Mur[1], J. M. Diego[2], C. Hernández-Monteagudo[3], D. Herranz[2], R. A. Lineros[4], E. Martínez-González[2] and J. A. Rubiño-Martín[5]**

[1] Observatori Astronòmic de la Universitat de València, C/ Catedràtic José Beltrán 2, 46980 Paterna, Spain
[2] Instituto de Física de Cantabria (CSIC-UC), Av. los Castros, s/n, 39005 Santander, Spain
[3] Centro de Estudios de Física del Cosmos de Aragón (CEFCA), Plaza San Juan, 1, Planta2, 44001 Teruel, Spain
[4] Instituto de Física Corpuscular (CSIC-UV), C/ Catedrático José Beltrán, 2, 46980 Paterna, Spain
[5] Instituto de Astrofísica de Canarias, 38200 La Laguna, Tenerife, Canary Islands, Spain

## Abstract

We briefly review the context of the SKA in the panorama of modern Cosmology. The SKA will undoubtedly be one of the most powerful tools for Cosmology during the first half of the XXI$^{st}$ century. Many of the fundamental questions of modern Cosmology, such as the nature of the dark energy and dark matter that dominate the dynamics of the Universe, will be answered by SKA-driven science. Moreover, SKA will shed light on many aspects of large-scale structure growth and galaxy formation as well as fundamental Physics regarding the early Universe, inflation and tests of General Relativity.

## 1 Introduction

One of the remaining challenges in cosmology is to close the observational gap in redshift space between $z \sim 1100$ (when the Universe was roughly $4\times10^5$ years old and the Cosmic Microwave Background, CMB hereafter, formed) and $z \sim 6$ (when the Universe was about 1 Gyr old and galaxies and quasars were common enough to have reionized most of the intergalactic medium). During most of the intervening period the only light in the Universe was the CMB itself (hence the name of the dark ages that has become popular in the literature). Dark neutral hydrogen (HI) and helium clouds thinned and cooled as the Universe expanded, started collapsing in overdense regions and eventually formed the first (yet unobserved) stars and galaxies. During the so-called Epoch of Reionization (EoR) the light from these first cosmic candles ionized the hydrogen gas in expanding bubbles around isolated galaxies until the bubbles eventually joined and percolated, reaching a state in which most of the hydrogen in the Universe was again ionized. After this moment, a small fraction of neutral hydrogen survived in dense regions in galaxies, where dust clouds are able to block the incoming ionizing radiation. Although this broad picture is almost universally accepted, the details of the cosmic history during the dark ages and the EoR are largely unknown. These unknown elements of the cosmic history contain the key for unveiling some of the most perplexing mysteries of modern Cosmology, such as the nature of Dark Energy and/or the equation of state of the Universe. Reaching the Dark Ages and the EoR is an overwhelming observational and technological challenge for modern cosmologists.

As the most common atomic species present in the Universe, hydrogen is a useful tracer of local properties of the gas. The simplicity of its structure -a proton and electron- belies the richness of the associated physics. The 21 cm line of the neutral hydrogen arises from the hyperfine splitting of the $1S$ ground state due to the interaction of the magnetic moments of the proton and the electron. This splitting leads to two distinct energy levels separated by $\Delta E = 5.9 \times 10^{6}$ eV, corresponding to a rest frame wavelength of 21.1 cm and a frequency of 1420 MHz. This frequency is one of the most precisely known quantities in Astrophysics having been measured to great accuracy from studies of hydrogen masers. The 21 cm line was theoretically predicted by van de Hulst in 1942 and has been used as a probe of astrophysics since it was first detected by [10]. Due to cosmological redshift, the observed frequency of the line depends on the redshift $z$ of the emitter as $\nu_{obs} = \nu_{em}/(1+z)$. This means that the 21 cm line is redshifted to frequencies $\nu_{obs} \sim 100 - 200$ MHz for sources in the approximate range of redshifts corresponding to the EoR, $z \sim 6 - 12$. To this cosmological redshift we must add the Doppler redshift due to the proper motion of the source. The optical depth of this transition is small at all relevant redshifts, yielding a differential brightness temperature of an hydrogen cloud when observed against a background source of light [28]:

$$\delta T_b \approx 27\, x_{HI} \left(\frac{T_S - T_R}{T_S}\right) (1+\delta_b) \left(\frac{\Omega_b h^2}{0.023}\right) \left(\frac{0.15}{\Omega_m h^2} \frac{1+z}{10}\right)^{1/2} \left[\frac{\partial_r v_r}{(1+z) H(z)}\right] \mathrm{mK}, \tag{1}$$

where $x_{HI}$ is the fraction of neutral hydrogen, $\delta_b$ is the fractional overdensity in baryons and the final term arises from the velocity gradient along the line of sight $\partial_r v_r$. The temperatures $T_S$ and $T_R$ are the *spin temperature* of the gas and the *brightness temperature* of the background radiation, respectively. Thus, the differential brightness temperature of the 21 cm is very sensitive to the environment and physical state of the intergalactic medium (IGM), as well as to fundamental cosmology, that enters in the last four terms of Eq. 1.

## 2 Status of modern cosmology and the role of the SKA

Modern cosmology rests upon the cosmological principle, which states that the Universe is isotropic and homogeneous at the largest scales, and on the validity of Einstein's theory of General Relativity. Both assumptions are roughly consistent with the current status of observations, but a number of observational and theoretical problems indicate that they might not be as solid as we previously thought. An example of observational tension between the cosmological principle and observations are the recently reported statistical anomalies of the CMB [26]. Regarding the conceptual problems of the relativistic description of the Universe, maybe the most difficult question is about the elusive nature of the 'Dark Energy' that drives the accelerated expansion of the Universe, and whether it can be described or not by the Einstein's equations. The third pillar of modern cosmology is the inflationary paradigm, which poses many interesting open questions by itself.

## 2.1 The Dark Side of the Universe

The analysis of the CMB allows us to measure the curvature of space and to model the matter and energy content of the Universe [24]. Together with other independent experiments and cosmological probes (supernovae, galaxy and galaxy cluster counts, Ly$\alpha$ forest, large-scale structure [LSS], etc) these recent observations have led to the so called *'Cosmological Concordance Model'*, according to which the Universe is spatially flat, in agreement with the prediction from cosmological inflation. But in this model only 5% of the Universe seems to be made out of atomic matter (and half of it has not been observed in the local Universe yet!). The vast amount in the energy budget of the Universe, about 69%, seems to stem from a 'dark fluid' with negative pressure (the so called Dark Energy) with accelerates the expansion of the Universe, and the remaining 26% is attributed to dark, non-baryonic matter. In other words, 95% of the Universe is made of exotic matter and energy that is not directly observable. One of the foremost challenges of fundamental Physics in the XXI$^{st}$ century is the understanding of this 'dark side' of the Universe [3].

### 2.1.1 Dark Energy

Some of the possible interpretations of the Dark Energy are the cosmological term $\Lambda$ in Einstein's equations (the so-called $\Lambda$CDM model), a scalar field rolling down its potential and giving rise to the model referred to as *quintessence* (QCDM models, [23]), an even weirder single cosmic fluid with an equation of state that causes it to act like dark matter at high densities and dark energy at low densities(Unified Dark Energy or Unified Dark Matter models, [8, 7]), or a breakdown of the laws of General Relativity (modified gravity theories, see for example [34]). A relatively large number of models have been proposed in the literature, but all of them suffer from conceptual problems that either require severe fine-tuning of fundamental parameters or involve exotic, not well understood physics. Moreover, the current status of observations is not enough to select among these competing models, either due to lack of sensitivity and redshift coverage or to dramatic degeneracy problems. The SKA data will be enough to rule out, or at least constrain, most of these models.

The Dark Energy equation of state (EoS) can be constrained, in principle, by matching existing models with observations at different cosmic times or, equivalently, redshift ranges. Interesting redshift regimes are $100 < z < 1000$ for the early Universe (CMB data), $10 < z < 100$ (linear growth of structures) and $0 < z < 10$ (SNeIa, radio-galaxies, etc). The problem is that for the moment only the low and high $z$ regimes have been explored, while most of the intermediate range of redshifts is still *terra incognita*. The SKA will improve this situation by opening a new window to redshifts up to $z \sim 30$.

### 2.1.2 Dark matter and the growth of the Large-Scale Structure

In the time interval between the formation of the CMB and the present day hierarchical, non-linear Universe populated by stars, galaxies and galaxy clusters, the tiny primordial density fluctuations have evolved due to the combined action of gravity, radiative transfer and cosmic expansion. The small density fluctuations have grown and eventually collapsed to form gravitationally bound structures of enormous extent: the LSS of the Universe. The evolution of the LSS is sensitive to the properties of the dark matter and neutrinos, as well as dark energy and the action of gravity. Going beyond the scales of clusters ($\geq$ few Mpc), the study of the LSS is one of the main observational probes able to provide constraints on the cosmological model (see e.g. [39] for a review). In particular, the main aim of the next generation of surveys (such as eBOSS, J-PAS, DESI or Euclid) is to analyse two features of the galaxy distribution: the Baryon Acoustic Oscillations (BAO), which track the geometrical expansion and its acceleration, and the Redshift Space Distortions (RSD), which provide a measure of the growth of structure able to discriminate between different theories of gravity. In addition, the clustering of galaxies gives us information about the relation between the galaxies and the dark matter haloes hosting them.

There is no doubt that around 80% of the total amount of matter of the Universe is in the form of Dark Matter (DM). Nowadays, the scientific debate is focused on the nature of this DM. Among many proposed particle candidates for DM, Weakly Interactive Massive Particles (WIMPs) are the most popular class of candidates that appear in many models beyond the Standard Model (SM) of particle physics. A main requirement for any candidate is to explain the observed DM relic abundance $\Omega_{\rm DM}h^2 \simeq 0.11$ [25]. A WIMP candidate can fulfill this condition if it is stable, electrically neutral, the thermally averaged cross section is $\langle \sigma v \rangle \simeq 3 \times 10^{-26}\,{\rm cm}^3/{\rm s}$, and its mass is in the GeV–TeV range.

It is expected that WIMPs, accumulated around galaxies and clusters of galaxies, can annihilate and act as source of cosmic rays and gamma rays. The production and propagation of cosmic rays through galactic magnetic fields leads to the emission of synchrotron radiation which, for cosmic ray electrons in the energy range of 100 MeV to 1 TeV, gives radiowaves in the MHz–GHz frequency range. The frequency dependence of this synchrotron radiation depends on the energy spectrum of the cosmic rays, which in turn depend on the nature of WIMPs, and has a spectral signature that is different to other sources of Galactic synchrotron. Moreover, the angular distribution of this synchrotron radiation will also be rather different to the one corresponding to astrophysical sources such as supernova remnants. The SKA's projected frequency range goes from 50 MHz to 14 GHz, which is in the right ballpark to study galactic WIMP dark matter.

Regarding extragalactic radio signal, SKA will largely contribute to unveil the origin of the isotropic radio background also known as *ARCADE2 anomaly* [12, 15]. This anomaly corresponds to an isotropic component

$$T(\nu) = (1.19 \pm 0.14)/{\rm K}\ (\nu/{\rm GHz})^{-2.62 \pm 0.04} \qquad (2)$$

that cannot be completely accounted by known radio source populations like AGNs and star forming galaxies. This phenomenon is also compatible with the contribution from WIMPs in far galaxies [13, 14] to the radio intensity. Extragalactic WIMP signal would appear like a faint population of radio sources. At 1.4 GHz, this population would emit with intensities $\sim \mu$Jy which one order of magnitud lower that available observations. However, it is expected that SKA1-survey may distinguish sources as faint as $\sim 2\,\mu$Jy after 10000 hours.

### 2.2 Reionization

We know from the Gunn-Peterson effect [18] and from analysis of the CMB that the Universe is almost fully ionized out to redshifts of about $z \sim 6$, but also that between $z \sim 1000$ and somewhere between 15 to 30 the Universe was filled with neutral hydrogen (the so-called Dark Ages). The history of ionization of the intergalactic medium is a key to reveal the origin of stars, galaxies and active galactic nuclei. We expect that the very first stars and galaxies produce X rays and UV light that eventually reionizes the Universe by a redshift around 10 (the 'Cosmic Dawn' that gives way to the Epoch of Reionization, EoR). However, there might also be other mechanisms of reionization like the injection of energy by annihilation or dark matter particles or by the decay of metastable particles[1].

### 2.3 Tests of inflation

The inflationary paradigm, which is often invoked as a solution to several conceptual 'coincidence' problems in Cosmology (the flatness problem, the horizon problem, the isotropy problem, the problem of the monopoles), is one of the milestones of the Concordance Model. But although the inflationary paradigm is widely accepted in the community, observations have not been yet sensitive enough to choose among the many different scenarios and hypothesis that have emerged in the literature during the last three decades. Conceptually, the mechanism of inflation is similar to many dark energy models: some kind of scalar or tensor potential that during the first instants of the Universe generated an exponential expansion of the scale factor. In principle, inflation models (single-field slow roll, multifields, curvaton, etc.) can be constrained using the same methods described for the EoS (section 2.1.1) and for the CMB: the study of the three-dimensional power spectrum, bispectrum and trispectrum of 21 cm fluctuations, as well as the study the Gaussianity of the same fluctuations, will add very useful information. We expect a significant progress in this topic from the combination of high accuracy and deep extragalactic source surveys achievable with SKA, the three-dimensional power spectrum of matter fluctuations at high $z$ also obtained with the SKA, and the CMB maps provided by *Planck* and from the future CMB missions.

### 2.4 The role of SKA

With its powerful redshift range ($z_{\rm max} \sim 20 - 30$), large sky area (approximately $2\pi$ sr) and broad frequency coverage (50 MHz to 14 GHz), the SKA will revolutionize the panorama of Cosmology. It will have the potential to perform competitive cosmological surveys, complementary to state-of-the-art optical/infrared surveys such as Euclid. One of its main char-acteristics is the possibility of covering very large areas, due to the large field of view and the fact that the selection of extragalactic radio sources is not affected by Galactic dust extinction. Together with its sensitivity, this will allow SKA to cover extremely large volumes much faster than any previous facility. As the uncertainties in the next generation of surveys will be dominated by cosmic variance, this represents an important advantage.

[1] These aspects will be discussed in the next chapter of this White Book.

Two observation modes will be of interest. On the one hand, SKA will perform surveys using the novel technique of HI intensity mapping (or HI density field mapping) [30]. As the relevant cosmological information comes from very large scales, it is not necessary in this context to identify individual galaxies. One can instead measure the integrated HI emission from all the galaxies in a given pixel in the sky and in frequency, with typical resolutions of $\sim 1^\circ$ and $\sim 5\,\mathrm{MHz}$. This technique will function similarly to how we map the CMB temperature anisotropies. But in contrast to the CMB, the SKA will resolve the HI density field in three dimensions (two angles and a redshift). This will add tomographic information (i.e. time evolution) about the growth of structures in the Universe as well as a direct measurement of the three-dimensional power spectrum $P_{\mathbf{k}}$ of density fluctuations. The 3D power spectrum $P_{\mathbf{k}}$ contains many more modes than the CMB angular power spectrum $C_\ell$, and therefore will serve better to discriminate among Dark Energy models. As this HI combined signal is much larger than that of individual sources, this approach allows the mapping of large volumes in short periods of time. This challenging technique has already been tested in small surveys [9, 20], and much research is in progress to address the problems related to foreground removal and calibration (see [30] and references therein). It is feasible, already in the SKA1 phase, to survey an area of $\sim 30000\,\mathrm{deg}^2$ with SKA-MID or SKA-SUR going down to $z \sim 3$ resulting in a volume of up to $V = 700\,\mathrm{Gpc}^3$, although the redshift range will depend on the frequency band(s) used. This SKA1 survey will produce BAO and RSD measurements a factor of 2-3 better than current constraints, and covering a much larger redshift range ($0.05 < z < 3$ if both frequency bands are used). This will be the only survey measuring the properties of the LSS over such a large period of cosmic time. The extremely large volume will make this survey optimal for testing the cosmological model at very large horizon-size scales, in particular measuring the primordial non-Gaussianity parameter $f_{\mathrm{NL}}$ with a precision of $\sim 2 - 3$.

Alternatively, the SKA will perform an LSS survey using as tracers galaxies detected through the 21-cm emission line of neutral Hydrogen (HI) [1]. This is an effective way of obtaining galaxy positions and redshifts over large volumes, and a complementary technique to optical spectroscopic surveys. According to the forecasts of [31, 2], a survey of this type with SKA1 will be able to obtain constraints on BAO and RSD comparable to those of the BOSS survey. The main contribution will arrive with the SKA2 phase, where a 10,000 hours survey with flux sensitivity $\sim 5\mu\mathrm{Jy}$ will be able to cover $\sim 30000\,\mathrm{deg}^2$, and detect $\sim 10^9$ galaxies up to $z \sim 2$. This will be the largest galaxy redshift survey ever performed, covering a volume $V \sim 400\,\mathrm{Gpc}^3$, a factor of $\sim 2$ larger than Euclid's spectroscopic survey [30]. The HI survey will select star-forming galaxies, which are low bias objects, presenting an additional advantage for RSD, as the signal is inversely proportional to the bias, and at the same time reducing the impact of non-linear effects. Recent works [6, 29] forecast that this survey will measure the BAO scale in redshift slices of $\Delta z = 0.1$ with a precision of $\sim 0.3\%$ in the range $0.4 < z < 1.3$, and constrain RSD to a precision of $< 0.5\%$ in that same range. These constraints are comparable to those predicted for Euclid, while extending to lower redshifts. These two surveys will be complementary, offering a unique opportunity for joint analyses.

A SKA-MID survey based on the radio continuum emission will be able to detect $10^8$ ($10^9$) extragalactic sources in SKA1 (SKA2) up to $z \sim 6$, although with no redshift information. The angular distribution of sources at very large scales will contain important cosmological information, such as constraints on the primordial non-Gaussianity, the accurate measurement of the cosmic dipole, and a way of testing the isotropy assumption. Moreover, it will be possible to combine this survey with data from overlapping surveys, either HI SKA surveys or others as Euclid, to use the multi-tracer technique [32, 17] and minimise the effect of cosmic variance, obtaining even tighter constraints on the cosmological parameters.

In addition to these cosmological volume surveys, SKA will be able to perform much deeper surveys over smaller areas. As explained in [5], a deep HI galaxy survey will be able to study the evolution of the clustering of star-forming galaxies up to $z \geq 3$ and its dependence on galaxy properties (in a similar way that optical and infrared surveys do at lower redshift to constrain models of galaxy evolution). In a similar way, deep radio continuum surveys in fields with multi-wavelength coverage will be able to study the clustering of different types of AGNs, providing valuable information to understand its formation and evolution.

Finally, new class of fundamental test will become possible by means of the superb frequency resolution of the SKA, which will allow us to directly see the expansion of the Universe based in the redshift drift of individual objects [21].

## 2.5 Cosmology with other hyperfine structure lines and SKA

Apart from the hydrogen, other atomic species show hyperfine structure lines that may be useful for cosmological studies with the SKA. This is the case of the hyperfine line of neutral deuterium at $\lambda = 91.6\,\mathrm{cm}$ [33]. Being more challenging to detect than the hydrogen 21 cm line due to its longer wavelenth and also to the smaller abundance relative to H, this line provides the cleanest possible way of measuring the primordial abundance [D/H], free from contamination by structure formation processes at lower $z$. This measurement provides an excellent indirect determination of the baryon-to-photon ratio $\eta = n_\mathrm{b}/s$. As big bang nucleosynthesis (BBN) [38, 35] is the only known natural production mechanism of deuterium, this in turn provides a measurement of the cosmic baryon abundance $\Omega_\mathrm{b}h^2$.

Another interesting line is the $^3\mathrm{He}^+$ $^2\mathrm{S}_{1/2}$ $F = 0$–$1$ transition at $\lambda = 3.46\,\mathrm{cm}$ [36, 4, 22]. Although again the primordial $^3$He abundance (relative to H) is very small ($\sim 10^{-5}$), in this case the spontaneous decay rate is 680 times larger than the 21 cm, and more importantly, for the same redshift $z$ the radio contamination is much lower because the observed frequencies are larger ($\nu_\mathrm{obs} = 8.666/(1+z)\,\mathrm{GHz}$). Intergalactic $^3\mathrm{He}^+$ absorption can be used as an observable of cosmological helium reionizations, both HeII and HeI. SKA could be used to measure this transition with two main goals. First, to constrain the primordial $^3$He abundance, providing an indirect measurent of the cosmic baryon abundance. Second, to use the absorption of $^3\mathrm{He}^+$ line along the line of sight of quasars to study the reionization of the first electron of helium, which is thought to occur at the same time as the reionization of hydrogen. This will provide valuable and complementary information to the 21 cm about the ionization state of the IGM and the reionization process.

### 2.6 Cosmology with SKA in combination with other observables

The 21 cm signal alone encrypts amounts of information that are potentially several orders of magnitude larger than other cosmological observables such as the CMB or the current galaxy surveys. For this only reason the SKA will be one of the most important tools for Cosmology in the first half of the XXI$^{st}$ century. But as it happens in many other fields of Astrophysics, the synergy between 21 cm Astronomy and independent observations in other wavelenghts will help to break degeneracies, remove possible systematic errors, tighten the constraints in cosmological models and lower the size of error bars in fundamental cosmological parameters.

CMB photons interact with hydrogen atoms in three main ways. Firstly, CMB serves as a thermal bath for the hydrogen atoms and a backlight against which the 21 cm signal can act as an absorber or as an emitter according to Eq. 1. Besides, CMB photons interact with the intergalactic medium (IGM) during reionization by Thomson scattering of free electrons or by interactions that may introduce spectral distortions like the Sunyaev-Zel'dovich (SZ) effect [37] or fine structure transitions associated to atomic, ionic and molecular species present in the IGM (resonant scattering collisional emission on the same transitions, and the Wouthuysen-Field coupling in the OI 62.3 $\mu$K transition [19]). Moreover, any energy injection process during the dark ages will distort the CMB spectrum at the same time that it will modify the observed 21 cm. Finally, clumps of hydrogen interact gravitationally with CMB photons through the gravitational lensing and ISW effects.

Similarly, and since the SKA will overlap at low redshift with current galaxy and quasar surveys, the joint study of the SKA galaxy redshift survey using the 21 cm line and the SKA continuum radio survey with optical, IR and radio surveys such as the SDSS, WiggleZ, BOSS, 4MOST, LSST, HETDEX, Pan-Starrs, or DES, PAU, J-PAS and Euclid will revolutionize our view of LSS and galaxy formation and evolution.

Other SKA synergies that will contribute to the advance of cosmology include: the radio observation with SKA of Gamma Ray Bursts, that will probe the star formation rate up to very high $z$; the interdisciplinary study of dark matter annihilation together with X-Ray observatories such as LOFT [11] and particle accelerators; General Relativity tests with precision SKA timing of objects in pulsar catalogues; the study of primordial magnetic fields by measuring the electron density and energy distribution in the direction of LSS features as observed by other experiments, etc.

## 3 Involvement of the Spanish Cosmology community

Spanish cosmologists have been deeply involved in radio, optical and CMB–related Cosmology for a long time. The Spanish community possesses wide expertise in both theoretical and observational Cosmology as well as in the development of astronomical instrumentation. Several members of our community have participated in both ground-based and space

CMB experiments such as COSMOSOMAS, the VSA, the *Planck* satellite and QUIJOTE. Spanish researchers are a reference in the statistical analysis of temperature and polarization CMB anisotropies and have led relevant projects, such as EPI (Exploring the Physics of Inflation), on microwave instrumentation, component separation, Gaussianity analysis, statistical characterization and the discovery and study of statistical anomalies of the CMB. Our cosmologists are expert on the statistical signal processing of large data sets. Moreover, several members of our community are leading the *Planck* science tasks regarding compact source catalogues, homogeneity and isotropy, cross-correlation with LSS through the integrated Sach-Wolfe effect, galaxy cluster follow-up, primordial magnetic fields and kinematic SZ effect. Spanish scientists are also involved in the preparation of future CMB experiments such as COrE+. All this experience will not only make Spanish researchers very fit candidates to lead SKA-CMB synergy research projects, but also put the Spanish community in a good position to tackle SKA-alone science cases, since most of the techniques, software and statistical tools we have developed for the CMB can easily be exported to 21 cm data.

In addition, Spain has a long tradition in optical, X-ray and radio galaxy surveys. Many members of the Spanish community have participated in most of the big survey projects of the last decades. Moreover, several Spanish researchers are currently leading specifically Cosmology-oriented surveys such as ALHAMBRA, PAU, DESI and J-PAS, and participating in other large current or upcoming surveys such as BOSS, eBOSS, DES, DESI and Euclid. With these surveys we intend to give answer to some of the many questions that plague modern cosmology, such as the kind of equation of state that better fits the observed behaviour of the dark energy.

Finally, Spain has also a deep interest in the theoretical/observational study of dark matter; MultiDark is an excellence project in which most of the Spanish research community working in the field of dark matter is involved. We also are experienced in the management and operation of the kind of supercomputing facilities that will be required by a project such as SKA. All of this expertise and previous work place the Spanish one as one of the most motivated and best situated communities for participating in the SKA.

## Acknowledgments

The authors acknowledge partial support from the Spanish MINECO through projects AYA2013-48623-C2-2, AYA2007-68058-C03-01, AYA2010-21766-C03-02, AYA2012-30789 and the Consolider-Ingenio projects CSD2010-00064 (EPI: Exploring the Physics of Inflation) and CSD2009-00064 (MultiDark), and from the Generalitat Valenciana through grants PROMETEOII/2014/060 and PROMETEOII/2014/084. RAL acknowledges the Spanish grant FPA2014-58183-P. CHM acknowledges the support of the Ramón y Cajal fellowship RyC 2011 148062 awarded by the Spanish MICINN and the Marie Curie Career Integration Grant CIG 294183.

# The Cosmic Dawn and the Epoch of Reionization

**J. M. Diego[1], C. Hernández-Monteagudo[2], D. Herranz[1], E. Martínez-González[1] and J. A. Rubiño-Martín[3]**

[1] Instituto de Física de Cantabria (CSIC-UC), Av. los Castros, s/n, 39005 Santander, Spain
[2] Centro de Estudios de Física del Cosmos de Aragón (CEFCA), Plaza San Juan, 1, Planta 2, 44001 Teruel, Spain
[3] Instituto de Astrofísica de Canarias, 38200 La Laguna, Tenerife, Canary Islands, Spain

## Abstract

In this chapter we briefly describe the physics behind the HI 21 cm line in terms of the interplay of the HI gas with the ionized plasma and the Cosmic Microwave Background, and the different phases the system undergoes as the ambient density and UV background evolve with cosmological time. We also address the problematics associated to the metal enrichment of the IGM and the implications for the models of galaxy formation and evolution. We briefly discuss possible synergies with other reionization probes like E-ELT, JWST and ALMA, and conclude by listing a number of cosmological scenarios describing some type of energetic injection in the Universe, scenarios to whose understanding SKA should be able to (at least partially) contribute.

## 1 Introduction

The so-called 'Cosmic Dawn', i.e. the age during which the very first sources of light (stars and galaxlies) kindled in the Universe, and the subsequent epoch of reinozation (EoR) during which most of the hydrogen of the intergalactic medium (IGM) returned to its ionized state, are not only two of the most fascinating and poorly understood phases of the evolution of the Universe, but also correspond to the range of redshifts that will be most effectively probed by the SKA. This is not an accidental coincidence. The SKA is being designed with the unveiling of the Cosmic Dawn and the EoR as one of its primary scientific goals.

In this chapter we will briefly describe the physics behind these interesting epochs. We first focus on the different states of the HI gas through different cosmological epochs, providing some insight on the corresponding behavior of the HI 21 cm line. We next address the problematic of the (early) enrichment of metals in the IGM, paying attention to the consequences and implications that the chemical enrichment has for models of galaxy formation and evolution. We also explore the possibility of combining different probes of reionization (like the European Extremely Large Telescope, or the James Web Space Telescope) to gain some more insight into the last episodes of reionization. Finally, at the end of this chapter, we enumerate a number of other cosmological problems that can be addressed (at least partially) with the future SKA radio data.

# 2 Dark Ages and the End of Reionization

## 2.1 The HI 21 cm brightness temperature fluctuations

The differential brightness temperature of an hydrogen cloud when observed against a background source of light is [1]:

$$\delta T_b \approx 27\, x_{HI} \left(\frac{T_S - T_R}{T_S}\right)(1+\delta_b)\left(\frac{\Omega_b h^2}{0.023}\right)\left(\frac{0.15}{\Omega_m h^2}\frac{1+z}{10}\right)^{1/2}\left[\frac{\partial_r v_r}{(1+z)H(z)}\right] \mathrm{mK}, \tag{1}$$

where $x_{HI}$ is the fraction of neutral hydrogen, $\delta_b$ is the fractional overdensity in baryons and the final term arises from the velocity gradient along the line of sight $\partial_r v_r$. The temperatures $T_S$ and $T_R$ are the *spin temperature* of the gas and the *brightness temperature* of the background radiation, respectively. Thus, the differential brightness temperature of the 21 cm is very sensitive to the environment and physical state of the intergalactic medium (IGM), as well as to fundamental cosmology, that enters in the last four terms of equation (1).

The interplay between temperatures $T_S$ and $T_R$ in equation (1) determines whether the differential temperature brightness $\delta T_b$ is seen as an absorption or an emission against the backlight CMB radiation. The spin temperature interpolates between the values of the radiation bath $T_S$ and the kinetic temperature $T_K$ of the IGM; therefore, there will be a non-null signal only when $T_S$ couples with $T_K$. This coupling can appear due to two physical processes: collisions, which are effective in the IGM at high redshifts, $z \geq 50$, or resonant coupling with a Lyman alpha background (Wouthuysen-Field coupling [2]), effective soon after the first sources (stars and galaxies) turn on at lower redshifts. Other second-order effects that may affect $T_S$ include heating by the decay of exotic dark matter particles or photon cascades originated by background X-rays. Although the exact timing of the cosmic epochs is uncertain, the relative order is robustly predicted [3]: the signal transitions from an early phase of collisional coupling to a later phase of Ly$\alpha$ coupling through a short period where there is little signal. Fluctuations after this phase are dominated successively by spatial variation in the Lyman $\alpha$, X-ray, and ionizing UV radiation backgrounds. After reionization is complete there is a residual signal from neutral hydrogen in galaxies. The net effect is a decrement of background brightness temperature during the dark ages and an increment during the EoR. Somewhere during the twilight of the first galaxies, the 21 cm signal virtually disappears. Though this qualitative behaviour is well understood, the quantitative details of $\delta T_b$ and the physics behind it are almost unknown. Figure 1 shows this qualitative behaviour as a function of time, with specification of the main milestones of cosmic history.

The Cosmic Dawn and the EoR are roughly marked by the redshift $z_*$ at which the first stars start forming, the redshift $z_\alpha$ at which the Lyman $\alpha$ coupling saturates, the redshift

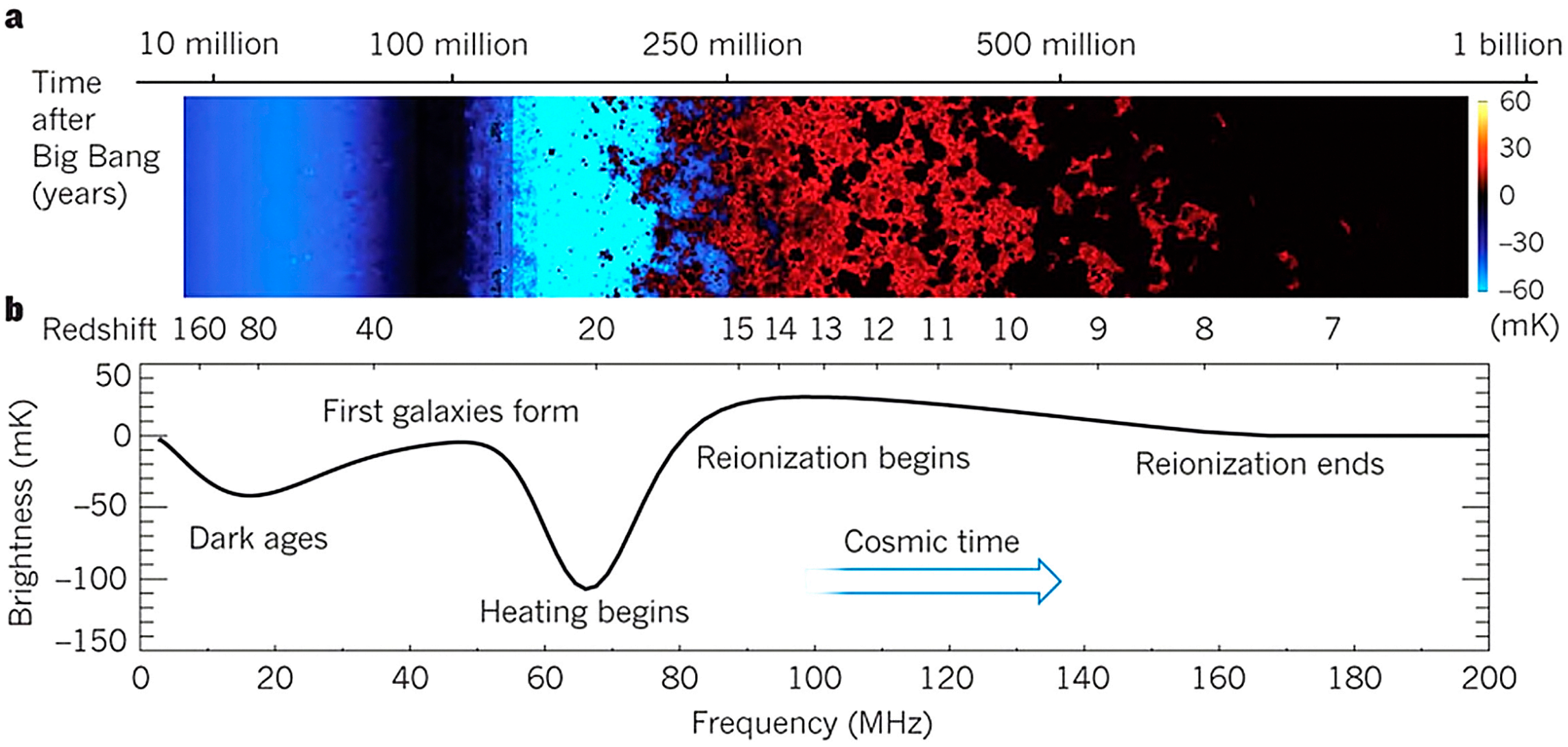

Figure 1: The 21 cm cosmic hydrogen signal. (a) Time evolution of fluctuations in the 21 cm brightness from just before the first stars formed through to the end of the reionization epoch. Coloration indicates the strength of the 21 cm brightness as it evolves through two absorption phases (purple and blue), separated by a period (black) where the excitation temperature of the 21 cm hydrogen transition decouples from the temperature of the hydrogen gas, before it transitions to emission (red) and finally disappears (black) owing to the ionization of the hydrogen gas. (b) Expected evolution of the sky–averaged 21 cm brightness from the Dark Ages at redshift 200 to the end of reionization, sometime before redshift 6. There is considerable uncertainty in the exact form of this signal, arising from the unknown properties of the first galaxies. Reproduced with permission from [4].

$z_h$ at which the gas temperature is heated to the temperature of the CMB, the redshift $z_T$ at which the 21 cm brightness temperature saturates, and the redshift $z_r$ at which the IGM is fully reionized. Most of these milestones are not sharply defined, and so there could be considerable overlap between them. In fact, our ignorance of early sources is such that we cannot definitively be sure of the sequence of events. Following [1] and references therein, we can summarize the approximate sequence of events and observational effects as:

$z_\alpha \leq z \leq z_*$ : the spin temperature is still coupled to cold gas and it is lower than $T_R$ and therefore there is an absorption signal. Fluctuations are dominated by density fluctuations and variation in the Ly$\alpha$ flux.

$z_h \leq z \leq z_\alpha$ : fluctuations in Ly$\alpha$ flux no longer affect the 21 cm signal and the spin temperature is driven by the kinetic temperature of the atoms. The signal is still in absorption, but as the temperature increases hotter regions may begin to be seen as islands in emission.

$z_T \leq z \leq z_h$ : after a transition of approximately no 21 cm signal anywhere, the spin temperature surpasses the value of $T_R$ and we expect to see a 21 cm in emission. Brightness temperature fluctuations are sourced by a mixture of fluctuations in ionization, density and gas temperature. By the end of this phase, the ionization fraction has likely risen above the per cent level.

$z_r \leq z \leq z_T$ : temperature fluctuations become unimportant. By this point, the filling fraction of HII regions probably becomes significant and ionization fluctuations begin to dominate the 21 cm signal.

$z \leq z_r$ : after reionization, any remaining 21 cm signal originates primarily from collapsed islands of neutral hydrogen (damped Ly$\alpha$ systems).

As noted by [5], the X-ray background due to the appearance of the first stars may give origin to a significant number of Ly$\alpha$ photons that could complicate the previous scenarios even more. Additionally, shocks could raise the IGM temperature very early and exotic particle physics mechanisms such as dark matter annihilation can also play a role. Clearly, there is still considerable uncertainty in the exact evolution of the signal making the potential implications of measuring the 21 cm signal very exciting.

## 2.2 The history of metal enrichment

It is during the epoch of reionization when metals are produced and ejected to the IGM. How this proceeds is intimately related to the process of galaxy formation and evolution, so whatever description is adopted must be consistent with the very different observational constraints existing not only on metallicities at high redshifts in different environments, but also in other properties of galaxies such as the fraction of visible matter or the metallicity inside galaxies.

By taking spectra along the direction of high redshift quasars, it is now well established that the metallicity of the IGM depends on the local gas density (and it tends to correlate with it, particularly in the denser environments, (see [6] and citations to this work). There is also evidence that metallicities remain flat versus redshift in the interval $z \in [2, 5]$, decreasing with increasing/decreasing redshifts outside this range, [6, 7, 8]. A satisfactory explanation for these observations must also explain why the fraction of visible matter in galaxies tends to decrease with increasing stellar halo mass, and why the metallicity in halos increases with stellar mass up to a *plateau* found at $M_\star \sim 10^{10}\,\mathrm{M}_\odot$.

A popular explanation nowadays invokes the *pre-enrichment* scenario, by which most of the reionization is due to Population III stars (with no metallicity, formed out of pristine gas) placed in small-mass galaxies ($M_\star < 10^7\,\mathrm{M}_\odot$) that were able to finish reionization by $z \sim 7$. Such galaxies would lose their metals early on due to supernova feedback, thus increasing the metallicity of the IGM while more massive galaxies above a given threshold, with a deeper gravitational potential well, would retain practically all their metals (in agreement with observations). As a consequence of this *feedback*, small halos would have a lower fraction of gas and visible matter too.

SKA observations will not provide measurements of the metallicities in the IGM and hence will not be able to confront this scenario directly. HI 21 cm observations will rather provide a tomographic view of the HI distribution, which constitutes *complementary* information. However, it is thought that by conducting intensity mapping (hereafter IM) on the HI 21 cm line it will be possible to identify those shielded, high density HI reservoirs placed at the cores of the ionized regions and hosting the ionizing sources. The definitive piece of information about these cores and the metal distribution may come from the synergy with other experiments probing reionization, either in the optical or infrared, or even in the millimeter (deferred to the chapter of this book by Diego et al. (Synergies with the Cosmic Microwave Background radiation)).

## 2.3 Synergies with optical and infrared probes of reionization

As of today there is a number of projects involving the construction of infrared and optical telescopes (some of them of gigantic size) such as the (optical) European Extremely Large Telescope (E-ELT, [9]), or the (infrared) James Webb Space Telescope (JWST, [10]). These facilities have been designed to detect bright galaxies placed as far as $z \in [10, 15]$ and so sample the last moments of the epoch of reionization. However, even when instruments of this size will only be able to detect a small fraction (the brightest tail) of the galaxies present at those redshifts, they should be able to identify and resolve the knots or the centres of the ionized regions sampled at the end of the EoR by HI 21 cm observations. This would provide a complementary description of the universe at those redshifts, since the smooth, large scale neutral and ionized regions would already be mapped by the HI 21 cm data.

Also complementary is the ability of the Atacama Large Millimeter/ Submillimeter Array (hereafter ALMA, [11]) to map molecular emission from clouds hosting young stars in the last stages of reionization ($z \in [8, 10]$). It is possible to combine the spectral and spatial information in SKA and ALMA data in order capture different lines observed at different frequencies but corresponding at the same redshift and to the same pixel on the sky (i.e., the same galaxy). Some of these molecular lines, such as those due to CO rotational levels, would set further constraints on the history of metal enrichment of the IGM.

## 3 Before the Cosmic Dawn: recombination lines

Another important predicted signal in the all-sky mean spectrum of the cosmic radio background is the presence of recombination lines from the epoch of cosmological recombination [27, 28] at redshifts $500 \leq z \leq 2000$ for HI, $1600 \leq z \leq 3500$ for HeII $\rightarrow$ HeI, and $5000 < z < 8000$ for HeIII $\rightarrow$ HeII recombination (see e.g. [29] and references therein). In contrast to the 21 cm signatures of reionization, the physical ingredients to describe the epoch of cosmological recombination are simple and well-understood. This fact allows us to find, within the standard model, potentially measurable consequences for the observed energy spectrum [30, 31, 32], as shown in Figure 2. These features could be used: (a) to measure in a model-independent way the redshift of the cosmological recombination; (b) to derive cosmological parameters (e.g. baryon fraction, Hubble parameter, etc.) in a new and completely independent way; and (c) to determine the pre-stellar abundance of helium in the Universe.

As shown in Figure 2, recombinational lines are expected to cover all SKA bands. However, the optimum frequency range for a SKA detection is at the top of the SKA-mid band, where the Galactic contamination is lower and the relative contrast of the distortions with respect to the un-distorted black-body spectrum is still of the order of $10^{-7}$. Being an all-sky (monopole) signal, this detection could be achieved with the total power spectra of the dish elements that form the SKA-mid interferometer [34]. Another possibility to measure all-sky signals interferometrically uses the presence of an occulting object (such as the Moon) that imposes spatial structure on an otherwise featureless all-sky signal [33, 34].

## 4 Other science cases: new sources of energy injection in the Universe

There are multiple physical processes that could inject energy into the IGM, whose effect could be potentially detectable via observations of the distortions of the CMB black body spectrum and the 21 cm line. In this section we focus on those detectable in the radio range. Among others, we can mention here:

- Silk damping of small-scale perturbations gives rise to CMB spectral distortions [12, 13, 14], providing information on the shape and amplitude of the primordial power spectrum at scales $0.6\,\mathrm{kpc} < \lambda < 1\,\mathrm{Mpc}$ (or CMB multipoles between $10^5$ and $10^8$), thus allowing us to probe 10 additional e-folds of inflation with respect to CMB anisotropies. Measuring this signature also provides an independent approach to study the scale-dependence of $f_{\mathrm{NL}}$, the non-Gaussian nature of the primordial fluctuations [15]. These features are again monopole signals, potentially observable in the SKA-mid band.
- Decay or annihilation of relic particles. The CMB spectrum can be used to set tight limits on decaying and annihilating particles during the pre-recombination epoch [16, 17]. The constraints are especially interesting for decaying particles with lifetimes $\sim 10^8$–$10^9$ s [18, 19], providing a unique test of the early-universe particle physics. Note that the CMB spectrum is less sensitive at particle masses higher than $m_{\rm DM} \sim 50\,$GeV and for decaying DM [20, 21], while the 21 cm line is more appropriate to explore this regime due to its sensitivity to smaller (and later time) energy injections [22].

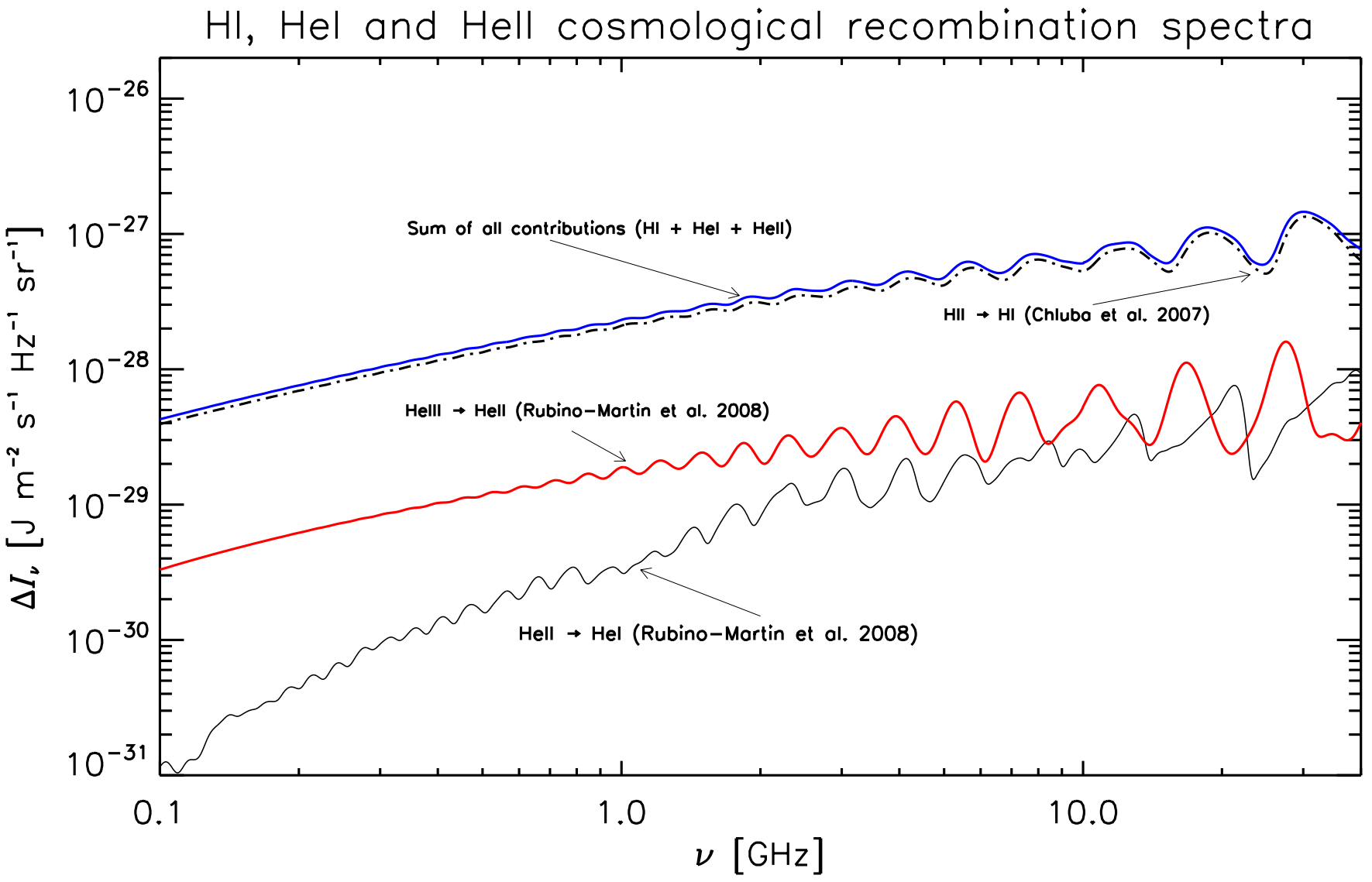

Figure 2: Full helium and hydrogen (bound-bound) cosmological recombination spectra, in the observing window of SKA. Details on the physical modelling of each spectrum can be found in [31, 32] and references therein.

- Evaporating black holes. Several inflationary scenarios predict the existence of primordial black holes (PBH) via originated after the collapse of overdense peaks in the initial density field. Those PBHs with masses in the range $5 \times 10^{13}$–$10^{17}$ g will evaporate due to Hawking radiation, and thus they will inject energy in the IGM. This excess of energy could be detectable as additional spin temperature fluctuations [23]. Note that for the larger masses, the expected signal will mimic that of a decaying dark matter species.

- Constraints on the radio-dim extragalactic population. There is currently some controversy on the interpretation of the results of the experiment ARCADE 2 [24, 25], since there is apparently an excess of radio background that cannot be explained with the CMB. This excess could be due to a high-redshift, faint radio source population placed at redshifts $z > 1$, but could also be due to other non-thermal mechanism of energy injection into the IGM. The SKA will be able to produce estimates of the radio source luminosity function down to very dim fluxes (5 to 200 nJy), and hence it should be able to confirm or discard the possibility of a discrete radio source population being behind the ARCADE 2 excess. At the same time, it should also set constraints on the free-free emission in halos throughout cosmic history [26], and improve current constraints on free-free induced distortion [25].

## Acknowledgments

The authors acknowledge partial support from the Spanish *Ministerio de Economía y Competitividad* (MICINN) through projects AYA2013-48623-C2-2, AYA2007-68058-C03-01, AYA2010-21766-C03-02, AYA2012-30789, and the Consolider-Ingenio project CSD2010-00064 (EPI: Exploring the Physics of Inflation). We also acknowledge the support of the *Ramón y Cajal* fellowship (RyC 2011 148062) awarded by the Spanish MICINN and the *Marie Curie Career Integration Grant* (CIG 294183).

# Synergies with the Cosmic Microwave Background Radiation.

**J.M.Diego[1], C.Hernández-Monteagudo[2], D.Herranz[1], E.Martínez-González[1], J.A.Rubiño-Martín[3], and P.Vielva[1]**

[1] Instituto de Física de Cantabria (IFCA, CSIC-UC), Avda. Los Castros, s/n, Santander,Spain
[2] Centro de Estudios de Física del Cosmos de Aragón (CEFCA), Plaza San Juan, 1, planta 2, E-44001, Teruel, Spain
[3] Instituto de Astrofísica de Canarias (IAC), C/Vía Láctea, s/n, La Laguna, Tenerife, Spain

## Abstract

The Cosmic Microwave Background (CMB), just like the HI 21 cm radiation, constitutes a privileged witness of the evolution of our universe during and after the epoch of reionization. This parallelism between the two radiation fields is also reflected in the fact that cosmological events leave different (but often complementary) signatures on both photon fields. In fact, there exists a number of physical channels of interaction between the CMB radiation with electrons, atoms, ions and gravitational potentials during and after the epoch of reionization: these strongly motivate a joint study of the CMB and future HI 21 cm data. In this context, we also highlight the existing parallelism between problems addressed in CMB analysis and problems foreseen in future HI 21 cm observations, in many of which the Spanish cosmological community has been deeply involved during the last decade.

## 1 Introduction

Just like the CMB photons, the HI 21 cm photons, emitted by neutral hydrogen some time between cosmological recombination and cosmological reionization, reach us after crossing a very significant volume of the observable universe. Both the CMB and HI 21 cm photon fields witness an evolving universe that becomes ionized by second time while the first stars are born, structure formation proceeds and metals are synthesized in stars and ejected to the Intergalactic Medium (IGM). All these physical processes leave their signature in the CMB and HI 21 cm photon fields in different (and often complementary) ways, and just for this reason a joint analysis in both radio and millimeter wavelengths is extremely interesting from the cosmological point of view. Furthermore, the evolution of the HI 21 cm field is linked to the evolution of the CMB photon bath itself, since the population of the level involved in the HI 21 cm transition is heavily influenced by the CMB. Depending on the dominant physical processes, the HI 21 cm line may be seen in emission or in absorption with respect to the CMB field, which acts as a background light.

In this chapter we describe those secondary anisotropies induced in the CMB during and after the epoch of reionization which can benefit from the combination with radio observations of the (redshifted) HI 21 cm line. We first address the interaction of CMB photons with metals and ions synthesized during the epoch of reionization, and discuss how this interaction may provide new insight on the physics of reionization. We next revisit the interaction of the CMB with free electrons in the IGM via inverse Compton scattering, and study its degree of complementarity with HI 21 cm observations. We also study how a deep galaxy survey built upon HI observations may also help to unveil the impact of the universal acceleration on the CMB by means of the integrated Sachs-Wolfe effect (ISW). Finally, we conclude by highlighting the similarity existing between the problems to be addressed in the analysis of HI 21 cm data and those related to CMB analysis, in which the Spanish cosmological community has extensively worked in the last decade.

## 2 Interaction of the CMB with atoms, molecules and plasma during the epoch of reionization

In this section we briefly describe the physics of the interaction of CMB photons with fine structure transitions associated to atomic, ionic and molecular species present in the IGM. Such species should be the result of the pollution induced by the explosion as supernovae of Population III stars, and hence these phenomena may allow to trace the chemical enrichment of the IGM at the redshifts of interest. We also make brief reference to those other processes injecting energy in the CMB and causing spectral distortions accessible by the next generation of CMB missions.

### 2.1 Resonant scattering on fine structure transitions during reionisation

It was shown in Basu et al. (2004) (hereafter B04) that a CMB photon presently observed at frequency $\nu_{\rm obs}$ may have interacted with a resonant transition $X$ of resonant frequency $\nu_X$ at a resonant redshift $z_X = \nu_X/\nu_{\rm obs} - 1$. For the range of frequencies for which the CMB can be clearly observed (say 10–300 GHz) , transitions in the fine structure range (20–600 $\mu$m) will fall naturally in a redshift range bracketing the reionization epoch ($z_X \in [5, 50]$). Since in resonant scattering the absorbed CMB photon is simply re-emitted in a different direction, the number of CMB photons is conserved, there is no distortion of the CMB black body spectrum and this process can only be seen in the angular anisotropies of the CMB.

B04 showed that the leading impact of this process on the CMB angular power spectrum is a frequency dependent *blurring* of the intrinsic CMB anisotropies generated at $z \simeq 1,100$. If a given species is generating an optical depth $\tau_X$ associated to a resonant transition $X$, then the ratio of the power spectrum of two CMB maps covering the same patch of the sky, but observed at two different frequencies $\nu_1$ and $\nu_2$, will equal

$$C_{l,1}/C_{l,2} = \frac{\exp -2\tau_{X,1}}{\exp -2\tau_{X,2}} \simeq \frac{1-2\tau_{X,1}}{1-2\tau_{X,2}}. \tag{1}$$

The optical depths $\tau_{X,1}$, $\tau_{X,2}$ are both associated to the same resonant transition $X$ but observed at different resonant redshifts $z_{X,i} = \nu_X/\nu_i - 1$, with $i = 1, 2$. We are assuming that both are much smaller than unity, as it is in practically all cases. The optical depth $\tau_{X,2}$ is proportional to the oscillator strength of the resonant transition $X$ and the *cosmic average* number density of resonant atoms/ions at redshift $z_{X,2}$. If $\nu_2$ is chosen so that $z_{X,2}$ is high enough to render the number of resonant atoms/ions negligible, then $\tau_{X,2} \simeq 0$ and from Eq. 1 above B04 find that

$$C_{l,1} \simeq C_l^{\rm CMB}\,(1 - 2\tau_{X,1}), \tag{2}$$

where the term $C_l^{\rm CMB}$ refers to the intrinsic CMB angular anisotropies generated during cosmological recombination. B04 concluded that by comparing CMB maps at different frequencies it may be possible to set constraints on the abundance of a resonant species at redshifts $z_{X,1}$. They also identified a number of resonant transitions (CI 370.4 $\mu$m, CII 157.7 $\mu$m, OI 63.2 $\mu$m, OIII 88.4 $\mu$m, etc) whose effect should be dominant and at worst close to the limit of detectability in future CMB missions. This pattern described so far is the predicted impact of the *average* number density of resonant atoms/ions at the resonant redshift $z_{X,1}$. The patchiness in the metal distribution will reflect in further anisotropies on angular scales comparable to the typical clustering of the bubbles containing the resonant species, but typical at much smaller amplitudes that will be neglected hereafter.

Given the small amplitude expected for realistic values of cosmic abundances of resonant species (the $\tau_X$'s are typically below $10^{-3}$), the level of foreground removal and control of systematics is highly demanding (see Hernández-Monteagudo et al. (2006) for details). Nevertheless, *a priori* this effect should show up in both intensity and polarization CMB anisotropies (Hernández-Monteagudo et al., 2007), and should lie within the range of detectability of foreseen future CMB missions of the type of PRISM[1].

The cross-correlation of HI 21 cm maps with future, high sensitivity CMB maps in both intensity and polarization would provide an unprecedented view of the history of cosmic enrichment of the IGM during the end of the Dark Ages and the epoch of reionization. The HI 21 cm observations will perform a tomographic study of distribution of HI during reionization. Similarly, accurate multifrequency CMB observations will provide a tomographic description of the pollution of the IGM with metals and ions during the same epoch. These metals and ions will be expelled by supernovae and thus are likely to follow the same spatial pattern of HII. Thus, a priori, it is expected that the blurring of CMB anisotropies induced by metals will be spatially anti-correlated to HI 21 cm emission, although an actual comparison or cross-correlation study will shed light on how metal pollution versus hydrogen ionization proceeded during reionization. Provided that systematics from radio and CMB observations are expected to be of very different nature, we can expect that this comparison would consti-tute a robust consistency test on the cosmological constraints derived for reionization from these two data sets.

[1]PRISM's URL site: `http://www.prism-mission.org`

### 2.2 Distortions induced in the CMB black body spectrum during reionization

Although the CMB is, to date, the most perfect black body that has ever been measured, increasing interest is arising around the study of deviations from this black body law. Possible future missions like PRISM or PIXIE (Kogut et al., 2014) should be sensitive to levels of spectral distortions of the CMB about 100–1000 times smaller than current upper limits. The presence of a re-ionized medium must introduce a floor distortion of the CMB spectrum at the level of $10^{-8}$–$10^{-7}$ But, addition to this, it is expected that other physical processes inject energy to the CMB photon bath, which are potentially detectable by the experiments just quoted above. Just to cite a few examples occurring during the epoch of reionization, such observations would shed light on the amount of energy released by shocks and the first supernovae during reionization (Cen & Ostriker, 2006; Miniati et al., 2000; Oh et al., 2003); they would also set constraints on the impact of non-linear bulk flows on the spectrum of CMB polarization (Renaux-Petel et al., 2014); and they may even detect how the first stars pollute the IGM with oxygen and distort the CMB spectrum via the Wouthuysen-Field effect on the OI 63.2 $\mu$m line (Hernández-Monteagudo et al., 2007, 2008).

Again, these observations would provide an alternative view of the onset and development of reionization, independently of that given by the HI 21 cm line. Therefore a joint study of both HI 21 cm and CMB distortion maps (suffering each from different systematics) would provide hints about the dominant physical processes which, during reionization, transferred energy and metals to the IGM.

## 3 The kinetic and thermal Sunyaev-Zeldovich effects during the epoch of reionization

The kinetic and thermal Sunyaev-Zeldovich effects (hereafter kSZ and tSZ, respectively Sunyaev & Zeldovich, 1972, 1980) describe the inverse Compton scattering of CMB photons off free electrons in the intergalactic medium. If there is no energy transfer between the CMB photons and the electrons (Thomson scattering), then the black body spectrum of the CMB is preserved and only the Doppler kick due to the relative motion of the electrons with respect to the CMB rest frame may introduce a (local) dipolar modulation of the CMB brightness. This is the kSZ effect. If instead the electrons are hot, then there exists energy transfer into the CMB photon bath that effectively distorts the CMB black body spectrum (tSZ effect).

On its way from the last scattering surface to us, the CMB photons propagate through a universe that, right after recombination, is filled with neutral hydrogen and helium. These *dark ages* come to an end by $\sim$ 30–50 as the first stars are formed (cosmic dawn), and the UV radiation emitted by these ionizes the intergalactic medium in bubbles that eventually percolate, leaving a newly ionized universe. How this *epoch of reionization* (EoR) proceeds is practically unknown, but both the tSZ and the kSZ effects can provide useful insight. Predictions about the tSZ are more uncertain, since the tSZ is proportional to the electron pressure which is highly dependent on the non-linear physics dominant during the EoR (but see Oh et al., 2003, and references to this work), whereas the kSZ depends solely on the electron density and velocity and thus predictions are less dependent of non-linear physics, see, e.g., Hernández-Monteagudo & Ho (2009) for a second order estimation of the all sky kSZ power spectrum and Zahn et al. (2005); Trac & Cen (2007) for studies based upon hydrodynamical numerical simulations. These studies showed that the kSZ would dominate over the tSZ during the EoR, and that the amplitude of the kSZ would depend on the duration and degree of patchiness of the re-ionization process. When the kSZ is not referred to electrons in collapsed structures, but rather to large scale, smooth electron distributions, it is also called the Ostriker-Vishniac (OV) effect.

At the same time, the EoR leaves also a trace in the radio part of the spectrum where the 21-cm line vanishes due to the absence of neutral hydrogen. SKA will probe this regime directly and be able to map in redshift slices the distribution of neutral hydrogen. Observations made with SKA will be crucial to understand the role of the kSZ effect in future CMB experiments. The cross-correlation of SKA data with future high-sensitivity, high-resolution stage IV CMB experiments (Abazajian et al., 2015), will help boost the signal from the EoR facilitating its study in greater detail (Slosar et al., 2007). In particular, Jelic et al. (2015) show how the cross-correlation between the cosmological 21 cm line and the kSZ is maximized for patchy reionization and peaks at scales $l \approx 100$ (correlation) and $l \approx 5000$ (anticorrelation). The positive correlation at large scales is due to the natural correlation of overdensities on large scales while the anticorrelation on small scales reflects the complementarity nature of the kinetic SZ and the 21 cm line. The same authors also consider the particular case of SKA and predicts a signal to noise of approximately 3 for an instantaneous reionization with 1 000 hours of integration in SKA and a *Planck*-like sensitivity for the kSZ (see also Tashiro et al., 2010).

The combination of kSZ and 21 cm line data from SKA opens the door also to novel analyses. Both the kSZ from the EoR and the 21 cm line signal from the same period can be used as a background for lensing studies. The combination of kinetic SZ and 21 cm line data can be used to increase the signal to noise of this background lensing plane at $z \approx 10$. Using this background for lensing studies has a variety of advantages that make it an attractive alternative for cosmological studies. This technique has been tested in the past with simulations Zahn & Zaldarriaga (2006); Diego & Herranz (2008); Metcalf & White (2009) with promising results. Similar techniques are nowadays starting to be applied but using the CMB as a background.

# 4 Combining CMB and HI 21 cm observations in the context of the Integrated Sachs Wolfe effect

The late integrated Sachs-Wolfe (ISW) effect (Sachs & Wolfe, 1967; Rees & Sciama, 1968; Martinez-Gonzalez et al., 1990) describes the gravitational interaction suffered by the CMB photons when passing through the time-evolving LSS. This secondary anisotropy of the CMB is given by an integral of the evolution of the gravitational potential $\Phi$, as a function of the conformal time $\eta$:

$$\frac{\Delta T}{T_{\rm CMB}} = -\frac{2}{c^3}\int_0^{\eta_{\rm CMB}} {\rm d}\eta\,\dot{\Phi}, \tag{3}$$

where $\eta_{\rm CMB}$ represents the conformal time at recombination, corresponding at a redshift of $z \simeq 1100$ in a $\Lambda$CDM cosmology. The ISW effect is very difficult to detect from CMB measurements, first, because it is a very weak signal and, second, because the CMB distortion caused by the time evolution of the gravitational potential redshifts (or blueshifts) the primary CMB photons, conserving its black body electromagnetic spectrum. However, it is possible to detect this effect by cross-correlating a map of the CMB anisotropies with tracers of the LSS, as original proposed by (Crittenden & Turok, 1996).

Because the ISW effect depends on the growth of structures, it is useful to constrain physics behind such evolution: a cosmological constant (e.g., Nolta et al., 2004), dark energy (e.g., Vielva et al., 2006), modified gravity (e.g., Zhao et al., 2010), or non-flat curvature (e.g., Li & Xia, 2010).

Since its first detection in 2004, through the cross-correlation of CMB data from WMAP and X-ray (HEAO-1) and radio (NVSS) catalogues (Boughn & Crittenden, 2004), several works (e.g., Fosalba et al., 2003; Nolta et al., 2004; Vielva et al., 2006; Giannantonio & Melchiorri, 2006; Cabré et al., 2007; Ho et al., 2008; Planck Collaboration et al., 2014) have confirmed the detection of the ISW effect (with significance levels ranging from $\approx 2\,\sigma$ to $4\,\sigma$). Recently, the Planck Collaboration has reported a $4\,\sigma$ detection, by the combination of NVSS, SDSS and WISE catalogues, and the *Planck* lensing map (Planck Collaboration et al., 2015).

For an ideal survey, i.e., full-sky, negligible shot noise, and mapping, at least, up to $z \sim 4$, the maximum signal-to-noise ratio for the detection of the ISW effect is $\approx 8$. Among the previous factors, the sky coverage is the most important limitation. Since SKA is expected to provide an almost half-sky coverage, it is possible to obtain a detection level $\approx 4\,\sigma$, similar to the one already obtained by the combination of multiple surveys, as mentioned above.

The possibilities that SKA offers to study the ISW effect are not limited to the signal detection, and the derived constraints on cosmological parameters. One the one hand, the precise redshift estimation of SKA galaxies by means of the HI 21 cm line will allow performing ISW-tomography, which can be very useful, in particular, to estimate a map of the ISW fluctuations, as function of redshift (using filtering techniques as the ones described in Barreiro et al., 2008). This can be used to probe further the nature of some of the CMB large-scale anomalies that puzzled the cosmological community (Planck Collaboration et al., 2014). On the other hand, the cosmic network that is mapped by the SKA galaxies will identify superstructures as clusters and voids, which can be used to prove the ISW effect through the stacking of the CMB fluctuation on the location of this structures. This kind of analyses has in fact revealed some anomalous signals, which are difficult to interpret in the context of the standard $\Lambda$CDM model (e.g., Hernández-Monteagudo & Smith, 2013; Ilić et al., 2013; Planck Collaboration et al., 2015).

## 5 Foreground removal: from CMB to 21 cm

The extraction of cosmological information from the analysis of the intensity mapping of the 21 cm requires of a precise process to disentangle the cosmological signal from foreground emission. According to its spatial properties and origin, these foregrounds can be classified as diffuse, anisotropic and galactic (as the synchrotron and free-free radiations) and compact, isotropic and extragalactic. The foreground emission is, in general, much higher than the cosmological 21 cm fluctuations and, therefore, its accurate removal is challenging.

The foreground removal in the 21 cm context has several similarities with the CMB component separation problem that has been widely studied during the last almost 20 years: maximum-entropy (MEM, e.g., Hobson et al., 1998; Barreiro et al., 2004), spectral matching (SMICA, e.g. Patanchon et al., 2005), internal linear combinations (Basak & Delabrouille, 2012; Fernández-Cobos et al., 2012), independent component analysis (Maino et al., 2002, FastICA, e.g.,), correlated component analysis (CCA, e.g., Bedini et al., 2005) ... However, one relevant difference between the component separation problem in the CMB and the 21 cm frameworks is that, while for the former the cosmological signal is the same among different observational frequencies, the source emitting the 21 cm signal is different across the frequency range, due to cosmological redshift. It is also worth mentioning that, as for the CMB case, the wealth of information encoded within the 21 cm measurements aim not only to obtain the cosmological signal, but also to separate and characterise the different foreground components, thus enabling an improved understanding of the physical processes behind them.

Another significant difference between HI 21 cm and CMB is that, while for the CMB case all signals (foregrounds and primordial) are highly smooth with respect to observational frequency, for the 21 cm the cosmic signal associated to neutral hydrogen exhibits much more structure. Despite of being a much younger field than the CMB, the study of the component separation problem in the context of 21 cm observations has already imported methods and techniques used in CMB analyses, (e.g., Ansari et al., 2012; Chapman et al., 2012, 2015; Alonso et al., 2015).

## Acknowledgments

The authors acknowledge partial support from the Spanish *Ministerio de Economa y Competitividad* (MICINN) through projects AYA2013-48623-C2-2, AYA2007-68058-C03-01, AYA2010-21766-C03-02, AYA2012-30789, and the Consolider-Ingenio project CSD2010-00064 (EPI: Exploring the Physics of Inflation). We also acknowledge the support of the *Ramón y Cajal* fellowship (RyC 2011 148062) awarded by the Spanish MICINN and the *Marie Curie Career Integration Grant* (CIG 294183).

# Gravitational waves with the SKA

**José A. Font[1,2], Alicia M. Sintes[3], and Carlos F. Sopuerta[4]**

[1] Departamento de Astronomía y Astrofísica, Universitat de València, Dr. Moliner 50, 46100, Burjassot (València)
[2] Observatori Astronòmic, Universitat de València, Catedrático José Beltrán 2, 46980, Paterna (València)
[3] Departament de Física, Universitat de les Illes Balears and Institut d'Estudis Espacials de Catalunya, Cra. Valldemossa km. 7.5, 07122 Palma de Mallorca
[4] Institut de Ciències de l'Espai (CSIC-IEEC), Campus UAB, Carrer de Can Magrans, 08193 Cerdanyola del Vallés (Barcelona)

## Abstract

Through its sensitivity, sky and frequency coverage, the SKA will be able to detect gravitational waves – ripples in the fabric of spacetime – in the very low frequency band ($10^{-9}$ – $10^{-7}$ Hz). The SKA will find and monitor multiple millisecond pulsars to identify and characterize sources of gravitational radiation. About 50 years after the discovery of pulsars marked the beginning of a new era in fundamental physics, pulsars observed with the SKA have the potential to transform our understanding of gravitational physics and provide important clues about the early history of the Universe. In particular, the gravitational waves detected by the SKA will allow us to learn about galaxy formation and the origin and growth history of the most massive black holes in the Universe. At the same time, by analyzing the properties of the gravitational waves detected by the SKA we should be able to challenge the theory of General Relativity and constraint alternative theories of gravity, as well as to probe energies beyond the realm of the standard model of particle physics.

## 1 Introduction

Einstein's theory of General Relativity (GR) predicted the existence of gravitational radiation. Gravitational Waves (GWs) are ripples in the curvature of spacetime produced by accelerating masses such as black hole collisions or supernova explosions. These spacetime waves reach the Earth fairly unperturbed, thus carrying valuable information about the astrophysical sources that produce them. Their direct detection, beyond providing yet another confirmation of the predictions of GR, will stir up a new revolution in the way we observe the Universe, comparable to those resulting from radioastronomy or other windows of observation of the electromagnetic spectrum. So far, there exists compelling evidence of the existence of GWs through the timing measurements of relativistic binary pulsars [1, 2].

At present, multimessenger astronomy is becoming a reality, thanks to the wealth of information we are gaining from the Universe through complementary cosmic messengers such as photons, neutrinos, and cosmic rays. Within such an exciting context, Gravitational Wave Astronomy has an enormous potential to provide fundamental contributions [3]. The GW *window*, as a result of the weakness of the gravitational interaction, is not affected in cosmological terms by the electromagnetic decoupling limit, thus allowing the access to information from the very early Universe, in a range of energies highly superior to that currently achievable through other approaches. In astrophysical terms, this window can provide new and complementary information about compact binary systems (formed by white dwarfs and/or neutron stars and/or black holes) from which information is already available through their electromagnetic emission, either directly or from their environment (i.e. accretion disks or other distributions of matter/energy around such compact objects). In the case of gravitational radiation we can anticipate a significant impact on topics of research at the forefront of physics, such as galaxy formation history and the physics of very compact objects, aside from unexpected surprises difficult to foresee. In any event there is an enormous potential as gravitational radiation will allow us to probe regions very close to black hole horizons (where the gravitational potential is strongest) and speeds very close to the speed of light which, in turn, will allow us to carry out definitive tests about the structure of black holes and even about the validity of GR and/or alternative theories of gravity.

## 2 Gravitational wave detection and the role of the SKA

After pioneering efforts with resonant bar detectors, present-day research to detect gravitational radiation is based on laser-interferometric detectors. Among the various current instruments two deserve special mention, the Virgo detector [4] (originally a French-Italian collaboration joined by Poland, Hungary, and the Netherlands), a 3 km long laser interferometer, and the Laser Interferometer GW Observatory (LIGO) [5] in the USA, which comprises two observatories in Hanford, WA and Livingston, LA with 4 km long arms. The formidable accuracy necessary to unmistakably detect GWs results in an unprecedented technological challenge on issues as diverse as laser stability, the suspension system of the test masses, the design of the lenses and mirrors, the seismic isolation, etc. The advanced LIGO detectors will start operating as early as mid 2015. They will be followed by advanced Virgo and the Kamioka Gravitational Wave Detector (KAGRA) [6], formerly the Large Scale Cryogenic Gravitational Wave Telescope (LCGT). KAGRA is a Japanese GW laser interferometer that will be located underground in the Kamioka mine and will use cryogenic technology for the first time. Furthermore, there are plans to place one of the LIGO detectors in India, which will greatly enhance the capabilities of sky localization of the GW sources once all of the detectors operate as a single network. The panorama of ground-based detectors should also include a mention to the proposed Einstein Telescope, a third-generation GW detector in Eu-rope, whose design study has already been undertaken by the European Commission [7]. The main astrophysical and cosmological sources for such ground-based detectors are compact binary coalescences and mergers, supernova core collapse, lumpy spinning stars, and stochastic cosmological backgrounds of diverse origin. The observations will provide key information to understand the formation and evolution of stellar-mass compact objects, the composition and equation of state of neutron stars, etc.

Current ground-based detectors operate in the high-frequency band (from 10 Hz to a few kHz) and are strongly limited in the low frequency band. The gravity gradient (Newtonian) noise turns impossible for them to operate below 1 Hz. Since below 1 Hz there are many different interesting sources of gravitational radiation, plans to build a GW observatory in space were naturally put forward during the last decades. In 2013, the European Space Agency (ESA) selected the science theme proposed by the eLISA Consortium in the white paper *The Gravitational Universe* [8] for its future large-class mission L3. The L3 mission will carry out a scientific program based on low-frequency Gravitational Wave Astronomy. The mission concept proposed in the white paper is eLISA, an all-sky monitor formed by a constellation of three spacecrafts in heliocentric orbits, each spacecraft separated from one another by a 1 million km baseline. Their relative positions will be monitored using high-precision laser interferometry (at the picometer level) using concepts and technology completely different to those of ground-based GW detectors. Since the core of these technologies can not be tested on ground, ESA developed a technology-demonstrator mission for a future GW space-based observatory, called LISA Pathfinder [9]. LISA Pathfinder is at the final stages of integration and tests and its launch is planned for September 2015. A space-based GW observatory like eLISA will be observing in the low-frequency band of the GW spectrum (from 0.01 mHz to 1 Hz), a band inaccessible to ground-based detectors, and will detect GWs from ultracompact stellar binary systems in our galaxy (which includes the so-called *verification binaries*, guaranteed known GW sources), collisions of (super)massive black hole binaries (within the mass range $10^{4-7}\,M_{\odot}$), capture of stellar-mass compact objects by (super)massive black holes in the centres of quiescent galaxies, and cosmological GW backgrounds of diverse physical origin.

In parallel to the design of GW detectors based on laser interferometry, there exists two completely different methodologies to detect (directly or indirectly) gravitational radiation. The first one is the search for B-modes originated by GWs of primordial origin in the Cosmic Microwave Background (CMB). These GWs are in the ultralow frequency range ($10^{-18} - 10^{-13}$ Hz) and they are the target of ground-based detectors like BICEP2 and space-based detectors like the ESA Planck mission (see [10, 11] for details). The second type of methodology, the one of interest for the SKA, is based on the use of an array of radiotelescopes for the accurate timing of a set of millisecond pulsars with known periods, a technique known as Pulsar Timing Arrays (PTAs) [12]. In this approach, when a GW passes across the region between the pulsars and the Earth it will perturb the local spacetime geometry producing minuscule but measurable changes on the times of arrival (ToAs) of the pulses (the effect being proportional to the amplitude of the characteristic strain of the GW [13]). Figure 1 illustrates this physical process. The differences between the predicted and measured ToAs are the so-called "timing residuals". Part of the residuals (irregular pulsar rotation, variability in the interstellar medium, etc.) are typically different between pulsars, while residuals due to GWs have a particularly simple functional form that only depends on the pulsar-Earth-gravitational wave angle [14]. PTA projects aim at extracting common signals present within the timing residuals for multiple pulsars in order to detect GWs and other correlated signals (e.g. fluctuations in atomic time standards or unmodelled effects in the Solar system ephemeris). With the appropriate technology (allowing timing precisions $< 100\,\mathrm{ns}$), a large enough number of pulsars ($> 20$) and a sufficiently long observational timing (about 10 years), the presence of GWs in the very low frequency range ($10^{-9} - 10^{-7}\,\mathrm{Hz}$) could be detected. It should be noted that a limit in the precision of the pulsar timing may exist due to various effects such as pulse jitter on short time scales (i.e. the intrinsic variability in the shape of individual pulses from a given pulsar), intrinsic pulsar timing noise on longer time scales (see [16] for an analysis of timing irregularities for 366 pulsars) and effects from the interstellar medium such as scattering [17].

Currently there exist three PTA consortia carrying out such kind of observations, one in Australia (the Parkes Pulsar Timing Array, PPTA [18]), one in Europe (the European Pulsar Timing Array, EPTA [19]) and one in the USA (the North American Nanohertz Observatory for GWs, NANOGrav [20]). All three Consortia have well established pulsar timing programmes on at least 20 millisecond pulsars with time baselines of 10 years or more [14]. Since 2008, the three PTAs have joined efforts in an international collaboration called the International Pulsar Timing Array (IPTA [21]). The sensitivity of the instruments has gradually improved through advances on instrumentation, software, and observing cadence and data span, to render increasingly accessible the PTA frequency range where suitable sources of gravitational radiation are expected to exist. We note that PTAs reach peak sensitivities at frequencies of $10^{-9} - 10^{-7}\,\mathrm{Hz}$ [14], therefore representing a unique complementary approach to that followed by GW observatories based on laser interferometry, either ground-based or in space. In Figure 2 we plot the sensitivity curves of the main GW projects that we have mentioned. As we can see, the SKA constitutes a significant improvement with respect to current PTA projects. We can also see in this figure the complementarity of PTAs, space-based GW observatories, and ground-based GW detectors, not only in frequency but also in the type of GW sources. Since the intersection of GW sources among all projects is essentially empty, the associated science to be extracted from each class of experiments is also quite different and complementary.

Likewise, owing to the frequency range where radiotelescopes are most sensitive, PTAs will be able to detect a unique class of GW sources out of reach for laser interferometers. The most likely source expected in their frequency band is the stochastic background of GWs produced by the superposition of a large amount of GWs from coalescing supermassive black hole binaries (with masses above $10^8 M_\odot$) in the far Universe at redshift $z \sim 1$ [22, 23]. In addition, cosmic strings [24] and inflation [25] are also prime candidates. To put bounds on these GW backgrounds it is assumed that their spectrum follows a power law defined by a characteristic (dimensionless) amplitude $h_c$ and a spectral index $\alpha$ ($-2/3$ for supermassive black hole binaries without environmental interactions, $-7/6$ for high emission modes from cosmic string cusps, and $-1$ for slow-roll inflation). The most stringent limit on $h_c$, from PPTA data [26], is $2.4 \times 10^{-15}$ at a reference frequency of $\mathrm{yr}^{-1}$. With the current bounds from PTA data, some extreme models for the coalescence rate of supermassive black holes as well as for the tension and loop size of cosmic strings have already been ruled out [14]. PTAs can also be used to search for individual supermassive black holes that produce continuous GWs, as the timing experiments are sensitive to sources emitting GWs with periods from a few weeks to around the total duration of the data set [14]. Finally, it has been argued that the merger of two supermassive black holes may produce permanent deformations in their surrounding spacetime that can be detectable as a memory event in the form of a "glitch" in the timing residuals [27, 28].

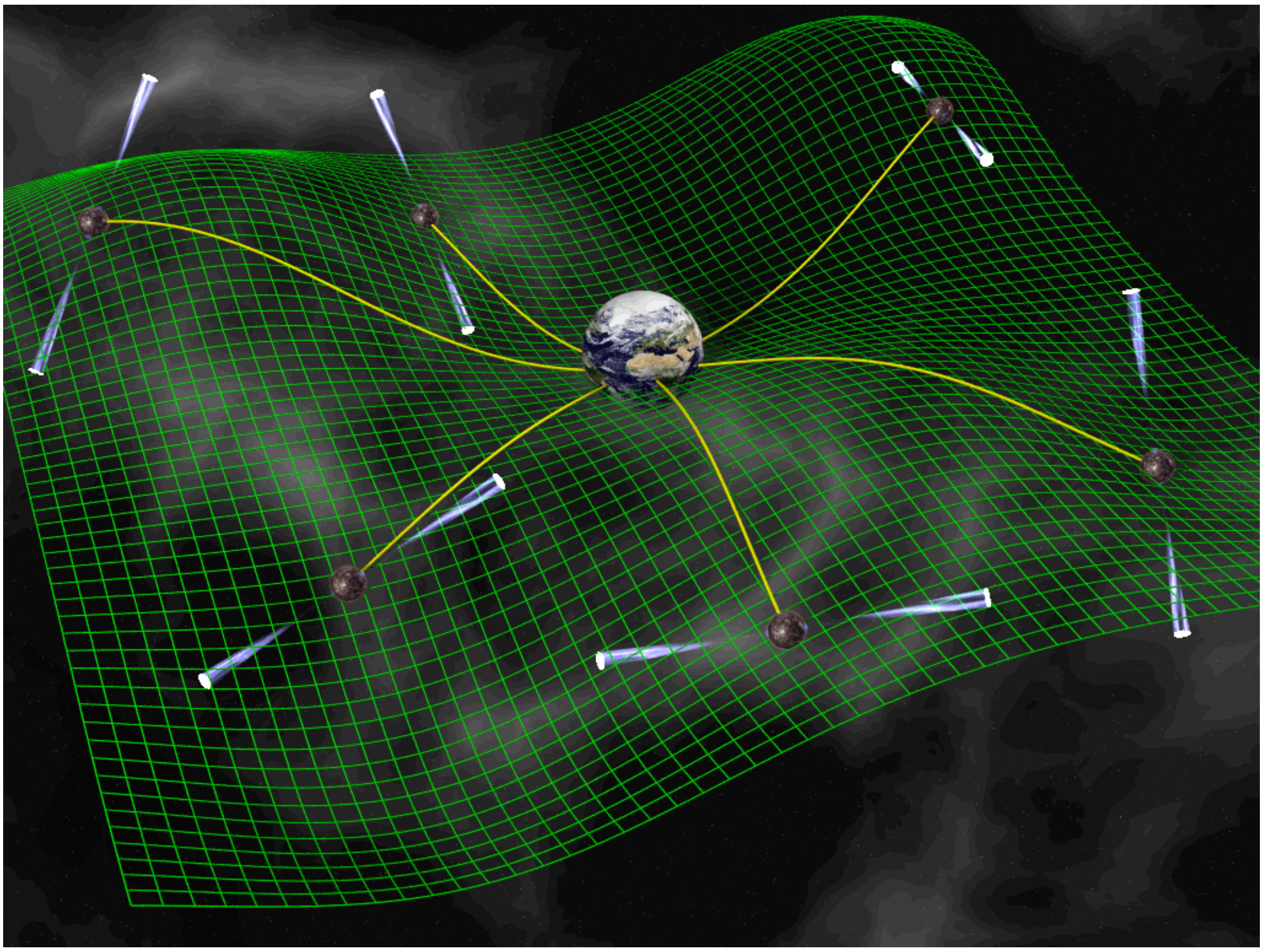

Figure 1: Artistic representation of the effect that GWs passing through our galaxy will produce in the arrival times of the pulses emitted by pulsars and received by radiotelescopes on Earth. Credit: Max Planck Institute for Radio Astronomy (D. Champion).

The SKA will no doubt constitute a major improvement of the ongoing efforts of the current PTAs. It is estimated that the SKA, as described in the baseline design document [29], will have a 50% probability of detecting a GW source after only five years of operation. However, as [14] note, this is a conservative estimate since the probability calculation does not include the pre-existing IPTA data sets. While the full IPTA reaches a nominal upper bound in the characteristic GW amplitude of $h_c \sim 4 \times 10^{-15}$ at $5 \times 10^{-9}$ Hz, the SKA is expected to reach levels of $h_c \sim 10^{-16} - 10^{-17}$ at a reference frequency of $\mathrm{yr}^{-1}$ [14]. Although current PTA experiments may succeed in detecting GWs by the time the SKA1 is commissioned, it is the SKA2 that will make Gravitational Wave Astronomy at nHz frequencies a reality. It is expected that the SKA will detect many more millisecond pulsars than is currently possible (or that are currently usable due to the different limiting noise processes mentioned before), and will allow to time them to very high precision ($< 100$ ns) every 10-20 days, making them very sensitive to the small space-time perturbations of GWs. Even if a red noise signal had already been identified by the existing PTAs, the SKA data sets could be used in conjunction with the longer PTA data sets to improve the sensitivity to GWs [30]. In the best case scenario where a GW detection had already taken place, the increased sensitivity of the SKA data sets could not only provide an independent confirmation of the detection but could also confirm the nature of the signal by distinguishing between the different spectral indices for the various sources. Finally, we briefly comment that information from Very Long Baseline Interferometry (VLBI) can significantly benefit GW searches by providing independent measurements of parameters that can be held fixed in the pulsar timing model [14].

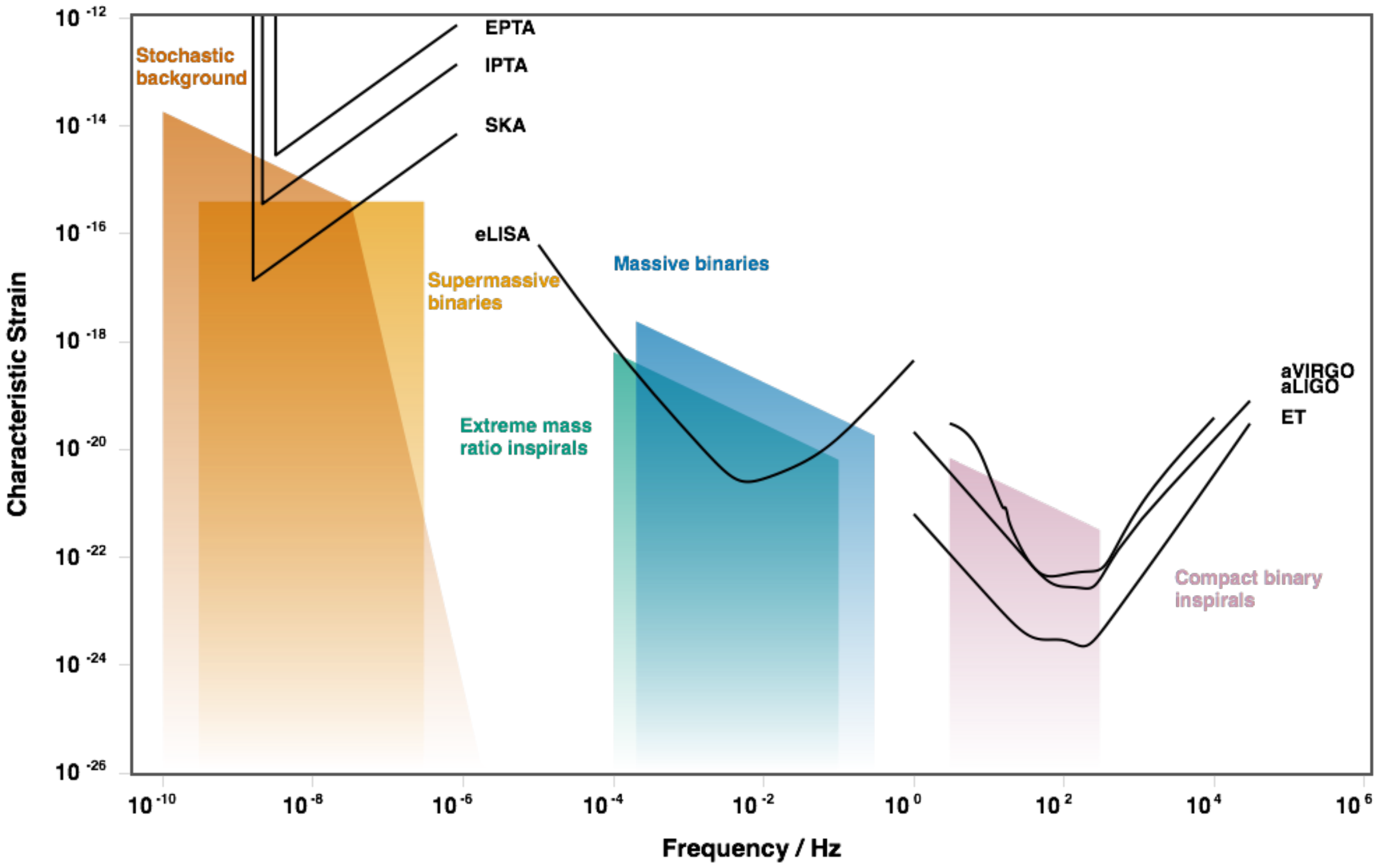

Figure 2: Sensitivity curves of PTA projects (EPTA, IPTA, and SKA), a space-based GW mission concept (eLISA), and ground-based GW detectors (LIGO and Virgo). These GW observatories represent the three more promising frequency bands: The very-low frequency band (PTAs), the low-frequency band (space-based detectors), and the high-frequency band (ground-based detectors). These sensitivity curves have been generated using the online tool described in [15].

## 3 Conclusion

To conclude, it is very likely that the SKA1 or the existing IPTA will make the first direct detections of GWs and also the first characterizations of the GW sources. With the upgraded version of the SKA1 (the SKA2) detailed properties of the GWs and their sources will be able to be studied opening the door for revolutionary discoveries in astrophysics, cosmology and fundamental physics. The most expected discoveries/scientific outcomes include: *Study of galaxy evolution.* The characterization of the properties of a GW background from coalescing supermassive black holes will provide constraints on models for galaxy evolution and black hole formation and growth. *Study of GWs from cosmological origin.* As we have already mentioned, cosmic strings could generate a GW background with different properties as compared to a GW generated by supermassive black hole binaries, in particular it will have a different spectral index. By identifying the GW background of cosmic strings we can have access to the physics at the energies of Grand Unification Theories or even beyond in the case such GW background would have been produced by cosmic superstrings. *Testing the theory of gravity.* The detection of GWs by PTA projects can provide information about the polarization states of GWs (two in GR but up to 6 in alternative theories of gravity) as well as constraints on the Compton wavelength of the GWs (or to the mass of the graviton) which affects the propagation of the different components of the GWs. Information of these two aspects of GWs will allow the study of the radiative sector of gravitation and/or help constraining alternative theories of gravity.

From the point of view of the Spanish scientific community, the GW science that the SKA has the potential to produce is of interest for existing groups working not only in GW astronomy but also in the astrophysics of supermassive black holes, cosmology, and fundamental physics of the Early Universe. The involvement of Spanish GW groups in the analysis of the data of the SKA is also possible and will benefit those groups by enlarging their expertise in GW data analysis.

## Acknowledgments

JAF is supported by the Spanish Ministry of Economy and Competitiveness (MINECO) through grant AYA2013-40979-P and by the Generalitat Valenciana (PROMETEOII-2014-069). AMS is supported by MINECO through grants FPA2013-41042-P and CSD2009-00064, by European Union FEDER funds and by the Conselleria d'Economia i Competitivitat del Govern de les Illes Balears. CFS is supported by the Spanish Ministry of Science and Innovation through grant AYA-2010-15709 and by MINECO through grant ESP2013-47637-P.

# Galaxy clusters with the Square Kilometer Array

**B. Ascaso[1], Y. Ascasibar[2], J.M. Diego[3], and S. Planelles[4,5]**

[1] GEPI, Observatoire de Paris, CNRS, Université Paris Diderot, 61, Avenue de l'Observatoire 75014, Paris France
[2] Departamento de Física Teórica, Universidad Autónoma de Madrid, E-28049 Madrid, Spain
[3] Instituto de Física de Cantabria (CSIC-UC), Avenida de los Castros s/n, E-39005 Santander, Spain
[4] Astronomy Unit, Department of Physics, University of Trieste, via Tiepolo 11, I-34131 Trieste, Italy
[5] INAF, Osservatorio Astronomico di Trieste, via Tiepolo 11, I-34131 Trieste, Italy

## Abstract

We review some science cases for galaxy clusters and the impact that the future SKA data will have in those analyses. We first describe how the search for galaxy clusters through radio-sources will be significantly improved through the detection of much fainter radio-sources in a big volume. Secondly, we bring out the benefits of using very sensitive radio data to study the thermal and non-thermal component of clusters and disentangle the main processes happening in the physics of their plasma. Moreover, we discuss the possibility of using the high frequencies of the SKA to separate the thermal Sunyaev-Zeldovich (SZ) effect from the radio halo emission and use the former as a mass proxy for galaxy clusters. Finally, we investigate how the very high sensitivity and spatial resolution of SKA will result into a great improvement in the lensing treatment, underlining the lensing distribution of the 21-cm intensity from the reionization period. As a whole, SKA will become an impressive window covering a significant wider range in redshift to look at an unknown radio universe and set constraints on different mechanisms happening in clusters.

## 1 Introduction

Within the standard model of cosmic structure formation, evolution proceeds in a hierarchical fashion. Clusters of galaxies, residing at the top of the cosmic hierarchy, are thought to be formed by mergers of smaller groups or clusters and the continuous accretion of matter and small galaxies ([33, 45] for recent reviews). Although dark matter is the main contributor to the mass in galaxy clusters ($\sim 70-80\%$), the cluster baryon content is shared between a hot and diffuse plasma ($T \sim 10^7 - 10^8$ K, $n_e \sim 10^{-4} - 10^{-2}$ cm$^{-3}$), called intra-cluster medium (ICM; $\sim 15-20\%$), and the stellar component ($\sim 3-5\%$). Cluster radio observations have also revealed the presence of a diffuse and extended synchrotron emission from the ICM ([25, 24, 9] and references therein) demonstrating that, besides the thermal plasma, there is also a contribution from non-thermal components, namely, magnetic fields ($\sim 0.1 - 10\,\mu$G) and cosmic rays (CR; in particular, relativistic electrons with $\sim$ GeV energies).

The SKA will become a superior instrument both in terms of sensitivity and resolution and therefore, will become very important at studying both the thermal and non-thermal component of the plasma. In this chapter, we aim to collect some of the interests regarding the understanding of physical properties of the clusters and illustrate the impact that the SKA data will signify for them. For a discussion of the potential of the SKA in the context of galactic evolution in dense environments, see Ascaso et al. (this proceeding).

The structure of the chapter is as follows. In §2, we focus on the expectations for SKA to detect galaxy clusters based on radio-sources and describe their selection functions. In §3, we summarize our current knowledge of the thermal and non-thermal components in clusters and describe their future significant improvement based on SKA data. We also focus on the benefits of using the high frequencies of SKA to measure the thermal SZ effect in clusters and calibrate their masses. Finally, section §4 is devoted to describe the new possibilities that the SKA data will offer for strong and weak lensing analysis, including the study of the lensing distribution of the 21-cm intensity from the reionization period.

## 2 Searching for galaxy clusters with SKA

Galaxy clusters, as structures consisting of dark matter, galaxies and ICM, can be traced by each of their components separately, providing different selection functions. Thousands of clusters have been found from systematic searches at different wavelengths, with a variety of techniques: modeling the properties of galaxies in the optical or infrared (IR) (e.g. [1, 2, 3] and references herein), using spectroscopic data to select the galactic population (e.g. [31, 32]), tracing the weak-lensing shear effect (e.g. [56]), detecting gas emission in X-rays (e.g. [49, 10]), analyzing the Sunyaev-Zeldovich (SZ; [53]) signature (e.g. [5, 52]), or searching for overdensities around different radio sources [6, 50, 30, 26, 15, 14, 8].

Radio-sources in galaxy clusters can be split into those associated with the ICM and those associated with Active Galactic Nuclei (AGN) [29]. Among the first group, we differentiate three main sources: (1) the (giant) radio haloes (RH), which are morphologically regular and extended sources ($\gtrsim$1Mpc), at the center of the potential well of merging clusters, (2) the mini-halo, smaller ($\lesssim$0.5 Mpc) sources, usually associated to cool-core systems featuring a central brightest cluster galaxy (BCG) or AGN, and (3), the radio relics or radio *gischt*, large elongated and strongly polarized sources, typically located in the outskirts of the clusters and also associated to merging clusters and/or energy injection from the AGN [25].

The ability of detecting galaxy clusters using RHs as a tracer has been investigated for the SKA1-LOW survey [13]. Based on their predictions, the SKA1-LOW would be able to detect $\sim$ 2600 RHs up to z$\sim$0.6, being the detection peak at z$\sim$0.2. This is a factor 6 times larger than previous surveys as LOFAR. Using scaling relations between $M_{500}$ and the minimum power of giant RH detectable in SKA1-LOW [12], the SKA cluster selection function as a function of halo mass was derived (see their Fig. 6), showing that SKA1-LOW will be able to detect galaxy clusters down to M $\approx 10^{14}$M$_\odot$ at local redshifts and down to M$\approx 4 \times 10^{14}$M$_\odot$ at $z \sim 1$. This selection function is comparable to the one obtained by Planck up to $z = 0.6$ or present X-ray surveys up to $z = 0.2$.

A similar analysis performed by [27] used instead the mini-haloes as galaxy cluster tracers. Assuming a given fraction of clusters with strong cool core clusters and deriving the observed radio-X-ray power correlation for mini-haloes, they estimated that around 600 and 2000 mini-haloes are waiting to be found with SKA1 and SKA2 respectively up to z$\sim$0.6. In addition, higher redshift mini-haloes might be detected with SKA.

Regarding radio relics, [41] derived the cluster halo mass function estimated from the radio relic number counts for different surveys, providing also a probability of discovery. As a tentative number depending on a variety of scaling relations based on a scarce sample of presently discovered radio relics and on the availably of complete galaxy cluster catalogues, SKA1 might be able to find up to $\sim$ 5000 radio relics

Additionally, AGN radio-galaxies are known to be found more likely in galaxy clusters than in the field (see Figure 13 in [20]). A variety of methods have been developed to search for overdensities around them (radio galaxy sources, [6, 50, 30, 26], FRI [15, 14], FRII, WAT [8], NAT and Bent-Tailed (BT) sources [17]. In particular, the latter method, based on a extrapolation from the results found in the ATLAS survey [17], has predicted to detect at least more than 1 million extended low-surface-brightness radio sources up to high-redshift ($\sim$ 2), out of which $\sim$ 80% will be BT sources. These number of detections are comparable to the number of clusters expected from next-generation survey such as LSST [35] and J-PAS [4] (Optical), Euclid [36] (Optical/IR), e-ROSITA [38] (X-Rays). Since many of these methods need deep IR data to search for overdensities around these radio sources, the data available from surveys such as Euclid will become an excellent tool to combine with the identification of such sources to detect galaxy clusters at high redshift ($z > 1.5 - 2$). Additionally, the combination with different datasets will also provide interesting scaling relations to calibrate the cluster mass and, therefore, to be usable for cosmological predictions.

Eventually, the SKA will be able to follow-up galaxy clusters using the SZ signature at $\sim$ 4 GHz frequencies (Band 5 of the SKA1-MID). In particular, this will be very useful to constrain the masses of the high-redshift ($z > 1$) clusters, since the SZ is redshift-independent. Indeed, [28] explored the ability for SKA1-MID to follow-up all the high-redshift clusters detected with e-ROSITA , finding that a 1000-hour program will be enough.

The upcoming SKA data is opening a new window to an unexplored universe and will undoubtedly deliver exciting results. The main properties of the radio-sources and their relation with the environment will be constrained within a wider range of redshift and halo masses. Our current knowledge is mostly based on the analysis of a few thousands of objects in the local universe and the extrapolation of the observed scaling relations between their physical properties. The SKA data ahead are likely to bring us interesting and unexpected findings that may force us to modify our expectations.

## 3 Thermal and non-thermal gas emission in clusters

As described in §2, the radio sources associated to the ICM (so far detected in only a few tens of clusters) are generally classified in halos, mini-halos and relics [29, 9]. Robust correlations between some physical properties of radio halos and their hosting clusters have been reported by current observational studies ([9] and references therein). The spectrum of radio emission in galaxy clusters provides crucial information on the energetics and physics of the relativistic particles involved as well as on the strength and distribution of magnetic fields. However, the sensitivity needed to perform such studies ($\sim m$Jy $-$ $\mu$Jy arcsec$^{-2}$ with frequencies from MHz to GHz) represents a challenge for current facilities, making it difficult to combine multi-frequency observations with different radio telescopes on a large number of radio sources. The increased sensitivity of the SKA (by a factor of $\sim 10$) to radio emission on cluster scales, will allow a detailed analysis of the spectra, polarization and brightness distribution of radio halos and relics for an unprecedented number of clusters, enabling a deeper understanding of the connection between radio sources and the dynamical history of the hosting clusters. In this regard, it is expected that a large number of new radio halos, those with very steep spectra which are not detected by current radio telescopes, will be unambiguously identified and associated to minor merging systems. The combination of these radio observations with X-ray or SZ-selected samples will help to study separately the thermal and non-thermal ICM components, and the combination of multi-wavelength data sets will provide invaluable constraints on the underlying physics.

In particular, the origin and evolution of the non-thermal component in galaxy clusters is still not understood. The steep radio spectra of the cluster radio sources suggest short life-times for the emitting electrons, indicating that, to account for the observed radiation, these particles need to be locally accelerated or injected. To explain the acceleration of relativistic particles within the ICM, two different models have been proposed: primary or re-acceleration models, in which relativistic electrons are accelerated by shocks and/or turbulence mainly through the process of Diffusive Shock Acceleration (DSA; [7]), and secondary models, in which relativistic electrons and $\gamma$-rays are thought to be released from the decay of pions generated in interactions between thermal ions and non-thermal protons in the ICM.

Numerical simulations have shown that the merger of two massive clusters produces shock and compression waves, turbulence and mixing, and amplification of magnetic fields in the ICM, contributing to the non-thermal emission in clusters (e.g. [11, 23]). Additional sources of shocks and turbulence, such as SN remnants or high-velocity jets from AGNs, can also provide relativistic particles to the ICM. The SKA, in conjunction with future $\gamma$-ray observations, will make possible to set strong constraints on the contribution of cosmic-ray protons to the relativistic electron budget in galaxy clusters. These studies, together with a precise measurement of Faraday rotation of cluster radio sources, will also shed some light on the origin, distribution and intensity of magnetic fields. Besides, they will help in understanding the role that turbulence plays in the acceleration and transport of relativistic particles and its influence on the ICM plasma physics. In addition, it will be possible to distinguish between the radio emission from different sources, such as AGN jets or SN remnants, and the intrinsic radio emission from the ICM.

Moreover, to quantify the impact of the process of cosmic structure formation on the acceleration of relativistic particles, it is important to map the strength and distribution of both shock waves and magnetic fields within the ICM. In this respect, numerical simulations have also reported a complex distribution of external and internal shocks within the cosmic web (e.g. [47, 44]) but the efficiency in the acceleration of CRs is still unknown. As for the cosmic magnetic fields, simulations suggest that cluster dynamics can contribute significantly to amplify their strengths from primordial values to the observed levels [23, 51]. In this regard, thanks to the spectral and spatial resolutions of SKA, a detailed analysis of the distribution and strength of shocks and magnetic fields throughout the cluster volume is expected.

The SKA will provide unique high-resolution and high-sensitivity radio observations of cluster-scale emission over a wide range of frequencies, allowing for a thorough investigation of the connection between thermal and non-termal cluster emission and the formation and evolution of the large-scale structure in the Universe. Forthcoming cosmological simulations, with improved resolutions and with a more precise modeling of the physics of cosmic plasmas, will also play a crucial role. Therefore, a coordinated effort between future cosmological simulations and the next generation of observing facilities at different wavebands will be essential to improve our understanding of the complex physical phenomena shaping the observational properties of galaxies and galaxy clusters.

In addition to the synchrotron radiation emitted by the relativistic component, all free electrons in the cluster interact with the Cosmic Microwave Background (CMB) photons through inverse Compton scattering. Since most electrons are moving at non-relativistic speeds, the small amount of energy imparted to the photons results in a weak distortion of the CMB spectrum, known as the SZ effect, whose amplitude is proportional to the electron density along the line of sight, the gas temperature (thermal SZ effect) and/or the bulk radial velocity of the cluster (kinetic SZ effect).

At the high frequencies of SKA ($\approx$ 20 GHz), the distortion in the spectrum due to the thermal SZ effect is seen as a decrement in the temperature (typically of the order of $\approx$ 0.1 mK in the Rayleigh-Jeans regime) with respect to the surrounding CMB radiation. The distortion due to the kinetic SZ effect is normally weaker than the thermal SZ effect, and it can be either positive or negative depending on the sign of the cluster radial velocity. In this frequency range, the SZ effect is expected to dominate over the radio halo emission in most clusters; even when they are comparable, the very different frequency dependence allows to separate both components.

The SKA will probe the physical conditions (density and temperature) of the hot intracluster plasma through the thermal SZ effect (see [28] for a recent discussion). The thermal SZ effect also provides a direct measurement of the electron pressure in the ICM, which can be used as an excellent mass proxy for galaxy clusters and makes possible to constrain the cosmological parameters (see e.g. [43]). The combination of high and low spatial resolution SKA data is particularly interesting, since the latter greatly improves the subtraction of contamination by discrete radio sources in the cluster.

Although most of the SZ effect with SKA will require sort baselines, the use of the longer baselines of SKA are also interesting for high resolution studies of brighter features in the SZ effect. Two phenomena producing small scale bright SZ features are particularly interesting. Cluster mergers produce a sharp enhancement in the thermal SZ effect at the collision point where the pressure increases due to the ongoing collision. In certain scenarios, the combination of SZ and X-ray data can be used to constrain the geometry of the collision [22]. Cluster mergers involve also large relative pair-wise velocities that can be studied through the kinetic SZ effect [39]. On smaller scales, subsonic motions of the cluster cores have been found to be present on several clusters. The kinetic effect distortion from these small scale ($< 1$ arc min) bulk motions is expected to reach a few tens of $\mu$K (or of the order of 100 Jy/sr at 20 GHz [18]).

## 4 Gravitational lensing from clusters with the SKA

Gravitational lensing is arguably one of the most powerful techniques to study the distribution of dark matter. The gravitational lensing effect is particularly obvious in galaxy clusters where the deep gravitational potentials produce large deflection angles, and often multiple distorted images of the same background galaxy. Massive galaxy clusters can be studied through the strong gravitational lensing (SL) and the weak gravitational lensing (WL) [34]. Detailed studies of the SL effect in galaxy clusters can be done with experiments that combine both, good spatial resolution and great sensitivity such as e.g. the Hubble Frontier Fields, HFF, program [16, 37].

A similar treasury program could be carried out by SKA which combines both capabilities (sensitivity and spatial resolution) and hence offers a competitive way of studying the dark matter in galaxy clusters through the gravitational lensing effect. Deep SKA observations of strongly lensed galaxies behind galaxy clusters offers also the unique opportunity to reach fainter fluxes that could not be observed otherwise. Current progress on the understanding of the magnification power of gravitational lenses establishes that in around 0.3 arcmin$^2$ per cluster, the ultra-faint galaxies can be magnified about a factor 3 or more [57, 48] for well studied gravitational lenses. Several clusters can be added together offering the possibility to statistically study the population of very faint sources below the detection limit of SKA.

At lower fluxes, the population of radio sources is largely undetermined with current estimates reaching approximately the 10 $\mu$Jy level at 1.4 GHz. In this regime, the density of radio sources is estimated to be $S^{2.5}dN/dS \approx 5\text{Jy}^{1.5}/sr$, with $S$ being the flux density [55, 40], resulting in a density of $N \approx 1$ source per arcminute$^2$ in the range $10 < S < 11$ $\mu$Jy for instance. Using these approximate densities and assuming the fraction at $z \geq 1$ to be about 30% of the total [42], it is possible to estimate the lensing effect for a realistic cluster. For this purpose we have used the deflection field derived by [21] and corresponding to the cluster MACSJ0717.5+3745 that was observed as part of the HFF program. After setting a typical deep SKA threshold of 1 $\mu$Jy at 1.4 GHz and assuming a beam of 1 arcsecond, we lensed the simulated background of sources above $z = 1$ by the cluster. A typical result is shown in Fig. 1 where several arcs and multiply lensed images can be appreciated above the limiting threshold of SKA. Even though this simulation did not account for magnification effects in the flux, it shows the power of SKA as a tool for gravitational lensing studies.

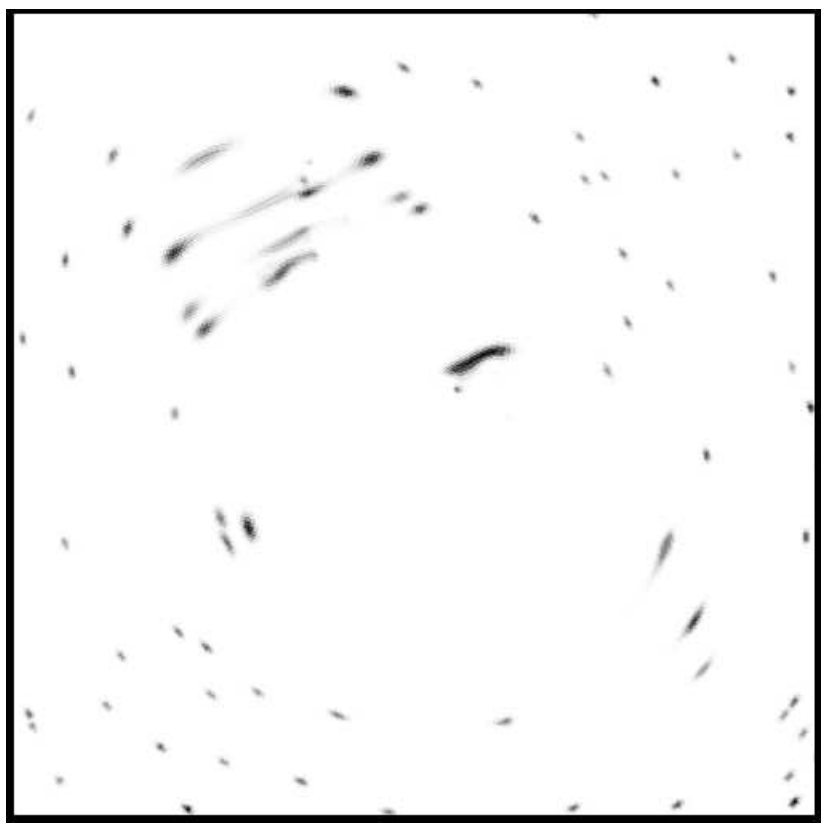

Figure 1: Simulated SKA observation through the lens MACSJ0717.5+3745. The field of view is 4 arcminutes on a side and all the radio sources have demagnified fluxes above 1 $\mu$Jy at 1.4 GHz. The density of sources is derived from the law $S^{2.5}dN/dS \approx 5\mathrm{Jy}^{1.5}/sr$ above 5 $\mu$Jy. Below 5 $\mu$Jy we adopt the law $S^{2.5}dN/dS \approx 1.0S_{\mu\mathrm{Jy}}\mathrm{Jy}^{1.5}/sr$ that follows the predictions. Both laws merge at S=5 $\mu$Jy. The lens model is taken from [21]

.

Since SKA can reach 1 $\mu$Jy fluxes (1.4 GHz), we should expect at least one strongly (or multiply) lensed galaxy per massive cluster ($M > 5 \times 10^{14}$ $\mathrm{M}_{\odot}$) allowing to estimate the masses of clusters from the SL data. The identification of the redshift of the background lensed galaxy through the 21-cm line will also eliminate one of the main sources of uncertainty in SL studies where redshifts are usually photometric. This will result in an improvement in the lensing reconstruction. A deep program carried out by SKA (below the 1 $\mu$Jy level at 1.4 GHz) on selected galaxy clusters could unveil a wealth of multiply lensed galaxies that would compete in quality with the optical data. This level of sensitivity is already planned for the SKA [40] and could easily be improved with a dedicated treasury program.

Furthermore, SKA data can be used to study the WL effect beyond the regime of the SL. The magnification bias does not rely on difficult shape measurements and it has been proven to provide useful information to constrain the cluster mass [54]. Studies of the magnification bias together with the classical WL analysis can be carried out with the SKA complementing the SL constraints. Radio data from SKA will be less affected than optical data by instrumental and astrophysical systematic effects. Radio telescopes have stable and well-understood point spread functions that simplify the corrections needed in optical-based data. This fact, together with the possibility of measuring redshifts for a significant fraction of the lensed galaxies through the HI emission lines, makes the SKA an attractive instrument for WL studies around galaxy clusters.

However, the unique contribution that SKA will do is on the lensing distribution of the 21-cm intensity from the reionization period. Anisotropies in the intensity due to the reionization are expected to produce a continuous background at high redshift. Lensing of this background can be studied with the SKA opening the door to novel techniques for mass reconstruction, some of which have been already implemented in the lensing of the CMB. In particular, lensing of a continuous background introduces correlations between the Fourier modes that can be used to reconstruct the mass distribution responsible for the lensing effect. However, while the CMB probes a particular redshift range, the 21-cm line probes a much wider redshift range that can be used to do tomographic studies of the universe by simply shifting to lower frequencies in SKA corresponding to higher redshifts. [58] shows that a large radio array such as SKA could measure the lensing convergence power spectrum and constrain the cosmological parameters. More recently, [46] focuses on the particular case of the low frequency of SKA and concludes that the SKA low frequency instrument could be used to study the lensing signal using the epoch of reionization as a background. With the mid-frequency instrument, SKA will survey a wider region of the sky and although not appropriate for studies of the epoch of reionization, fluctuations in the intensity from galaxies will form also a background of sources that can be used for lensing studies as described above. In the particular case of clusters, when a cluster is known to lie along the line of sight, spatial filters can be used to isolate the lensing signal as shown by [19] although it is difficult the evaluate the feasibility of such direct studies as the signal depends on the amount of anisotropy in the background signal which is controlled by the type of reionization (patchy, or instantaneous).

## Acknowledgments

This work has been supported by a grant funded by the "Consorzio per la Fisica di Trieste". SP also acknowledges support by the PRIN-INAF09 project "Towards an Italian Network for Computational Cosmology", by the PRIN-MIUR09 "Tracing the growth of structures in the Universe", and by the PD51 INFN grant. Partial support is also provided by *Spanish Ministerio de Ciencia e Innovación* (AYA2010-21322-C03-02). JD acknowledges support from the Spanish Ministry of Economy and Competitiveness (MINECO) through grants AYA2010-21766-C03-02, AYA2012-30789, and the Consolider-Ingenio project CSD2010-00064 (EPI: Exploring the Physics of Inflation). YA is financially supported by the Spanish Ramn y Cajal programme (RyC-2011-09461) and grant AYA2013-47742-C4-3-P (MINECO), as well as the 'Study of Emission-Line Galaxies with Integral-Field Spectroscopy' (SELGIFS) exchange programme, funded by the EU through the IRSES scheme (FP7-PEOPLE-2013-IRSES-612701). BA acknowledges financial support for a postdoctoral fellowship from the Observatory of Paris.

# HI and galaxy evolution

**L. Verdes-Montenegro[1], J. M. Solanes[2], R. Domínguez-Tenreiro[3], A. Gil de Paz[4], J. H. Knapen[5,6], U. Lisenfeld[7], Y. Ascasibar[3], A. del Olmo[1], S. García-Burillo[8], B. García-Lorenzo[5,6], J. Iglesias-Páramo[1], J. Perea[1], E. Pérez-Montero[1], M. A. Pérez-Torres[1,9], V. Quilis[10], J. Sabater[11], S. Sánchez[12], M.C. Toribio[13], B. Ascaso[14], J. Blasco[1], C. B. Brook[3], J. Font[5], R. García-Benito[1], I. Márquez[1], P. Ramírez-Moreta[1], and J. Vílchez[1]**

[1] IAA-CSIC
[2] Departament d'Astronomia i Meteorologia and Institut de Ciències del Cosmos, Universitat de Barcelona, C. Martí Franqués, 1, E-08028 Barcelona, Spain
[3] Departamento de Física Teórica, Universidad Autónoma de Madrid, Madrid 28049, Spain
[4] Dept. Astrofísica, U. Complutense, Avda. Complutense s/n, CC. Físicas, Madrid, 28040, Spain
[5] Instituto de Astrofísica de Canarias, E-38200 La Laguna, Tenerife, Spain
[6] Departamento de Astrofísica, Universidad de La Laguna, E-38206 La Laguna, Tenerife, Spain
[7] Departamento de Física Teórica y del Cosmos, Universidad de Granada, Spain; Instituto Universitario Carlos I de Física Teórica y Computacional, Facultad de Ciencias, 18071, Granada, Spain
[8] Observatorio Astronómico Nacional (OAN)-Observatorio de Madrid, Alfonso XII, 3, 28014, Madrid, Spain
[9] Centro de Estudios de la Física del Cosmos de Aragón (CEFCA), 44001 Teruel, Spain
[10] Departament d'Astronomia i Astrofísica, Universitat de València, E-46100 Burjassot, Valencia, Spain
[11] Institute for Astronomy (IfA), University of Edinburgh, Royal Observatory, Blackford Hill, EH9 3HJ Edinburgh, UK
[12] Instituto de Astronomía,Universidad Nacional Autonóma de Mexico, A.P. 70-264, 04510, México, D.F., México
[13] Netherlands Institute for Radio Astronomy (ASTRON), Postbus 2, 7990 AA Dwingeloo, The Netherlands
[14] GEPI, Observatoire de Paris, CNRS, Universite Paris Diderot, 61, Avenue de l'Observatoire 75014, Paris France

## Abstract

Observations of atomic hydrogen (HI) are a fundamental component of SKA science, with the Hydrogen Array being part of the initial concepts of the project. The atomic gas is key in central processes of galaxy evolution and structure formation, such as the buildup of stellar mass in galaxies, star formation (SF) in and beyond the optical disc, galaxy dynamics or in probing environmental influence. The study of HI in nearby galaxies is part of one of the five Key Science Projects identified by the community as key science drivers for the SKA: *Galaxy Evolution, Cosmology and Dark Energy.*
SKA will make HI imaging of galaxies possible with unprecedented sensitivity and resolution, both characteristics crucial to address these questions. The first stage of SKA (SKA1, first science expected in 2020) will reach column density limits 3 times deeper than current facilities for the same integration time and angular resolution. At column densities similar to the ones reached by current interferometers, a $3''$ resolution will be achievable ($\leq$150 pc for galaxies out to 10 Mpc, i.e the Local Volume, starting hence to resolve individual H I clouds). The full SKA will provide mapping of individual galaxies at $1''$ resolution out to z = 1, one order of magnitude deeper than current interferometers, and reach the same angular resolutions now achieved for the Local Group out to 10 Mpc. This will be complemented by a survey speed that will allow detecting HI emission from around a billion galaxies over 3/4 of the sky, out to a redshift of z$\sim$2 [123].
In this chapter we summarize the main areas of HI studies of galaxies in which SKA will have an impact, with special emphasis on those areas of particular interest for the Spanish astronomers, both observers and theoreticians.

## 1 Introduction

The HI component of galaxies, the main phase of their neutral gas content, informs us about the different processes that shape them, from their very formation to their internal evolution, as well as the way they are connected with their environments. However the 21cm HI line is intrinsically faint, hence limiting the extent of the studies performed so far. HI mapping of galaxies has been feasible till now at only moderate redshifts (z < 0.25). Nearby spirals have been the subject of extensive HI-mapping (Sections 2 and 4), although with angular resolutions limited to values an order of magnitude higher than $1''$, the typical one for standard optical observations. On the other hand HI mapping of early-type galaxies (ETGs) has been scarce till recently, when the picture of red and quiescent ETGs has started to change (Section 3). HI stripping has been proposed to inhibit SF in extreme environments, although the removed gas is suspected to exist at column densities below the present detection limits (Section 4.1). The same applies to cosmological cold gas accretion, suspected to maintain SF in galaxies: unfortunately models predict it to occur at too low column densities to be detected by current interferometers (Section 4.2). In summary, as nicely expressed by T.J. Cornwell and R.A. Perley in [16], in order to motivate a km$^2$ Hydrogen Array, '*A volume of the 'Encyclopaedia of the Universe' is written in 21 cm typescript. Unfortunately the printing* *is rather faint and we need a large 'lens' to read the text!*".

This Chapter focuses on those aspects of galaxy evolution in which HI studies play a relevant role and that are of particular interest for the Spanish astronomical community. This includes synergies with nearby galaxy surveys (Section 2), formation history of early type galaxies (ETGs, Section 3), environmental effects (Section 4), SF in the outskirts of disks and cold gas accretion (Section 4.2) or theoretical models (Section 5). We remark that covering all angles of galaxy evolution in which this component plays a relevant role, such as, for instance, galactic fountain processes, high-velocity clouds, kinematical studies of galaxies and their dark halos, the Tully Fisher relation or the HI mass function, is beyond the scope of this chapter. Detailed discussions of these and related topics can be found in the revision of SKA capabilities for the study of the history of HI that the HI and Galaxy evolution Science Working Group has performed ([4, 18, 85, 76]).

## 2 Nearby galaxy surveys

A number of studies have focused in the last years on interferometric HI studies of nearby galaxies, in order to investigate how HI relates to SF, galaxy morphology, mass and environment, complementing blind single dish studies as HIPASS [2] and ALFALFA [37]. THINGS [119] has performed HI observations with the VLA (Very Large Array) of 34 galaxies at distances $3 < \mathrm{D} < 15$ Mpc, reaching the highest angular resolution so far ($6''$), and LittleTHINGS observed 21 dwarf irregular and blue compact galaxies [48]. VLA-ANGST [79] is an NRAO Large Program to observe 35 gas-rich and actively star forming (D $<$ 4 Mpc) galaxies in HI, selected from the HST ACS Nearby Galaxy Survey Treasury (ANGST) performed with the HST. FIGGS is a Giant Metrewave Radio Telescope (GMRT) based HI imaging survey of a systematically selected sample of extremely faint nearby dwarf irregular galaxies [3]. LVHIS [57] is a project comprising deep HI line observations with the Australia Telescope Compact Array (ATCA) for all nearby, gas-rich galaxies (D $<$10 Mpc, $\delta <$ -30) that are detected in the HI Parkes All Sky Survey. For $\delta >$ +20 degrees the Westerbork Radio Synthesis Telescope was used (WSRT-LVHIS project), with which the WHISP HI-survey [107] of spiral and irregular galaxies was performed as well. SHIELD [10] aims at measuring the total halo masses of the extremely low-mass galaxies detected by ALFALFA by means of deep, high spatial and spectral resolution HI observations with the VLA. Other more specific studies are described in the next sections.

Spanish astronomers play a major role in several studies that comprise either HI observations and/or studies of galaxy evolution with important synergies with the SKA. In the following we summarize those that would benefit from the improvement in spatial resolution and/or sensitivity that SKA will provide. Studies focused on specific environments, like isolated galaxies, compact groups or clusters, will be presented in Section 4.

The Calar Alto Legacy Integral Field Area survey (CALIFA [93]) is obtaining integral-field spectroscopic data in the optical of a diameter selected sample of 600 galaxies of all Hubble types in the local universe ($0.005 \leq \mathrm{z} \leq 0.03$) with the 3.5 m telescope at the Calar Alto observatory. CALIFA provides very precise measurements of the stellar and ionized gas kinematics and metallicities, as well a spatial resolved derivation of the full SF history and the properties of the ionized gas mapped over the entire optical extent of galaxies at an angular resolution of $2''$. A complementary survey of HI at a similar spatial resolution will provide with unique information about the gas duty cycle, since it will connect the atomic gas, with the SF processes and the SF and chemical enrichment histories of galaxies, and allow to answer, among others, key questions on the origin of the gas or the angular momentum and dark matter distribution in galaxies. In the first phase of the SKA, SKA1-MID will reach similar depths as e.g. THINGS but at $3''$ resolution in 10 hours, hence starting to resolve individual HI clouds. Reaching $1''$ resolution would require 1000h per galaxy with SKA1-MID to map the main HI component of a disc, hence the full SKA will be required for this aim.

The kind of observations that SKA will be able to perform are also of interest for the Spitzer Survey of Stellar Structure in Galaxies ($S^4G$), a representative, volume-limited (D < 40 Mpc) survey of nearby spiral and dwarf star-forming galaxies, elliptical, and dwarf galaxies at 3.6 and 4.5$\mu$m with the IRAC camera on board of the Spitzer Space Telescope, a project with significant participation of Spanish researchers. Synergies with SKA HI studies will be most important in the outer regions of the galaxies for which deep mid-IR and often optical [56] imaging is available. Reaching low HI column densities will provide vital complementary information, allowing us to progress in the study of important issues such as radial HI profiles in relation with truncation/antitruncation studies (e.g. [71, 74]) detection of tails, shells (e.g. [63, 64]), or presence of tidal debris in ETGs (e.g. [55]; see Section 3).

Complementarities are as well foreseen with the IRAM key project NUclei of GAlaxies (NUGA; see [32]), established to systematically tackle the issue of nuclear fueling, and later extended to the outer discs (HI-NUGA) using the VLA, covering scales from a few tens parsec to the outer few tens kiloparsec. NUGA showcases the complementarity between ALMA and SKA, which will lead to a revolutionary improvement in our understanding of SF and nuclear activity. Mapping the various spatial scales of the different instabilities (HI: outer, CO: inner) is essential to establish whether nuclear activity depends on the host galaxy properties. The dissipative nature of the gas makes HI ideal to trace dynamical disturbances, both internal, such as non-axisymmetric potentials, or external, such as tidal interactions. The first results of NUGA showed a slightly larger HI extent and less prevalent disturbed HI disks for Seyfert hosts than for LINER host galaxies [40]. In addition, there is the recent study by [92], based on global properties of more than $250,000$ galaxies drawn from the Sloan Digital Sky Survey, suggesting that the level of nuclear activity depends primarily on the availability of cold gas in the nuclear regions of galaxies —similar results were obtained by [26] by measuring the average HI content of a volume-limited sample of 1871 Active Galactic Nuclei (AGN) host galaxies—, while large-scale environment and galaxy interactions only affect AGN activity in an indirect manner, by influencing the central gas supply. In contrast, the conclusions from the NUGA project were based on a sample of only 16 galaxies, so a detailed HI mapping study with a larger dataset may reveal more relations with respect to the AGN type and would certainly improve the significance of these early results.

However, studies of the 2D distribution of HI in galaxies using samples much larger than that seem nowadays sheer impossibility. For instance, THINGS required typically 10h per galaxy to produce a complete observation with a $5\sigma$ sensitivity of $2.7 \times 10^{20}$ cm$^{-2}$ ($6''$ and 5 km/s of spatial and spectral resolution respectively, [119]). SKA Australian pathfinder ASKAP will anticipate the capabilities of the full SKA ,with an instantaneous field of view of $\sim$ 30 deg$^2$, allowing to map a large number of galaxies together with their environments as part of WALLABY large program (Koribalski et al 2012).

Another complementary strategy will be the observation of the 21cm HI line in absorption, which is mainly a function of the strength of the background continuum source. This sort of observations are hence useful to increase our understanding of the spatial distribution and kinematics of the circumnuclear disks that inhabit the central parsecs of radio sources, thus providing an important test of unified schemes for AGN (see Perucho et al. Chapter).

## 3 Early type galaxies

Resolved HI studies of ETGs are in a more preliminary stage than for spiral galaxies, the traditional target of HI studies. Searching for cold gas in ETGs started in the 1960s, with single dish detections reaching about a hundred in the 1990s, detecting HI in 15% of ellipticals and 25% of lenticulars for a detection limit of a few times $10^8 M_\odot$ (e.g [46] and references therein). ALFALFA data, with a detection limit of around $10^8 M_\odot$ for nearby galaxies, show a very low detection rate of ETGs ($\sim$ 1%) in the region of the Virgo cluster [33]. High-z observations show that intermediate-mass ETGs actually finished forming stars much more recently than the most massive ones [13], which have traditionally biased our view on the properties and evolution of ETGs. Our picture of ETGs no longer sees them merely as red, featureless and quiescent. Rejuvenation signatures in ETGs are often detected, including HI gas (e.g. [78, 39] - ALFALFA) and even SF in galaxy outskirts, as shown by GALEX [88]. According to [78], the fraction of ETGs with HI detections depends strongly on the environment, being very common in field galaxies, and extremely uncommon in clusters, suggesting that very different mechanisms are at play. Slow and fast rotators seem to have similar HI mass and rates of discs/rings ([99], ATLAS3D), and the weak relation between HI and stellar disc is confirmed by their frequent kinematical misalignment in fast rotators. More recent results, based on CALIFA observations, indicate that most if not all ETGs present a diffuse ionized gas emission, which is most probably ionized by post-AGBs and other ionization sources different than SF [100, 80]. This indicates that ETGs are not totally gas free objects, and the lack of SF has to be explained by other reasons. Measuring the amount of HI would clarify the nature of this gas, and why it is not forming stars nowadays.

This complex picture highlights a diversity of ETG formation histories. A correlation between HI mass and morphology with the density of the environment has been shown by [99], and tidal debris has been identified in S$^4$G ETG targets [55]. However it is not clear yet whether cold gas in ETGs is always due to gravitational interactions/mergers or is leftover gas being directly accreted from the halo, since preliminary studies of isolated galaxies from the AMIGA sample (Analysis of the interstellar Medium of Isolated GAlaxies, Section 4) reveal that HI detections are not rare in isolated ETG galaxies. Such low density environments are hence ideal to disentangle the origin of cold gas in ETGs. Unfortunately the above mentioned studies have shown that HI surface density is generally much lower in E+S0 than spirals, hence a high sensitivity is needed for their study. ATLAS3D reached $5\sigma$ sensitivities with the WSRT of 0.6 - 3.1 $\times$ $10^{19}$ cm$^{-2}$ with typical resolutions of 35″ and 16 km/s in 12h observations. SKA1-MID will be able to go down to 2.9$\times10^{18}$ cm$^{-2}$ at similar or better resolutions (30″ and 5 km/s) in a 10h integration. These data, combined with ALMA data in order to characterize both the atomic and molecular components of the interestellar medium, are key to understand the origin and SF history of ETGs.

# 4 Galaxies and their environments

## 4.1 HI as a priviledged tracer of galaxy interactions

HI, as the most extended cold component of the interstellar medium, is a very sensitive tracer of interactions, hence playing a major role in disentangling internal from environmental processes. The inventory of detailed HI observations of galaxies in environmentally selected samples is growing, ranging from isolated galaxies to groups or clusters. Having a sufficiently large HI data set in and around many galaxies in different environments will be a prime goal for the SKA, as it will provide very valuable information in order to determine to which extent are galaxy properties governed by initial conditions (nature) or driven by environmental effects (nurture). In this section we will refer to works on isolated galaxies and compact groups, while clusters are discussed in detail in the Chapter by Ascaso et al in this book.

While it is easy to recognize a rich cluster, definitions of low-density environments can be confusing. In recent years there has been an increased emphasis on identifying low density or isolated galaxy populations. The AMIGA program ([111]) was designed to provide a sample with strict isolation criteria [116, 117] implying that its galaxies have likely been unperturbed for $\sim$ 3 Gyr. Variables expected to be enhanced by interactions have lower values in AMIGA galaxies than in any other sample, even field galaxies: $L_{FIR}$ [66], radio continuum [65], AGN rate [91], a lower fraction of dense gas which could be more diffuse [67, 69, 77], or no radioexcess above the radio-FIR correlation [90]. Redder (g-r) colours and larger discs [28] coexist with red pseudobulges having formed most of their mass at an early epoch [21, 29]. Having minimized environmental effects, a key finding in a single dish HI study of isolated galaxies [24] was that AMIGA shows the lowest values of the HI profile asymmetry of any studied sample. Some significant asymmetries (involving 25 - 50% HI mass) are still found in AMIGA sample, and in an effort to discover their cause interferometric HI data are being gathered for a significant subsample. Neither tidal features nor gas rich companions are found in the most sensitive interferometric HI studies of AMIGA galaxies, reaching N(HI)$\sim$5x$10^{19}$ cm$^{-2}$ even for the most asymmetric cases (e.g. [23, 25, 86, 98, 97]), with the signatures of HI asymmetry stronger in the resolved morphology than in the velocity field. Those studies indicate that at least some of their HI asymmetries were generated recently ($\sim10^8$ yr) and were not produced by interactions with currently observed minor companions. If asymmetries can only be generated by tidal interactions, lopsidedness in isolated galaxies should not be observed. Asymmetries in the accretion of cosmological cold gas would be an obvious candidate to generate them (see Section 4.2). In 100 hours per pointing, SKA1-MID will achieve brightness sensitivities equivalent to the best current datasets (N(HI)$\sim$2e18 at/cm2, 10 km/s and 15″ resolution) over 0.46deg$^2$ [85].

ALFALFA data was used by [43] to study a sample of $\sim$ 740 groups of 3 or more members (HI mass sensitivity $\sim 10^9$-$10^{10}$M$_\odot$), that led the authors to prove the evolution in the spatial distribution of the HI content of galaxy groups as a function of their group membership. They find strong evidence that HI is being processed in galaxies as a result of the group environment and that the infall of gas rich objects is important for the continuous growth of large scale structure at the present epoch. Finding out which physical mechanisms are responsible for processing or removing the cold gas reservoir of galaxies requires high resolution and time consuming HI observations, currently feasible only for a few nearby groups (e.g. ALFALFA Leo region, [102]). SKA1-MID will be able to perform routinely deep observations of large galaxy fields in a modest amount of time, allowing a search for intragroup HI clouds, tidal debris, or swept back HI disks.

Hickson compact groups [44] (HCG) constitute a more extreme kind of environments, with densities similar to cluster cores, which should favor pre-processing through extreme tidal interactions. In general their average SFR and molecular gas content [108, 49] are similar to control samples, but in HCGs there exists a higher fraction of young late type spirals compared with isolated galaxies, and most of them belong to groups in which the majority of their members are also young and actively forming stars [82]. According to [73], the low frequency (20%) of star forming nuclei supports that there is no SF enhancement in HCGs. Although 42% of galaxies in HCGs present nuclear activity , most corresponds to low luminosity AGN. In fact in this kind of environment there is a remarkable deficiency of broad-line AGNs, consistent with gas stripping through tidal interaction [72]. As we explain below, recent studies suggest that their intragroup medium might be refueled with diffuse processed cold gas, but current radiotelescopes are unable to confirm this. Single dish observations of HCGs showed a large HI deficiency ([121, 47]). Interferometric VLA observations of the HI in 26 Hickson Compact Groups ([112, 103, 109, 120, 110, 113, 114, 20]) have shown that they seem to be evolving from a phase where the gas is located in the galaxy discs, to intermediate cases where the HI is still detected as a net of high surface brightness tidal tails and bridges (see example in Fig. 1), and finally into a stage where almost no HI is detected, neither in the galaxies nor in the intragroup medium. What physical processes give rise to this extreme deficiency is still a mystery, despite of the numerous studies performed since this evolutionary model was proposed. Comparison of VLA imaging with high-quality single dish observations with the GBT ([5, 6]) has provided evidence for the existence of a diffuse HI component missed by the VLA that increases with evolutionary stage, more consistent with tidal stripping than with ram-pressure (see as well [89]), and spread over a velocity range of more than 1000 km/s. This suggests that slow evolution of tidal debris may lead to a final stage where the HI becomes faint and extended – hence escaping detection by current interferometers – being returned to the IGM. Sensitivities in the above referred interferometric studies of HI in HCGs reach a maximum value of N(HI) = $5\times 10^{19}$ cm$^{-2}$ (e.g. HCG 92 study with VLA C+CnB+D, $20''$ resolution, [103, 120]). SKA1-MID will provide the required field of view ($30'$) with a highly improved sensitivity (e.g. $1.3\times10^{18}$ cm$^{-2}$ at $30''$ and 5 km/s in 10h) able to provide key information on the fate of gas in HCGs.

The unprecedented high sensitivity and resolution observations that will be achieved by SKA will also help us to better characterize the environment of galaxies in a more unbiased and quantitative manner. The environment of galaxies is usually characterized by means of the definition of isolation criteria and cluster/group/field membership or other quantitative measurements of the density of galaxies. As mentioned above, samples of galaxies with properties as less affected by environment as possible are used as a reference for environmental studies. In that respect, isolating the HI emission of galaxies in close proximity in combination with galaxy redshift measurements and intracluster medium observations at other wavelengths, will provide us with the most accurate information on both the HI content and environment of galaxies. With SKA we will be able to significantly improve our current standard measurements of the HI content of galaxies (e.g., [41, 101, 106]) by overcoming confusion effects and any potential biases that limit the studies carried out to date.

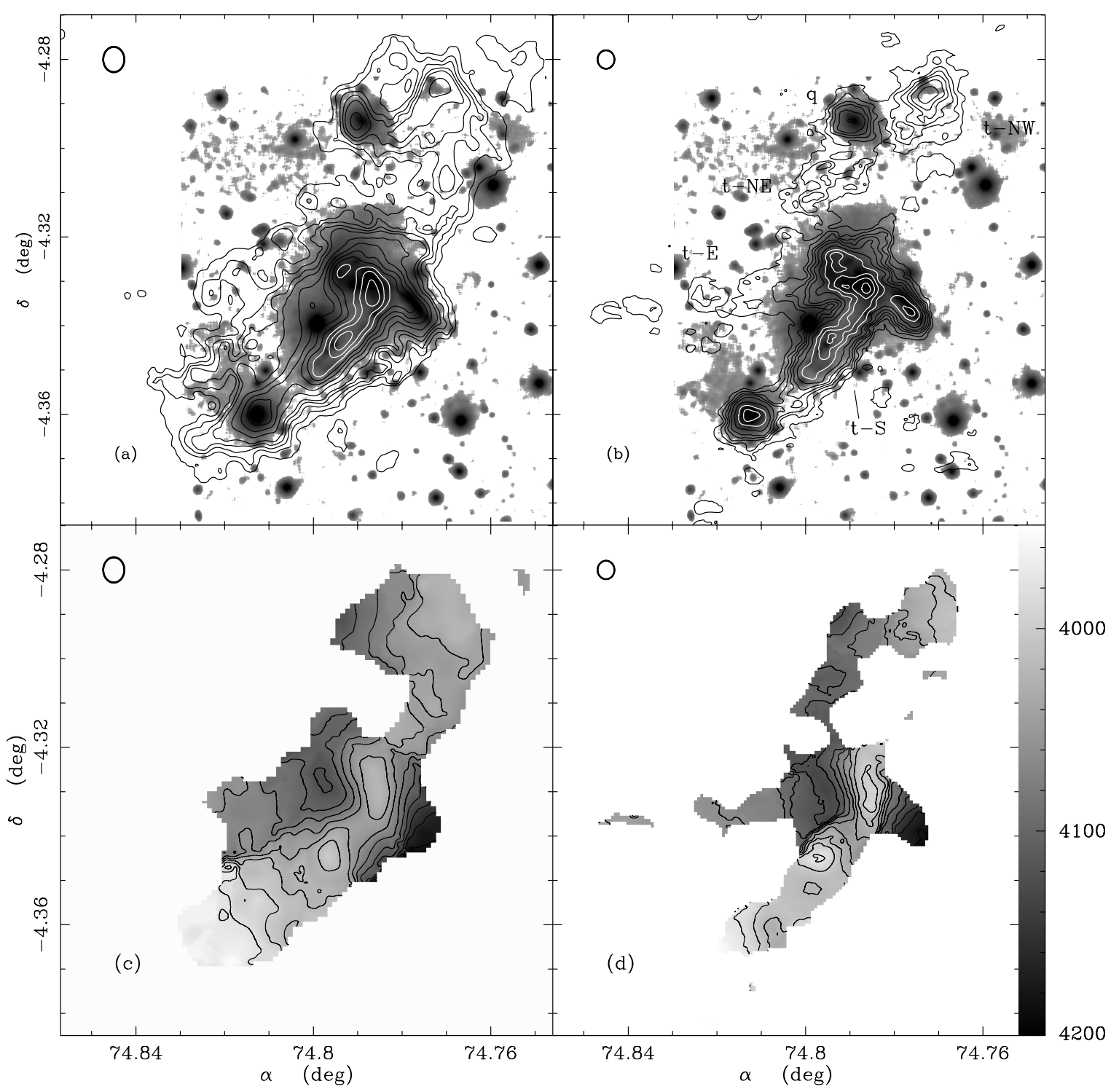

Figure 1: (a) Map of the HI column density distribution in HCG 31 obtained with the VLA DnC configuration, superimposed on a B-band image. (b) The same as (a) for VLA CnB array data. (c) Map of the first moment of the HI radial velocity in HCG 31 obtained from the VLA DnC configuration data showning both with iso-velocity contours and gray-scale. (d) The same as (c) for the CnB array data. The gray scale in panels (c) and (d) represents heliocentric velocities in km $s^{-1}$ as shown in the wedge. See [110] for more details.

## 4.2 Star formation beyond the optical disc and cold gas accretion

For three decades it has been known that about one third of galaxies show unusually extended HI distributions (e.g., [7, 45]). It was not until recently, with the advent of the Galaxy Evolution Explorer (GALEX) [68], that we could observe easily SF far beyond the optical radius of galaxies ([104, 35, 22]). The star-forming regions generally show a good spatial correlation with the HI component (the main phase of neutral gas at large radii) in such extended HI discs (e.g [25]). Comparing UV and HI in them allows the study of the Kennicutt-Schmidt (KS) law, i.e. the SFR per area versus gas surface densities ([96, 53]), in the extreme low-density and often low-metallicity environments of the outskirts of galaxy discs, which can elucidate the limiting conditions for gas cloud and SF ([105, 70]). In this respect, SKA observations will provide valuable insight into the fundamental question of whether it is possible or not to trigger SF in very low-density gaseous environments. In addition, the gas reservoirs observed in galaxies seem to be insufficient to explain sustained SFRs over a Hubble time, both at high redshift ([15]) and in the local Universe ([30]). It is therefore essential to unravel whether the availability of HI in the outer disks of galaxies bears any connection with the depletion time scales of molecular gas in their inner regions due to SF.

Apart from this mechanism, several other alternatives which also require the study of HI at low column densities have been proposed to sustain SF over cosmological times. Gas accretion from the intergalactic medium seems necessary, although its origin is controversial. It could be partly explained by SNe and AGN feedback (galactic fountain models, e.g. [31]). Additionally, cosmological pristine gas might be accreted into the galaxy, as simulations suggest [27], along what is called the Cosmic Web. However, the observed cold gas accretion seems to be an order of magnitude too low to explain the measured SFRs [95]. Besides, the flat, non-primordial chemical abundances measured in the ionized-gas in outer discs suggests that either this gas has been transported from the inner diks (metal transport) or it is IGM gas pre-processed by galactic outflows [8, 36]. If gas accretion from the IGM is the dominant process by which galaxies acquire their gas, then it must happen below the current observational limit for the HI column density N(HI)$\sim 10^{18}$ cm$^{-2}$. While gas features have been detected at such column densities using the GBT antenna ([122]), deep interferometric observations cannot reach beyond a few $10^{19}$ cm$^{-2}$, where ubiquitous clouds and streams were not found (see deep HI WSRT data of nearby edge-on galaxies performed as part of HALOGAS survey, [42]).

The picture that emerges from HI observations is consistent with the study of cosmological hydrodynamic simulations by [84]. They find that, in the high over densities where the galaxies and clusters are, the gas is fully neutral. However, in the filaments N(HI) decreases well below $10^{18}$ cm$^{-2}$, making them only accessible to SKA. The issue is complicated by the fact that galaxies live their lives embedded in a variety of environments. Hence, lonely galaxies could constitute privileged witness of cold gas accretion effects till SKA comes to a reality. The same applies to the study of the compact dwarf star-forming galaxies in the Local Universe [125], as these objects have probably the same nature as the building blocks from which larger galaxies formed in a more primitive Universe. Understanding the role of the inflow of pristine neutral gas from the cosmic web to trigger the intense bursts of SF in these systems is mandatory to explain the rhythm in the formation and evolution of these objects. SKA will provide HI maps in these dwarf star factories with an unprecedent spatial scale. The full SKA is expected to reach N(HI) = $2.8{\times}10^{17}$ at/cm$^{-2}$ in 10h with spatial and spectral resolutions of respectively $30''$ and 4 km/s.

## 5 Synergies with theoretical models

Numerical simulations are key in order to define in detail many of the here summarized programs, as they can play a double role either motivating new observations or trying to look for an explanation for the new results.

From the numerical point of view, two complementary approaches are available. The first one are the so called semi-analytical models of galaxy formation, a very successful technique to study galaxy formation in cosmological context [14] although limited by the over-simplified description of the physical phenomena involved. One of the most widely used is GALFORM, where the authors [59, 60] have implemented a more general treatment of SF, which is able to track the HI and $H_2$ content of galaxies individually. The second approach is based on hydrodynamical simulations in a cosmological context, (HSs hereafter) where not only DM but also gas processes are followed in detail through basic, well known physical laws. In implementing physical processes, different approaches are possible depending on how large the box size is and how the initial conditions are implemented. Fully cosmological hydrodynamical simulations, with high resolution regions in large periodic boxes, can self-consistently include a detailed modellization of the physical ingredients involved in the growing of cosmic structures (see for instance [118]). Both, semi-analytical models and these fully cosmological simulations, will make possible a statistical assessment of the relation between HI and molecular gas, SF, stellar masses and luminosities, providing a physical explanation of observed local and high-redshift relations (e.g. [60, 34, 61, 62]). This substantial progress still requires significant improvement in the gas accretion models inside simulations of galaxy formation. Besides, detailed simulations of the formation and growth of galaxy discs indicate that changes in the gas volume density of the outer discs seem to be the cause of the evolution in the position of SF threshold radius [94].

On more general grounds, the outcome of simulations will allow a direct comparison of the virtual simulated Universe with the new data produces by SKA, leading to detection of cosmic shock waves associated with galaxy formation [81], the understanding of the nature and nurture of the galactic halos [87], and the description of the SF processes in galaxies [58].

Hydrodynamical simulations can be based as well on the so-called constrained initial conditions, in particular for the local Universe (CLUES project, http://www.clues-project.org/, [38]). In this case, the local structures at scales larger than $\sim$ 5 Mpc have been fixed according to observational data on the density field and velocity flows, with random fluctuations seeding smaller structures. This enables us to analyze in detail our Local Universe formation, including the Local Group, its intra-group gas, its constituent galaxies of different masses, with an even increasing resolution (some 100 pcs to date). Another widely used approach are the zoomed-in simulation technique, where a subox is re-simulated with very high resolution, allowing us to obtain galaxies (in general, disk ones) and their (varying) environments at high resolutions and in deep detail.

These two last approaches show a particular synergy with expected high resolution HI SKA maps of galaxies (and environments). Some of their predictions crossed to SKA capabilities are: 1) The gas within the disk and at the outskirts of galaxies (both HI and $H_2$), its space and velocity structure with high spatial resolution, and correlations with SFR, stellar mass and element abundance [19, 75]. 2) Inflows and outflows within and around forming galaxies, as a function of environment and cosmic time. Cold, j-rich gas reservoirs are predicted by HSs around disk-forming galaxies ([11, 9, 17]) with messy kinematics. 3) Subhalo abundance (e.g. [124]), that may allow SKA discriminating between the CDM and WDM models. Also, HI detection in satellites, in particular those that are devoid of stars but may have HI. 4) Improved measurements of the distribution of dark matter profiles within galaxies, providing crucial constraints for dark matter detection experiments (e.g. [83]) .

A different approach is based on fully analytic chemical evolution models, useful to test (subgrid) modeling in simulations. They have recently highlighted that the observed relation between metallicity and stellar-to-gas fraction [126, 1] provides invaluable constraints on the amount of cold gas infall and galactic winds throughout the history of the galaxy.

Summing up, SKA maps of the radial distribution of neutral hydrogen, combined with the information provided by ALMA on the molecular gas, as well as present and forthcoming integral-field optical spectroscopic surveys (which measure the physical properties of the stellar and ionized components), will provide, for the first time, all the data that is required in order to break the degeneracies present in current theoretical scenarios, as well as those required to test the predictions of hydrodynamical simulations, thus leading to an unprecedented leap in our understanding of galaxy formation and evolution.

# 6 Conclusions

SKA will provide breakthrough observations, revolutionary in the combination of parameters sky coverage, spatial resolution, and sensitivity, which will allow us to tackle many important questions in the field of galaxy structure and evolution. We have highlighted here a few examples of specific topics where Spanish scientists are currently leading or are playing a major role, with the expectation that they can extend their lead if Spain has access to SKA. This will allow our community to build on and further enhance the strengths we have developed over the past decades.

## Acknowledgments

LVM, PRM and JB have been supported by Grant AYA2011-30491-C02-01, co-financed by MICINN and FEDER funds, the Junta de Andalucía (Spain) Grant TIC-114, and BIOSTIRLING4SKA project, funded under FP7 ENERGY.2012.2.5-1, project number 309028; YA by the Spanish Ramón y Cajal programme (RyC-2011-09461), grant AYA2013-47742-C4-3-P (MINECO), and the 'Study of Emission-Line Galaxies with Integral-Field Spectroscopy' (SELGIFS) exchange programme, funded by the EU through the IRSES scheme (FP7-PEOPLE-2013-IRSES-612701); BA by a postdoctoral fellowship from the Observatory of Paris CBB by the Spanish Ramón y Cajal programme; AO by Grants AYA2010-15169 and AYA2013-42227-P; RDT and CBB by AYA2012-31101 from the PNAyA; SGB by the Spanish Ministry of Economy and Competitiveness (MINECO) through grants AYA2010-15169, AYA2012-32295, and program CONSOLIDER INGENIO 2010, under grant Molecular Astrophysics: The Herschel and ALMA Era-ASTROMOL (ref CSD2009-00038). SGB acknowledges support from the Junta de Andalucía through TIC-114 and the Excellence Project P08-TIC-03531; BGL by the Spanish MINECO grants AYA2012-39408-C0-02 and AYA2013-41656-P; AGP by AYA2012-30717 and AYA2013-46724-P; JHK by the DAGAL network from the People Programme (Marie Curie Actions) of the European Union's Seventh Framework Programme FP7/2007-2013/ under REA grant agreement number PITN-GA-2011-289313, and by the Spanish MINECO under grant number AYA2013-41243-P; UL by AYA2011-24728 from the Spanish Ministerio de Ciencia y Educación and the Junta de Andalucía (Spain) grants FQM108; IM by Grant AYA2013-42227-P; JP and JMS by AYA2013-40609-P; JIP, JVM and EPM by the Spanish MINECO under grant AYA2010-21887-C04-01 and from Junta de Andalucía Excellence Project PEX2011-FQM7058. MAPT by the Spanish Ministerio de Economía y Competitividad (MINECO) through grant AYA2012-38491-C02-02; VQ by MINECO, AYA2013-48226-C3-2-P; SFS by the CONACYT-125180 and DGAPA-IA100815 projects.

# Galaxy evolution in clusters with the Square Kilometer Array

**B. Ascaso[1], M.C. Toribio[2], M. González-García[3], M. Huertas-Company[1], J. Iglesias-Páramo[4], J. Sabater[5], L. Verdes-Montenegro[4], A. Vidal-García[6], and J. Vilchez[4]**

[1] GEPI, Observatoire de Paris, CNRS, Université Paris Diderot, 61, Avenue de l'Observatoire 75014, Paris France
[2] Netherlands Institute for Radio Astronomy (ASTRON), Postbus 2, 7990 AA, Dwingeloo, The Netherlands
[3] Group of Molecular Astrophysics, ICMM, CSIC, C/Sor Juana Inés de La Cruz N3, E-28049 Madrid, Spain
[4] Instituto de Astrofísica de Andalucía (CSIC), Glorieta de la Astronomía s/n, E-18008 Granada, Spain
[5] University of Edinburgh, Royal Observatory, Blackford Hill, EH9 3HJ Edinburgh, UK
[6] UPMC-CNRS, UMR7095, Institut dAstrophysique de Paris, F-75014 Paris, France

## Abstract

Galaxy clusters, the largest systems in the universe, result excellent laboratories for studying the properties of galaxies under similar environmental conditions. The general properties of the galactic population in clusters remarkably differ to their field counterparts. Consequently, the mechanisms responsible for the formation and evolution of the galaxies are believed to be closely tighten to the environmental conditions. In this chapter, we review the state-of-art of radio observations in galaxy clusters. We first mention the role that the neutral gas plays in tracing the properties of galaxies in high-density environments and highlight the impact that the SKA data will have in sampling deeper and wider samples. Secondly, we present our knowledge on the active galactic nuclei (AGN) mechanism in clusters and their impact on the brightest clusters galaxies (BCG). We then explore the power that the SKA data will have in sweeping a large range of redshift and host masses to study the relation of the AGN with the cluster host properties and to constrain the main evolutionary mechanism of BCGs. Finally, we investigate the presence of the diffuse emission in galaxy clusters and describe the impact on the characterization of their properties through a larger range of redshift than previous surveys that the SKA will achieve.

# 1 Introduction

Galaxy clusters are the largest structures gravitationally bound in the universe. Along with a large quantity of dark matter and very hot gas known as the intra-cluster medium (ICM), they contain several tens to thousands galaxies with tangential properties to the galaxies located in the field. Preliminary works (e.g. [51, 70, 31, 94]) discovered the so-called morphology-density relation, for which early-type galaxies are concentrated in the central part of the clusters and their density decreases with the cluster-centric distance. Reversely, the late-type galaxy fraction increases progressively as function of the distance to the center. Many works have later investigated the evolution of this relation with redshift and in different environments (e.g. [73, 87, 6]), a finding that holds at least up to redshift z$\sim$1 with the early-type galaxies becoming more spatially concentrated with redshift.

In addition to the different mix of morphological types, many differences have been detected in the properties of the spiral population itself: late-type galaxies inhabiting in clusters systematically differ from their field counterparts regarding their neutral gas content, star formation activity, molecular gas content, metallicity, cold dust content, kinematic perturbations, and radio continuum synchrotron emission (see [13] for a review on the effects of environment on late-type galaxies). The origin of these differences is still unknown and the debate about nature or nurture is still ongoing.

Using radio sources both associated to the ICM and to the Active Galactic Nuclei (AGN) [58] as cluster tracers, the SKA is expected to detect thousands of galaxy clusters up to redshift $\sim$0.6 and several hundreds at redshift z>1.5 (see Ascaso et al. in these proceedings). These cluster detections, in combination with other overlapping datasets such as X-ray, optical and infrared surveys and Cosmic Microwave Background (CMB) Planck maps, will provide a measurement of some of their overall properties such as masses, luminosities or temperatures. This cluster set will become several times superior to the previous radio surveys in terms of depth and redshift range. Therefore, it will constitute a unique sample to analyze the radio properties of the galaxies inhabiting there. Furthermore, it will play a crucial role in disentangling the main mechanisms involved in the formation and evolution of the Brightest Cluster Galaxies (BCGs), particularly at high redshift. The SKA then, will help unveiling unanswered questions and set constraints on the effect of the environment on the processes transforming galaxies.

The structure of the chapter is as follows. In §2, we focus on the detection of neutral hydrogen in cluster galaxies and highlights the advantages that the SKA will offer. §3 discusses the possibilities that the SKA will open for the study of AGN and the star formation of cluster galaxies. Finally, §4 reviews the diffuse emission in galaxy clusters and explores the improvement that will result in the understanding of the processes related with infra-red and radio emission thanks to the sensitivity and resolution of the SKA.

## 2 Neutral gas in galaxies in clusters with SKA

The content of neutral hydrogen, H I, in galaxies has been studied for a wide variety of galaxy samples. Here we focus on the studies regarding the galactic population in clusters, but a more general discussion on the topic can be found in Verdes-Montenegro et al. in these proceedings.

Since the pioneering observations of [30], the investigations on the H I content of cluster galaxies generally agree on the fact that on average, cluster spirals tend to have less neutral gas than their field counterparts. They also found evidence for correlations between the H I deficiency and the cluster-centric distance, with H I-poor disks typically situated close to the cluster cores and galaxies removed from those regions showing normal gas content (e.g., [43, 80]).

While the selection of reduced size samples has been reported to lead to biased results [1, 2, 72], plenty of these studies on galaxy clusters focused their attention on few nearby clusters, mainly Coma [15] and Virgo (e.g., [47, 19, 20, 81, 21, 86]), due to the current difficulty in obtaining deep H I observations. In particular, a very impressive example is the VIVA (VLA Imaging of Virgo in Atomic Gas [22]) survey, an imaging survey at $15''$ resolution of 53 late-type Virgo cluster galaxies located throughout the cluster, covering more than a factor of 20 in mass. Detailed imaging studies carried in all these nearby clusters show spirals with truncated gas disks, smaller than their undisturbed stellar disks, as well a trend of the extent of the H I disks with location in the cluster, being the most reduced those located near the core (e.g., [20, 2, 22]). Additionally, BUDHIES (Blind Ultra-Deep H I Environmental Survey of the Westerbork Synthesis Radio Telescope; [55, 56]) is a deep H I survey of galaxies in two clusters at z$\sim$0.19 and $\sim$0.21 and the large-scale structure around them. The survey aims at understanding where, how, and why star-forming spiral galaxies get transformed into passive early-type galaxies. The unique aspect of this study is that it has obtained measurements of the H I content of galaxies in different environments at intermediate redshift. This is the first time that optical properties and gas content are combined at a redshift where evolutionary effects begin to show, and in a volume large enough to sample all environments, ranging from voids to cluster cores. As part of this project, [54] found that the H I gas and the SF correlate with morphology and environment at z$\sim$0.2. Their results suggest that the H I gas gets removed and SF suppressed progressively, from the lowest mass galaxy groups to cluster-sized structures, as smaller structures get assembled into larger structures.

The measurements of H I content have been made extensive to other density regimes, showing that gas-depleted disks are not exclusive of cluster spirals: H I-deficiency has also been detected in Hickson Compact Groups (e.g. [90] and see Verdes-Montenegro in these proceedings), in X-ray bright groups (e.g., [78]), as well as in loose groups [77, 61]. In these less dense environments, in which the relative velocities of the galaxies are low, the tidal forces in close galaxy encounters can result in the removal of significant portion of a galaxy H I mass. Precisely, the efficiency of this mechanism in groups of galaxies has been used to favor the idea that galaxies suffer a pre-processing before entering the cluster environment (e.g., [37]). Several pieces of observations support this scenario, such as the compact group falling towards Abell 1367 [25], or of the Eridanus group [71].

All these observational results indicate that, at least in the densest regions of the cluster, gas-sweeping occurs. Since the 1970s, when the first models of tidal interactions between galaxies [85] and interactions of the inter-stellar medium (ISM) of galaxies with the dense ICM were proposed [45], theoretical models and numerical simulations have provided a wealthy variety of possible scenarios (see reviews by [46, 88, 13]). In general, the results of most of the aforementioned H I studies point towards ram pressure stripping (RPS) of the interstellar gas of galaxies by the hot ICM as the most plausible mechanism acting on the spiral galaxy population in today clusters. However, simple models of RPS are only partially consistent with galaxy data. For some galaxies, more than one mechanism appears to be necessary to explain the observations (e.g. [92]), whereas in other cases the results suggest that the effects of the ICM reach further than expected from simple ICM-ISM models (e.g. [28, 29, 65, 16]). Another mechanism that might affect cluster disks is galaxy harassment [68], proposed as a mechanisms to transform late-spirals into lenticulars (S0), and recently claimed to be responsible for the morphology of NGC 4254, as unveiled by the Arecibo Legacy Fast ALFA (ALFALFA, [44]) detection of this system [48].

We are currently limited when studying galaxies in clusters by two factors: they are scarcely detected in H I (high sensitivity and resolution is needed to detect cluster galaxies in H I, because they are gas-poorer than field galaxies) and the environment is not properly characterized in a quantitative manner. The SKA resolution and sensitivity will allow us to isolate the emission from galaxies in close proximity and therefore will help to mitigate confusion effects in dense environments over a large redshift range. This, in combination with optical redshift samples and observations of the ICM at other wavelengths (e.g., X-rays), will allow us to better quantify the effects of environment of galaxies.

The RPS has also been observed in other components of the interstellar medium such as the molecular gas. Indeed, several studies have reported no $H_2$ deficiency in galaxies associated to clusters, but a HI deficiency [59, 17, 12]. These results have been interpreted as a consequence of the larger concentration of the molecular gas towards the center of the galaxies and more recent works point towards the confirmation of molecular gas deficient galaxies in some nearby clusters [38, 39]. Additionally, it has been found that HI deficient Virgo galaxies show also truncated dust disks [26, 14], implying that RPS not only affects the neutral gas.

The HI deficiency is related to the star formation history (SFH) of galaxies in clusters, in the sense that HI deficient galaxies show truncated SFHs compared to non HI deficient galaxies [41]. This is explained by the fact that the star formation activity is regulated by the gaseous content of galaxies [42]. Supporting the link between the HI content of spiral galaxies and their star formation activity is the finding that many Virgo spirals show truncated H$\alpha$ profiles [63] indicating that RPS is at the basis of this effect.

Historically, the SFH has been represented in a simple way, the exponentially decreasing parametrization being the most common one. This parametrization is able to explain present-day galaxies of all morphologies [60]. Some other more difficult parametrizations have also been explored in order to find a typical SFH of higher redshift galaxies. Unfortunately, the search for a typical SFH at higher redshifts has been unsuccessful and the relation might not be as simple as the local-universe one. To constrain the SFH of the galaxies, the ongoing star-formation needs to be measured, together with the mass of the stellar populations that has already been formed. Radio wavelengths offer the possibility of observing obscured and non-obscured star formation at the same time.

The SKA will be able to provide constraints on the SFH since the reionization epoch through a tiered survey strategy whereby enough volume is sampled at each cosmic epoch: an ultra-deep tier, a deep one and a wide one. As shown by [23], the deep SKA1 survey will be capable to recover the ∼85% of the star-forming galaxies detected by Euclid [64] up to $z = 3$ with ∼2000 hours. Furthermore, the ultra-deep tier SKA1-MID will detect the faintest star-forming populations to the highest redshifts (SFR $\sim 10 M_{\odot} yr^{-1}$ out to z∼4, and SFR$\sim 100 M_{\odot} yr^{-1}$ for z∼8). Finally, the SKA Wide Tier (1000-5000 deg$^2$ survey with $0.5''$ resolution with a $5\sigma$ detection limit of 5 $\mu$Jy/beam) will relate the findings in the higher universe with those obtained locally. Since galaxies in clusters tend to populate the gas-poor regime, they are hardly detected in Hi. A particularly powerful technique to study their properties with SKA will be the stacking of the Hi spectra (e.g. [91, 32]). This technique will provide constraints thanks to the large statistics that will be available with SKA observations in combination with independent measurements of galaxy redshifts.

Likewise, the SKA will perform a breakthrough in the study of the relation between gas content and the star formation rate (SFR) of galaxies in clusters given its combined unprecedented sensibility and angular resolution. Thousands of galaxy clusters and groups will be mapped up to redshift $z = 1$ at least and even higher (see Ascaso et al. in these proceedings) with SKA2 facilities. Hence, a complete spatial coverage up to the infall regions ($r \approx 5R_{200}$) of the Hi component will be ready to be fully exploited for a sample of clusters spanning a wide range in mass, from loose groups to Coma-like clusters, and different evolutionary states according to their substructure. This will result in a complete view of the effect of the local and global density on the gaseous component of galaxies. The resulting dataset will contribute to shed light on the role of the pre-processing of galaxies in small groups before entering the cluster environment.

Moreover, this study will benefit of the multiwavelength data available for different cluster samples in order to address other related topics like the relations between star formation and metallicity and the gas content of galaxies (via UV/IR spatially resolved imaging and integral field spectroscopy) in different environments. Combining SKA 21-cm with measurements and data in other wavelengths will be key in this respect. Optical and NIR imaging (e.g, J-PAS [7], LSST [53], Euclid), millimeters (e.g., ALMA [95]), Far Infrared (e.g. SPICA [84]) and X-ray (e.g. e-ROSITA[67]) imaging and spatially resolved optical spectroscopy will be available at the time SKA operates. These datasets will provide excellent follow-up samples to fully resolve the chemistry and dynamics of cluster galaxies: star-formation, metallicity, stellar population ages, angular momenta and stellar dispersions.

## 3 AGN in clusters and BCGs with SKA

Faint radio sources in clusters are currently an unexplored population that can become crucial to set constraints on the main scenarios explaining the evolution of galaxies in clusters. In particular, AGNs seems to play a fundamental role on the control of star formation of galaxies [18, 33, 49]. AGNs can be divided in two types depending on their accretion mode: a) quasar-like AGN or radiative mode AGN with high accretion rates ($\geq 0.1$ per cent of the Eddington rate) and, b) radiatively inefficient or jet-mode AGN, with low accretion rates and radio jets carrying the bulk of their energetic output. The feedback of jet-mode AGN is suggested as the main mechanism that can maintain massive galaxies (mainly found in clusters [31], e.g. the BCGs, see below) "red and dead" (see [9] and Perucho et al. in these proceedings). This radio-mode feedback evolves in time increasing up to a redshift 0.5 and decreasing at higher redshifts [10]. Galaxies in clusters are less likely to harbor a radiative-mode AGN and more likely to harbor a jet-mode one [75]. The most probable explanation is the presence and physical properties of the supply of gas that feeds the activity [76]. In this way, AGNs can shape the properties of the gas of the cluster and vice versa.

By contrast, while the AGN is generally believed to suppress the star-formation activity in galaxies, several observational [11, 69, 62, 27, 52, 5, 82] studies claim that AGN might also induce star formation. This result has also been supported by theoretical [79] and simulation [40] works claiming that the energy released from the jets can indeed trigger star-formation by allowing the collapse of over-dense clouds. This might be a crucial issue for the early universe, where the density of the gas was much higher in galaxies and when AGN activity was much more prevalent.

The SKA will go deep enough (e.g. the SKA1 is expected to detect about 500 million radio sources at a limiting noise level of 2 $\mu Jy$ rms [74]) to distinguish between the different populations of AGN and star forming galaxies at high distances. This will allow to trace the evolution of the radio-mode feedback up to high redshifts, quantifying its role across the history of the universe. It will also show enough resolution ($\sim 0.2'' - \sim 1''$) to decompose local galaxies into its compact individual radio sources, and will help to clarify the feedback processes that can trigger or quench the star formation. Indeed, with only <0.5 arcsec angular resolution continuum imaging capability, the SKA will study the individual components within a distance of <100 Mpc in the local universe [57]. Besides, its wide coverage will provide the necessary reference frame to compare the physical processes that drive the evolution of galaxies in clusters with those found in other environments.

In addition to the AGNs, the BCGs, the most luminous galaxies of the cluster lying usually on the centre of the cluster potential well, are interesting objects that will enormously benefit from the SKA potential. The BCGs have been explored in large samples drawn from optical and X-ray cluster samples. Two main mechanisms have been proposed to explain their evolution since $z \sim 1$: major/minor mergers (e.g. [8, 93, 66, 4]) or adiabatic expansion through AGN/quasar feedback (e.g. [34, 24, 83, 3]). According to numerical simulations predicting the latter mechanism (e.g. [50]), a galaxy would lose a fraction of its central mass, therefore producing an increase of its size but keeping a similar surface brightness profile. The observational data available at present do not allow to rotundly give an answer of the preference of one scenario versus the other. However, this scenario will change with the arrival of SKA. Since SKA will be able to map very faint radio-galaxies up to z=1 at least, these radio sources will be easily correlated with the already-identified BCGs in next-generation optical and X-ray samples (e.g. Euclid, J-PAS, LSST, eROSITA) overlapping with the SKA area. The correlation of the optical/X-ray properties of the host cluster (e.g. the presence of cooling-flows) with the type of radio-source identified in the BCG within a wide range of redshift and halo masses will definitively provide a new window on the type of mechanism that has helped forming the BCGs, particularly at $z \sim 1$, range where striking differences with their local counterparts start being appreciated (e.g [4]).

# 4 Diffuse emission in galaxy clusters with SKA

Clusters mergers produce cluster-wide shock waves that can be detected as faint elongated structures at their borders [35, 89]. These structures are the signature of large scale magnetic fields interacting with the ICM. Several theories have been proposed to explain the origin of these magnetic fields (see Battaner et al. in these proceedings). There are also radio halos linked to the X-ray emission of the ICM which are usually found around radio-loud BCGs and are probably related to AGN feedback processes. We can also find radio relics associated to the old jet emission of radio- galaxies in the cluster. Their morphology and physical properties can trace the formation and evolution of the cluster. The morphology of these relics is shaped by the interaction with the ICM tracing their relative movement, on the other hand, the spectral index of the source is related to the age of the synchrotron emission of the jet. Finally, there are AGN radio relics, whose emission is boosted by the adiabatic compression of merger shock waves, known as radio phoenixes (see Ascaso et al. in these proceedings). Fig. 1 reproduces a simulation of a galaxy cluster at $z = 0.5$ (see [36] for a wider explanation) where a diffuse radio halo can be seen together with several tailed radio galaxies and point sources. All these faint radio sources can be used to trace the formation and evolution of clusters.

The wide spectral coverage of SKA, from 50 MHz to 13.8 GHz, will allow us to determine the age of the synchrotron emission of the jets that can be used to constrain the AGN duty cycle. The sensitivity of SKA-LOW to low surface brightness radio emission will allow the detection of faint extended radio sources. It is expected to detect ~2600 radio haloes up to z ~0.6, almost an order of magnitude more than with current state-of-the-art instruments [74]. The SKA all-sky survey at band 2 with a noise level of $2\mu Jy$ and a resolution of 2 arcsecs will provide more than two orders of magnitude increase in the number of radio sources currently known, including radio relics in clusters. The sensitivity of SKA to low surface brightness radio emission combined with its resolution make it an ideal instrument to study these faint radio sources, which trace the evolution of the cluster, at high distances.

Radio continuum emission is widely used as a dust-unbiased tracer for measuring the recent massive star formation activity in galaxies. The observed tight relation between infra-red and radio emission is used to convert the radio flux density into a star formation rate measurement. However, as discussed in [13], such an indirect tracer should be used with caution for cluster galaxies because the involved quantities might not be reliable indicators of the star formation activity due to dust removal or magnetic field compression. In that respect, the sensitivity and resolution of future SKA observations will allow us to gain a better understanding on the underlying processes that give rise to the infra-red and radio emission relation itself and the origin of the differences for galaxies inhabiting clusters.

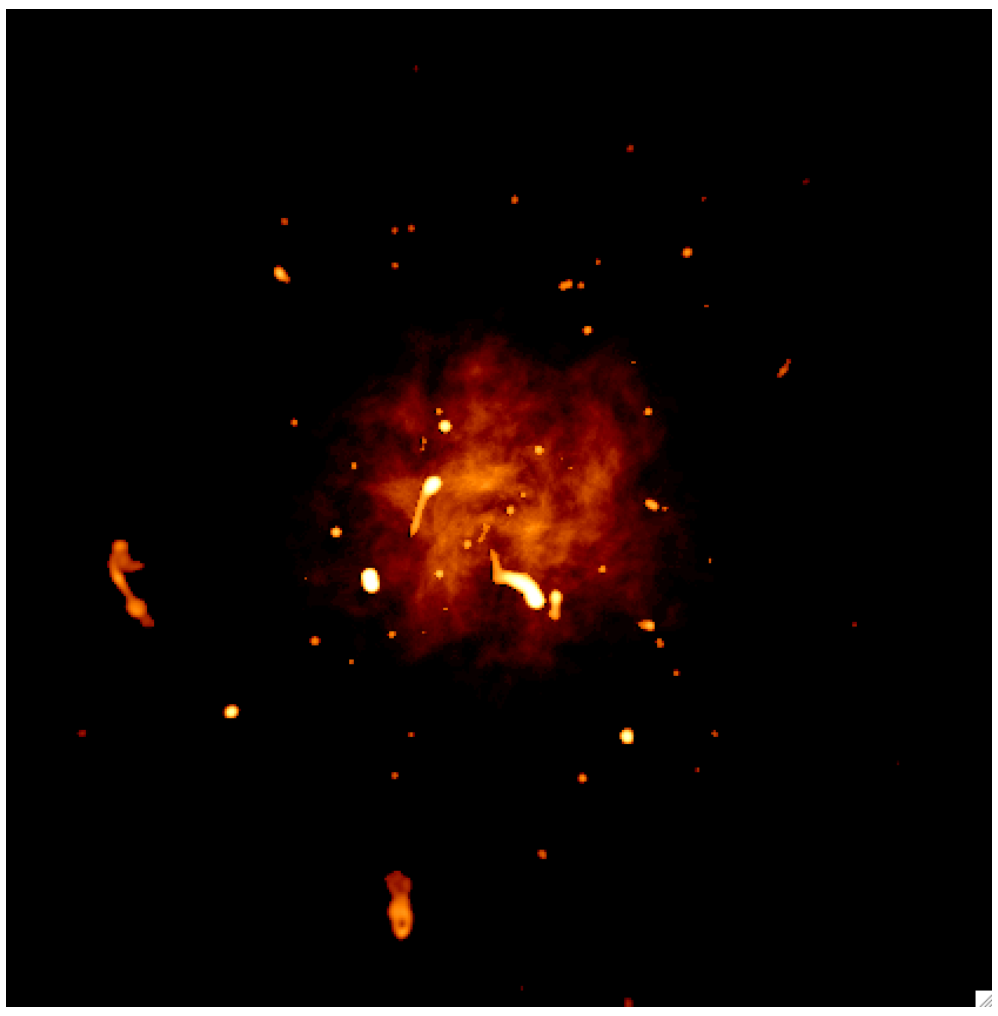

Figure 1: The population of radio sources of a local cluster (A2255) is simulated and then redshifted. In addition to the central halo, the disruption of the shape of extended radio jets due to the interaction with the ICM is clearly visible. SKA1 will observe at relatively high resolution the diffuse emission of clusters like this at high redshifts. Source: [36]

.

In summary, the future SKA facilities will provide a great advantage in a variety of topics related with the main formation and evolutionary mechanisms happening in galaxy clusters. These data, with unprecedented quality in radio, will explore an uncharted range of redshift and clusters masses which, will immediately shape the evolution of the radio properties of galaxies in clusters and their relations with the environment.

## Acknowledgments

LVM has been supported by Grant AYA2011-30491-C02-01, co-financed by MICINN and FEDER funds, the Junta de Andalucía (Spain) Grant TIC-114, and BIOSTIRLING4SKA project, funded under FP7 ENERGY.2012.2.5-1, project number 309028. AVG acknowledges support from the ERC via an Advanced Grant under grant agreement no. 321323-NEOGAL. JIP and JMV acknowledge financial support from the Spanish MINECO under grant AYA2010-21887-C04-01 and from Junta de Andaluca Excellence Project PEX2011-FQM7058. BA acknowledges financial support for a postdoctoral fellowship from the Observatory of Paris.

# Active Galactic Nuclei at kiloparsec scales and their cosmological evolution

**M. Perucho[1,2], I. Agudo[3], A. Alberdi[3], S. García-Burillo[4], J.I. González-Serrano[5], I. Márquez[3], J.M. Martí[1,2], I. Martí-Vidal[6], M. Mezcua[7], P. Mimica[1], M.A. Pérez-Torres[3,8], A. Prieto[9], V. Quilis[1,2], C. Ramos Almeida[9,10], E. Ros[11,1,2], J. Sabater[12], A. Usero[4]**

[1] Dept. d'Astronomia i Astrofísica. Universitat de València, Burjassot, València, Spain
[2] Obs. Astronòmic. Universitat de València, Paterna, València, Spain
[3] Instituto de Astrofísica de Andalucía (CSIC), Granada, Spain
[4] Obs. Astronómico Nacional (OAN)-Obs. de Madrid, Madrid, Spain
[5] Instituto de Física de Cantabria, Universidad de Cantabria, Santander, Spain
[6] Onsala Space Obs., Chalmers University of Technology, Onsala, Sweden
[7] Harvard-Smithsonian Center for Astrophysics (CfA), Cambridge, MA, USA
[8] Centro de Estudios de la Física del Cosmos de Aragón, Teruel, Spain
[9] Instituto de Astrofísica de Canarias, La Laguna, Tenerife, Spain
[10] Departamento de Astrofísica, Universidad de La Laguna, La Laguna, Tenerife, Spain
[11] Max-Planck-Institut für Radioastronomie, Bonn, Germany
[12] Institute for Astronomy (IfA), University of Edinburgh, Royal Obs., Edinburgh, UK

## Abstract

Active galactic nuclei (AGN) play a fundamental role in the evolution of the host galaxies and the clusters in which they are embedded. The interplay between the most compact regions in these galaxies and their environments occurs via outflows and can be studied at kiloparsec scales, where those outflows generate shocks and get mixed with the interstellar and intergalactic media. The SKA represents a unique opportunity to study this process in detail, not only in the Local Universe, but also at high redshift, allowing us to study how the AGN properties vary with time and their role in galaxy evolution. In this chapter we collect a number of AGN processes taking place on kiloparsec scales in which the SKA will certainly provide revolutionary advances, namely, those related with gas distribution around galaxies that host active nuclei, molecular and relativistic outflows and their impact on the host galaxy and intracluster medium, and the cosmological evolution of AGN populations.

# 1 Introduction

Besides being among the brightest objects in the universe, AGN are further characterised by being strong emitters at any range of the electromagnetic spectrum. The power of these sources is based on the availability of material in their immediate surroundings, host galaxy, and/or circumgalactic medium. Different processes cause the fall of matter onto the center (see Sect. 2), eventually fueling the black hole through an accretion disk (e.g., [10]). Yet, the source of this material, its mere nature and origin remain big questions that make the unification of the various AGN classes more difficult, in particular of the two more extreme groups: the radio loud, low luminosity AGN - kinetically very efficient but poor radiators - and the radio quiet AGN - extremely efficient radiators [57]. Neutral HI gas is the most abundant element and thus the most important matter reservoir to form molecular complexes, stars and to fuel supermassive black holes (SMBHs). Most of what we know from HI comes from single dish aperture with limited angular resolution. Large HI surveys, e.g. ALFALFA, done with the 300 m Arecibo radio telescope [26], show that HI is ubiquitous in galaxies (e.g., [24]). Yet, these surveys provide an integrated emission for most cases. Our knowledge about the precise spatial location, morphology and kinematics of this pervading gas within a galaxy and its circungalactic medium is limited to a few cases. One of the reasons is the relatively low column density of this gas, which makes HI mappings with suitable angular resolution difficult and expensive. Combining its high angular resolution and sensitivity, the SKA shall produce unprecedented HI spectral 3-D distributions of the AGN circungalatic medium where massive HI disks, presumably captured from a merger event, seem to orbit around the host of radio quiet and radio loud AGN (e.g.,[68, 75], see Fig. 1).

In the cm domain, AGN present the most spectacular morphologies, associated with synchrotron (non-thermal) or bremsstrahlung (thermal) plasma emission and extending from tens to hundreds of kiloparsecs away from the optical galaxy. These structures, which carry a non-negligible fraction of the AGN power to be deposited in the galactic and intergalactic medium (IGM), are of paramount importance because of the role they may play in galaxy evolution [52]. They will greatly benefit from the continuous frequency coverage from 350 MHz to 20 GHz offered by the SKA. This, complemented with superior angular resolution, will allow us to get extremely well sampled spectral energy distributions of the various components, core, jet-knots, hot spot, lobes, without inter-contamination and high sensitivity. It will open the opportunity to test more physically-motivated models of jet production, propagation, their interaction with the interstellar medium (ISM) and IGM, and in turn, a sound determination of the total power injected by the jet in the medium.

The studies of these regions in the low frequency range, which the SKA offers with much higher sensitivity than current facilities, will open a new window for discovering previous periods of AGN activity and cycles. This is possible because the low energy particles suffer much smaller synchrotron losses. This in turn makes it possible to observe them at low frequencies, millions of years after their acceleration (which presumably happened in the jet regions that underwent a surge of activity). The sensitivity of the SKA will allow us to catch them at their final stage.

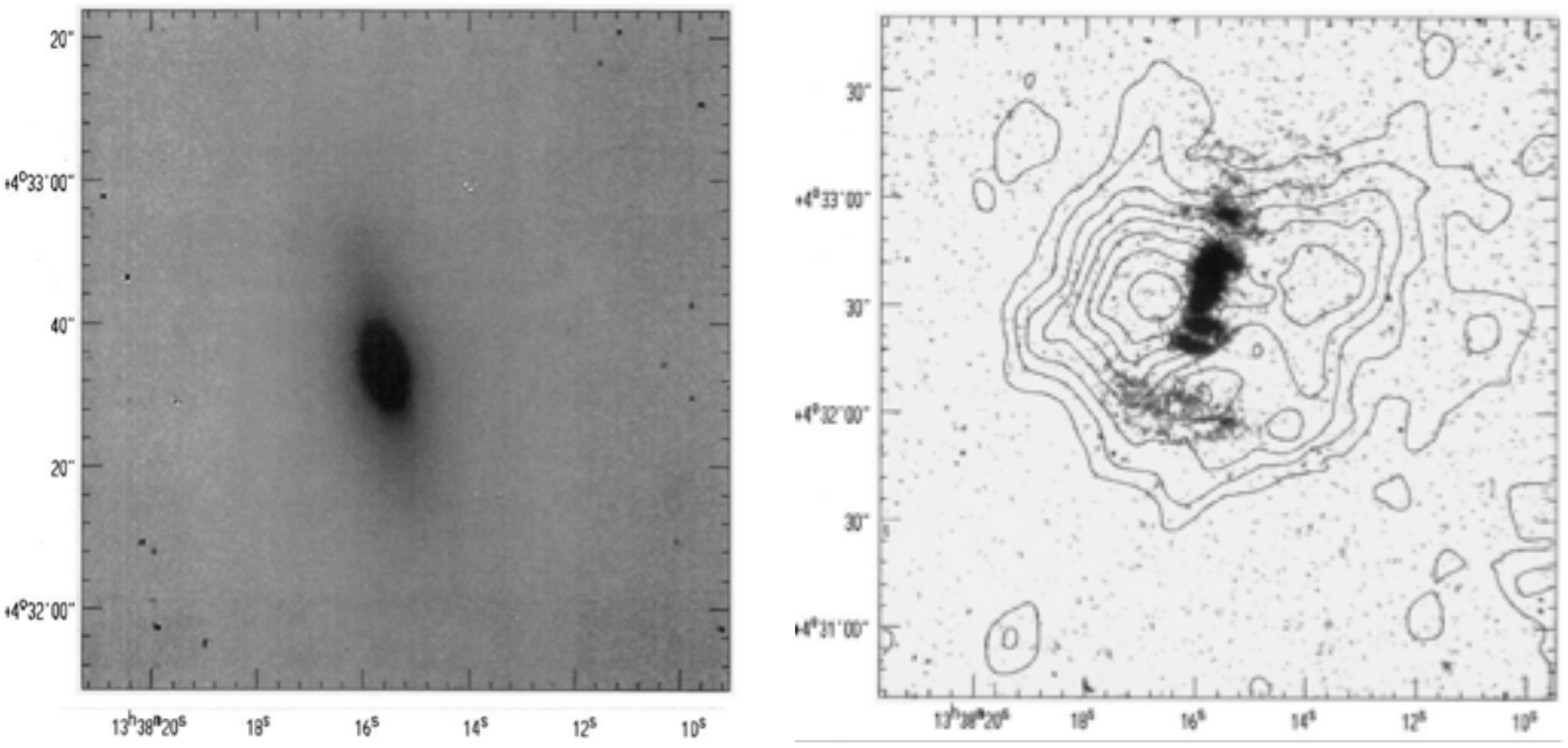

Figure 1: The HI circungalactic medium of a radio quiet AGN, the Seyfert galaxy NGC 5252. The optical image is shown above; in the right panel, the same field traces the ionised gas (H$\alpha$ in dark), and the HI in contours. Both gas phases constitute a single rotating disk with equal column desnity and velocity. The HI mass disk is $10^8$ $M_\odot$. The sensitivity and angular resolution of the SKA will permit us to investigate the frequency and origin of these massive HI disks. From [68].

## 2 Gas distribution around galaxies: AGN triggering

The highly luminous optical, and sometimes also radio, emission of quasars first led to their use as probes of the high-redshift Universe. It is now widely accepted that SMBHs power AGN activity and that feedback during the AGN phase may play an important role in regulating galaxy evolution. However, we still know very little about how and when quasars are triggered as part of the hierarchical growth of galaxies (see [2] for a review).

The tidal torques associated with galaxy bars, disk instabilities and galaxy interactions between galaxies are efficient mechanisms to transport the cold gas required to trigger and feed AGN and star formation to the centres of galaxies. This gas has to lose ~99.9% of its angular momentum to travel from the kpc-scale host galaxy down to ~10 pc radius (see e.g. [39]). Interferometric CO maps of nearby AGN have been used to quantify this angular momentum transfer budget. For example, the NUclei of GAlaxies (NUGA) project observed a sample of 25 nearby low-luminosity AGN in CO with a maximum resolution of 0.5" (~50–100 pc) and found that only ~1/3 of the sample showed evidence for fuelling [30]. However, there has to be gas fuelling into the nucleus to trigger and maintain nuclear activity and thus, it becomes necessary to look at smaller scales (<50 pc) to study the gas kinematics and quantify the gas content of the inflow [47, 50, 29].

Recent studies of samples of luminous radio-quiet and radio-loud quasars ($L_{bol} > 10^{45}$ erg $s^{-1}$) revealed a high incidence of tidal features in their host galaxies (see Fig. 3) compared to control samples of non-active galaxies [69, 6, 87]. This indicates that galaxy mergers/interactions likely play a role in the triggering of luminous AGN. On the other hand, for less luminous AGN ($L_{bol} \sim 10^{42-45}$ erg s$^{-1}$, including Seyferts and LINERS), it has been found that the incidence of disturbed morphologies is not significantly enhanced over the general population of galaxies (e.g., [12]), leaving secular processes as the most likely triggering mechanism at such luminosities [48]. Thus, it seems that AGN of different luminosities are triggered in different manners [86], although other factors such as their environment and/or radio power can also play a key role [70, 76].

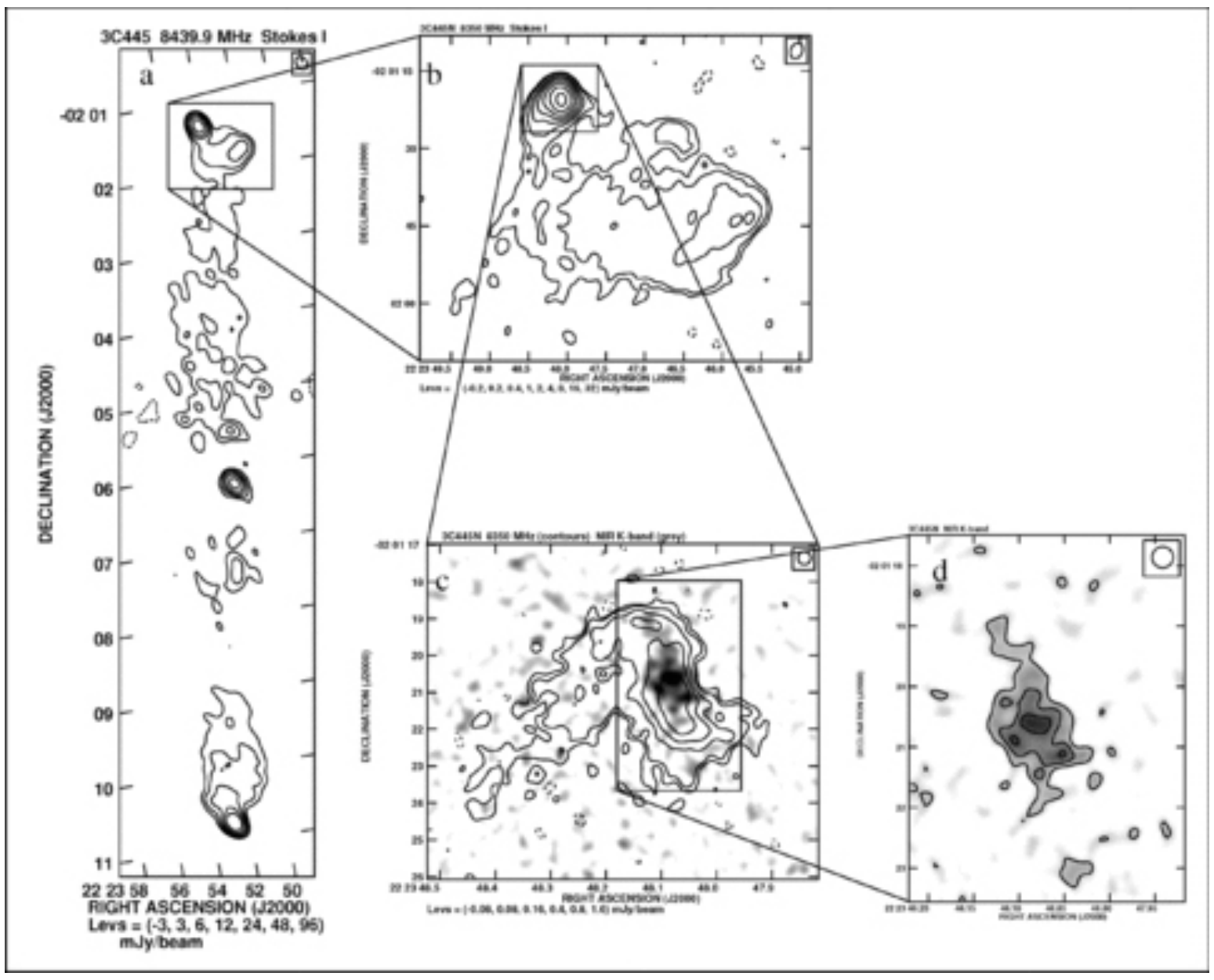

Figure 2: (a) 3C 445: entire source, 8.4 GHz, VLA C-array; (b) 3C 445 North hot-spot: 8.4 GHz, VLA B-array; (c) superposition of VLA A-array data (contours) and VLT K-band image (grey), both smoothed to a common resolution of $0.35 \times 0.35$ arcsec$^2$; (d) 3C 445 North hot-spot: VLT near-infrared emission in original resolution (0.31 arcsec). From [46].

Detecting the sometimes faint morphological signatures resulting from galaxy interactions is challenging, and it requires deep and high spatial resolution optical observations with 8-10 m telescopes. These studies are therefore restricted to intermediate-redshift samples (up to $z$ ∼0.5–0.8), for which cosmological dimming and spatial resolution are not critical. Alternatively, these tidal features can be easily detected in HI observations [77]. Performing such studies in HI will bring clear advantages with respect to optical surveys. For example, they will provide kinematic information that can be used to reconstruct the dynamics of the interaction. Besides, the tidal features resulting from the interaction between galaxies, which are sometimes difficult to detect in the optical, will be clearly visible in HI observations.

The good spatial resolution and excellent sensitivity of the SKA will provide key information about the distribution of HI in AGN samples at high and intermediate redshift. Thanks to the SKA, we will not only be able to understand how the cold gas is accreted in active and non-active galaxies and how important mergers are in galaxy evolution, but also to further investigate the triggering of star formation in these merging systems. The SKA will allow us to trace the amount of cold gas present in those systems and therefore to clarify whether gas-rich mergers are more prevalent than dry-mergers in the case of luminous AGN. Using far-infrared Herschel observations of radio galaxy samples at intermediate redshift, it has been hinted that minor mergers are capable of triggering luminous quasars [85]. The SKA will definitely confirm if minor mergers provide sufficient fuel for such powerful AGN, and if so, whether this is also the case at high redshift.

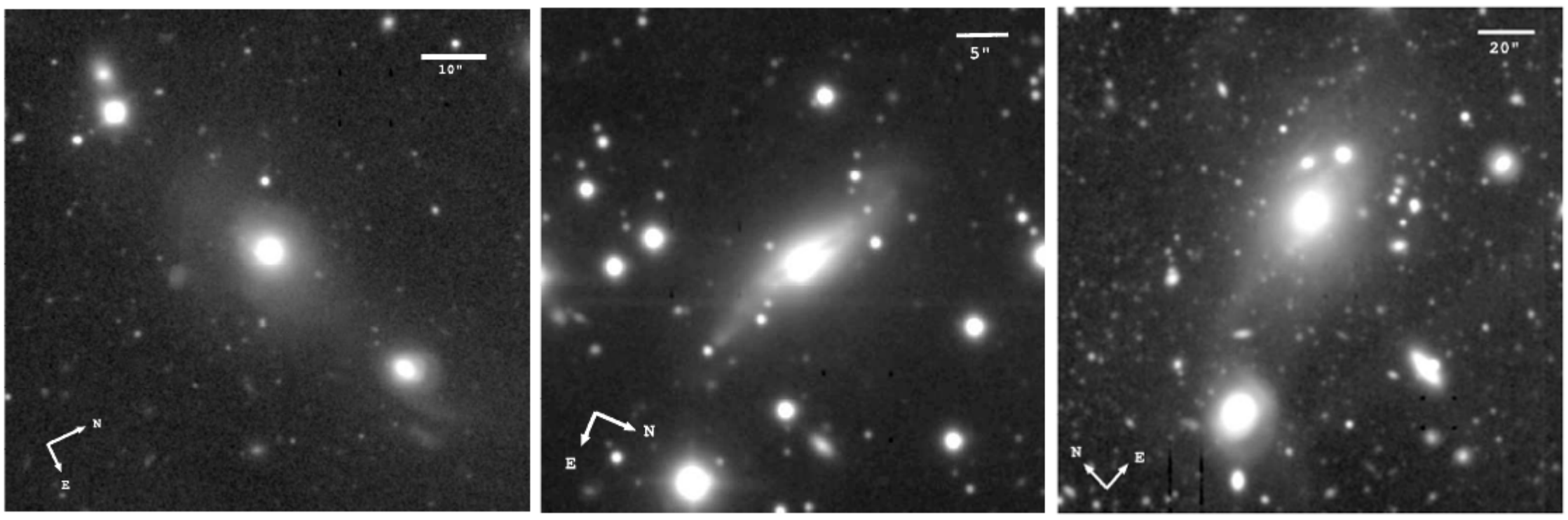

Figure 3: Optical morphologies of powerful radio galaxies in the 2Jy sample [69]. The morphological features detected include bridges, tidal tails, shells, etc.

# 3 Neutral gas outflows

Radiative and mechanical feedback is invoked as a mechanism of self-regulation in galaxy evolution. Outflows could prevent galaxies from becoming overly massive and help regulate the fueling of both the star formation and the nuclear activity (e.g., [40, 16, 81, 21, 38, 89]). Observations of AGN where the distribution and kinematics of atomic and molecular gas can be spatially resolved are a key to our understanding of how gas accretion can self-regulate in galaxies. There is growing observational evidence of the presence of gas outflows in a wide variety of starbursts and active galaxies. Outflows are found in ultra luminous infrared galaxies (ULIRGs), radio galaxies, quasars and Seyferts, and they are detected in nearly all the phases of the interstellar medium (ISM), including the ionized gas (e.g., [36, 13, 49, 3]), the neutral atomic medium (e.g., [58, 74]), and the molecular gas phase (e.g., [23, 59, 14, 11, 31]). This multi-wavelength based evidence suggests that the global impact of outflows can only be evaluated if all of their ISM phases are scrutinized. The quantitative balance between the different ISM phases of outflows is seen to change dramatically depending on galaxy type for reasons that remain to be fully understood. The dominant phase of the ISM in the central kiloparsec regions of active galaxies resides in the neutral gas. As argued below, the SKA will be able to make a significant progress in the study of galaxy outflows by observing the distribution and kinematics of atomic and molecular gas in different populations of active galaxies.

The SKA will permit studying the neutral hydrogen content of galaxies to cosmological significant distances ($z \sim$2). A key tool will be the observation of this line in absorption in front of radio continuum sources. The chances of detecting the HI line at 21 cm in absorption is virtually redshift-independent, since it is mainly a function of the strength of the background continuum source. This makes that HI absorption studies become an excellent tool to probe the neutral hydrogen reservoirs in distant galaxies where the HI emission line would be way too weak to be detected. The atomic gas producing the absorption can be used to probe the distribution and the kinematics of gas in circumnuclear disks (CNDs) of radio-loud AGN on very small scales. In particular, HI profiles can unveil the presence of fast ($> 1000$ km s$^{-1}$) atomic outflows, a signature of ongoing mechanical feedback from the AGN (e.g., [58]). The work presented in [32] and [33] has been instrumental in defining the expected detection rate of HI absorption associated with radio-loud AGN, which amounts to $\sim 30\%$ for radio-continuum flux sources of $\geq 50$ mJy. As extensively discussed in [60], about 10% of the HI 21cm absorption detections would show high-velocity components corresponding to atomic outflows. Furthermore, to trace around $\sim 600$ atomic outflows in a survey of radio continuum sources of $\geq 100$ mJy, covering $\sim 10,000$ deg$^2$ of sky, the rms goal required would approach the 30–150$\mu$Jy level in channels of 30 km s$^{-1}$, taking into account that HI outflows are characterised by line opacities that typically range from 0.001 to 0.005.

The SKA will also probe the emission of extragalactic mega-masers from a number of molecular species, which include OH and $H_2O$ lines (see also the Chapter by Pérez-Torres, García-Burillo et al., this book). These lines, seen in emission, are able to trace compact structures in the CNDs of active galaxies, where they can be used as the best tools to probe the 2D kinematics of molecular gas down to spatial scales of microarcseconds. Therefore, these SKA observations represent an ideal complement to interpret HI absorption data of radio-loud AGN, which are to some extent limited in their ability to give, mainly, 1D information along the line of sight in front of the radio continuum sources. It has been discussed if OH emission can trace the jet-ISM feedback action in the CNDs [28]. The dynamical status of the CNDs could be heavily influenced by a significant AGN feedback action. AGN feedback can explain the onset of nuclear warp instabilities or the existence of massive outflows recently uncovered by the Atacama Large Millimeter Array (ALMA) on similar spatial scales [14], [31]. As an indication of feasibility, current estimates suggest that a detection experiment for OH would need a few–to–20 hours per source for modest maser luminosities up to $z \sim 2$ [8].

# 4 Relativistic outflows at kiloparsec scales

Relativistic outflows from radio-loud AGN propagate from the nuclear regions of the galaxy to hundreds of kiloparsecs away. At the largest scales, radio jets are divided into FRI and FRII populations [20], depending on their morphology: The former have a lower radio brightness and an irregular, decollimated morphology, whereas the latter show collimated jets and bright hot-spots at the terminal shock of the jet flow, where the interaction with the ambient takes place. In principle, this dichotomy is related to jet power (see, e.g., [71]), the most powerful showing typically an FRII morphology, but there is a number of open issues related to this dichotomy. Whereas FRII jets are thought to have high internal mach numbers, FRI jets are decelerated along the first kiloparsecs of evolution. There are two main proposed decelerating mechanisms: entrainment of denser, slower ambient gas (e.g., [42, 64] and references therein) or entrainment of stellar wind gas from stars embedded within the jet flow (e.g., [66] and references therein). Another ingredient that has been proposed as causing the dichotomy is the magnetic field structure and intensity (e.g., [56] and references therein).

The velocity profile of the jets -along the transversal direction to the jet axis- at these scales can give clues about which of the processes proposed to explain FRI jet deceleration is more relevant (see, e.g., [43]): If the deceleration is caused by internal entrainment, no velocity profile across the jet is expected, whereas if the mass-load of the jet is external, there would be a velocity profile, with slower layers at the jet boundaries. Determining the magnetic field structure will also be relevant to understand its role on jet dynamics and stability. These studies require a resolution of the order of 0.1 arcsec at GHz frequencies, which will be provided by the SKA1-MID, Bands 3-5 (see [44, 1] and the Chapter by Agudo, et al., this book). The characterisation of the velocity profiles in narrower jets corresponding to FRII morphologies will require the SKA2 capacities (resolution of the order of 10 mas, see [44]). In the case of wide FRI jets, the SKA will allow us to obtain Faraday rotation measures, which can unveil the possible intrinsic helical topology of the magnetic field in these jets at kiloparsec scales, using the SKA1-VLBI, Band 5.

A significant fraction of FRII galaxies present two pairs of misaligned radio lobes, the high-brightness and the loss-brightness luminosity ones, forming an X shape [45]. The peculiar radio morphology of these X-shaped radio galaxies was suggested to be caused by a recent merger of two SMBHs (e.g., [54]), where the low-brightness lobes constitute the relic emission of the past radio jets. This scenario was supported by the finding that X-shaped radio galaxies are hosted by early-type galaxies and present statistically higher black hole masses and an enhanced star formation and nuclear activity compared to a sample of normal double-lobed galaxies [55, 56]. The high sensitivity of the SKA will permit us to detect these sources at higher redshifts ($z$ >0.3), to perform polarimetric and spectral aging studies of their radio lobes, and thus to study these peculiar sources in great detail.

Finally, the sensitivity of the SKA will allow us to address another of the fundamental questions in extragalactic jet astrophysics, namely, the existence of a cosmological evolution of FRII-FRI populations, as it has been suggested by different authors (e.g., [9, 88, 4, 5]). The detection of high-redshift sources and the possibility to resolve their large-scale structure –for instance, the presence or absence of hot-spots– will be crucial to develop statistical studies of these populations depending on redshift. In this same line of research, the SKA will also help to quantify the intermittency of jet injection along the evolution of an AGN [80, 17], by making possible to observe radio relics from ancient outflows in currently active radiogalaxies or even the relics of present radio-quiet active galaxies (see Sect. 5), at the 100-600 MHz bands.

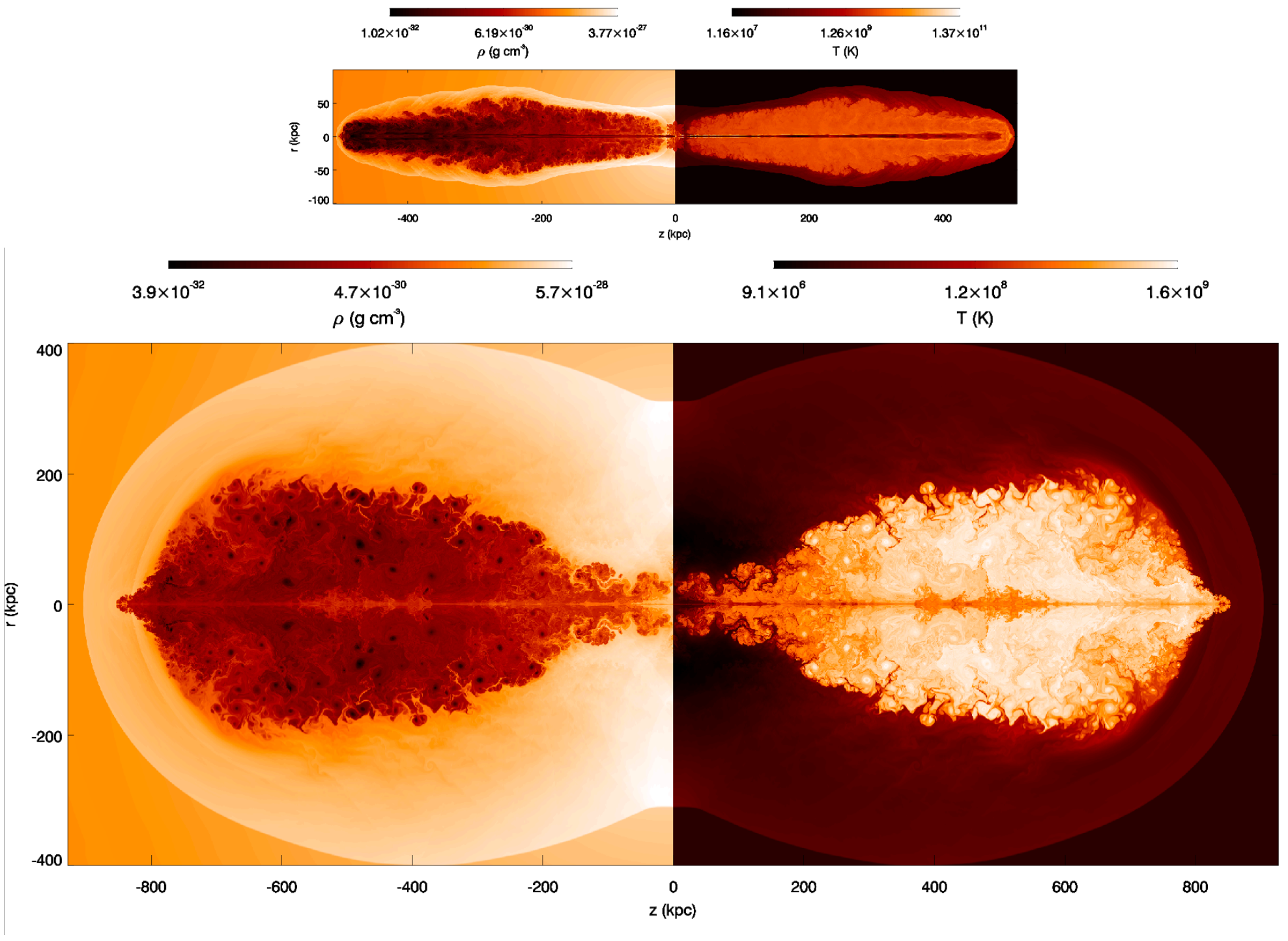

Figure 4: The impact of relativistic outflows from AGN on their environment up to hundreds of kiloparsecs is shown by long-term numerical simulations. The image shows two snapshots of the rest-mass density and temperature distributions of an evolved two-dimensional axisymmetric jet. The upper panel shows the jet-cocoon-shell structure after $\simeq 2 \times 10^7$ yrs, whereas the bottom panel shows the relic cocoon-shell structure after $\simeq 2 \times 10^8$ yrs. The jet injection was stopped after the time shown at the top panel, thus the bottom panel represents the long-term evolution after the jet active phase. The jet is surrounded by a region (cocoon) of low-density and high-temperature shocked jet gas mixed with shocked ambient. The *cocoon* is the region that corresponds to X-ray cavities in typical *Chandra* observations of clusters. The dense shell of shocked ambient density surrounds the jet and the cocoon. This simulation was performed at Mare Nostrum (BSC) within the frame of the Spanish Supercomputing Network (RES) [67].

## 5 Feedback and impact on galaxy evolution

A fundamental problem in Astrophysics and Cosmology is the role that AGN can play on the evolution of the host active galaxy and its host cluster. One of the ways in which nuclear galactic activity affects its environment can be characterised by the so-called *cooling-problem*: the gas among galaxies in clusters is X-ray bright owing to thermal emission and cools down at a rate that implies relatively short cooling times with respect to the age of the Universe; therefore flows of cooled gas onto the galactic potential well should be expected ($10 - 100$ $M_\odot$/yr), as the gas loses energy. However, X-ray observations of clusters with *XMM-Newton* and *Chandra* revealed a lack of such flows in many cases (see e.g., [52, 18], and references therein). As a consequence, it was postulated that a heating mechanism was necessary. The main mechanisms that have been proposed are the so called *quasar mode* and the *radio mode*. The former, also known as radiative mode, consists of the cold gas being pushed away by winds triggered by the nuclear activity [18]. The latter is justified by the observed anti-correlation between the radio lobes formed by jets and the X-ray emission from the cluster gas surrounding active galaxies: the interaction between the jets and their environment could be responsible for the heating. The heating process has been typically associated to turbulent mixing caused by the gentle buoyancy of bubbles inflated by the jets. Nevertheless, evidence of weak shocks (Mach numbers $M \simeq 1 - 4$) surrounding the radio lobes in powerful sources has been collected (see, e.g., [61, 83, 34, 15, 84]). The relevant role of these shocks in the heating process has been confirmed by recent numerical simulations of long-term evolution of relativistic jets [65, 67]. In addition, those shocks can displace large amounts of gas from the host galaxy, possibly triggering star formation during the initial stages [27] and quenching it in the long term [65, 67], see Fig. 4. This whole process has important implications for the star formation rates in the galaxy and the growth of the central black hole (e.g., [10]), i.e., regarding the evolution of the host galaxy and its environment. Moreover, the recurrent activity of radio jets is considered to play a fundamental role in the lifecycle of the most massive galaxies maintaining them "red and dead" [7], thus providing a possible explanation for the shape of the luminosity function at the high mass end (e.g., [82]).

On the one hand, aged particles injected by jets in the radio lobes and relics (in the case of exhausted nuclei) and having emitted most of their gained energy at the jet terminal shock emit at low radio frequencies and can be very dim, depending on the source age and power. Thus, at the lowest frequencies, the radio data from the aged synchrotron emission will trace back the history of feedback for 100-300 Myrs. On the other hand, detecting their emission is necessary to understand the interaction between extragalactic jets and the intracluster media (see also [35]). The SKA offers the high-sensitivity and resolution at hundreds of MHz to unveil these regions and show the details of those interactions: SKA1-MID and SKA1-LOW represent a perfect observational tool for this aim: Angular resolutions on the order of the arcsecond ($\sim$ 10 arcseconds for the SKA1-LOW), and sensitivities of the order of the $\mu$Jy/beam in wide band continuum mode, in 30 min. integrations. Furthermore, low frequency radio interferometric observations have revealed the presence of extended structures in galaxy clusters, suggesting the presence of large scale shocks with enhanced magnetic fields that play a relevant role in the particle acceleration. The SKA, through the measurement of the Faraday Rotation in background and embedded radio sources, will permit to study the magnetic fields in the clusters of galaxies and their exact role in the feedback process.

# 6 Cosmological evolution of Active Galactic Nuclei

The superb sensitivity and survey speed of the SKA is also perfectly suited to produce an actual revolution on cosmological studies of large samples of radio loud AGN up to redshifts $z \sim 10$. At such high redshifts, the Universe is still completely unknown, as well as the evolution of bright radio loud AGN from those early epochs to the Local Universe. There is therefore still an almost entire new field to explore regarding the cosmological evolution of AGN. This field needs to answer relevant questions such as: when were the first AGN formed? How were they formed? What were the properties of their environments and their SMBH and accretion systems? How did they evolve with time up to $z \sim 0$?

A relevant issue regarding the possibility to detect radio loud AGN at very high redshifts (beyond $z \sim 8$) was raised because of the strong influence of the dense Cosmic Microwave Background (CMB) at those redshifts [22, 72, 41], which is expected to produce a dramatic loss of relativistic electrons on the AGN jets via enhanced inverse Compton scattering, therefore decreasing the output synchrotron emission of the AGN/jet by several orders of magnitude. The huge cosmological distance of these early AGN sources makes the possibilities of detection very small. However, the group of G. Ghisellini has recently shown, through numerical simulations of jet emission, that the detection of powerful relativistically beamed radio loud AGN (i.e., blazars) is feasible for LOFAR up to redshifts between 3 and 7 [25], and of course for the SKA even for $z > 8$ at detection level $\sigma > 10$ [1]. For the case of the SKA, a single 30 min. observation suffices to detect such dimmed and far away radio sources. This will not only allow for unprecedented cosmological studies of radio loud AGN, but also for new estimates of the CMB itself, by modeling the impact of the CMB on the spectral energy distribution of blazars.

Concerning the detection and possible characterization of the first AGN formed in the Universe, and therefore the characterisation of the basic properties of their putative black holes, their appearance is highly speculative and there are lots of uncertainties regarding predictions so far [73, 37, 78]. However, numerical simulations predict that dense clouds are capable to form massive compact objects (of $\sim 10^5 M_\odot$) containing accretion disks [79]. These scenarios would naturally lead to the formation of relativistic jets, but the dense environments on the early-Universe galaxies would frustrate the development of large scale jets. Instead, small scale radio structures, similar to the GHz-Peaked-Spectrum (GPS, [62]) sources known in the Local Universe, would form. This is actually favourable in terms of the detection with the SKA. For small scale relativistic jets, the radio emission is not strongly affected by IC looses by the dense CMB as it happens for large scale radio lobes from AGN. Falcke et al. [19] claimed that, for central black hole masses of $\sim 10^7 M_\odot$, the compact radio structures corresponding to this early type of GPS should be visible even at $z > 10$. Deep searches on the 100-600 MHz bands of the SKA at arcsecond and sub-arcsecond resolution [19], combined with ultra-high sensitivity SKA-VLBI [63] (see also the Chapter by Ros, Alberdi, et al., this book), are ideal for the eventual detection and study of these sources. This would open a promising window for studies of the first radio loud AGN and SMBHs, and their evolution along cosmic time.

## Acknowledgments

MINECO grants AYA2013-40979-P (MP -member of the work team-, JMM, VQ and PM) and AYA2013-48226-C3-2-P (MP -member of the work team-, JMM and VQ), AYA2013-40825-P (IA), AYA2012-38491-C02-02 (AA and MAPT), AYA2010-15169 and AYA2012-32295 (SGB), and AYA2012-38491-C02-02 (ER) are acknowledged. IA research is supported by a Ramón y Cajal grant of the Spanish MINECO. SGB acknowledges support from program CONSOLIDER INGENIO 2010, under grant Molecular Astrophysics: The Herschel and ALMA EraASTROMOL (ref CSD2009-00038), and the Junta de Andaluc ía through TIC-114 and the Excellence Project P08-TIC-03531. PM acknowledges support from the European Research Council (grant CAMAP-259276), and GVPROMETEOII2014-069. CRA is supported by a Marie Curie Intra European Fellowship within the 7th European Community Framework Programme (PIEF-GA-2012-327934). ER acknowledges GVPROMETEOII/2014/057.

# Active galactic nuclei on parsec scales

**Iván Agudo[1], Manel Perucho[2,3], Antxon Alberdi[1], Almudena Alonso-Herrero[4], José L. Gómez[1], Isabel Márquez[1], José María Martí[2,3], Iván Martí-Vidal[5], Josefa Masegosa[1], Mar Mezcua[6], Petar Mimica[2], Almudena Prieto[7,8], Cristina Ramos Almeida[7,8] and Eduardo Ros[9,3,2]**

[1] Instituto de Astrofísica de Andalucía-CSIC, Granada, Spain
[2] Dept. d'Astronomia i Astrofísica, Universitat de València, Burjassot, València, Spain
[3] Observatori Astronòmic, Universitat de València, Paterna, València, Spain
[4] Instituto de Física de Cantabria, CSIC-UC, Santander, Spain
[5] Onsala Space Observatory, Chalmers Univ. of Technology, Onsala, Sweden
[6] Harvard-Smithsonian Center for Astrophysics, Cambridge, USA
[7] Instituto de Astrofísica de Canarias, La Laguna, Tenerife, Spain
[8] Departamento de Astrofísica, Universidad de La Laguna, Tenerife, Spain
[9] Max-Planck-Institut für Radioastronomie, Bonn, Germany

## Abstract

The nuclei of active galaxies are the most powerful long-lived sources of radiation in the Universe. They often outshine the host galaxy in which they reside and are able to eject outflows or jets of relativistic plasma that emit all the way from radio waves to the highest energetic gamma rays. To understand the mechanisms that govern AGN we have to go down to parsec or sub-parsec scales, where a central engine formed by a supermassive black hole and a surrounding accretion disc produces helical magnetic fields in which jets are thought to originate. The exact role of the magnetic field and its structure, the composition and dynamics of the ejected jets, as well as the feedback effect of the jets on the gas and dust that surrounds the central engine are however still far from understood. The SKA, with its superb sensitivity and polarization capabilities, will allow for a relevant advance on the field of relativistic jets and in active galactic nuclei and their connexion with the fueling material and star formation present in the vicinity of the central supermassive black hole.

## 1 Introduction

More than six decades after the discovery of active galactic nuclei (AGN) by Carl Seyfert [49], these objects still represent one of the major puzzles in astrophysics. AGN are currently well known as the most luminous long-lived sources of radiation in the cosmos. They are powered by gas falling from an accretion disc onto a super-massive black hole (SMBH, $\sim 10^6$ to $\sim 10^9\,\mathrm{M}_\odot$) at the center of a large fraction of galaxies [32, 50]. Indeed, AGN are able to emit enormous amounts of rapidly variable radiation, within time scales from minutes to years, all along the electromagnetic spectrum (from the longer radio wavelengths to the most energetic TeV gamma-rays).

When considered at the smallest (parsec and sub-parsec) scales, AGN are ideal objects to further our understanding of relevant physical processes surrounding SMBHs. Among all AGN components, one of the most intriguing ones is the central engine. Relativistic jets in radio-loud AGN have not still revealed the detailed mechanism for their formation. In models of magnetically driven outflows (pioneered by Blandford-Payne [9], but see also [31]), the rotation of magnetic fields anchored at the base of the accretion disc generate a poloidal electromagnetic flux of energy (Poynting flux) that accelerates the magnetospheric plasma. Energy can also be extracted from rotating black holes with similar efficiencies by the Blandford-Znajek mechanism [10]. In general, it is thought that both mechanisms operate simultaneously to produce stratified jets in both composition and speed, with an electron-proton (slow) disc wind shrouding an inner electron-positron relativistic jet. The current paradigm of jet formation, and in particular the one for the jet composition and density stratification, its magnetic field intensity and configuration, as well as the one for the non-thermal particle distribution responsible for jet emission will be efficiently tested by future sensitive radio facilities as SKA, see Sections 2 and 3.

The existence of the dusty torus (a toroidal structure of dust and molecular gas that, depending on the viewing angle, obscures the inner AGN regions, in particular the broad line region) is the basis of the unified model for AGN [7]. Interferometric infrared (IR) observations show that the torus of Seyfert galaxies is rather small ($r \leq$10 pc at 12 $\mu$m, [12]), and therefore it cannot be directly imaged with current instrumentation. Thus, the only methods available to infer torus properties are IR interferometry, which is restricted to the most nearby and brightest AGN, and the fitting of nuclear spectral energy distributions (SEDs) with torus models. The peak of the torus emission in these models occurs in the mid-IR and the most realistic dust distribution is not homogeneous, but clumpy (e.g., [42]). Another important process that can be probed is the accretion of cold gas from kiloparsec to sub-parsec scales. While this gas is important for the accretion process that ultimately leads to relativistic jets, it can also be used to produce new stars. Both the basic geometrical properties of the dusty torus and those of the nuclear star formation regions will be accessible by SKA thanks to its unprecedented sensitivity and its good angular resolutions. How SKA will impact our current knowledge of these systems is outlined in Section 4.

The very large, several (hundreds of) kiloparsec-size, radio structures seen in radio galaxies are in general not observed in Seyfert galaxies and in many low luminosity AGN (LLAGN). In particular, large scale collimated jets are rare. However, collimated radio structures, at scales of parsec to hundreds of parsecs, appear to be common in LLAGN when observed with sufficient angular resolution and sensitivity [38, 41, 35]. These jets are the best laboratories to study galaxy-AGN feedback because of their frequent interaction with the innermost regions of the AGN [19, 52], see also the chapter by Perucho et al.

(this book). Typically associated with Seyfert/LINER optical spectra, LLAGN have typical radio luminosities from $\sim 10^{19}$ to $\sim 10^{21}$ W/Hz [40] and can be studied only in the Local Universe with current radio facilities. Systematic statistical studies including thousands of LLAGN up to the most distant Universe will be allowed by the SKA through both large area and deep surveys (see Section 5), which will help to solve the long-standing problem of the understanding of the ultimate reason for the radio-loudness of a fraction of the AGN population.

From the spectroscopic point of view, HI studies with SKA (Section 6) will provide unprecedented three-dimensional information of the gas distribution of the central regions of a large number of AGN, therefore allowing for reliable statistical studies over large populations. This will in turn allow for additional tests of the AGN paradigm, in particular on which regards to the properties of their tori.

## 2 The central engine

The basics of the central engine in radio-loud AGN have been outlined in Section 1. Two of the major challenges on the astrophysics of the inner regions of radio-loud AGN are the determination of the composition of the jet and the observational confirmation of the jet/disc-wind paradigm. If the jets are formed around the black-hole corona (Blandford-Znajek model, [10]), their composition should be dominated by electrons and positrons, whereas the load of particles from the accretion disc [9] would imply an electron-proton composition [51]. Full polarization observations (including circular polarization), using SKA-VLBI [44, 3, see also Chapter by Ros et al. (this book)] and the SKA alone [3], will allow to determine the jet composition (see [26] and references therein) and the presence or not of slower and denser winds originated at the accretion discs surrounding a jet fast spine. Furthermore, knowing the jet composition and the energy distribution of the particles is fundamental to measure the dynamical role of the magnetic field. SKA-MID on Band 5 using the SKA as a VLBI station will permit high-frequency observations of polarized light (4.6 - 13.8 GHz) that will help understanding the evolution of the magnetic field intensity along the jet, in the same line. Therefore, Blandford-Payne type models (where the jet is anchored to magnetized rotating discs and helical fields are generated in a natural way) should result in an edge-brightened polarization structure up to the base of the jet with a high degree of polarization. In contrast, Blandford-Znajek models, where the jet is driven by the SMBH spin, result in a more compact footprint of the jet, which implies a higher opacity and a smaller polarization degree. The observations will help to discriminate between the models and will therefore also represent an extra independent test of the jet formation mechanism that can add information relevant for studies of jet composition (see above).

The violent variability in many radio-loud AGN, in the form of high energy flares accompanied by the ejection of superluminal radio components (e.g., [33, 34, 1, 2]), implies strong non-stationarity of the central engine on time scales of one year. The theory governing this variability, most likely related to the change in the black hole magnetic field and accretion rate, and affecting the whole electromagnetic emission spectrum, has to be developed. In the currently accepted acceleration models by Vlahakis & Koenigl [55], jets change from Poynting dominated to particle dominated along the acceleration phase. One of the key parameters in these models is the initial magnetization of the wind, determining the asymptotic speed of the flow. It is expected that this relevant parameter for jet formation models can also be inferred from the previously outlined full-polarization observations with the SKA-MID and SKA-VLBI at the highest possible (range of) frequencies.

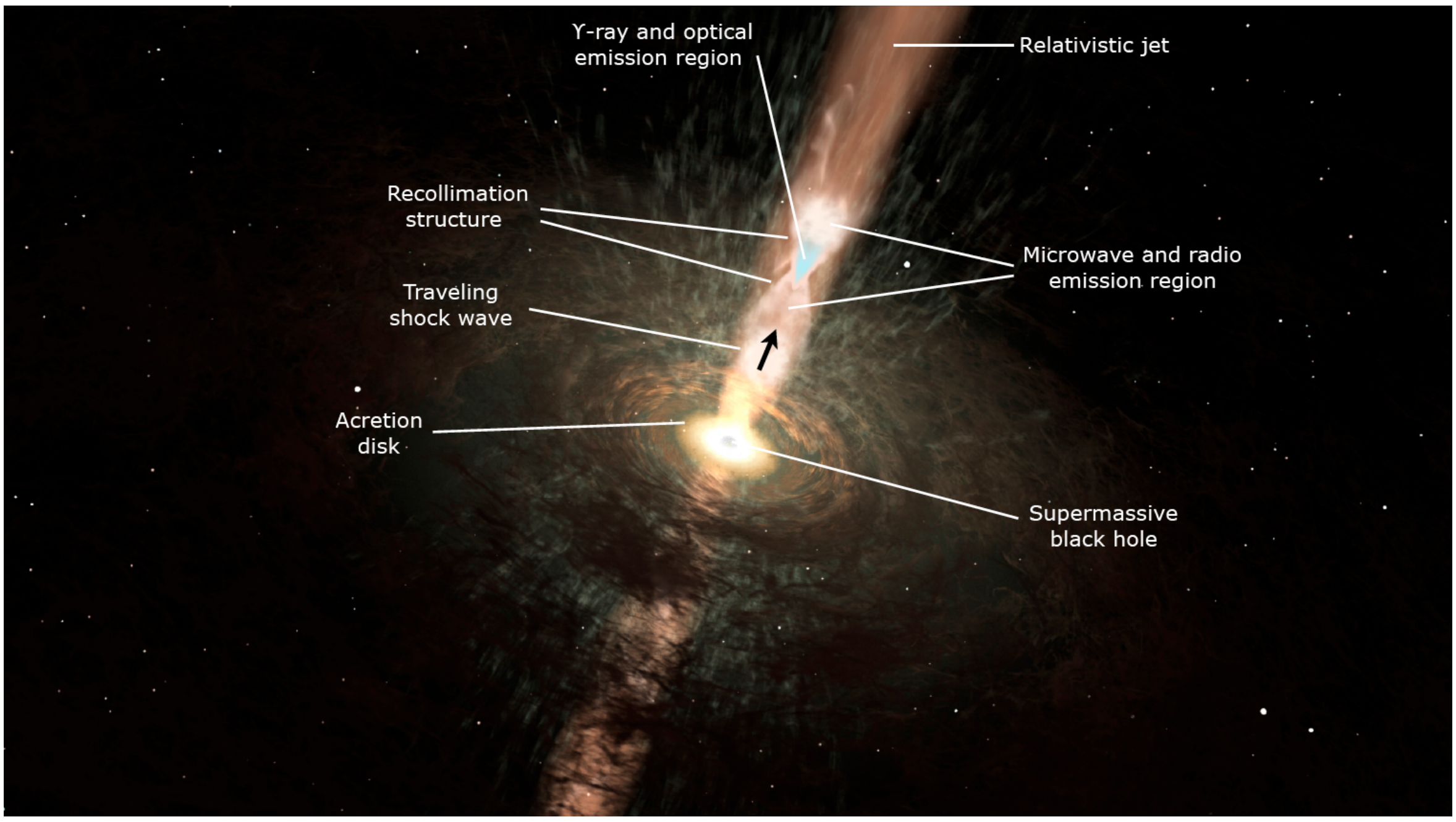

Figure 1: Conceptual representation of the multi-wavelength signature of the interaction of a moving shock with a stationary recollimation structure in the innermost regions of a radio-loud AGN, see [2] for an example of the observational basis of this model. Image credit: W. Steffen (UNAM).

Another open issue is the division of AGN in radio-loud and radio-weak objects. Current observations seem to be more consistent with a smooth single distribution of radio loudness rather than a bimodal one. The superb sensitivity and survey speed of the SKA is expected to produce massive data sets of AGN in the entire domain of luminosities up to the nano-Jansky level, which will provide key answers regarding this long standing problem through the statistical study of unprecedentedly large samples [3].

## 3 Physics of jets and their magnetic fields

At parsec scales, several relevant open questions remain unanswered regarding the study of relativistic outflows from radio-loud AGN. These include the actual structure of the magnetic field, the jet composition, its possible transversal stratification, and the role of the magnetic field on the jet dynamics after the acceleration of the jet flow. This jet acceleration is thought to be produced by conversion of magnetic energy into kinetic energy of the particles (e.g. [55] and [29]). The unprecedented sensitivity of the SKA will allow to perform detailed studies of the interaction regions between traveling perturbations and standing shocks at the compact regions of jets within their radio-cores, which have been claimed to be the origin of multi-wavelength flares (see e.g., [34, 1, 16, 17], and Fig. 1) in blazars and radio galaxies that are observed at small viewing angles. In particular, SKA-VLBI at the highest possible observing frequencies will grant a detailed study of a number of jets in which these flares are produced and help determining the exact regions and processes by which they are produced. Polarization observations will also allow to determine the exact behavior of the magnetic field in this whole process and will help constraining the parameter space of these events, which can be related to the SMBH/accretion-disc system [33] and to the acceleration mechanisms that end up in the production of X-rays and gamma-rays from these sources.

The excellent sensitivity and polarization purity of the SKA (see the SKA1 Baseline Design and [3]), both on VLBI mode [44, see also Chapter by Ros et al. (this book)] (for non local radio-loud AGN) and with SKA-alone observations (for nearby AGN), will certainly impact our knowledge of the physical properties of the parsec scale region of extragalactic relativistic jets and their immediate environments. The SKA will allow, for the first time, to resolve a large number of parsec-scale jets on the transversal direction to their axis, and even to image the counter-jets in a fraction of them.

If the parsec-scale jet is threaded by a magnetic field with a helical structure, as expected from currently accepted jet formation models (see Section 2) and inferred from VLBI observations [8], detectable stratification in the total and linearly polarized emission is expected across the jet axis [30, 6]. This, together with Faraday rotation studies, would allow for a determination of the three dimensional structure of the magnetic field and flow velocity [21]. The case where both the jet and counter-jet can be observed is particularly interesting. In this case, measurements of the flux density ratio between the jet and counter-jet features, as well as their proper motions, provide a direct determination of the jet viewing angle and bulk flow velocity. These studies would require polarimetric multi-frequency SKA-VLBI observations at the highest possible frequencies, therefore achieving the necessary angular resolution to image the innermost regions of AGN jets.

Although there are indirect indications for the existence of a significant toroidal component in the parsec-scale magnetic fields of AGN jets [20, 8], only two direct cases (involving polarimetric imaging of the actual magnetic field) have been reported so far, see [56, 36] and Fig. 2. These latter two cases are still under debate though, because of unavoidable uncertainties on the actual nature of the source of polarimetric emission. Such definitive direct evidences require deep and transversely-resolved SKA-VLBI observations of the parsec-scale emission at a broad range of high frequencies from a statistically significant sample of jets. This will allow (with the help of detailed Faraday rotation images and profiles) to shed light on the actual magnetic field strength and its three dimensional structure, as well as on the low-energy particle properties.

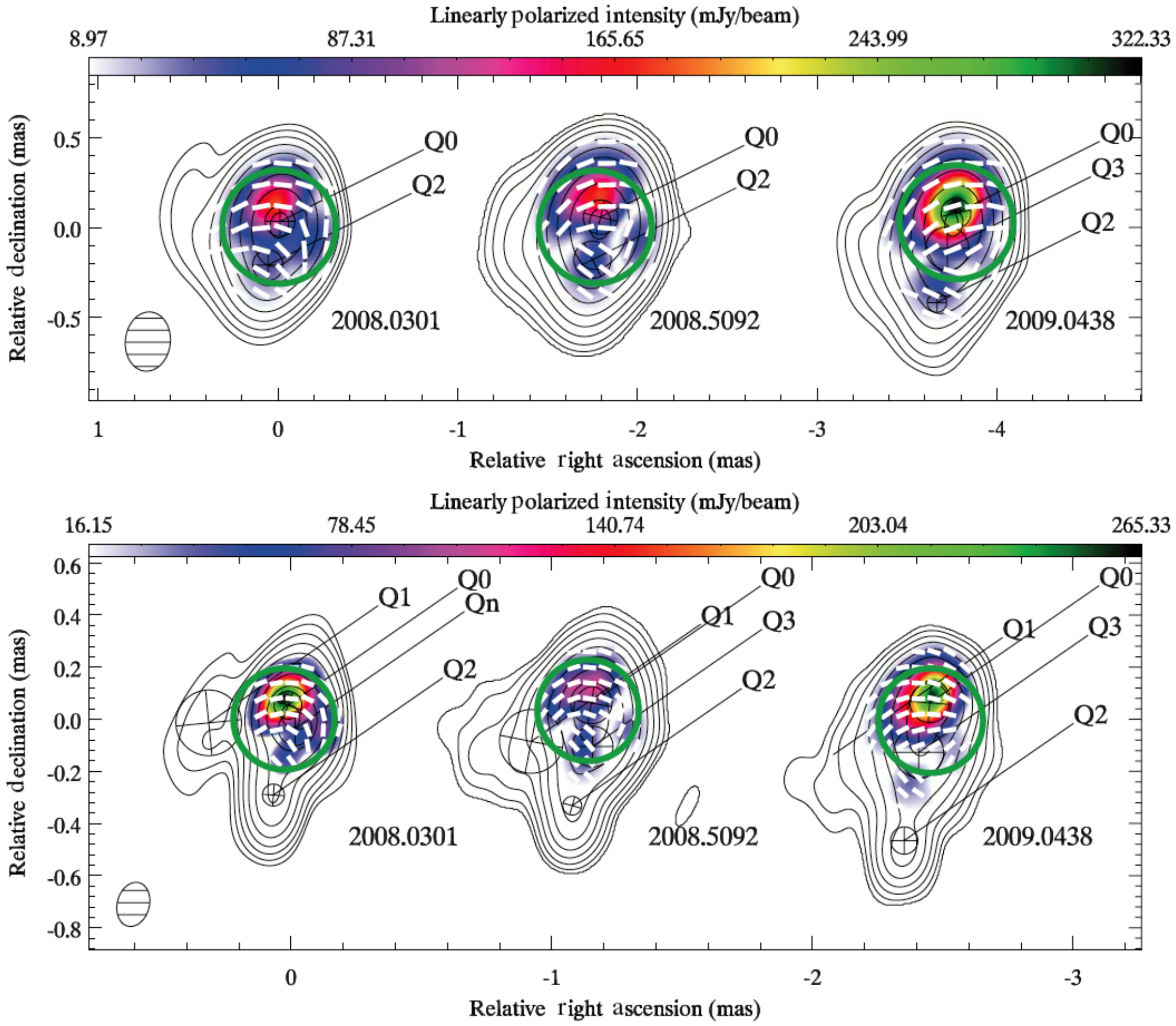

Figure 2: VLBA sequence of total intensity images (represented by contours), linearly polarized intensity images (represented by the color scale), and magnetic vector polarization angle distribution (symbolized by short white sticks), of the radio loud quasar NRAO150. The top three images were obtained at 22 GHz, whereas those on the bottom side correspond to 43 GHz observations. Assuming that the the observed structure corresponds to the strong radio jet on NRAO150 as seen face-on (i.e. pointing at $\sim 0°$ to the line of sight), the green line would represent the toroidal component of the magnetic field that would be observed in this case. The images show a good match with the observed magnetic vectors. Reproduced from [36].

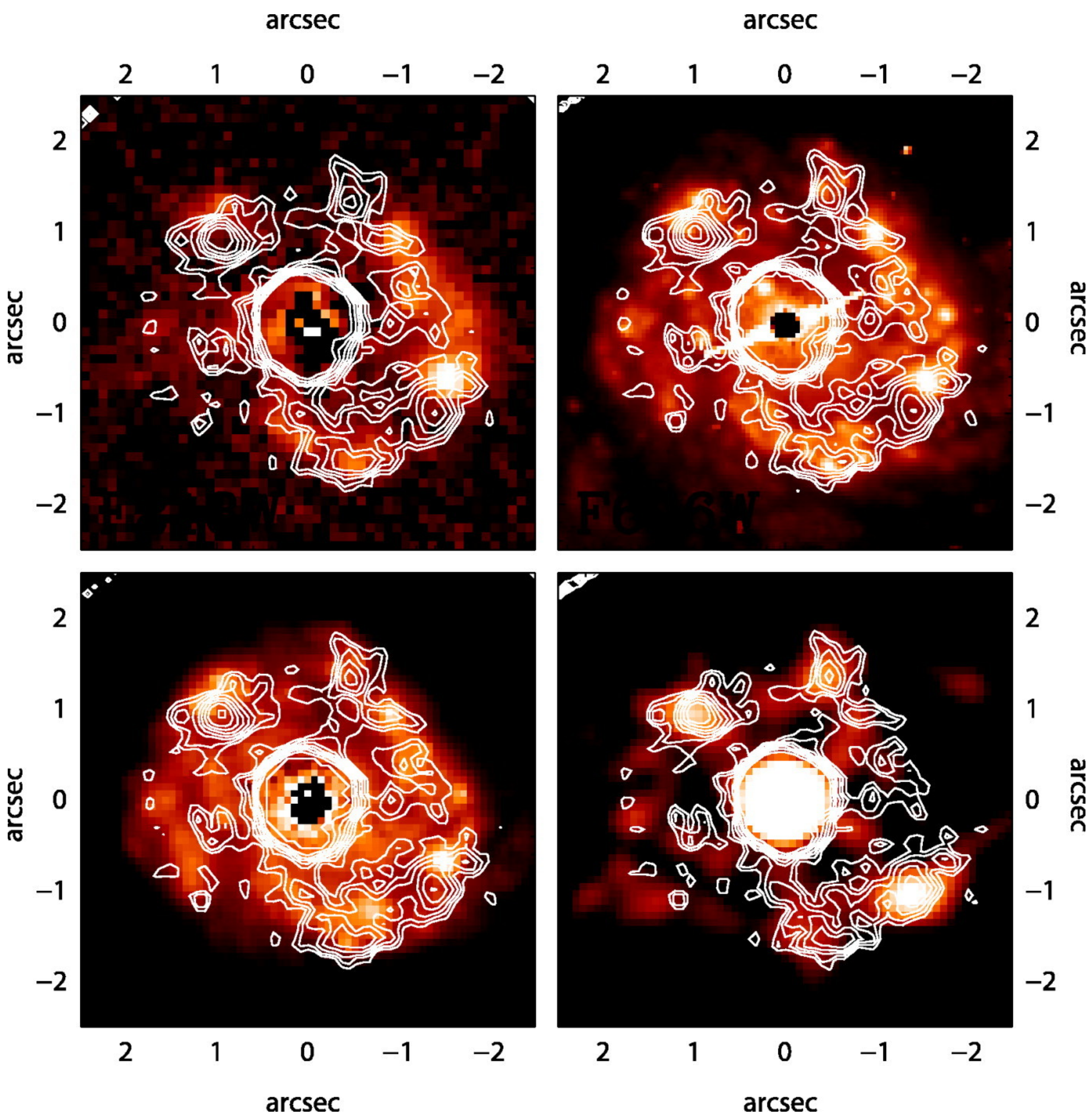

Figure 3: Multifrequency observations of the central region in NGC 7469, which contains a bright Seyfert 1 nucleus and a circumnuclear ring of star formation. Contours are the mid-IR 12.5 $\mu$m emission superimposed on the ultraviolet, optical, radio 8.4 GHz, and near-IR continuum maps (from top left clockwise). There is an excellent spatial correspondence between the extended mid-IR emission probing the on-going star formation activity in the ring and the radio continuum emission. Figure from [14].

# 4 The dusty torus and nuclear star formation

In the last few years we have started to constrain the torus properties of radio quiet AGN using sub-arcsecond resolution infrared (IR) data. These properties include, for example, the torus size, inclination, and covering factor [54, 46, 47, 4, 12]. However, those results are not statistically significant because the samples studied are not complete, due to the paucity of both interferometric and sub-arcsecond imaging and spectroscopic IR observations. Besides, the torus model parameters that describe the torus geometry and properties are highly degenerated. Therefore, to reduce these degeneracies it is paramount to have observational constraints on the torus properties such as its inclination angle. For a handful of galaxies, we can infer the torus inclination from the orientation of the ionization cones, but this is definitely not sufficient for a statistical study. High angular resolution and sensitive radio observations of nearby AGN have the potential to solve this problem. Parsec-scale jets have been detected in low and intermediate-luminosity radio-weak AGN [28, 48, 39, 35] and the orientation of these jets, together with other observations, can be used to constrain the torus inclination. In addition, radio observations of water and OH masers are unique tools for investigating the structure and kinematics of the gas in the immediate vicinity of the AGN central engine (e.g., [11]), and they can also provide constraints on the torus inclination angle.

The fueling of AGN requires the cold gas to be driven from the kiloparsec-scale host galaxy to physical scales of less than a parsec. As this cold gas is also the fuel to form new stars, nuclear star formation (on scales of less than 100 pc) appears to be an inevitable consequence of gas accretion processes. Moreover, numerical simulations predict that the star formation rate and black hole accretion rate should be tightly correlated on nuclear scales in AGN (see e.g., [27]). There is plenty of observational evidence for nuclear star formation in the inner ∼100 pc of nearby Seyferts using high angular resolution near and mid-IR observations [13, 53, 15, 5], as well as very long baseline interferometric (VLBI) radio observations [45]. However, star formation indicators such as ultra violet emission, hydrogen recombination lines, and polycyclic aromatic hydrocarbon emission are not straightforward to use in the nuclear regions of AGN, as they can be easily contaminated or diluted by the bright AGN emission and additionally obscured in type 2s. Radio continuum observations can be used to trace star formation activity in AGN as they are not affected by extinction, although observations at different frequencies are necessary to distinguish between thermal and non-thermal processes in the nuclear environment of galaxies (see [25] and references therein). Until a few years ago, sensitive radio observations could only trace star formation activity on kiloparsec-scales (see Fig. 3) in nearby Seyfert galaxies. However, recent high-angular resolution ( $\lesssim$ 10 mas) radio observations with the high sensitive European VLBI Network has proven to be an excellent tool for scrutinizing the inner $\approx$ 100 pc region of local ultra luminous IR galaxies. For the case of galaxy Arp299-A (D= 45 Mpc), an extremely prolific supernova factory was discovered in the central 150 pc, together with a low luminosity AGN at a mere distance of 2 pc of a recently exploding core-collapse supernova [45]. More recently, it has been reported evidence for the existence of nuclear discs ( $\lesssim$ 100 pc in size) in starburst galaxies from their radial distribution of supernovae [24].

The good angular resolution and superb sensitivity of the SKA will allow to determine the orientation of the parsec-scale jets as well as to characterize nuclear masers, when present, in complete samples of low and intermediate-luminosity radio-quiet AGN. These observations will be key to set constraints on the torus models and study the kinematics of the nuclear gas, and will include not only nearby AGN but also intermediate redshift radio-quiet AGN where such constraints are lacking. The large field of view of the SKA and its good angular resolution will also allow to characterize both the nuclear and circumnuclear star formation activity of large samples of radio-quiet AGN.

## 5 Low luminosity AGN

Most of the SMBH at the center of galaxies are thought to go through phases of activity during their evolution. However, the duty cycle of such phenomenon is currently not well constrained. The detection of compact jet emission in the low power counterpart of double size radio sources, Seyferts and LLAGN is a key test for the AGN unification scenarios. LLAGN are thought to be characterized by radiatively inefficient accretion (frequently modeled as advection-dominated accretion flows). They sometimes show small scale and low luminosity compact radio jets that can be useful to explain properties of the system, but observational parameters like intensity and polarized flux density, spectral index, size and variability are only available for a few very bright and nearby cases. Most AGN would not be radio silent at their radio quiescence phase. Therefore, radio observations will be paramount to detect LLAGN and probe the physics of their accretion flows. Their radio detection rate is relatively large, about 67% of LLAGN on the Palomar sample show extended jet emission at sub-arcsecond resolution, indicating that most LLAGN are energetic enough to power parsec-scale jets, see Fig. 4 and [45, 35]. The reason why their radio emission often appears compact is that previous radio observations were not sensitive enough or did not have enough angular resolution. The large collecting area, good angular resolution and contiguous frequency coverage of the SKA will allow to unveil these mini-jets, their morphology, and spectrum in full detail in representative samples of AGN covering a much wider range of luminosity and radio loudness than feasible before. SKA surveys will become an efficient tool to search for faint (intrinsically weak or not) AGN. Latter, SKA continuum observations in the 1 to 3 GHz range will be ideal to improve our current knowledge, since many more sources will be revealed, both fainter ones in the Local Universe, and with much more distant galaxies. Variability studies are currently limited by sensitivity, which will not be a major limitation in the SKA era. The comparison between the jet structure and the surrounding ionized gas at similar angular scales will allow a direct assessment on the jet power dumped into the medium (see for example the study in the X-rays by [22, 23]).

## 6 HI studies of AGN

One of the relevant observing opportunities with the SKA will be HI imaging of galaxies with unprecedented detail, see the chapter by Verdes-Montenegro et al. (this book). The SKA is an optimum instrument for measurements of the HI 21 cm line, in the region around 1.4 GHz where SKA-MID will deliver excellent sensitivity and angular resolution (almost an order of magnitude better than any other current instrument, e.g. a factor of six better than JVLA, see the SKA1 Baseline Design).

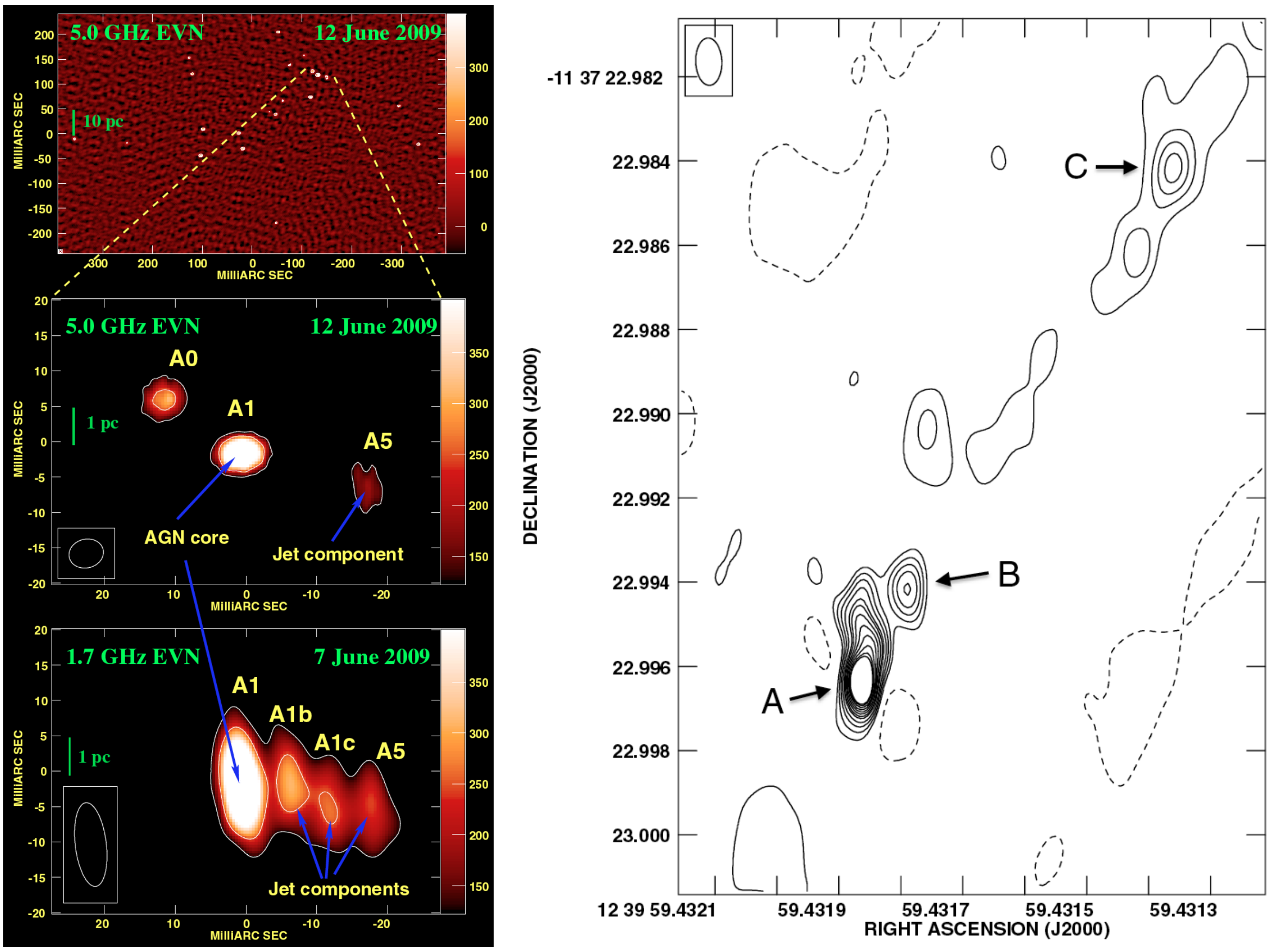

Figure 4: *Left:* European VLBI Network observations of the Arp299-A galaxy from the 100 pc to the parsec scale. The zoomed-in images provide clear evidence of the presence of a jet associated to the LLAGN at the center of the galaxy. Reproduced from [45]. *Right:* VLBA image of NGC 4594 at 23.8 GHz (beam $1.13 \times 0.62$ mas), uncovering a 0.6 pc-size mini jet at the core of prototype LLAGN the Sombrero galaxy. Reproduced from [35].

Combining its good angular resolution, nominally $\sim 0.25$ arcsec at 1.4 GHz, and large collecting area, the SKA will produce HI three dimensional mapps of low redshift AGN, all the way from the central parsecs. This will allow to directly trace the central obscuring material where large HI-equivalent column densities are inferred from X-rays, to the central few hundred parsecs where HI gas discs of these scales. Promising reservoirs for circumnuclear star formation and black-hole fuel are inferred from HI in absorption in some Seyfert galaxies [38].

The excellent sensitivity of the SKA will allow to expand these observations to large samples of AGN at all possible luminosities in the Local Universe. Therefore, the SKA will provide a direct confirmation of large HI-equivalent column densities inferred from X-rays, and will help to asses on the nature of the absorbers. This will provide an stringent test of AGN-tori. The SKA will establish how frequently the central HI discs are present in active and non active nuclei, and uncover the interplay between the neutral gas, the molecular gas structures, and the ionized gas outflows seen at comparable resolutions with ALMA, adaptive-optics assisted IR, and HST observations [52, 43, 37].

# 7 Relevance of the SKA for AGN studies at parsec scales

Previous sections give an advance of the impact that the SKA will have in the field of AGN, in particular on the study of the parsec-scale region of these objects. The fabulous sensitivity and full-polarization precision (and purity) of the SKA, together with its participation in dedicated high sensitivity VLBI campaigns (see Ros et al. [this book]), will allow to provide (for the first time) robust observational evidence about the actual composition of extragalactic relativistic jets in AGN, as well as the possible existence of the accretion-disc wind (that is postulated to produce most of the initial collimation of the jet) and the actual structure of the magnetic field involved in the jet formation process. It is expected that SKA will implement policies to respond to interesting transient phenomena also observed by facilities covering other spectral ranges. These policies may include extreme flaring states of AGN jets observed with SKA-VLBI. If so, SKA will be an excellent tool for understanding the relation of the different radiation processes in the innermost (and less understood) regions of AGN relativistic jets.

Another of the relevant problems on physics of AGN is about the understanding of the mechanism driving the radio loudness (i.e. the ability to form powerful relativistic jets). The excellent sensitivity and survey speed of the SKA will allow to build huge data sets including millions of AGN at all ranges of radio luminosity to make adequate statistical studies. These are expected to provide definitive answers about this long standing problem. Later, dedicated observations of relevant sources will help constraining the properties of the different components of the AGN (e.g., jets, dusty torus, circumnuclear star formation region, and broad line region), for which high resolution spectral observations of the HI and OH lines will be crucial on the understanding of the three dimensional structures of cold gas in the close vicinity of the AGN. Observations of masers in the innermost nuclear regions of AGN will certainly provide key information about the geometry and dynamics of the accretion discs in AGN at all radio luminosities.

## Acknowledgments

IA research is supported by a Ramón y Cajal grant of the Spanish MINECO. CRA is supported by a Marie Curie Intra European Fellowship within the 7th European Community Framework Programme (PIEF-GA-2012-327934). MINECO grants AYA2012-38491-C02-01 (AA), AYA2012-13036-C02-01 (AA), AYA2012-31447 (AAH), AYA2013-40825-P (JLG & IA) AYA2013-40979-P (JMM & MP), AYA2013-48226-C3-2-P (JMM & MP), and AYA2012-38491-C02-02 (ER) are acknowledged.

# The far-infrared-radio correlation in galaxies

**U. Lisenfeld[1], M. A. Pérez-Torres[2], A. Alberdi[2], L. Colina[3], S. García-Burillo[4], and A. Alonso-Herrero[5]**

[1] Departamento de Física Teórica y del Cosmos, Universidad de Granada, E-18071 Granada, Spain
[2] Instituto de Astrofísica de Andalucía (IAA-CSIC), E- 18008 Granada, Spain
[3] Astrophysics Department, Center for Astrobiology (CSIC-INTA), Torrejon de Ardoz, 28850 Madrid
[4] Observatorio Astronmico Nacional (OAN)-Observatorio de Madrid, Alfonso XII, 3, 28014, Madrid, Spain
[5] Instituto de Física de Cantabria, CSIC-UC, E-39005 Santander

## Abstract

The tightness and universality of the far-infrared (FIR) to radio continuum (RC) correlation is still not completely understood. This correlation is followed by all star-forming galaxies not dominated by an Active Galactic Nucleus, both globally as well as locally within the disks. There is a general consensus that star formation (SF) is the ultimate driver of the relation, in the sense that the bulk of dust emission in the FIR is powered by young stars ending their lives as supernovae which are the main sites of Cosmic Ray (CR) acceleration. Although this simplistic view is correct, it neglects many of the additional parameters that affect the correlation. Thus, a detailed understanding is still missing which is crucial in order to correctly use the RC emission as a tracer of recent SF with the important advantage not to be affected by extinction. Furthermore, a detailed understanding of the correlation will lead to a deeper understanding of dust heating, the interstellar medium (ISM) and propagation of CRs.

The capabilities of the SKA are needed to make progress in our understanding of the correlation. In particular, they will allow us to (i) extend the study of the correlation to low-luminosity dwarf galaxies which are expected not to follow the correlation so well, (ii) extend the correlation to high-z objects and test whether the correlation is still fulfilled, and (iii) study the properties of CR propagation in galactic halos via changes in the spectral index in order to be able to compare the relative relevance of propagation, escape and energy losses.

# 1 Physical basis and relevance of the correlation

The tight correlation between the radio continuum (RC) and the far-infrared (FIR) emission of galaxies [13, 8] holds over 5 orders of magnitude in luminosity with a very low scatter of about 0.26 dex [42] and is followed by basically all star-forming galaxies unless they have a radio-loud Active Galactic Nucleus (AGN). Discovered from data at 60 $\mu$m and 100 $\mu$m from the IRAS satellite, it has since afterwards also been tested with data from the satellites Spitzer (24 $\mu$m, 70 $\mu$m, and 160$\mu$m) and Herschel (70 $\mu$m, 100 $\mu$m, 160$\mu$m, and 250$\mu$m).

There is a general agreement that the basic underlying driver of this correlation is massive star formation (SF). Massive stars are the main heating sources of the interstellar dust, both locally in HII regions and also, mainly through their non-ionizing UV radiation [43], of the diffuse dust distributed in the ISM in galaxies. Massive stars are at the same time responsible for the radio emission in two ways: The ionizing photons that they emit produce the thermal (free-free) radio emission in HII regions and they end their lives as supernovae (SN) whose remnants accelerate Cosmic Rays Electrons (CREs) emitting radio synchrotron emission on their way through the interstellar magnetic field.

However, although this basic relation is clear, the step from massive SF to the RC emission on one side and FIR emission on the other side is not completely direct but involves instead a considerable number of parameters. For the FIR emission the main additional parameter is the dust opacity in a galactic disk which can vary from well below one in dwarf galaxies to completely optically thick as in dust-enshrouded starburst galaxies. The RC emission is based on even more complex processes. First of all, it consists of two radio components, each of which is based on a completely different process: free-free or bremsstrahlung emission from thermal electrons in HII regions (the thermal radio emission) and synchrotron emission from CREs. Generally, in the GHz observing range, the synchrotron emission is dominant with a typical thermal-to-total radio flux ratio of 10% [6]. Synchrotron radiation is emitted from CREs propagating away from their acceleration sites (SN remnants), suffering at the same time energy losses and emitting synchrotron radiation. The energy losses are predominately inverse Compton and synchrotron losses in spiral galaxies, but in starburst galaxies also bremsstrahlung, ionization and adiabatic energy losses can become important [38, 28]. Therefore a large number of parameters come into play when describing the synchrotron emission of a galaxy, the most important ones being the magnetic field, energy density of the radiation field, gas density, diffusion and convection velocities and the halos size. Considering this complex relation between the origin (massive stars) and the results (FIR and RC emission) it is a priori hard to understand how the FIR-radio correlation can be so tight and universal.

The FIR-radio correlation has had, apart from challenging our understanding of the processes of dust heating and CR propagation, two important applications. First of all, the relation of the RC emission with massive SF that it implies show that the RC emission can be used as a SF tracer [6, 15]. The chapter of Torres et al. (this book) deals with the topic of RC emission as SF tracer in more detail. Other commonly used SF tracers, as the H$\alpha$ or UV emission have the disadvantage that they are extincted by dust and that the correction for dust-extinction is difficult and results in a considerable uncertainty. The dust emission in the mid-infrared (typically taken at 24 or 70 $\mu$m) which is also frequently used as a SF tracer on its own or in combination with H$\alpha$ or UV (e.g. [18, 4]) is not affected by dust extinction. However, these observations need to be done from space and come therefore only from satellite missions (Spitzer, Herschel, WISE) and have a limited spatial resolution of $6''$ to $36''$. The RC emission avoids both disadvantages since it is not affected by dust and interferometers from the earth surface can obtain sub-arcsec resolutions. Both properties are particularly important when observing distant galaxies and will be crucial for studies of the cosmic star formation history once sensitive observations with SKA will be possible. A second application of the FIR-radio correlation has been the use of the radio-submillimeter ratio as a photometric redshift indicator [3]. This relation could be derived based on the tightness of the FIR-radio correlation.

Thus, a deep understanding of the detailed physical processes that drive the FIR-radio correlation is crucial not only for our knowledge of the ISM but also in order to be able to apply the RC emission as a SF tracer and the radio-submillimeter ratio as a redshift indicator. Both applications are of particular relevance for studies of the high-redshift universe where unknown extinction corrections can lead to enormous errors.

## 2 Observations of the FIR-radio correlation

The FIR-radio correlation has been tested for many galaxy samples, and found to hold for all galaxies irrespective of their morphological type or star formation rate (SFR). The only exception are galaxies where CRE acceleration outside SN remnants (SNRs) takes place, as is the case in galaxies with a radio-loud Active Galactic Nucleus (AGN), galaxies in clusters where interaction with the Intragroup Medium might produce shocks accelerating CRs [40, 25] or between interacting galaxies that collided face-on [30]. The correlation is roughly linear and holds over 5 orders of magnitude [42]. At low luminosities, the situation is less clear. Some studies indicate that the correlation holds also for dwarf galaxies [5, 31], whereas in the largest sample studied up to date of 26 dwarf galaxies, Kitchener et al. [17] find a lack of RC emission compared to brighter objects. The discrepancy is probably due to the faintness of the objects which lead to a high fraction of upper limits in the samples [5] and the need to stack data [31].

Apart from samples of local galaxies, the FIR-radio correlation has also been tested for high-z galaxies. So far, no clear indications for a deviation at high redshifts have been found. Several studies confirm the validity of the FIR-radio correlation out to redshifts of about 1-2 [12, 1, 32, 33] with only a few giving tentative evidence for a deviation (lower radio emission) at higher z (z $\sim$ 3-6) [26, 35]. However, due to their faintness, reliable conclusions for these high redshift galaxies are difficult.

The correlation does not only hold globally, but also locally within galaxies. It has been shown to hold down to about 20-50 pc in the Milky Way [41] and the Large Magellanic Cloud [16], a scale which is dominated by thermal radio emission from HII regions. At larger scales, where synchrotron emission becomes dominant, the scale where the correlation breaks down is expected to depend on the dust opacity and, most importantly, on the diffusion length of CREs in galactic disks. Since the arrival of Spitzer and Herschel data, studies of the local FIR-radio have been possible at scales on the order of 10″. Murphy et al. [22, 23, 24] has shown for galaxies from the Spitzer Infrared Nearby Galaxy Survey (SINGS) that the RC emission agrees very well with a spatially smoothed version of the FIR emission showing that CRE propagation away from the sites of acceleration is important. Detailed analysis of the spatial scales with a wavelet analysis have shown that the correlation holds down to scales of between several hundred parsec to 1 kpc ($<$ 0.4 kpc for M33 [37], 0.7 kpc for M51 [11], $\sim$ 1kpc for M31 [37] and $\sim$ 2 kpc for NGC 6946 [36]). The different scales might be related to properties of the magnetic field determining the CR propagation [37].

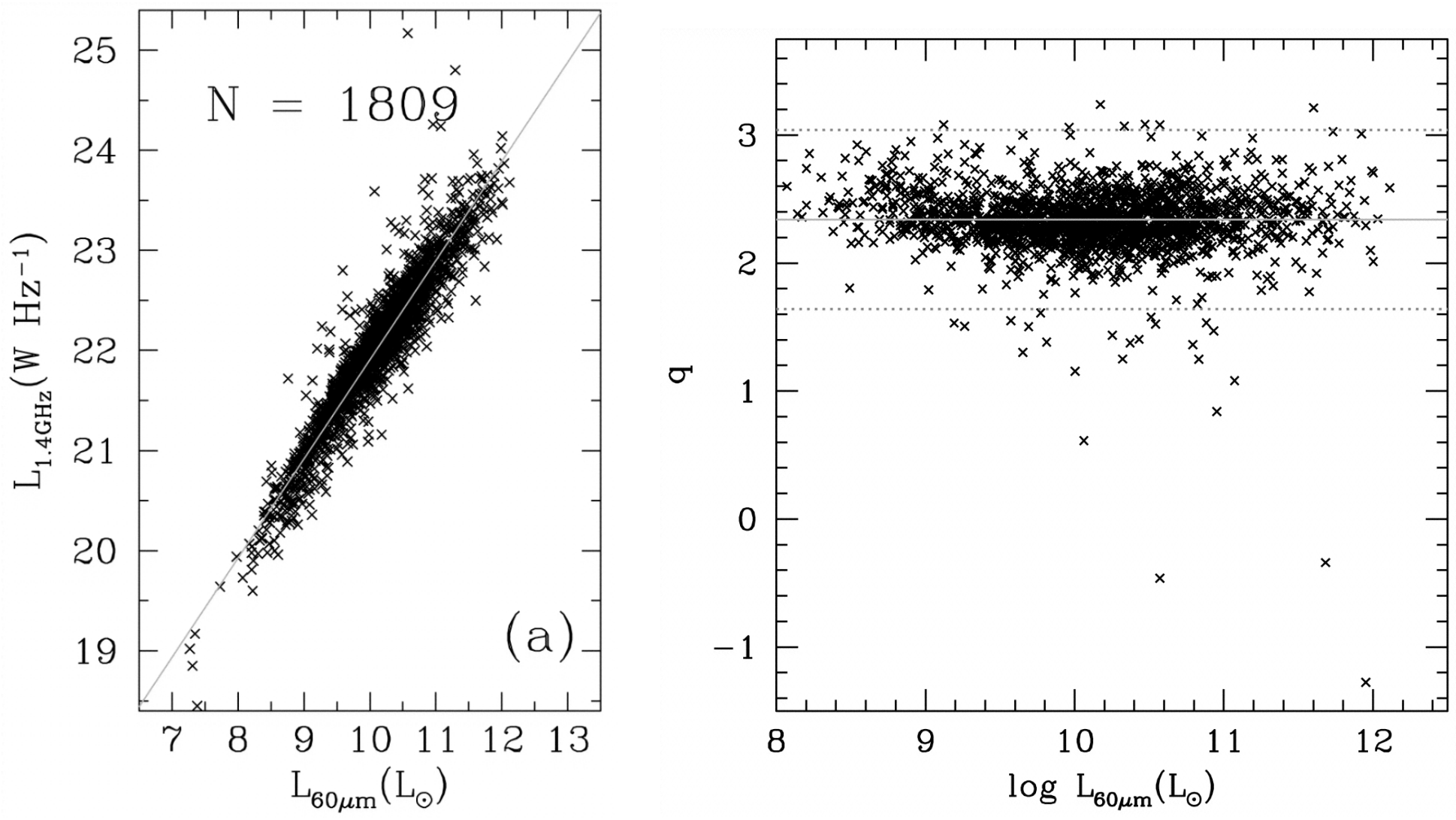

Figure 1: Example of the FIR-radio correlation for a sample of 1809 galaxies with $S_{60\mu m} > 2$ Jy, taken from Yun et al. [42]. **Left:** The radio luminosity at 1.4 GHz vs. the luminosity at 60 $\mu$m from IRAS data. The solid line corresponds to a linear relation with a constant offset. **Right:** Distribution of q-values ($q = \log\left(\frac{FIR}{3.75\times10^{12}\mathrm{Wm^{-2}}}\right) - \log\left(\frac{S_{1.4\mathrm{GHz}}}{\mathrm{Wm^{-2}Hz^{-1}}}\right)$, with $FIR = 1.26\times10^{-14}(2.58S_{60\mu m} + S_{100\mu m})$, where $S_{1.4\mathrm{GHz}}$ is the observed 1.4 GHz flux density and $S_{60\mu m}$ and $S_{100\mu m}$ denoting the IRAS 60 and 100 $\mu$m band flux densities in units of Jy), plotted as a function of the IRAS 60$\mu$m luminosity. The solid line marks the average value of $q = 2.34$, while the dotted lines delineate a limit of 5 $\times$ the standard deviation. The radio-excess objects (below the line) are likely to be dominated by an AGN.

## 3 Models

The tightness and the universality of the correlation has inspired many researchers to model the correlation in order to draw conclusions about processes taking place in the ISM. In general, models agree that the fundamental reason for the existence of the correlation is massive SF that is ultimately responsible for both the dust heating as well as for CR acceleration and thus synchrotron emission.

A crucial question for our understanding of the FIR-radio correlation is whether galaxies are optically thick ("calorimeters") to their UV radiation and to the energy of their CRE, or optically thin. In the calorimeter case it can be shown that the massive SFR is basically the only parameter, and that the dust opacity and the magnetic field play only a very minor role [39, 19]. In this case the tightness of the correlation can thus be easily understood. In the optically thin case, models relating dust opacity and the CRE escape length [13] or the gas density and magnetic field [27] have been proposed.

A calorimeter situation is certainly present in dense starburst galaxies where the energy loss time-scale of CREs becomes very short [38]. However, in this regime both gas density, the energy density of the radiation field and most likely the magnetic field is much increased [38, 28]. This has several consequences: (i) Secondary electrons and positrons, which are the result of the decay of pions that were produced in collisions from CR protons with the interstellar gas protons, contribute considerably to the RC emission. (ii) Energy losses due to the inverse Compton effect, ionization losses and bremsstrahlung dominate over the synchrotron losses, producing a "synchrotron dimming". Lacki et al. [28] present a model taking these effects into account. They show that all these effects are proportional to the gas surface density and that for a certain choice of the parameters, the different effects compensate and produce a rather constant FIR-to-radio ratio.

On the other side of the luminosity spectrum, dwarf galaxies are certainly no calorimeters because they are characterized by low dust opacities and CR escape leading to a suppression of both the FIR and the RC emission compared to other SF tracers [2]. To which extent the decrease of both emissions "conspires" [2] to preserve the FIR-radio correlation is unclear. The model of Lacki et al. [29] predicts a break-down of the correlation for low surface brightness dwarfs who are expected to be radio synchrotron weak. A firm observational confirmation of this prediction is still missing, partly due to the difficulty to separate synchrotron and thermal radio emission.

At which luminosities the transition between optically thin and optically thick takes place exactly is still uncertain, and in particular the question of whether normal spiral galaxies are calorimeters or not. The only way to answer this question is via radio spectral index maps of the galactic halos because a steepening of the spectral index away from the disk is a unique sign that energy losses dominate over escape [20]. So far, the sensitivity of the radio data only allows this measurement in bright starburst galaxies and has generally shown a steepening of the spectral index [9, 10, 14]. Measurement for normal spiral galaxies will only be possible with the SKA.

All these models have been developed for nearby galaxies. From theoretical considerations, the FIR-radio correlation is expected to break down at high redshifts. The main reason is the increase of the energy density of the Cosmic Microwave Background which leads to an increase of inverse Compton losses compared to synchrotron losses and therefore to an increasing radio synchrotron dimming [26, 29]. The break-down of the correlation is expected to occur first in galaxies with a moderate SFR (at $z \sim 2$) and later in starburst galaxies (at $z \sim 10-20$) if the general morphologies of the galaxies are similar to local ones [29].

## 4 The need for the SKA

In order to make progress in our understanding of the correlation, observational data of higher sensitivity are required which can only be provided by the SKA. Already in the first phase, SKA1-mid will provide an improvement of a factor of 5 in sensitivity and at the same time an improvement of a factor of $\sim 5$ in spatial resolution compared to the JVLA which is the most powerful instrument at present. The large band width of 500 - 700 MHz in the GHz range making an in-band spectral index determination possible is an additional feature that will help to solve the open questions. making galaxy surveys very efficient. At full performance, the SKA is expected to improve a further factor 10-20 in sensitivity in the GHz frequency range. The data that can be acquired with the SKA1 (and later SKA) will help us to answer the following questions:

- Does the FIR-RC correlation hold in dwarf galaxies with the same low scatter as in spirals and starbursts? If yes, this would require a fine-tuning between parameters relating the dust emission and the CR escape. So far, only a modest amount of RC data for dwarf galaxies is available and indicate a lack of RC emission compared to the FIR. Data for a larger sample are needed in order to further quantify this trend and draw statistical conclusions. The largest sample of 26 galaxies observed by Kitchener et al. [17] contains galaxies out to a distance of up to 10 Mpc with total SFRs down to $10^{-3}$ to $10^{-4} M_{\odot}$ yr$^{-1}$. The improvement in sensitivity of a factor of 5 of SKA-mid compared to the JVLA will allow us to observe similar galaxies as in this survey with the same observing time out to about twice the distance with about twice the linear spatial resolution. This will allow us to obtain high-quality data for a sample of about 10 times more galaxies, i.e. hundreds of objects, in the same observing time.
- What are the properties of radio halos of galaxies? Are they loss-dominated or escape-dominated? We need to know this in order to be able to understand the FIR-to-radio correlation and in particular, find out whether there is a direct relation between RC emission and the SFR which is assumed when using the RC emission as a SF tracer. Radio halos are intrinsically faint and extended which makes their observation difficult. They are most likely powered by star formation [9, 10] so that a higher SFR can power larger and brighter halos. With the current instruments, radio halos have only been detected in a handful of starburst galaxies, with NGC 253 [14] (SFR of 5 $M_{\odot}$ yr$^{-1}$) being one of the least luminous objects observed. The large field of view, high sensitivity, high resolution (necessary to resolve also smaller halos) and multi-frequency capacity of SKA1 will allow such studies in galaxies similar to the Milky Way (SFR $\approx 1$ $M_{\odot}$ yr$^{-1}$) in a few hours.
- Does the FIR-radio correlation hold for galaxies at high redshift? The correlation is expected to break down at some point and the precise value of the redshift where this happens will allow us to draw conclusions about the other energy losses that CRE suffer apart from invers-Compton losses. Furthermore, we need to know the range of validity of the FIR-radio correlation with redshift in order to limit the range where the RC can be used as a SF indicator. Surveys up to date, like VLA-Cosmos [34] which achieved in 275 hours of observations a rms of 17 $\mu$Jy at a resolution of 2.5$''$ have been able to detect galaxies out to $z \approx 1 - 2$ with its field of 2 deg$^2$. The instrument SKA1-mid is able to observe in the same time the same area down to a 25 times lower rms (0.7 $\mu$Jy) at a high spatial resolution ($\sim 0.2''$), which is important to avoid confusion. Following the estimation of Murphy et al. [26] (his Fig. 3), adapted to the sensitivity of SKA1-mid, we find that this instrument will be able to detect Luminous Infrared Galaxies (LIRGs, L $\geq 10^{11}$ L$_\odot$) out to $z \sim 1.5$ and Ultraluminous Infrared Galaxies (ULIRGs, L $\geq 10^{12}$ L$_\odot$) out to $z \sim 3$. With the factor 10 improvement expected in the future, the final SKA survey instrument will be able to detect ULIRGs at all redshifts and LIRGS out to $z \sim 3$.

## 5 Summary and conclusions

Observations with the SKA will allow to answer open questions about the FIR-radio correlation that are beyond the reach of present instruments. They will improve our knowledge about the CR propagation and properties of the ISM (magnetic field, gas density, radiation field) and will allow us to reliably calibrate the RC emission as an extinction-free SF indicator. Such an extinction-free SF indicator is highly needed, especially when studying the high-z universe which will be an increasingly accessible topic in the future.

## Acknowledgments

UL acknowledges support by the research projects AYA2011-24728 from the Spanish Ministerio de Ciencia y Educación and the Junta de Andalucía (Spain) grants FQM108. MAPT and AA acknowledge support from the Spanish Ministerio de Economía y Competitividad (MINECO) through grant AYA2012-38491-C02-02. AAH acknowledges support from the Spanish Plan Nacional de Astronomia y Astrofisica under grant AYA2012-31447, and LC from grant LC from grant AYA2012-32295.

# How is star formation fed and quenched in massive galaxies at high redshift?

**P. G. Pérez-González[1], J.I. González-Serrano[2], and A. Fernández-Soto[2,3]**

[1] Departamento de Astrofísica, Facultad de CC. Físicas, Universidad Complutense de Madrid, E-28040 Madrid, Spain
[2] Instituto de Física de Cantabria (Universidad de Cantabria-CSIC), E-39005 Santander, Spain
[3] Unidad Asociada Observatorio Astronómico (Universitat de València-IFCA), E-46980 Paterna, Spain

## Abstract

Observations of the location and kinematics of the atomic gas (HI) and the continuum radio emission from high redshift galaxies would mean a huge step forward in our understanding of galaxy evolution. We now have a secure global picture of the stellar content of massive galaxies and their precursors up to z∼4. But we still have to understand why star formation in these systems started early and quenched some time after, a scenario known as downsizing which, at face value, conflicts with the predictions from the current hierarchical galaxy formation paradigm. SKA will provide the missing piece to solve the puzzle: information about the amounts of gas falling into galaxies to form stars, as well as data to measure when and how the star formation turns off as the gas stops cooling due to still to be understood feedback mechanisms, such as (radio mode) obscured nuclear activity.

## 1 Introduction

Looking at early-type galaxies (ETGs), as well as at the bulges/halos of spiral galaxies such as the Milky Way, we are able to find some of the oldest stars known. These stars can be over 14 Gyr old, so they formed when the Universe was very young. In fact, ETGs not only typically harbor the oldest stellar populations, but they also present a variety of metallicities (from very low to quite high) and $\alpha$-element enhancement, pointing out to early and also rapid starburst events that formed a significant fraction of their stellar mass (see, e.g., [54, 25, 39]).

The most probable progenitors of nearby ETGs and bulges of spirals have been identified in the last 15 years out to redshifts of z∼4 (12 Gyr ago). Indeed, based on optical and especially near-infrared surveys of galaxies, we have detected a numerous population of distant massive galaxies [65, 24, 17, 47, 10]. A good fraction of them (about half) are already dead at z∼2 [48, 23], but also many are experiencing very intense bursts of star formation, with SFRs of hundreds or even thousands of solar masses per year, and large amounts of dust [56]. In fact, about half of the stars that we see today were formed in the first 6 Gyr of cosmic time, and a fair fraction of the most massive galaxies ($M>10^{11}$ $M_\odot$) were already in place and evolving passively (with some further merging events, mostly dry) by z$\sim$1, with the bulk of the star formation in the Universe progressing to less massive systems as we move to lower redshifts [51, 41, 44, 37]. The epoch when the most massive galaxies were forming the bulk of their stellar content is then z=1–4. This is a formation scenario known as *downsizing* [15, 33, 28, 5, 50, 2, 51].

The observational evidence for early and rapid assembly of stars in ETGs is now very compelling. But the theoretical explanation of this behavior is not well understood, and modelers have difficulties reproducing the number density and the properties of high redshift (dead and alive) massive galaxies [41, 31]. This is also an observational problem: we do not have data to characterize in detail and with certainty relevant properties of this early assembly such as the timescale of the star formation and the amount of gas available to transform into stars, as well as its physical properties (i.e., whether the fuel can actually feed star formation), as a function of time. Very remarkably, the current paradigm of galaxy formation is hierarchical, with baryons following the behavior of the CDM halos, which merge and evolve from the smallest to the largest scales as we advance in cosmic time ([63]; see also [6, 61, 11, 62, 55]). The downsizing formation scenario is, in some way, antihierarchical (large galaxies assemble first), and that is nowadays a major problem in our understanding of galaxy formation.

The difficulty found to simulate the formation of galaxies, especially in the case of the most massive systems, is twofold. First, the physics governing the collapse, recombination, and cooling of ionized and atomic gas (the main form of baryonic matter in the early Universe) to form molecular gas and then turn this into stars is still poorly constrained. And second, the fuel consumption while forming stars must stop for some reason (still to be fully understood), i.e., there must be some negative feedback mechanism(s), given that we see massive galaxies stop forming stars and start evolving passively at high redshift, but we know that there are still large amounts of gas available in the intergalactic medium.

The most advanced simulations for the formation of galaxies now typically introduce (more or less *ad hoc*) some kind of feedback mechanism to regulate star formation (see, e.g., [45, 36]). For the most massive galaxies, the main suspects for the job are Active Galactic Nuclei (AGN). Moreover, many simulations now identify a *radio mode* as the most probable form of nuclear activity responsible for the quenching of the star formation, as well as for the dissemination of metals far into the inter-galactic medium, tens or hundreds of kpc away, the typical size of radio jets [57, 16]. Indeed, from the observational point of view, we have identified several evidences of this interplay between nuclear activity and star formation in galactic scales, e.g.: the intra-cluster medium in the local Universe seems to be shocked and heated by AGN radio jets [22]; the AGN activity peaks in the same epoch when the Universe was forming stars most efficiently [50, 34, 32]; AGN evolution models reproducing, among other observables, the X-ray background, predict many more obscured AGN than we have actually detected [27]; and, last but not least, many massive galaxies at high redshift present X-ray emission and/or dust properties compatible with the presence of intense heating sources such as dust-enshrouded AGN [20, 21, 18].

In this context, SKA will be a key facility to understand how gas feeds star formation in massive galaxies as well as why the job is abruptly stopped. It will provide two of the key pieces, probably the most relevant, to solve the puzzle of galaxy evolution: how much atomic gas (HI) falls into the center of the DM halos to form stars and the dynamics of this accretion, and why that cooling is stopped in spite of the availability of more gas in the IGM.

In this white paper we discuss an experiment to understand the earliest phases in the formation of massive galaxies by studying the amounts of atomic gas in the outskirts of high-z massive galaxies, as well as its dynamics, in a sample of massive galaxies in various stages of star formation (and probably nuclear) activity at $1<z<4$, when they formed the bulk of their stars. We also present a scientific case for the study of feedback mechanisms linked to the activity of obscured AGN presenting intense (and even not so intense) radio emission.

## 2 Atomic hydrogen at z>1

SKA will provide excellent data to understand the formation of massive galaxies from the earliest stages of their evolution. In Figure 1 we show how SKA1-MID and SKA2 surveys will be able to detect Damped Lyman-$\alpha$ systems up to $z\sim3$. These observations would tell us how much gas is falling into galaxies at the peak of their star formation history, z=1–3, when the Universe was most efficient forming stars. SKA would also allow to study the kinematics of the gas, so we can understand how gas stops collapsing into the center of the dark matter halo, eventually resulting in the quenching of the star formation. We want to emphasize that these observations would provide direct measurements of the effects of (supernova and/or AGN) feedback. We still have very limited understanding about feedback since, up to now, we have only been able to study its effects for nearby systems (mostly in clusters) and a few high redshift sources, based mainly on the study of the kinematics of ionized gas, as well as on the X-ray emission from hot gas in clusters. SKA will allow us to characterize the dynamics of the main component of intergalactic gas feeding star formation in galaxies.

In Figure 1 we also show the expected HI column densities for the compact star-forming galaxies (also called *nuggets*) in [3]. These galaxies have been identified as the progenitors of the compact quiescent massive galaxies at $z\sim2$, which are believed to be the cores of nearby elliptical galaxies [35]. The compact star-forming nuggets exhibit large SFRs, very compact sizes, and many of them have X-ray emitting AGN, being in the way of quenching their star formation and starting evolving passively [4, 49]. Even if their HI content were relatively small and similar to nearby quenching early-type galaxies, $M(HI)=2–40\times10^8$ $M_\odot$ (a very conservative assumption, since they are forming the bulk of their stars and probably have larger gas contents than nearby galaxies), they would be readily detectable by SKA in just 100 hours.

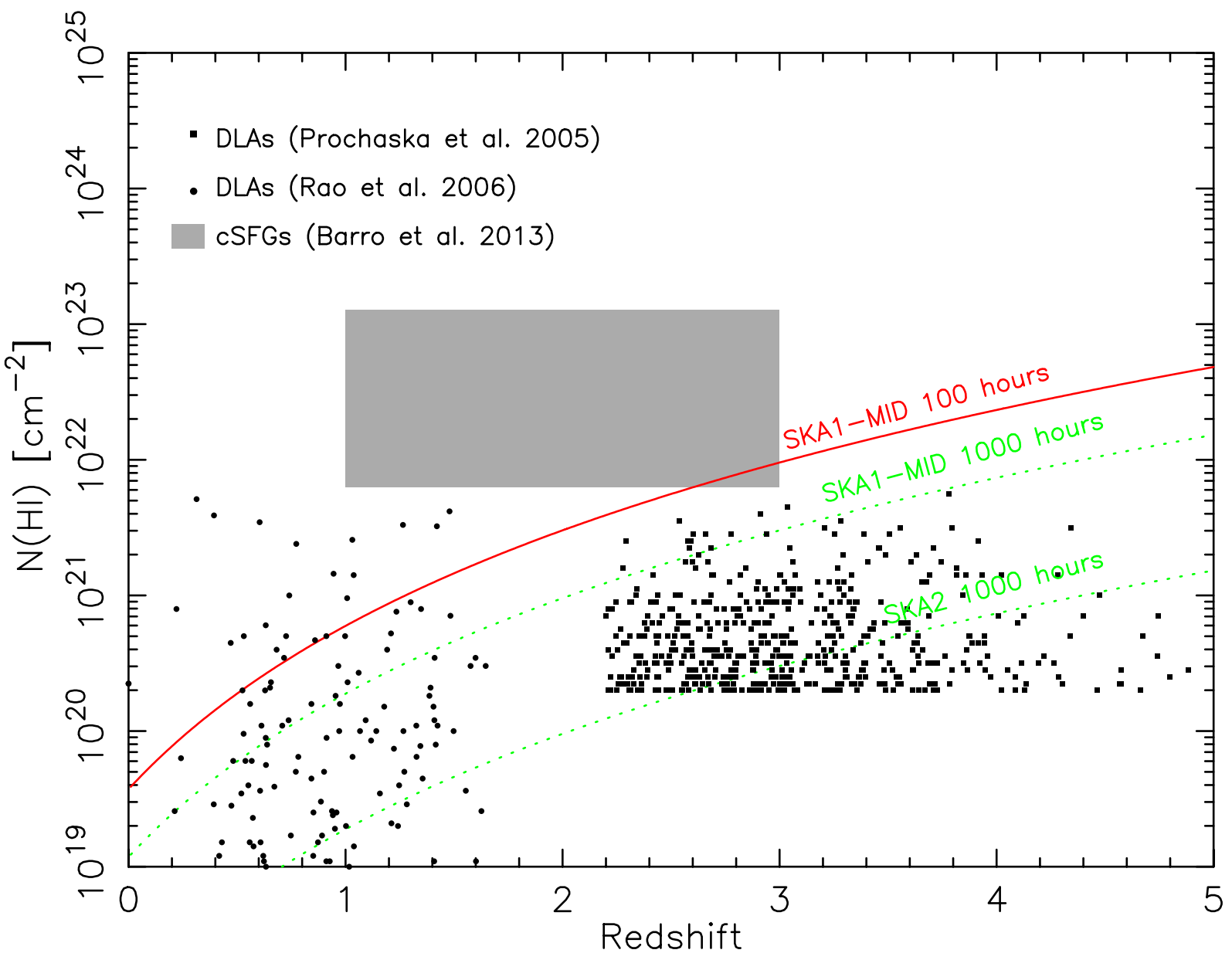

Figure 1: Sensitivity of the SKA in terms of the detection of neutral hydrogen at high redshift. We depict detection limits for HI 21 cm line surveys integrating for 100 and 1000 hours in the SKA1-MID configuration, and 100 hours with SKA2. We plot the column densities for the z>2 DLAs in [52], as well as for the z<1.65 DLAs in [53]. The shaded gray region corresponds to the HI column density expectations for the 1<z<3 compact star-forming nuggets in [3]. We have assumed that their typical size is 2 kpc, they would present HI masses similar to those for quenching nearby early-type galaxies in [64], and the density would be constant throughout their optical extent. Given that the high-z star-forming nuggets are forming the bulk of their stars and they are quite massive, probably they will present larger HI contents and the SKA observations will reveal HI gas far beyond the zone dominated by stellar emission, allowing the most detailed characterization of the gas in-fall to feed star formation in high-z galaxies.

## 3 AGN-star formation separation

The radio sky is dominated by extragalactic sources at metre to cm wavelengths. At flux densities above 1 mJy (at 1.4 GHz) most are AGNs while radio sources associated with starburst galaxies dominate the source counts below that flux [14, 26, 19]. However, results by Gruppioni et al. [29] indicate that early-type galaxies at $z > 1$ dominate at sub-mJy levels. It has also been suggested that the flattening of radio counts below 1 mJy may be caused by radio-quiet quasars and type 2 AGNs [59].

It is crucial to disentangle the star-forming galaxies from the AGNs, although the expected fraction of AGN vs. star-forming galaxies is expected to be $\sim 17\%$ at $z < 5$ below the 10$\mu$Jy level, and $\sim 7\%$ at the 1$\mu$Jy level [38]. A diagnostic tool to distinguish between AGN and star formation is the known correlation between FIR and radio continuum emission from star-forming galaxies (e.g. [12]). A criterion is that galaxies with radio to far-IR flux ratio more than three times higher than the mean for star-forming galaxies are classified as AGN [13], although Mauch and Salder have found a disagreement at the $\sim 10\%$ level between spectroscopic and radio/FIR classification [42].

Optical spectra can be used to identify the origin of the radio emission. However, optical AGN spectra do not always imply that the radio emission is of nuclear origin. In fact, there is evidence that nuclear activity and star formation are connected. The widespread detection of blueshifted rest-UV ionic absorption lines demonstrates that most star-forming galaxies (SFGs) at $z \sim 1-3$ exhibit nuclear AGN-driven gas outflows (e.g., [58]). Multiwavelength SED fitting (e.g. [1]) and X-ray and infrared information are useful to discriminate AGNs from star-forming galaxies. For instance, using three bands from WISE ($3.4, 4.6$, and $12\mu$m) it is possible to separate AGNs from ULIRGS and normal galaxies [40]. Other methods are based on the finding that the NUV-NIR SEDs of galaxies are well correlated with line strengths and hence with their position on the BPT diagram [60].

Thermal and synchrotron self-absorbed spectra are expected to show different properties. Observations covering the range $\sim 1-10$ GHz will be useful to discriminate thermal and non-thermal radio emission. A flat spectrum up to high frequency is evidence of thermal emission, while a turnover around a GHz supports synchrotron radiation from an AGN. The broad-band radio spectrum may be combined with polarization information in order to study the nature of the radio emission from different regions. A jet structure is expected to show a polarised steep-spectrum, while a diffuse star-forming region should be highly depolarized (see [46]). With 3-4 measurements in the spectral range $1-10$ GHz at a resolution of $\sim 100$ MHz it should be possible to discriminate AGN from star-forming galaxies, reaching flux densities levels of around $10-20\mu$Jy in one hour. High resolution radio observations can be used to distinguish between compact AGN cores and inner jets from $\sim 1-10$ kpc-scale star-forming disks [43, 7, 30, 9]. Resolutions of at least 0.05 arcsec at 1.4 GHz (i.e. 1000 km baselines) are necessary to separate individual sources. However, to image sources with sufficient detail to determine if the emission comes from the disk, star formation, or an AGN will require higher resolutions.

Radio morphological identification can be also useful to reveal obscured AGN activity otherwise undetected at other wavelengths [8].

## 4 Synergies with other facilities

The observations put forward in the previous sections should be accompanied by observations in separate wavelengths taken at different facilities. The main thrust of the science here presented is the analysis of the neutral hydrogen component associated to moderate-to-high redshift normal galaxies, its dynamics and its relationship to the star formation and galaxy evolution processes.

In order to complete this objective it will be important to add other pieces of information. Many of them are treated in detail in other chapters of this work, but we will only mention here some of the most obvious:

- VLT, GTC: Optical/near infrared imaging and, specially, spectroscopy of the observed sources will be crucial to determine the properties of the galaxies being observed. Whereas we cannot realistically expect to study the same galactic components that will be measured by SKA, because their surface densities and emission levels will be too low, we can study the metallicities and star formation histories of the galaxies that are receiving (or, in the case of very strong winds or AGN activity, expelling) the HI flows that will be observed in the SKA data. VLT will be perfectly suited in terms of sky coverage, while GTC will be able to observe areas close to the celestial equator–there is still a significant areal overlap between both facilities.

- ALMA: Observations in other radio wavelengths in the millimeter range will allow for the analysis of abundances of other molecular species, unreachable to SKA. This study will contribute to our knowledge of the physical state of the neutral gas (in particular its temperature), and also to analyse its metallicity and, from a different point of view, its dynamics. A wider coverage of the radio spectrum should also become useful for the separation of the star formation and AGN components.

- Space: By the time SKA will be productive, we expect JWST to be the most important tool for the kind of science we are presenting here. In particular, observations with the MIRI instrument will cover the wavelength range 5-25 $\mu$m, both in imaging and spectroscopy. At the redshifts we intend to study, this will fully cover the rest-frame near infrared bands, allowing for detailed measurements of the stellar mass and adding critical data to analyse the past star formation history of the surveyed galaxies.

## 5 Acknowledgements

PGP-G acknowledges support from Spanish Government MINECO AYA2012-31277 Grant. JIGS acknowledges support from the Spanish Ministry for Economy and Competitiveness AYA2011-29517-C03-02. AFS acknowledges support from the Spanish Ministry for Economy and Competitiveness and FEDER funds through grants AYA2010-22111-C03-02 and AYA2013-48623-C2-2, and Generalitat Valenciana project PROMETEOII/2014/060.

# Nearby Normal and Luminous Infrared Galaxies with the SKA

**M.A. Pérez-Torres[1,2], S. García-Burillo[3], A. Alberdi[1], A. Alonso-Herrero[4], L. Colina[5], A. Prieto[6], J. Knapen[6] and J. A. Acosta Pulido[6]**

[1]Instituto de Astrofísica de Andalucía (IAA-CSIC), 18008 Granada, Spain
[2]Centro de Estudios de la Física del Cosmos de Aragón (CEFCA), 44001 Teruel, Spain
[3]Observatorio Astronómico Nacional (OAN), Alfonso XII, 3, 28014 Madrid, Spain
[4]Instituto de Física de Cantabria (CSIC-UC), 39005 Santander, Spain
[5]Centro de Astrobiología (CSIC-INTA), Torrejón de Ardoz, 28850 Madrid, Spain
[6]Instituto de Astrofísica de Canarias (IAC), 38200 La Laguna, Spain

## Abstract

The SKA will routinely provide $\mu$Jy sensitivity and sub-arcsecond angular resolutions at radio wavelengths. Planned SKA surveys will image vast numbers of nearby galaxies, which are expected to provide a cornerstone in our understanding of star-formation and accretion activity in the local Universe. Here, we outline some of the key continuum and molecular line studies of local galaxies, where the SKA will have a significant scientific impact and where the Spanish astrophysical community is particularly active.

## 1 Introduction

Star-formation (SF) and accretion onto a supermassive black hole are the two key processes in the evolution of galaxies in our Universe. Further, the SF history within individual galaxies still remains a crucial physical parameter that observations are only now beginning to accurately characterise. Radio observations provide by far some of the best diagnostics of these two processes, allowing a direct view of SF even in dusty environments, as well as the detection of AGN and the measurement of their accretion rate at bolometric luminosities far below anything detectable at higher energies. The high sensitivity, resolution and imaging fidelity capabilities of SKA, will ensure that the SKA will become a dominant instrument for the detailed study of nearby galaxies over the next decades. In this chapter, we outline the scientific motivation for undertaking large surveys and deep observations of galaxies in the local Universe across all available frequency bands of SKA, both continuum and spectral line studies.

# 2 Radio continuum as a star-forming tracer in nearby galaxies

—*Thermal and non-thermal radio emission as star-formation tracers*

Both continuum thermal free-free radio emission and synchrotron radio emission can be considered as SF tracers. Thus if the presence of an AGN can be excluded (or its contribution neglected), one could then obtain independent estimates of the SF rate, and thus check the linearity of the RC(thermal)–FIR and RC(sync)–FIR correlations. RC and FIR emission both depend on (recent) SF and will thus be correlated. Surprisingly, this RC–FIR relation of galaxies holds over 4 orders of magnitude in luminosity, irrespective of galaxy type [29, 18, 7] and has been observed to hold out to a redshift of about 3 [25, 3]. A well calibrated RC–FIR relation offers a powerful method to probe the cosmic SFR out to intermediate redshifts, initially with SKA pathfinders and precursors, and eventually with the SKA. For further details, we refer the reader to the chapter by Lisenfeld et al. (this book), which is exclusively devoted to discuss the RC–FIR correlation. However, we note that a well calibrated RC-FIR relation is an essential prerequisite for deep extragalactic surveys with the SKA.

An alternative to estimating SFR from the FIR, which at best provides modest angular resolution and relies on the availability of suitable satellites, one can use the RC to determine directly the current SFR in galaxies [16]. The thermal and non–thermal emission are the result of fundamentally different processes, with different RC–SFR relations expected for each component. The thermal RC, due to the ionized flux from massive stars, is expected to be directly proportional to the SFR. This makes it an ideal, virtually extinction–free proxy for SF [40]. The non–thermal RC depends on the magnetic field strength as well as the cosmic–ray energy density, unless one assumes an electron calorimeter which is unlikely, particularly for dwarf galaxies, or in general those galaxies with large–scale outflows. Usually one assumes energy equipartition between the CRe and the magnetic field, so that the RC–SFR relation is closely connected to a relation between the magnetic field and gas.

SKA1–MID will transform what can be achieved. Rather than an essentially monochromatic information, observations of about 1 hr per band will sample the entire radio continuum spectrum between 1.67 and 10 GHz (about 4 hr total per target), down to $\mu$Jy sensitivity. This will allow to put the above relations on a much more robust footing, as we will be able to separate off the thermal contribution, leaving just the non–thermal fraction. The spatially resolved spectral index distribution will come within reach, thus providing a lever on the propagation and aging of CRe as they diffuse from their sites of origin. At the $\mu$Jy level of sensitivity, we will be able to trace all star formation activity in local galaxies down to star formation rates as low as (see, e.g., equation 6 in [6])

$$SFR \lesssim 6.62 \times 10^{-4} \left(\frac{S_{1.4\mathrm{GHz}}}{\mu Jy}\right) \left(\frac{D}{100\,\mathrm{Mpc}}\right)^2 \mathrm{M}_\odot\,\mathrm{yr}^{-1}$$

—*The constituent parts of local galaxies*

Sensitive and critically high angular resolution radio images of nearby galaxies such as will be provided by the SKA1-MID ( $\lesssim 0\farcs22$) and by the complete SKA ( $\lesssim 0\farcs01$), provide one method by which observations can directly probe SF in a way which is independent of complex physical emission mechanisms. Whereas lower resolution radio observations of normal and star-forming galaxies trace the diffuse radio emission, sub-arcsecond angular resolution observations are required to systematically characterise the populations of individual compact SF products on a galaxy by galaxy basis by resolving away the diffuse emission. This population census can hence be used to directly infer the levels of SF.

Observed with $<0\rlap{.}''22$ angular resolution, each individual galaxy can be considered as a laboratory containing a large sample of discrete radio sources, all at essentially the same distance, which can be studied in a systematic way. At $\mu$Jy and sub-$\mu$Jy sensitivities and between frequencies of 1 and 7 GHz this source population, with the exception of accretion dominated objects and in particular AGN, will consist exclusively of sources related to various key phases of the stellar evolutionary sequence. This population will be a mixture of sources from the early stages of SF, such as compact HII regions, through SSCs, and stellar end-points like X-ray binaries, planetary nebulae, SNe [48] and their SNR.

By first detecting and then identifying the physical nature of these objects using a combination of radio morphologies and spectral indices, alongside extensive multi-wavelength ancillary data, high angular resolution radio observations will provide the first detailed extinction-free census of SF products within nearby galaxies. The majority of core-collapse SNe evolve to form long-lived radio SNR, hence this statistically well-constrained census combined with information regarding the sizes and hence canonical ages of SNR (for the nearest galaxies observed with sub-arcsecond resolution available in higher bands with SKA1–MID), can be used to directly infer levels of SF in individual galaxies. This will provide a further obscuration-independent SFR tracer and, because it detects massive stars which form CCSNe, will preferentially probe the top part of the IMF. When combined with other SFR measures such as IR/UV or global RC free-free and synchrotron emission (e.g., the case of the circumnuclear region of NGC 1614; see [30] and Fig.1), this will provide constraints on the universality of the IMF as a function of galaxy type, evolution and environment within the local volume.

Importantly, such a local galaxy radio survey will also identify populations of sources that trace earlier stages in stellar evolution, such as HII regions and SSCs, placing useful constraints on the levels of SF at various phases in the evolution of individual galaxies. When compared with other wavelength tracers, which probe different ranges of SF age and different spatial regions, these radio diagnostics will provide significant new insights.

Whilst this will be achievable on a galaxy-by-galaxy basis, the power and importance of an SKA survey of nearby galaxies arises from its large size and the available complementary multi-wavelength data-sets. By combining these direct radio tracers of SF products, with other multi-wavelength SF proxies (e.g. IR), significant constraints will be placed on their calibration and interpretation, with important implications across a wide range of observational astrophysics. Large-area SKA1 continuum surveys with both SKA-MID and SKA-SUR with sensitivities of $\sim 4\mu$Jy will be capable of detecting all local galaxies thus spanning a complete range of types and levels of both historical and ongoing SF, and allowing this census of SF products to be applied over the wide range of luminosity and environment parameter space inhabited by galaxies. Such a survey will provide the radio benchmark for studies of local galaxies with application to all observational astronomers.

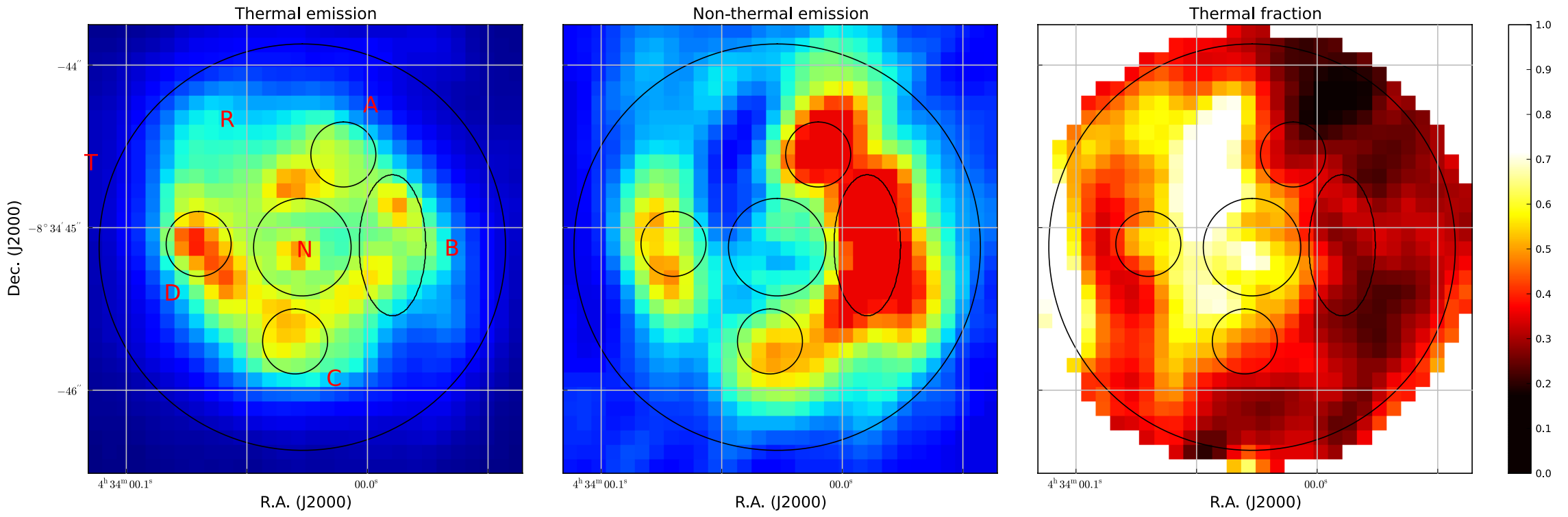

Figure 1: Decomposition of the 3.6 cm flux for the central kpc of the luminous infrared galaxy NGC 1614 into thermal (left) and non-thermal (middle) radio emission. The right panel shows the relative contribution of the thermal emission. Note that regions A and B are dominated by synchrotron non-thermal emission, in contrast with regions C and D. (Taken from [30].)

—*Resolving extragalactic star forming regions in clusters with the SKA*

A sizable fraction of the emission from young star forming clusters comes in the radio domain. This emission mainly arises in SNR of thermal or non-thermal nature and bresmtrahlung due to the cooling gas. In the extragalactic domain, the characterization of young star forming regions is difficult in the radio domain because of limited angular resolution and sensitivity, particularly at the lowest radio frequencies. However, the key advantage of studying starformation in radio is the transparency to dust, contrarily to the studies in the UV, optical and even in the near-IR where suitable angular resolutions are at reach but dust is the major drawback. Yet, with the present instrumentation, detecting these regions and spatially resolving them into their building blocks, the star clusters, are possibilities limited to the nearest, brightest starburst galaxies using facilities as the VLA in its larger configuration, e.g. NGC 253 (see Fig. 2 and [19]).

The large collective area and angular resolution of SKA will allow us to individually characterize these stellar building blocks not only in prototype objects like NGC 253, but also in most star forming regions in galaxies in the local Universe. Detailed mapping of extragalactic young star clusters is accessible today from the UV- to the optical- bands with the HST, in the near-IR band using Adaptive Optics in 8-10 m class telescopes, and in the millimetre/submillimetre with ALMA. The SKA will provide the complementary information in the cm range at comparable scales to those provided by all the above facilities. The ensemble of all these data will allow us to produce extremely well sampled spectral energy distributions. Modelling of these SEDs will deliver a more consistent picture of the onset and evolution of star formation in a wide range of environments within a given galaxy - e.g. star formation in the disk vs. that in the spiral arms - and between galaxies of different type and activity level (e.g. [41]).

—*From local U/LIRGs to high-redshift star-forming galaxies*

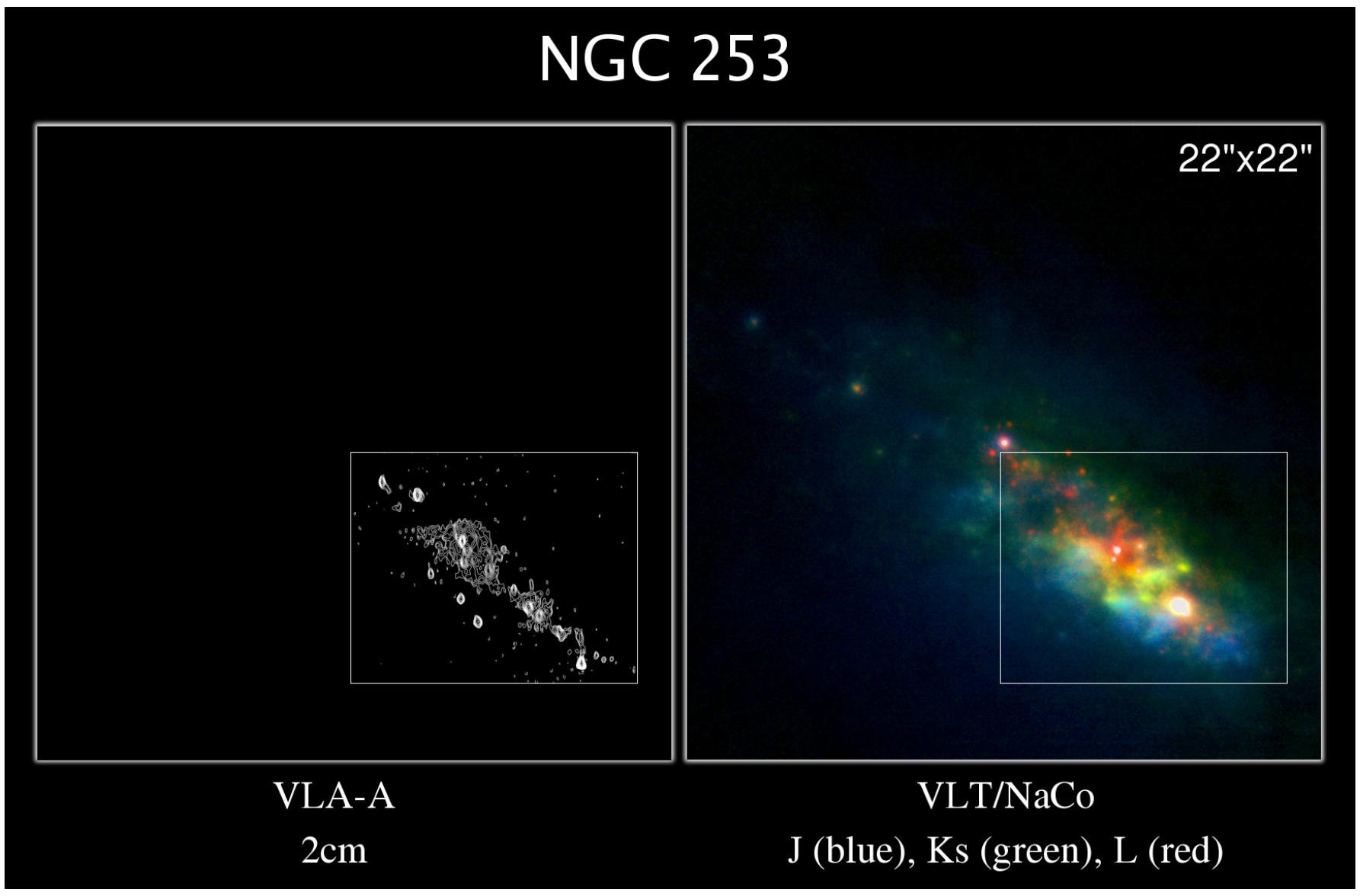

Figure 2: The central 150 pc of the nearest starburst NGC 253 (D=3.9 Mpc) spatially resolved into stellar clusters. The right panel shows a color composite VLT adaptive image taken at 1.2 $\mu$m, 2.2$\mu$m, and 4.6 $\mu$m. The inside square resolves into 37 young clusters, each with size of 3 pc (adapted from [19]). The left panel shows the same region at 2 cm collected with the VLA-A. Sensitivity limits restrict the detection to $\sim$15 clusters, all with a counterpart in the near-IR VLT image (adapted from [54].

Local ($z \lesssim 0.3$) Luminous Infrared Galaxies (LIRGs) are thought to be nearby scaled-down versions of high-z star-forming galaxies. Since a major science goal for the SKA and its pathfinders is the study of SF across cosmic time, it is crucial to i) have a detailed and accurate knowledge of local star-forming galaxies, and ii) test the radio-infrared (radio-IR) relation. Since radio emission is a dust-unbiased SF tracer, an accurate calibration of the radio-IR relation will be needed to determine the SFRs at high-z. Indeed, at $z \geq 1$, 1 arcsec corresponds to 8 kpc, so that disentangling AGN from star-forming activity is challenging, unless angular resolutions better than $\sim 0\farcs1$ are provided, so that one starts to separate a putative AGN from a compact starburst at essentially any redshift. This capability will be provided by the full SKA.

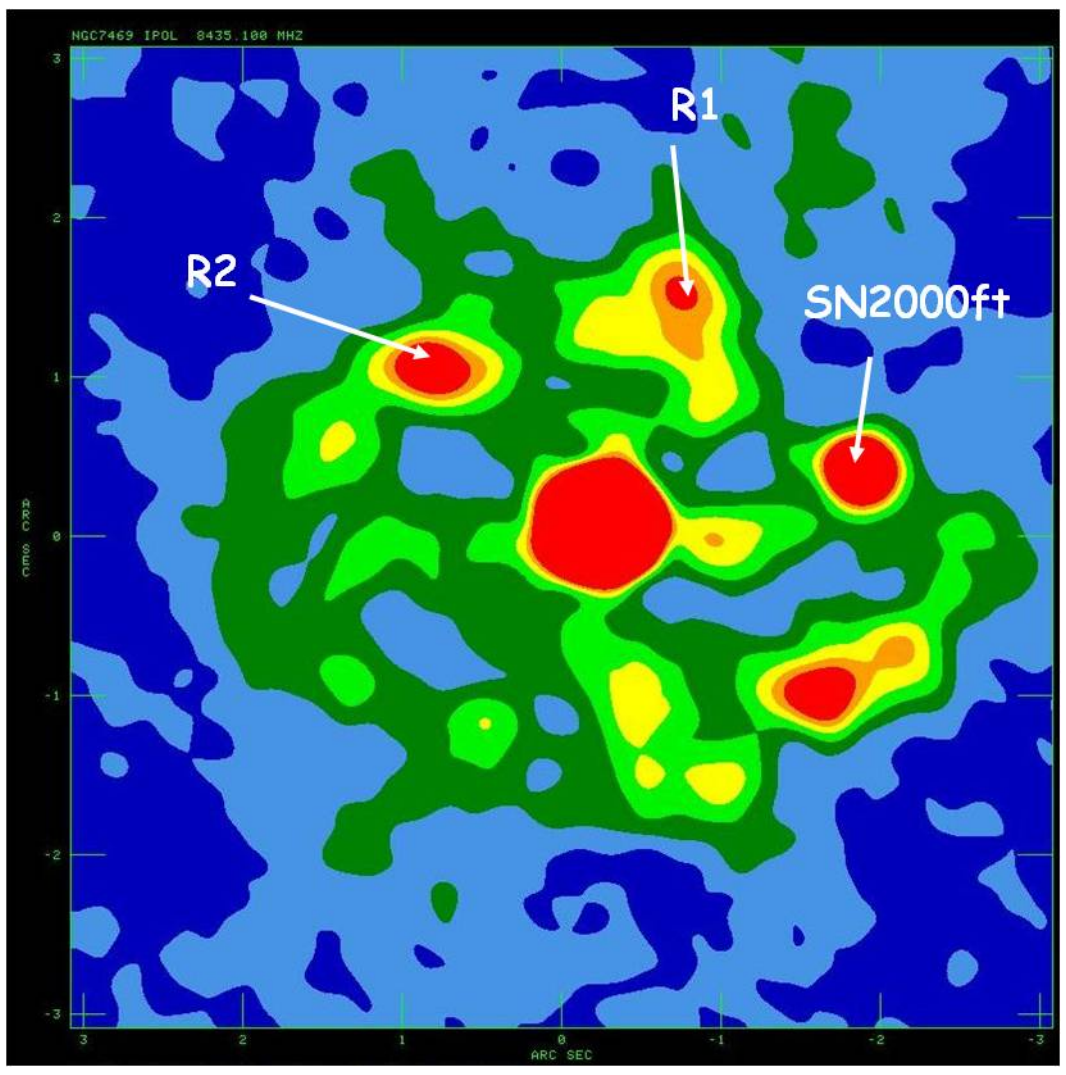

Figure 3: 8.4 GHz VLA image of SN 2000ft in the Luminous Infrared Galaxy NGC 7469, 70 Mpc away from us. The SN was discovered in 27 October 2000, about a few months after the explosion. The SN exploded at a deprojected distance of about 600 pc from the nucleus, whose radio emission is very prominent. Sub-arcsecond angular resolution radio observations are thus a very useful tool to detect extragalactic CCSNe. (Image taken from [1].)

A large fraction of massive SF at both low- and high-$z$ has taken place in (U)LIRGs. Their implied high SFRs are expected to result in CCSN rates a couple of orders of magnitude higher than in normal galaxies. Therefore, a powerful tracer for starburst activity in (U)LIRGs is the detection of CCSNe, since the SFR is directly related to the CCSN rate. However, most SNe occurring in ULIRGs are optically obscured by large amounts of dust in the nuclear starburst environment, and have therefore remained undiscovered by (optical) SN searches. Fortunately, it is possible to discover these CCSNe through high-resolution radio observations, as radio emission is free from extinction effects. Furthermore, CCSNe are expected (unlike thermonuclear SNe) to become strong radio emitters when the SN ejecta interact with the CSM that was ejected by the progenitor star before its explosion as a supernova. Therefore, if (U)LIRGs are starburst-dominated, bright radio SNe are expected to occur. Given their compactness and characteristic radio behavior of radio SNe they can be pinpointed with high-resolution, high-sensitivity radio observations (e.g., SN 2000ft in NGC 7469 [14, 1, 45], see Fig.3; or the supernova factories in Mrk 273 [11], Arp 299 [46, 12], or Arp 220 [43, 5]). However, since (U)LIRGs are likely to have an AGN contribution ([49, 2]) , high-sensitivity, high-resolution radio observations are required to disentangle the nuclear and stellar (mainly from young SNe) contributions to the radio emission, thus probing the mechanisms responsible for the heating of the dust in their (circum-)nuclear regions.

In view of the importance of (U)LIRGs in tracing the SF history across cosmic time, a targeted survey of local (U)LIRGs using SKA1–MID, up to a distance of 100 Mpc, will be essential. The sub-arcsecond angular resolution in band 2/3 will be well-matched to current (J)VLA-A images at higher frequencies, permitting the thermal and non-thermal contribution to be disentangled in the very centres of galaxies. Considering the continuum sensitivity provided by SKA1–MID, as little as 2 minutes per source would be sufficient to produce 1.7 GHz images of a similar depth to those currently provided by the JVLA at 8.4 GHz in 1 hour. A census of all local (U)LIRGs could be obtained in just a few hours with SKA1–MID. Even if the specifications deviate by as much as 30% in terms of sensitivity, this science will not be severely affected. However, baseline lengths of at least 200 km are required, to provide the necessary angular resolution. An SKA-MID survey providing an angular resolution of $0\farcs5$ corresponds to a physically interesting resolving linear scale of $\sim$250 pc at 100 Mpc.

Similarly, SKA1–MID will be a game–changer when it comes to providing a benchmark study for relating the CCSN rate to the SFR in both star-forming and normal spirals, and down to dwarf irregulars. With sensitivities of over an order of magnitude better than current instruments, and sub-arcsecond resolutions providing linear resolution scales of a few tens of pc within nearby galaxies, SKA1–MID will be well matched to spatially separate CCSNe from their surrounding diffuse emission, thus enabling a complete census of radio supernovae, provided the SKA1–MID survey is made of several visits of each field, so that the flux density and spectral variability of each individual source can be ascertained.

The full SKA will reach an angular resolution of about 10 mas at 1.67 GHz, i.e., a 20-fold increase with respect to SKA1. At this angular resolution, it will be possible to locate any individual core-collapse supernova (or supernova remnant) within the nuclear region of any local starburst galaxy, similarly to detailed studies in e.g., Arp 220 and Arp 299 (see Fig. 4, taken from [45, 47]), but with the potential of unveiling the much more numerous, fainter population of radio supernovae and supernova remnants. In turn, this will allow us to test scenarios of SN/CSM-ISM interaction, including estimates of the energy budgets in particles and magnetic fields, and determine the SNR luminosity vs. size relation for essentially all local (U)LIRGs. In addition, we will be able to extend the study described above to essentially all redshifts, as the 10 mas beam at 1.67 GHz will yield spatial resolutions of 80 pc, or better, at all redshifts. The limitation will be dictated by the sensitivity. In fact, even Type IIn supernovae will be realistically detected up to $z \lesssim 0.5$ (3-$\sigma$ detection after 1-hr with SKA1-MID against a source-free background). Even if lensing is taken into account, e.g., a factor of 5 increase, the maximum redshift to which we can expect a detection is $z \lesssim 1.2$. While exciting, the detection of such events will require specific, deep searches, rather than simply making a commensal use of the moderately deep surveys discussed here.

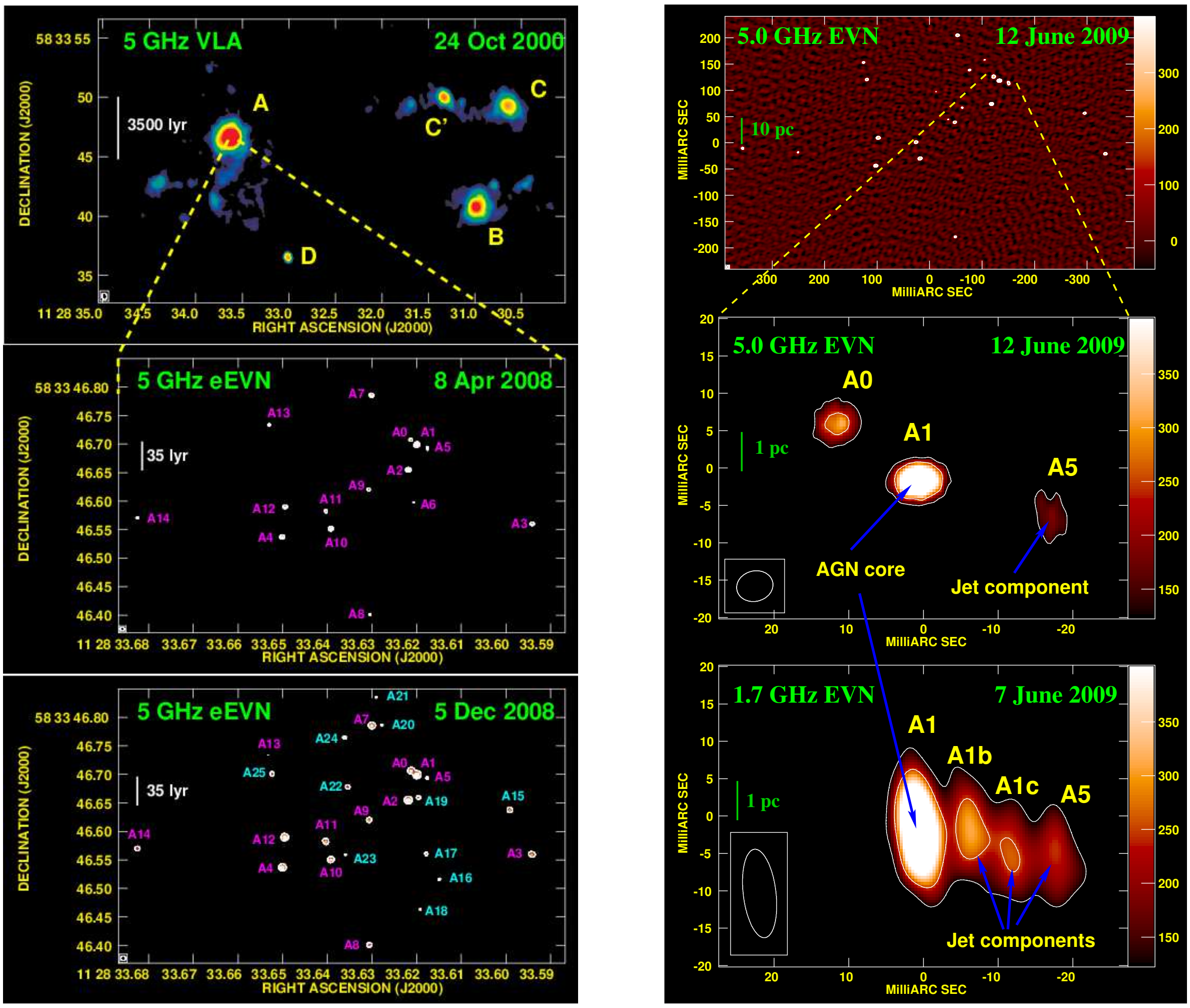

Figure 4: *Top left:* 5 GHz VLA archival observations of Arp 299 on 24 October 2000, displaying the five brightest knots of radio emission in this merging galaxy. *Middle and bottom left:* Contour maps drawn at five times the r.m.s. noise of our 5 GHz eEVN observations of the central 500 light years of the luminous infrared galaxy Arp 299-A on 8 April 2008 and 5 December 2008, revealing a large population of bright, compact, non-thermal emitting sources. To guide the reader's eye, we show in cyan the components detected only at the 5 December 2008 epoch. *Top right:* 5.0 GHz full EVN image of the central 150 parsec region of the luminous infrared galaxy Arp 299-A, displaying a large number of bright, compact, non-thermal emitting sources, mostly identified with young RSNe and SNRs. *Middle and bottom right:* Blow-ups of the inner 8 parsec of the nuclear region of Arp 299-A, as imaged with the full EVN at 1.7 and 5.0 GHz. The core-jet morphology, spectral index and luminosity of the A1–A5 region clearly revealed the location of the long-sought AGN in Arp 299A.

## 3 Molecular line probes of the fueling and feedback of activity

We list below a number of scientific questions in the study of the fueling and the feedback of activity in galaxies that can be addressed by using the capabilities of the SKA to probe the emission and absorption of a set of molecular lines in different environments:

—*Mega-masers:*

Extragalactic mega-masers can trace compact structures in the circumnuclear disks (CND) of active galaxies. They are the best probes of the molecular gas kinematics down to angular scales of $\mu$-arcseconds. Specifically, mega-masers trace gas within the narrow range of densities and temperatures at which the IR radiation is able to invert the population of the energy levels.

OH-lines at rest frequencies 1.6-1.7 GHz rank among the most important extragalactic mega or kilo-masers. They are up to 3 orders of magnitudes brighter than observed in Galactic regions of massive star formation. The OH luminosity tightly correlates with $L_{\rm IR}$ in the more than 100 ULIRGs where its emission has been detected (e.g.,[58]). This lends support to a picture where OH masers are favoured in merging systems, presumably thanks to the intense radiation from warmed-up dust ([34]). The OH-IR correlation opens the possibility of measuring the merger rate as a function of redshift. In a closer detail, high-spatial resolution observations of a dozen of nearby active galaxies have shown that OH mega-masers originate in compact rotating disks of up to a few hundred parsec size (see, e.g.,[4, 24, 39]). Therefore, further higher-quality observations with SKA could address two key questions of black-hole fueling mechanisms: whether these CND structures feed supermassive black holes and if nuclear star formation is also a coeval phenomenon. In addition, it has also been discussed if OH emission can trace AGN-feedback phenomena like the jet-ISM interaction at the CNDs ([24]). AGN-driven feedback might distort the dynamical status of the CNDs, which would explain the onset of nuclear warp instabilities or the existence of massive and energetically relevant outflows that are starting to be imaged by the Atacama Large Millimeter Array (ALMA) on similar spatial scales ([15, 22] and Fig. 5). The SKA in phase 1 has the high-spatial and spectral resolution capabilities required to image the emission of OH in different types of active galaxies at different redshift ranges. More specifically, resolving an OH megamaser disk of size $\sim$500 pc with SKA1-MID will be possible up to a distance of about $\sim$160 Mpc. With SKA, the maximum distance is not set by the resolution element, but rather by the sensitivity limit. Estimates indicate that a typical detection experiment for OH will require a few–to–20 hours per source for modest maser luminosities up to $z \sim 2$ ([8] and references therein).

The emission of extragalactic $H_2O$ mega-masers at 22 GHz, detected to date in more than 100 type-2 AGNs, is known to come from the inner pc or sub-pc regions around the putative tori of active galaxies. The gas kinematics inferred from these observations has provided the most accurate estimates of supermassive black hole masses in these sources. Combined with measurements of their proper motions, these observations, done in a statistically significant sample of galaxies at different redshift ranges, are the basis of the Mega-Maser Cosmology project, which aims at determining the Hubble constant (see [32] and references therein). As for OH masers, water mega-masers can probe the shocked jet-ISM interface in the CNDs of nearby AGNs (e.g., [44]). While the SKA in phase 1 will not have access to the 22 GHz range, making impossible the imaging of the $H_2O$ line in local galaxies, it is expected that the SKA will be sensitive enough to detect its emission at high redshift. The prospects of this type of project are good in view of the predicted higher occurrence of $H_2O$ masers in the early Universe ([31]).

—*Diffuse molecular gas:*

One of the main drivers of the SKA resides in its ability to detect the emission of the neutral hydrogen content of galaxies to cosmological distances. This goal can be reached thanks to the significant gain in sensitivity: the SKA will have a collecting area ∼two orders of magnitude larger than current radiotelecopes. With these sensitivity goals at hand, the SKA will be able to detect the emission of atomic hydrogen in Milky Way-like galaxies out to z∼1–1.5 and, also in the disks of large spiral galaxies like M 101 out to z∼2.5, after a typical integration time of 12 hours (see [55]). The detection of atomic and, also, molecular gas emission in the halos of isolated spirals, like the edge-on spiral NGC 891, is a representative example of the type of science that the SKA will routinely address in a large number of galaxies ([20, 42]). This neutral gas reservoir may be continuously falling onto the disks of galaxies and thus feed star formation activity on long time scales. The detection of massive atomic and molecular halos in a significant number of galaxies may provide an answer to the long-standing problem of the missing baryons.

It is expected that molecular gas at high galactic latitudes will be preferentially diffuse with typical densities $n(H_2) \sim 10^2 - 10^3$ cm$^{-3}$. This particular phase corresponds to the transition between the atomic medium, traced by the HI line at 21 cm, and the condensed molecular medium typically found in Giant Molecular Clouds (GMC) of galactic disks. While the CO (1–0) and (2–1) emission will be detected in the molecular thick disks and halo components with ALMA, the translation of line intensities into column densities is handicapped by the highly uncertain CO–to–$H_2$ conversion factor at these low densities ([35]). The emission of molecular species like CH, which has a micro-maser line at 3.3 GHz, is extremely useful in the low density phase ([26, 35, 33, 36, 37, 38]), which is expected to be prevalent in the halo gas components. First, the abundance of CH in diffuse molecular gas is high ($\sim 10^{-8}$). Furthermore, the typical densities required to excite the line are low $n(H_2) \sim 10^2 - 10^3$ cm$^{-3}$. Last, but not least, the conversion factor from CH–to–$H_2$ in the diffuse molecular gas medium is much more accurate than the corresponding factor for CO in diffuse gas, partly because the CH line is expected to be optically thin.

Extragalactic observations of the 3.3 GHz line have proved to be feasible in nearby galaxies like the LMC, NGC 253, NGC4945 and NGC 5128, though they are scarce due to the limited sensitivity of the first single-dish radio telescopes used in this experiment ([57, 13]). The improved capabilities of the SKA will make possible to map out the emission of CH in the disks but also in the halos of a significant sample of nearby galaxies in a few hours.

—*CO lines at high redshift:*

Current millimeter interferometers have been able to map the emission of carbon monoxide (CO) and that of other more complex molecular species (like HCN, $HCO^+$, CN or CS), and thus trace the content, distribution and kinematics of molecular gas in a growing number of galaxies. The study of CO emission in Local Universe targets has been crucial to studying star formation laws in galaxies ([9]). Furthermore, high-spatial resolution CO observations have been key in the study of AGN feeding mechanisms in the Local Universe ([21]). The high-sensitivity of current mm-interferometers have also allowed us to detect the emission of molecular gas in galaxies situated at redshifts beyond z$\sim$1. This includes extreme starbursts/mergers outside the Main Sequence (like Sub-Millimeter Galaxies (SMGs)), but also QSOs, Lyman Break Galaxies(LBGs), and normal Star Forming Galaxies (SFGs) ([27, 56, 52, 17, 53]). It is worth noting the case of SDSS J1148+5251, the highest redshift quasar currently known detected in CO at z$\sim$6.4 ([56]). Current models of galaxy formation and evolution require the observation of the *normal* population of SFGs at different redshift ranges.

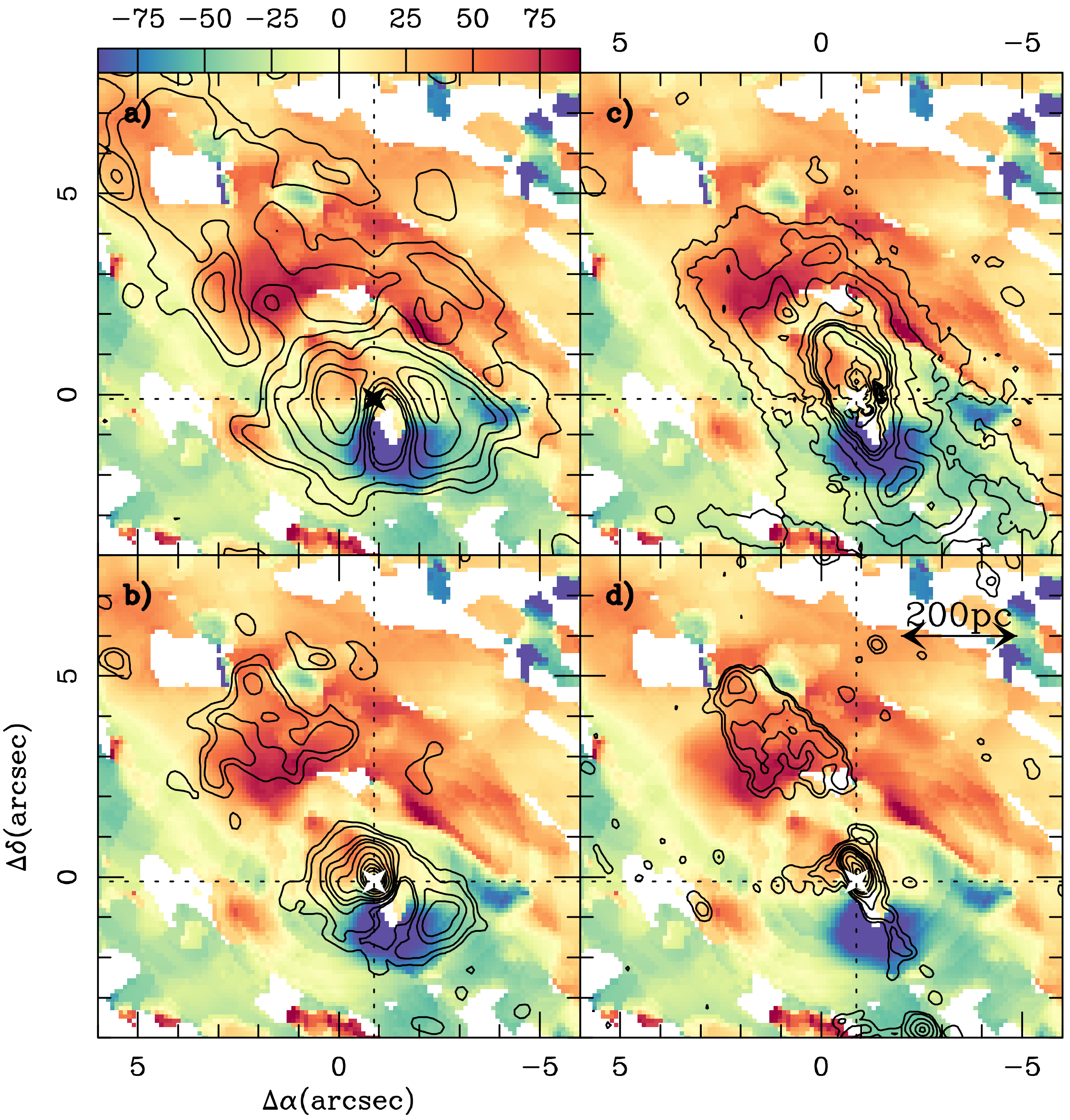

Figure 5: Overlay of the CO(3–2) velocities (in color scale) of the massive molecular outflow, as derived from the high-resolution observations done with ALMA in the CND of the Seyfert 2 galaxy NGC 1068 (adapted from [22]), with the contours representing: the integrated intensity of CO(3–2) (**a**) *upper left panel*; contours), the 349 GHz continuum emission (**b**) *lower left panel*; contours), the HST Pa$\alpha$ emission (**c**) *upper right panel* contours), and the 22 GHz VLA map of [23]) (**d**) *lower right panel*: contours).

Besides being a powerful HI radiotelescope, the SKA will also be able to probe the distribution of molecular hydrogen in galaxies to cosmological distances. As an illustration of the SKA capabilities, the cumulative number of CO(1–0)-detected galaxies at $\nu \sim$22 GHz would amount to $\sim 10^3$ per square degree only after one hour of integration time ([10]). In this context, it is worth noting that the SKA telescope may be a nice complement to the capabilities of ALMA. While ALMA will detect and image the emission from the high rotational lines of CO and other molecules in the high-redshift universe, the SKA will give information on the lower–J rotational lines of these species. The combination of SKA and ALMA observations is a prerequisite to characterize the spectral line energy distribution (SLED) of different molecular species in a statistically significant sample of galaxies and, also, for a wide range of redshifts. Modelling the SLEDs will allow us to derive the physical conditions and the chemical abundances in different extragalactic environments.

—*Other molecular species:*

Ongoing pilot studies carried out with Arecibo and the Jansky Very Large Array (JVLA) at the 1–10 GHz frequency range have revealed a surprising richness of strong molecular lines in the spectrum of the prototypical ULIRG Arp 220 ([51, 50]). The catalog of detected lines includes pre-biotic molecules, like methanimine ($CH_2NH$), in emission, three vibrationally excited $v_2 = 1$ direct l-type absorption lines of HCN and several line transitions of the OH radical seen in absorption against the continuum source. The high signal-to-noise ratio of the spectra opens the possibility of extending this type of survey with the SKA to a much larger sample of galaxies. In particular, it is expected that some of the molecules with lines in the 1-10 GHz range may also have strong maser transitions in the so far unexplored lower frequency range around 0.3 GHz in these sources.

## 4 Conclusions

The SKA will yield transformational science across a wide range of astrophysics for the next decades. In particular, the use of SKA for studies of the local Universe will enable an extremely wide range of science covering many areas of astrophysics, and will form the bridge between the detailed studies of objects in our own Galaxy and the distant high-redshift Universe. Deep and moderate-resolution (few arcsec) continuum and spectral line surveys of large numbers of nearby galaxies will be provided by projected SKA1 "all-sky" surveys at frequencies of ∼1-2 GHz. Such large area surveys will provide an essentially full radio atlas of the local Universe, and allow detailed studies of the non-thermal radio component of galaxies. Yet multiple, pointed observations covering a wide range of frequency bands, in particular higher band SKA1–MID (band 5), will be required to characterise the non-thermal components of local galaxies. This will allow us to obtain an independent measure of the SF within galaxies covering the full range of type and environments, which will be critical to our understanding of galaxy evolution and SF through cosmic time.

The use of both continuum and spectral line facilities of SKA will not only allow to study the physics of SF and extreme physics in accretion-dominated sources on an individual basis in nearby galaxies, but also allow statistical properties of these sources to be investigated, how they interact with the ISM and how they affect galaxy evolution.

## Acknowledgments

MAPT, AA, and RHI acknowledge support from the Spanish Ministerio de Economía y Competitividad (MINECO) through grant AYA2012-38491-C02-02. S.G.B. acknowledges support from the Spanish MINECO through grants AYA2010-15169, AYA2012-32295, and program CONSOLIDER INGENIO 2010, under grant Molecular Astrophysics: The Herschel and ALMA EraASTROMOL (ref CSD2009-00038). S.G.B also acknowledges support from the Junta de Andalucia through TIC-114 and the Excellence Project P08-TIC-03531. A.A.H. and LC acknowledge support from the Spanish MINECO through grants AYA-2012-31447 and AYA-2012-32295. JHK acknowledges financial support to the DAGAL network from the People Programme (Marie Curie Actions) of the European Unions Seventh Framework Programme FP7/2007-2013/ under REA grant agreement number PITN-GA-2011-289313, and from the Spanish MINECO under grant number AYA2013-41243-P.

# Spectral line mapping of the Milky Way

**José F. Gómez**[1], **and Emilio J. Alfaro**[1]

[1] Instituto de Astrofísica de Andalucía, CSIC. Glorieta de la Astronomía s/n, 18008 Granada, Spain

## Abstract

SKA will give the opportunity to map our Galaxy in thermal lines of HI, OH, and recombination lines, tracing all three phases of the gas content of the Galaxy (atomic, molecular and ionized hydrogen) with the same angular resolution. These studies would deliver a comprehensive spatial and kinematical view of the interstellar medium, its evolution, and the interrelationship between its phases. Moreover, observations of OH maser emission will provide us with a complete catalog of sources undergoing energetic mass loss, covering star-forming regions and evolved stars, and with detailed information about their internal motions and the strength of their magnetic fields.

## 1 Mapping our Galaxy

Understanding our own Galaxy is obviously of great importance. Our location within it gives us certain observational advantages, but also imposes some restrictions and challenges. The most favorable aspect is that we can study the spatial and kinematical characteristics of different structures with far better detail than in any other external galaxy. This means that we can directly observe the interplay between different phases of the interstellar medium, as well as feedback between the gas dynamics and key physical processes such as mass-loss from young and evolved stars. In summary, it is possible to study evolutionary processes in our Galaxy and then use this detailed knowledge to understand Galactic evolution in other galaxies, where many of these physical phenomena are not accessible.

Spectral line studies are key in our knowledge of the general Galactic structure, as well as in the study of star formation and the last stages of stellar evolution. This is mainly because these lines provide information on the gas kinematics (determined by their Doppler shifts) and on its temperature and column densities. The line of atomic hydrogen (HI) at 1.4 GHz [12] has been instrumental to determine the structure of the spiral arms in the Galaxy, and the location of the Solar System in our Galaxy. The millimeter CO lines, as well as a plethora of other molecular lines at different wavelengths (such as CS, $NH_3$, SiO, and $NH_2^+$) are good tracers of molecular gas (under different density and excitation conditions), and have provided fundamental insights in the processes of star formation [21]. On the other hand, maser emission from different molecules (OH, SiO, $CH_3OH$, $H_2O$) are useful signposts of energetic phenomena in the circumstellar environments of young and evolved stars [9].

Specific aspects of the study of molecular lines in star-forming regions and evolved stars with SKA are treated elsewhere in this document. Here we will focus on some of the information that we can obtain by carrying out a large-scale mapping of our Galaxy in spectral lines. Considering the frequency coverage of SKA1, the most relevant spectral lines for such a mapping are the 1.4 GHz line of HI, and the ground-state OH lines at 1.6 − 1.7 GHz. Thermal emission from OH is a tracer of diffuse molecular gas, while OH maser emission traces energetic processes with extremely high spatial accuracy. Other interesting lines are the maser transitions of methanol (e.g., 6.7 and 12.1 GHz) and excited OH (4.7, 6.0, 13 GHz).

This mapping of spectral lines in our Galaxy is challenging in several aspects. First, Galactic emission covers a large fraction of the sky. For instance, a mapping of Galactic latitudes $|b| < 3^\circ$ of the plane south of declination $40^\circ$ would cover $\simeq 1500$ square degrees. For such a huge area it is not enough to be able to make observations with high sensitivity. It is also necessary to do so fast enough to cover the whole Galactic emission in a reasonable time. Therefore the relevant figure of merit in Galactic mapping is not sensitivity, but "survey speed", i.e., how fast one can map a large area of the sky at a given sensitivity level, and it is governed by the immediate field of view, in addition to the collecting area. Survey speed will be one of the strongest points of SKA1. For instance, SKA1-mid will have a survey speed around two orders of magnitude higher than the Very Large Array in its A configuration. This means that a mapping of the Southern Galactic plane with high sensitivity and angular resolution will be feasible in reasonable integration times with SKA. Note also that the field of view of the SKA will be smaller as we move to higher frequencies, so a large-scale mapping of the Galaxy is more efficient at lower frequencies. This, and the possibility of simultaneously observing several lines (see section 2), suggest that spectroscopic surveys of our Galaxy should first focus on the lines around 1.4-1.7 GHz, for which the field of view will be $\simeq 0.49$ square degrees (a mapping of the southern Milky Way with $|b| < 3^\circ$ and dec $< 40^\circ$ requires $\simeq 3000$ individual pointings). Other lines at higher frequencies would be better observed commensally with radio continuum surveys, in a similar way as the MeerGal survey will do at $\simeq 14$ GHz using MeerKat [29].

An additional challenge when carrying out a spectral line mapping of our Galaxy is not only that the total size to be covered is large, but also that there are relevant structures at different scales, from several degrees down to the resolution limit of the telescope. However, interferometers have a limit to the largest angular scale they can sample, determined by the length of its shorter baseline. Although SKA will probably be designed to minimize such losses, this is an inherent limit that cannot be completely avoided. A complete picture of the Galaxy will require the combination of SKA data with data from single-dish telescopes, such as the Parkes Galactic All-Sky Survey (GASS [17], Fig. 1) in HI and the Southern Parkes Large-Area Survey in Hydroxyl (SPLASH [6] in OH), so that we can properly sample both large and small scales. An important step forward will be provided by the Galactic ASKAP Survey (GASKAP [8]), which will map our Galaxy in HI and OH with the Australian Square

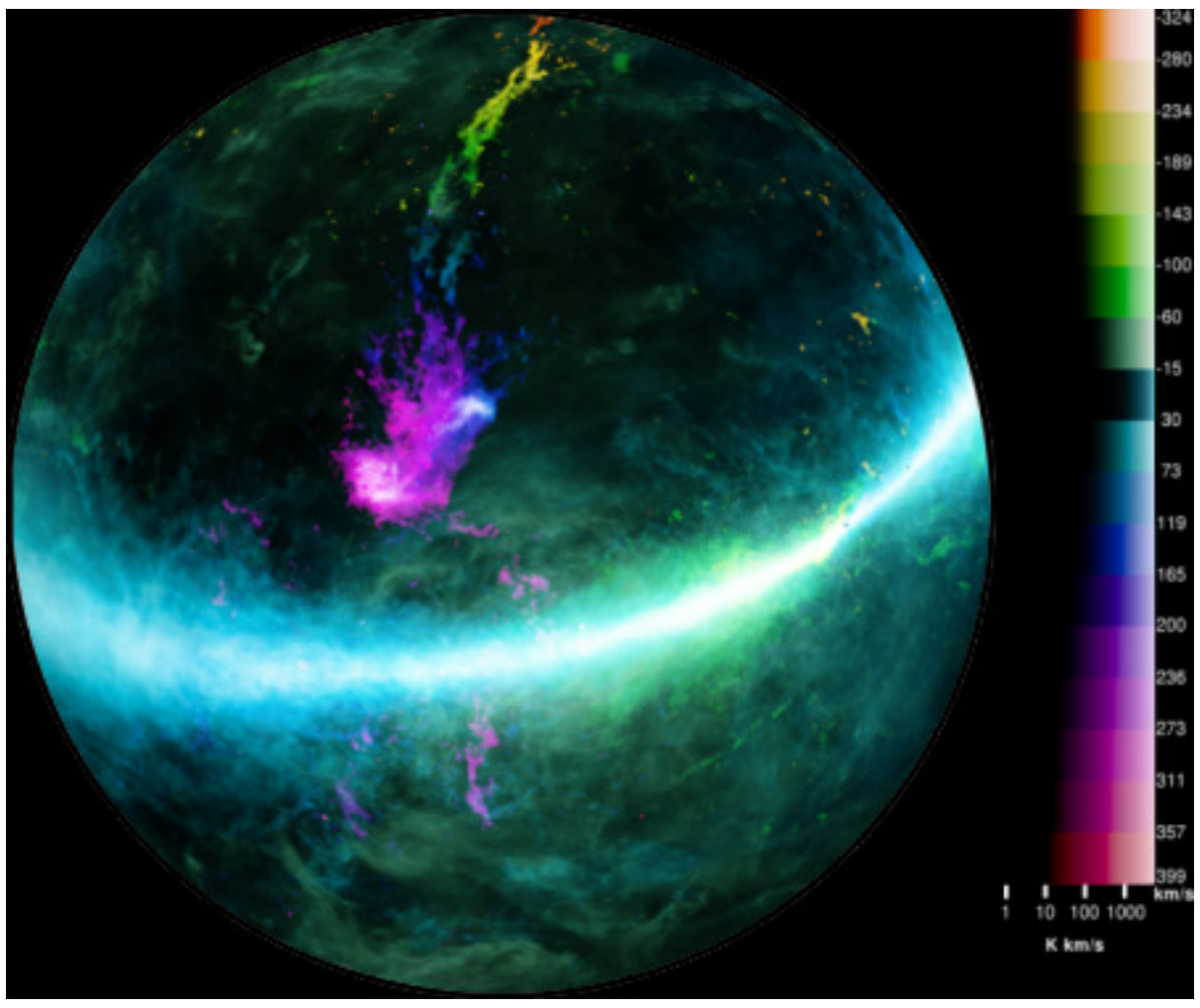

Figure 1: Map of the HII emission obtained with the Parkes antenna for the GASS project (image from [17]). The South Celestial Pole is at the center. Right ascension zero is at the top and it increases counter-clockwise. Different colors represents the velocity with respect to the Local Standard of Rest, in intervals of 40 km $s^{-1}$. The color intensity is proportional to the integrated brightness temperature over each velocity interval.

Kilometre Array Pathfinder with $10'' - 20''$ angular resolution. SKA will add another piece to the puzzle, to increase the detail and sensitivity provided by previous spectral line surveys.

## 2 Thermal lines: the three phases of the interstellar medium

SKA1 will allow us to study thermal spectral lines from the three phases in which the gas can be found in the insterstellar medium. In particular, the HI line at 1.4 GHz emitted by the atomic component, four OH transitions at 1.612, 1.665, 1.667, and 1.720 GHz tracing molecular gas, and several recombination lines of hydrogen and helium around those frequencies (e.g, H$n\alpha$, with $n$ between 155 and 163) for the ionized phase. The rest frequencies of all these lines are within $\simeq$ 400 MHz, so it will certainly be possible to observe them simultaneously with the appropriate correlator setup and *with the same angular resolution*, which would be an extremely efficient way to carry out studies of the physical conditions and evolution of the interstellar medium in our Galaxy. In a single pass we will have a picture of the whole mass content of the Galaxy and with unprecedented detail. The angular resolution of SKA1-mid at these frequencies ($0.22''$) would correspond to a linear scale of only $\simeq$ 1800 AU at the distance of the Galactic Center, so we could reach circumstellar scales.

A spectral line mapping of our Galaxy will give valuable information on each of the phases of the interstellar medium. For instance, recombination lines from ionized gas will provide a complete census of regions of massive star formation and planetary nebulae, even at the farthest reaches of the Galaxy, without being affected by interstellar extinction, and including kinematical information of each system in its Galactic motion and in its internal outflows. However, even more important will be the availability of a coherent picture of all phases of the interstellar medium, to study their interrelationship. This includes answering questions on how the atomic gas evolves into molecular gas, creating the molecular clouds from which new stars will form. The energy input from young stars will create winds and photoionized (HII) regions that will influence the subsequent generation of stars. At the end of the stellar life cycle, planetary nebulae and supernovae will send material back to the interstellar medium, enriching it with heavier elements.

By combining data from SKA1 and sensitive single-dish surveys, we will have access to the whole hierarchy of scales. Starting from the kpc scales (flocculent spiral arms and gas supercomplexes), new structures and substructures could arise [16], indicating how spiral arms drive processes down to smaller scales. The largest hierarchical scales of gas with coherent velocity patterns are relevant, since the largest complexes of coherent HI gas can be related to how star formation will proceed, and to what extent it can be triggered by large-scale Galactic processes and turbulence [10]. Down in the hierarchical ladder, we find shell and chimneys created by stellar winds and supernovae (at scales of 10-100 pc), direct witnesses of the energy exchange between stars and the interstellar medium [3]. Smaller structures ($< 1$ pc) can be created by instabilities in shell walls [5] or in turbulent flows [32]. Some of these smallest structures can evolve into molecular clouds with induced star formation.

The comparison between HI and OH will give us the fraction of molecular gas (with respect to total gas) in the whole galaxy, with an unsurpassed angular resolution. This fraction is thought to depend on pressure, metallicity, and ultraviolet radiation [11]. The detailed variation of the fraction of molecular gas with galactocentric distance and height above the plane will give us information about the physical processes acting at different locations and about the history of star formation in our Galaxy. Sampling multiple scales will allow us to study the characteristics of turbulence, by studying the fractal structure of atomic and molecular clouds [24], or the relationship between the velocity distribution and the mass and size of clouds [15] or of clumps within clouds. By doing this in a complete sample of clouds, we could see how turbulence evolves, and whether there are particular cloud characteristics that will determine the subsequent process of star formation.

At high Galactic latitudes, we can obtain a full count of high-velocity clouds in the galactic halo [33], together with their detailed structural and kinematical characterization. This could reveal the origin of these clouds (which is still uncertain), how they interact with both the halo and the Galactic disk along their path of motion, and whether they are a significant source of mass input into the Galactic disk, which would in its turn determine the star formation rate in our Galaxy.

Diffuse HI and OH emission does not have a high brightness temperature. Therefore, its mapping at the full angular resolution of SKA1 would in principle require prohibitively long integration times. However, the antenna configuration of SKA1-mid is centrally condensed, so the visibilities can be tapered (downweighting the longest baselines) without losing much overall (point-source) sensitivity. If we set a target sensitivity in terms of brightness temperature, increasing the integration time per field would then allow us to improve the angular resolution of the final maps. As a benchmark, a dwell time of 1 hour per field with SKA1-mid would already provide maps with the same sensitivity as those in the low-latitude component ($|b| < 2.75°$) of the GASKAP survey at $20''$ angular resolution (dwell time of 50 hours), but with a factor of two better angular resolution ($10''$). Such a mapping would require a total of $\simeq 2800$ hours of SKA1-mid time. Subsequent visits over the lifetime of the instrument would then allow obtaining maps with increasingly higher final angular resolution.

## 3 Maser emission: signposts of energetic processes

Maser emission requires a very energetic process (either shocks or intense radiation fields), so that the population between two energy levels of an atom or molecule is "inverted" (i.e., the upper level is overpopulated over a thermal situation). Background photons with the energy corresponding to the energy difference between these states rapidly stimulate the energy decay. The result is a bright, and usually compact ($< 1$ AU) emission.

At the frequencies covered by SKA1, the most widespread maser lines are those of the ground state of the OH molecule at 1612, 1665, 1667, and 1720 MHz. Although different types of sources can emit in all these transitions, there are some identifiable trends. The satellite line at 1612 MHz is dominant in the circumstellar envelopes of oxygen-rich evolved stars, especially those in the Asymptotic Giant Branch (AGB) that show characteristic double-peaked OH maser spectra with peak separations of $\simeq 30$ km s$^{-1}$ [26]. The main-line transitions at 1665 and 1667 MHz are prominent in regions of massive star formation [2]. The presence of the 1720 MHz line alone is usually associated to shocks at the interactions of supernova remnants with molecular clouds [13]. Thus, the relative intensities of OH transitions are useful to discriminate among young and evolved objects.

Maser spots will appear as point sources with the angular resolution of SKA1. Therefore, observations will not suffer from any loss of extended emission and no complement from single-dish data will be needed. Moverover, maps can be obtained at the full angular resolution provided by the antenna configuration, while retaining the maximum signal-to-noise ratio. We also note that, if there is a velocity gradient in a source, different spots of maser emission will appear in separate spectral channels. This allow us to measure spatial and kinematical structures at scales much smaller than the resolving power of the telescope, since the relative positional accuracy within two point sources in a map [19] is given by $\sigma = \theta/(2 * sn)$, where $\theta$ is the synthesized beam, and $sn$ is the signal to noise ratio of the emission. The sensitivity of SKA will be the key factor in improving this relative positional accuracy. As an example, this accuracy is $\simeq 0.15$ milliarcsec for maser emission at a level of 0.5 Jy with SKA1-mid with only one hour of observing time and a velocity resolution of 0.2 km s$^{-1}$.

Another useful aspect of OH emission is that its levels undergo Zeeman splitting in the presence of a magnetic field [1] (see chapter by Girart in this book). Thus, it is straightforward to measure circular polarization with these lines, which is related with the magnetic field strength along the line of sight. In sources where internal Faraday rotation does not have a strong effect, the orientation of the linear polarization vectors [14], and the degree of linear polarization can give us information about the overall spatial orientation of the field.

SKA1-mid will have enough sensitivity to detect almost all OH masers in our Galaxy with just 45 minutes per field of view, assuming that their typical luminosity is similar to those detected near the Galactic center [28]. So we will obtain a complete catalog of Galactic OH masers, including a detailed measurement of the spatio-kinematical structure within each source, and the strength and orientation of their magnetic field (in $\simeq$ 2300 hours, to cover $|b| < 3°$). This will be a powerful tool when compared with large scale surveys at other wavelengths, since we can obtain statistically significant correlations about different physical properties both in young and old stars. Moreover, the high sensitivity of SKA could unveil a new population of weaker OH masers, pumped in circumstellar environments different from those we now know. Even known OH-emitting objects could reveal a wealth of weaker maser components, tracing kinematics and magnetic fields over a more extended region. These observations will also uncover the whole catalog of some particularly interesting OH-emitting evolved objects, of which very few members are known so far, such as post-AGB stars undergoing collimated mass loss [7] or extremely young planetary nebulae [31]. In the case of star-forming regions, OH masers in star-forming regions, not directly associated with ultracompact HII regions, can pinpoint the location of high-mass protostars, in their very first stages of evolution.

The structures traced by OH masers show a significant diversity. In particular, these masers can be found in jets, circumstellar disks, or in uncollimated winds, and this is the case both in young and evolved objects [22, 4]. A sensitive OH survey in our Galaxy would allow us to determine the relative incidence of these different structures, and under what conditions either is produced. It is possible that this is an evolutionary effect, and the structure traced by OH masers will depend on the relative age of the object. A complete morphological catalog of OH masers in our Galaxy can give us a better understanding of the processes leading to the formation and death of stars, since the presence of disks and collimated outflows are key ingredients in models of star formation via accretion [27] and to explain the shaping of planetary nebulae [23]. However, the earliest phases of massive star formation may be characterized by short episodes of uncollimated mass-loss [30]

Previous studies of Zeeman splitting of OH masers suggest that the orientation of the field traced by this emission may be consistent with that found at larger scales in our Galaxy [20]. This persistence after collapse in molecular clouds is surprising, since masers trace scales and densities that differ several orders of magnitudes from those of the diffuse interstellar medium. With a large and unbiased data base of Zeeman splitting of OH masers in star-forming regions we can make a more precise comparison with the Galactic magnetic field [18]. In evolved stars, we could study the evolution of magnetic fields, and whether there is any correlation betwee their strength and orientation with the degree of collimation observed in post-AGBs and PNe.

The radial velocities of OH masers can also be used to study Galactic dynamics [25]. Specially in the case of AGB stars, whose OH maser spectra typically consist of two well-defined peaks, the radial velocity of the system is very easy to obtain (it is just the mean velocity between the peaks). Thus, we will have a perspective of Galactic dynamics from the population of these old stars, which can be compared with that provided by the gaseous component (from thermal HI and OH emission) or from different stellar populations sampled in the optical by GAIA.

## Acknowledgments

The authors are supported by Ministerio de Economía y Competitividad (MINECO), grants AYA2011-30228-C03-01 (JFG) and AYA2013-40611-P (EJA), including FEDER funds.

# Cosmic Magnetism

**Eduardo Battaner,[1,2], Iván Agudo,[3], Antxon Alberdi,[3], Estrella Florido,[1,2], Ana Guijarro, [4], Beatriz Ruiz-Granados,[5], and José Alberto Rubiño-Martín[6,7]**

[1] Departamento de Física Teórica y del Cosmos, Universidad de Granada, Spain
[2] Instituto Carlos I de Física Teórica y Computacional, Universidad de Granada, Spain
[3] Instituto de Astrofísica de Andalucía (CSIC), Granada, Spain
[4] Centro Astronómico Hispano Alemán, Calar Alto, Almería, Spain
[5] Instituto de Física de Cantabria, CSIC-Universidad de Cantabria, Spain
[6] Instituto de Astrofísica de Canarias, La Laguna, Spain
[7] Departamento de Astrofísica, Universidad de La Laguna, Spain

## Abstract

High–sensitivity, linear and circular polarization, large frequency range and good angular resolution are characteristics of the SKA which will allow a much deeper insight in all topics of Cosmic Magnetism. More specific goals will be: magnetism and cosmology, nearby galaxies, the so called fan region, the physics of AGN jets and cosmic ray anisotropies.

## 1 Introduction

Considering the possibilities of SKA, we have two types of objectives: general and particular. On one hand we are interested in participating in the general goals of the SKA Working Group *Cosmic Magnetism*. We would address general topics considered in the collaboration as are the 3D distribution of magnetic fields in the Milky Way as well as fields in SNR and HII regions. We aim to study the turbulent power spectrum, mainly in the small angular scales, which is largely unknown, and for which SKA will provide unprecedented high resolution. Existing data of the Faraday depth of extragalactic sources will be highly improved with the great capability of SKA. The Milky Way can be considered a foreground for observations of magnetic fields at the epochs of Recombination and Reionization, thus overlapping our interests with those of the cosmological working group. We do not comment on the general purposes as basically coincide with those outlined by the international SKA Collaboration. We take the presentations in AASKA14 and [1] as the basic information about the general objectives of SKA in the field of cosmic magnetism. In order to understand how the magnetic fields are built up, wide band polarization spectral imaging will be a major contributor. SKA will make fundamental contributions to the study of magnetic fields in galaxies and clusters, thanks to its superb sensitivity, high polarization purity, large frequency range, high frequency agility and the possibility to observe large bandwidths with high spectral resolution.

On the other hand, our particular interests are focused on more specific goals which are outlined here.

## 2 Magnetism and cosmology

A precise description of the Faraday rotation in our Galaxy would permit us to better identify the regions where the detection of the rotated angle of polarization could have measurable effects to assess magnetic fields at the Recombination epoch ($z$=1100) and at the Reionization epoch ($z \sim$10). An accurate separation of the different components is crucial for this task. An upper limit of the magnetic strength at Recombination of 4 nG (comoving field) has been established by Planck [2], but future polarization data can give more precise upper limits. We could in turn obtain better constraints with the large capability of SKA. With respect the Reionization epoch ($z <$10) the possibility of detecting a direct Faraday signal cannot be disregarded yet. In this case, a correlation with the angular distribution obtained by SKA and LOFAR would provide observational clues to study the role of magnetic fields in the formation of the patched structures of ionized regions. At least, interesting upper limits could be established about this influence. This search would be related to the possible influence of magnetic fields in the formation of the large-scale structure.

After decoupling, magnetic fields may influence the thermal and ionization history of the Universe in two ways [3]. First, they can dissipate energy via ambipolar diffusion, due to the existence of a residual ionized component in the almost neutral plasma after decoupling. In addition, and for small enough scales, they can generate decaying magnetohydrodynamic turbulence, due to non-linear effects. These processes may affect the ionization history of the Universe, producing potentially measurable effects in the redshift evolution of the measured 21 cm line, but also creating distortions of the black-body spectrum of the cosmic microwave background (CMB) [4], and generating additional "Thomson optical depth" to CMB photons, which affect the the temperature and polarization power spectra at high multipoles [5, 6].

## 3 Fan Region

The study of the magnetic field distribution in the Milky Way is one of the main objectives of SKA. The high angular resolution, the high sensitivity and the inclusion of polarization of this large sky area would clearly permit the study of many particular features of subsystems in the Milky Way. A puzzling area is the so called *fan* region, which is probably a complicated structure (see [7]). This region is especially interesting because it is near the anti-center region, it is affected by a spiral arm, it has a very strong synchrotron emission and the order of the polarization vectors is highly noticeable. It contains a SNR but its size is small compared with the dimensions of the fan. A 3D dissection carried out by SKA could permit a better description of this region and to assess the role of magnetic fields in its structure and dynamics. It cannot be disregarded that this region is enriched with extragalactic cosmic rays in the 10-100 TeV spectral region.

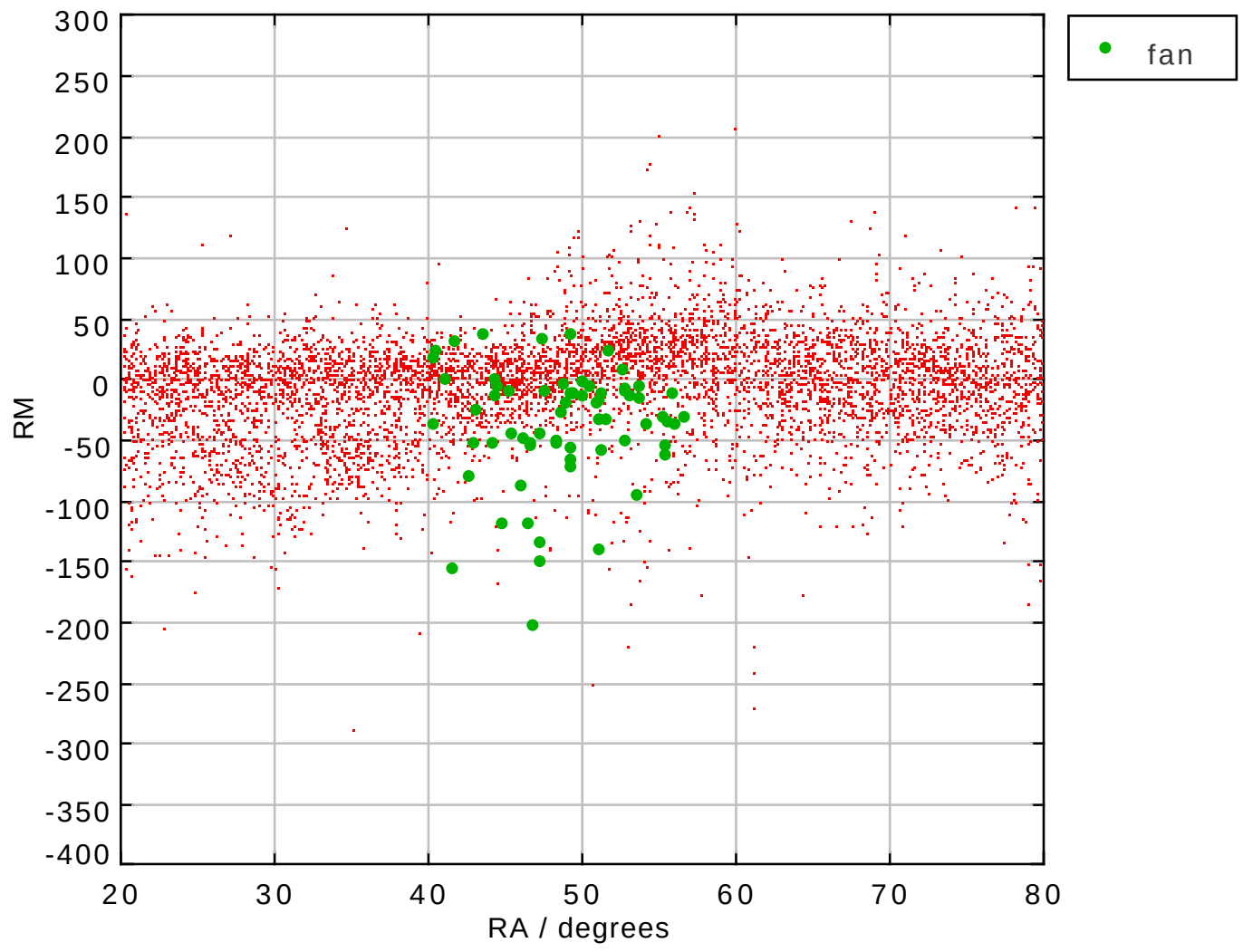

Figure 1: Faraday rotation measurement of pulsars in the Galaxy (red small points) and in the fan region (green large dots) showing that this region has peculiar magnetic properties. The data source was the ATNF Pulsar Catalog http://www.atnf.csiro.au/research/pulsar/psrcat/.

## 4 Other nearby galaxies

The objective of SKA covers about 30 nearby galaxies that could be observed in 3D, thanks to the different Faraday depths at different wavelengths. We would be particularly interested in assessing the magnetic field patterns in these galaxies, as they are clue to know the balance between pregalactic distributions (bisymmetric) or dynamo driven distributions (axisymmetric) [8]. Toroidal patterns could be explained by both mechanisms. The evolution of magnetic fields in galaxies is a controversial subject of which the observations of SKA could permit deeper understanding. Of particular interest is the role of magnetic fields in bars, because the strength is higher than in normal spirals, reaching values of about 40 $\mu$G [9]. The properties of the field in the upstream and the downstream regions divided by the bar shock are very different and the influence on star formation is far from being understood. Dynamical effects, including the feeding of the AGN are important but ignored.

## 5 Active Galactic Nuclei Jet Polarization Studies

As it has been described in detail by Agudo et al. (this book), and [10], SKA will allow to make a step forward in our understanding of extragalactic jet physics in general, and in particular of their magnetic fields: i) thanks to the high precision in the determination of the linear and circular polarization in AGN jets, it will be possible to characterize the relativistic-jet composition (electron-positron vs. electron-proton), particle acceleration and magnetic field configuration; ii) thanks to the determination of the magnetic field configuration, discriminate between different models for the relativistic jet formation, disk-launched jets or black hole launched jets; ii) thanks to the SKA multifrequency capabilities, it will be possible to estimate the plasma velocity, density and magnetic field for hundreds of relativistic jets; iii) thanks to the superb sensitivity of SKA, it will be possible to understand the physical differences between radio-loud and radio-quiet AGN populations; iv) thanks to multi-epoch polarimetric observations, it will also be possible to establish the location of the high energy bursts within the relativistic jets at a much wider range of radio luminosities than currently feasible.

## 6 Anisotropies in cosmic rays and magnetic fields

Regular toroidal fields in disky structures could constitute magnetic lenses, for charged Cosmic Ray (CR) particles [11]. The dimensions of the lens should depend on the field strengths and mainly on the energy of the incoming cosmic rays. The theoretical study of lenses and their properties compared with optical and gravitational lenses is simple but the identification of lenses in the real sky, with more or less degree of geometrical symmetry requires a more detailed description with a very large angular resolution. The interpretation of large and small scale anisotropies observed by CR telescopes such as Milagro or Hawk requires a precise knowledge of magnetic structures and their possible lensing effects [12].

## 7 Dynamical effects of magnetic fields at the rim of galaxies

Magnetic fields in the outer part of galaxies will be measured with unprecedent precision, because of the very large amount of foreground polarized extragalactic sources that can be measured. In principle, we could have a value of the strength for each extragalactic source, due to the Faraday Rotation. The outer part of spiral galaxies are expected to be more affected by the dynamic action of fields. One of the problems to be considered is the role of magnetic fields in the rotation curve. If ordered magnetic fields have strengths of the order of $\mu$G inside galaxies and of the order of $\mu$G outside, in the intergalactic medium, they must be of the order of $\mu$G in the most external galactic rim. Therefore, the Alfvén speed must increase exponentially, eventually reaching the order of the rotation velocity [13]. Other galactic features in the outer disk with a possible dynamical contribution of magnetic fields [14] (e.g., warps [15] and stellar truncations [16]) will be revisited with the high resolution of SKA.

## Acknowledgments

EB, EF, AG, BRG and JARM acknowledge support from the Spanish Ministry of Economy and Competitiveness (MINECO) through grants AYA2011-24728 and the Consolider-Ingenio project CSD2010-00064 (EPI: Exploring the Physics of Infation), and from Junta de Andalucía (CEIC) through the FQM-108 project. IA research is supported by a Ramón y Cajal grant of the Spanish MINECO. IA also acknowledges funding support by MINECO through grant AYA2013-40825-P.

# Zeeman observations: Measuring magnetic fields in the atomic and molecular ISM.

**Girart, J.M.**[1]

[1] Institut de Ciències de l'Espai (CSIC-IEEC). Campus UAB, C/ de Can Magrans, S/N, 08193 Cerdanyola del Vallès, Spain

## Abstract

Magnetic fields are believed to play an important role in the dynamics of the ISM. The Zeeman effect in the 21 cm HI line, as well as in several molecular lines, can be used to characterize the magnetic fields in the interstellar medium and in circumstellar envelopes. Feasible Zeeman measurements are presently restricted to maser observations and few cases of strong thermal lines. The significant increase of sensitivity of the SKA telescope will allow us to carry out Zeeman measurements in different environments in a systematic way. However, special care should be taken to characterize and correct the instrumental polarization.

## 1 Introduction

In Astrophysics, the only way to directly measure magnetic fields is through polarimetric observations. In the case of interstellar medium (ISM) studies, the polarization can be detected from the following mechanisms (see [27] for a description of these mechanisms): synchrotron radiation, thermal dust emission, interstellar absorption (by dust particles) of background stars, Faraday rotation of ionized radiation, and spectral line emission. Under the presence of a magnetic field, the spectral transitions of molecules and atoms split into magnetic sub-levels (this is the so-called Zeeman effect). There are two different processes generating polarized emission of spectral lines as a consequence of this splitting. One is the so-called Goldreich-Kylafis effect [22] that produces linear polarization in molecular rotational transitions. This polarization arises when the magnetic sub-levels of the rotational have an unequal population. In order to have significant polarized emission, some special conditions should be fulfilled such as: anisotropic radiation, moderate optical depths, the excitation of the observed transition should not be dominated completely by collisions. By measuring the Goldreich-Kylafis effect, one obtains the direction of the magnetic field projected in the plane of the sky, but it does not provide information of the magnetic field strength. The second process is related with the fact that the different magnetic sub-levels have slightly different energies. Observationally, the resulting atomic and molecular transitions are partially circularly polarized. The circular polarized emission depends on the magnetic field strength along the line-of-sight and on the magnetic dipole moment. Only H I and molecules with an unpaired electron in the outer layer (i.e, radicals such as OH, CN, $C_2S$, $C_2H$) have magnetic moments high enough to the circular polarization to be detectable at radio wavelengths under the typical ISM magnetic field strengths. Masers are the exception, where their emission is so strong that the circular polarization can be relatively easily detected.

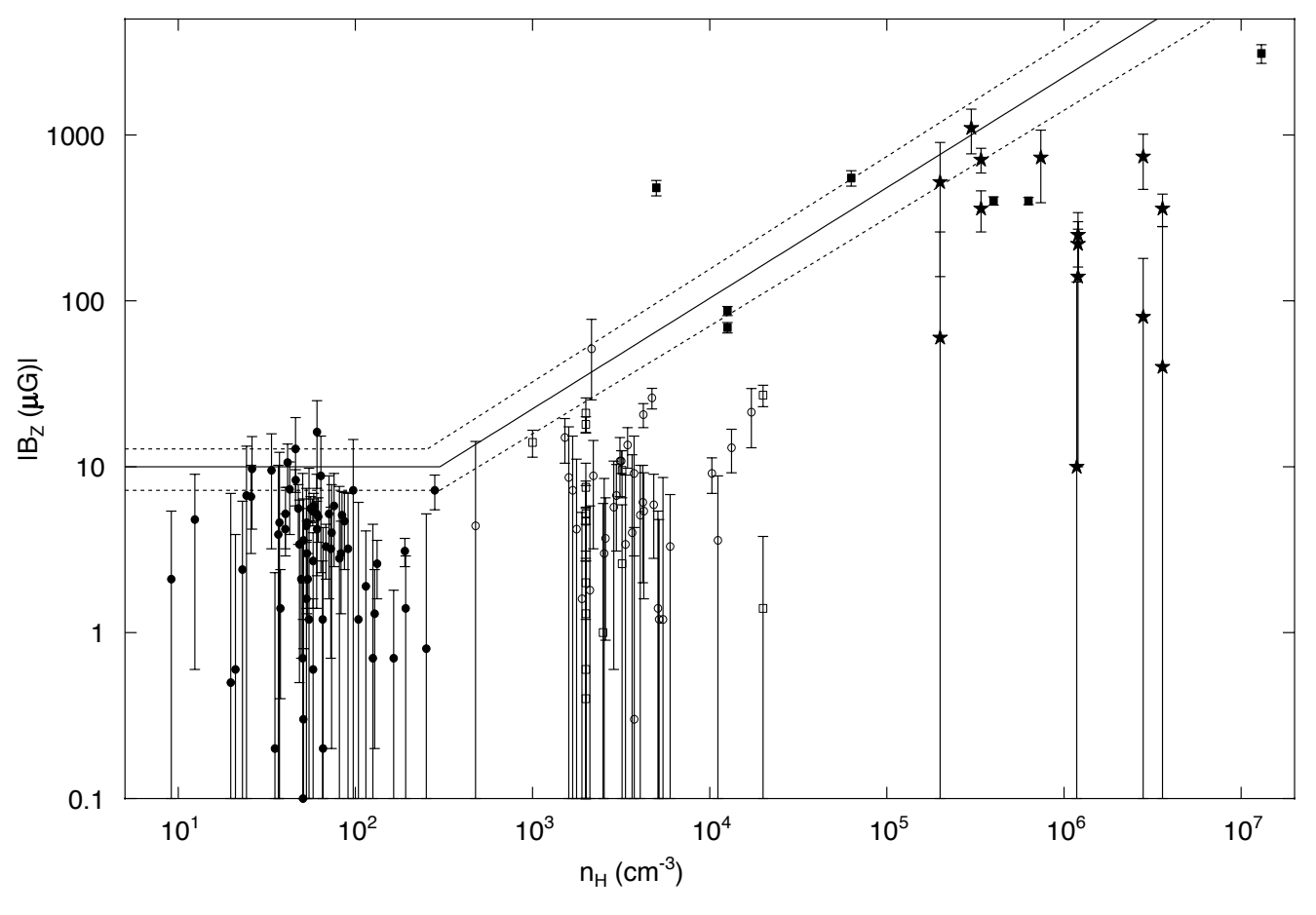

Figure 1: Set of diffuse cloud and molecular cloud Zeeman measurements of the line-of-sight component $B_z$ of the magnetic field strength, plotted as a function of the volume density of H I (Figure 1 from [11]).

Here on, we focus on the science goals that can already be achieved with the SKA1 phase. The relevant feasible polarization mechanisms measurable with SKA1 are synchrotron radiation (see the chapter by Perez-Torres et al., this book), Faraday rotation (see the chapter by Battaner et al., this book) and the Zeeman effect. We note that we cannot discard that the Goldreich-Kylafis effect could be detected with SKA1 in some molecules with rotational transitions within the SKA1 frequency range (e.g. $HC_3N$). The polarization level of the interstellar radiation is, in most cases, very low (a few percent of the total intensity). These low polarization levels are comparable (and in some cases even lower) to the instrumental polarization produced in the radio telescopes (including aperture synthesis arrays). This requires an accurate determination and removal of this instrumental contribution from the telescope, making polarization observations difficult. As a result, and although the magnetic field is one of the main ingredients in the ISM, the studies of the interstellar magnetic fields have been relatively limited [23].

At radio wavelengths the Zeeman splitting has been measured in a small number of atomic and molecular lines (e.g., H I 21 cm line, OH 18 cm lines). The Zeeman observations provide a direct measurement of the magnetic field strength in the neutral atomic and molecular phases of the ISM. The studies based on the different polarimetric techniques have shown that magnetic fields are dynamically important in the evolution of the atomic and molecular

ISM, as well as in the star formation process (e.g., [23, 11, 10, 17, 18]). Figure 1 shows a recent compilation of magnetic field strengths as a function of the ISM volume density (measured from Zeeman observations of H I, OH and CN, [11]). Note that most of the measurements are upper limits and done with time-expensive observations, mainly with single-dish radio telescopes. The enormous increase of sensitivity and the angular resolution of SKA will allow us to obtain a significant improvement in this field. [31] has also recently described the science goals that can be achieved from Zeeman observations with SKA.

## 2 The Zeeman effect and SKA

For the typical magnetic field strengths measured in the ISM, the Zeeman splitting produced in spectral lines is extremely small, much smaller than their line width. Therefore, the effect of the Zeeman splitting can only be measured through circular polarization observations (with the exception of a few cases of maser observations at high angular resolution: e.g., [21]) and requires high spectral resolution ($\simeq$ 0.1–0.5 km s$^{-1}$). Figure 2 shows a typical example of a Zeeman detection in circularly polarized emission.

Within the SKA1 frequency coverage (specially at $\nu \geq 1$ GHz) there are a number of atomic and molecular transitions with strong Zeeman splitting factors (see presentation by Robishaw et al. at AASKA14). However, in many of them, the Zeeman effect has not yet been detected. The SKA1 sensitivity may allow us to expand the feasibility of the Zeeman effect detection to a larger sample of spectral lines. These will allow us to widen the sample of physical environments where the magnetic field can be measured. As previously stated, a problem of the radio astronomical polarimetric observations is that the instrumental polarization is of the same order of the polarized signal (in the case of Zeeman observations, polarized signal could be significantly smaller than the instrumental polarization). The present aperture synthesis radio telescopes correct the instrumental polarization on-axis. The large field of view of SKA implies that a special care would be necessary to characterize and correct the off-axis instrumental polarization (e.g., [33]).

## 3 Diffuse H I Medium

In the diffuse atomic ISM, the H I mass is distributed approximately half in the so-called Cold Neutral Medium component (CNM) and half in the so-called Warm Neutral Medium component (WNM). The Zeeman observations of the 21 cm H I line typically target regions with strong radio continuum sources. In these cases, the H I lines appear in absorption. This implies that Zeeman observations trace the magnetic fields in the CNM (the H I line opacity is inversely proportional to the gas temperature, so the H I absorption features are more sensitive to the colder H I diffuse component; [24]). The Zeeman observations of the H I line in emission, although technically more difficult (because of the instrumental polarization issues, [25]), can provide interesting results since it allows the mapping of large areas [25]. However, no positive detection of the Zeeman effect in the H I line in emission has been made with an aperture synthesis radio telescope [31]. SKA is going to provide a major step in this field: SKA1 is expected to detect in a few hours of observation the Zeeman splitting of H I lines corresponding to (line-of-sight) magnetic fields strength of $\sim 1\ \mu$G, which in addition to the great imaging coverage (field of view and angular resolution) will improve the understanding of the magnetic field properties in the CNM (Robishaw et al. at presentation at AASKA14).

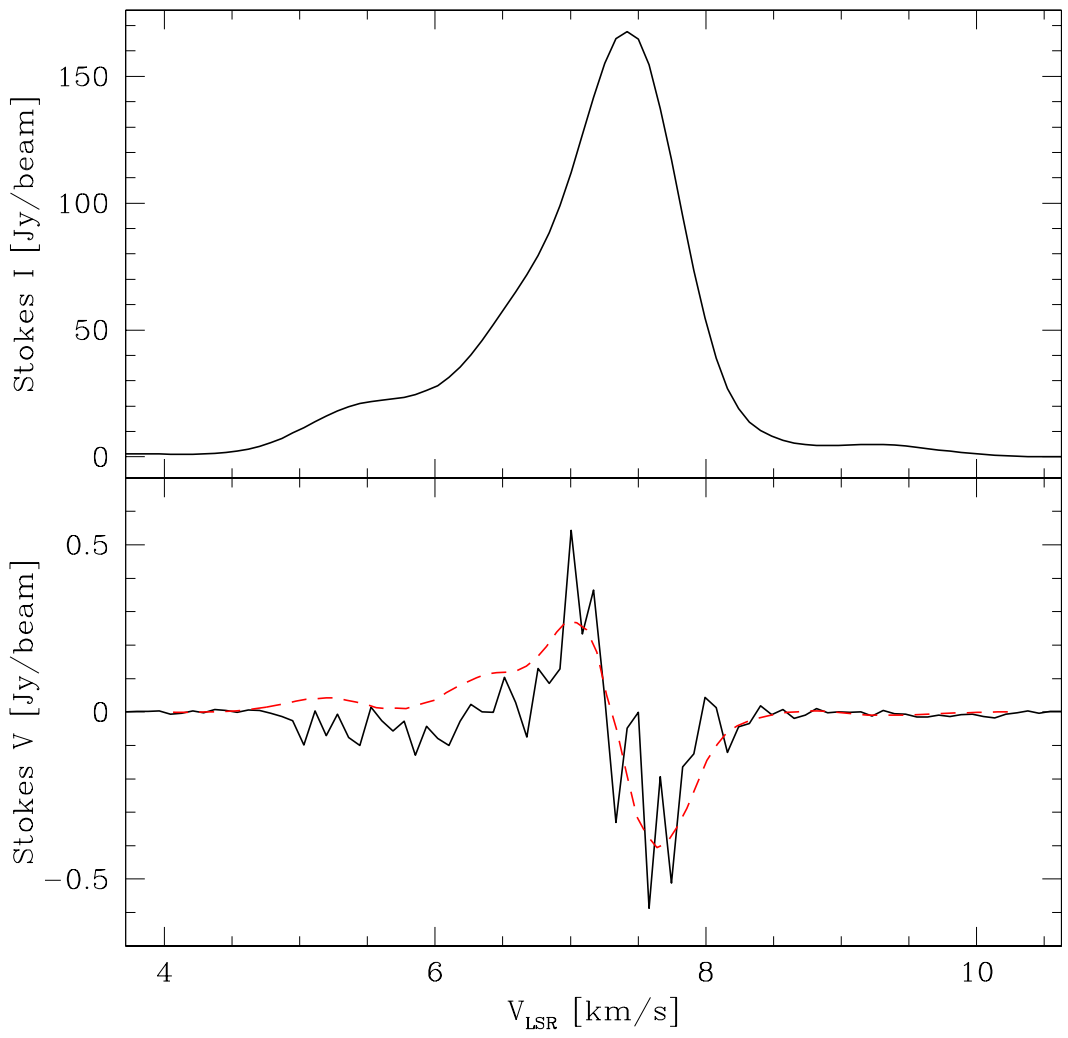

Figure 2: Typical spectra in Zeeman observations: Stokes $I$ (total intensity, upper panel) and Stokes $V$ (circular polarization, lower panel) spectra of the water maser emission at 22.23 GHz. The red dashed line is the scaled derivative of the total power Stokes $I$ in order to guide the eye (Figure from [3]). Note that the water maser frequency will not be available for SKA1, but it will possibly be within the observing frequency for SKA2.

# 4 Molecular clouds and Star formation

The last decade has seen a significant progress in the characterization of the magnetic field distribution (more specifically the projected component on the plane of the sky) in molecular clouds and toward star forming regions (e.g., [16, 8, 40, 26]). However, there is a lack of direct measurement of the actual magnetic field strength [15, 32]. The most effective way to measure the magnetic field strength in molecular clouds at centimeter wavelengths is observing the 18 cm OH lines in absorption with respect to a strong continuum source, typically H II regions generated by massive young stars (e.g., [9], [6]). Observations of the Zeeman effect of the 18 cm OH lines in emission is also feasible, but because, in general the emission is weak, long observing times are required to obtain meaningful upper limits of the field strength, specially toward dark molecular clouds (see [32]). SKA will allow us to measure the magnetic field strength through OH observations of the 18 cm lines (regardless of the presence of H II regions) toward star forming molecular dense cores. However for the proper interpretation of the results, the chemistry of the OH should be well characterized in the observed regions, as its abundance may change significantly (specially toward the densest part of molecular clouds: e.g., [34]).

The OH masers (in the 18 cm lines) associated with high mass star forming regions are also used as a tracers of the Galactic magnetic field [21]. The Galactic ASKAP Survey (GASKAP, [12]) will already provide an unbiased, flux-limited survey of OH masers associated with high-mass star forming regions. This survey will serve as base to carry OH Zeeman observations with SKA1 towards a large amount of sources.
In addition, the 18 cm OH masers are also an excellent probe to study the ISM in nearby galaxies, but also in high redshift galaxies [28, 31].

Other type of masers that need a higher volume density to be excited (e.g., $CH_3OH$ line at 6.7 GHz, OH line at 6.0 GHz: e.g., [14], [2] ) can be used to more specifically study the role of the magnetic fields at the scales of few thousand AU in very dense and hot molecular environments around massive prostars.

A more complete information on the study of molecular clouds and star formation in the context of SKA to the can be found in the chapters of this book by Anglada et al., Osorio et al. and Martin-Pintado et al.

# 5 Circumstellar envelopes

An interesting phenomenological issue in the last phases of the stellar evolution is the morphological change from expanding spherical envelopes in the AGB phase to the mostly aspherical and very varied shapes of the PNs [5, 38, 4]. Among the interesting observational features is the presence of fast collimated outflows, with properties resembling those of the star forming regions and that cannot be produced by the stellar radiation pressure [7]. The origin of the asphericity is attributed to the influence of a binary companion, a disk, a magnetic field, or a combination of these [30, 37]. Most of the information on magnetic fields in circumstellar envelopes and proto-PN comes from polarization observations of maser emission, mainly from SiO, $H_2O$, OH (e.g., [1, 29, 20, 35]). Recently, polarization has also been detected in non-masing molecular lines [36, 19].

SKA1 is going to provide a significant step in understanding of the magnetic field in the last phases of stellar evolution through the observations of the maser emission of the 18 cm OH lines. Note, that a few thousand of OH masers have already been detected in different phases of evolved stars (e.g., AGB stars, young Planetary Nebulae: [13]). A high fraction of the OH masers exhibit polarized emission [39] . As in the case of high-mass star forming regions, GASKAP [12]) will obtain an unbiased, flux-limited survey of OH masers associated with evolved stars. The catalogue obtained could be used to carry out extensive studies to characterize the role of magnetic fields in the mass-loss processes of evolved stars, and how these depend on mass and evolutionary stage. Finally, we note that more information about the scientific contribution from SKA to the study of circumbinary envelopes can be found in the chapters of this book by Alcolea et al. and by Gomez et al.

## Acknowledgments

JMG acknowledge support from the Spanish Ministry of Economy and Competitiveness (MINECO) through grant AYA2011-30228- C03-02.

# The Earliest Stages of Star Formation: Massive Protostars and the SKA

**Mayra Osorio[1], Asunción Fuente[2], Guillem Anglada[1], Alejandro Báez-Rubio[3], Gemma Busquet[1], Javier R. Goicoechea[4], Izaskun Jiménez-Serra[5], Jesús Martín-Pintado[3], Juan M. Mayen-Gijon[1], Aina Palau[6], Álvaro Sánchez-Monge[7] Mario Tafalla[8], and Belén Tercero[4]**

[1] Instituto de Astrofísica de Andalucía (CSIC), Glorieta de la Astronomía s/n, E-18008 Granada, Spain
[2] Observatorio Astronómico Nacional (OAN IGN), Apdo 112, E-28803, Alcalá de Henares, Spain
[3] Centro de Astrobiología (CSIC/INTA), Ctra. de Torrejón a Ajalvir, km 4, E-28850 Torrejón de Ardoz, Madrid, Spain
[4] Instituto de Ciencia de Materiales de Madrid (ICMM-CSIC), Sor Juana Ines de la Cruz 3, E-28049 Cantoblanco, Madrid, Spain
[5] University College London, Department of Physics and Astronomy, 132 Hampstead Road, London NW1 2PS, UK
[6] Centro de Radioastronomía y Astrofísica, Universidad Nacional Autónoma de México, P.O. Box 3-72, 58090 Morelia, Michoacán, México
[7] I. Physikalisches Institut, Universität zu Köln, Zülpicher Str. 77, 50937, Köln, Germany
[8] Observatorio Astronómico Nacional-IGN, Alfonso XII 3, E-28014 Madrid, Spain

## Abstract

The high spatial resolution and sensitivity provided by the SKA will indubitably increase our knowledge on the processes that are taking place during the formation of high-mass stars. Free-free emission, recombination lines and molecular emission will be easily detected and mapped even in remote regions where these stars are located. This will enable us to investigate star formation in both clustered or isolated mode. In this Chapter, we describe the potential of SKA for studying the main processes of the early stages of massive star formation: cloud fragmentation, kinematics and chemistry of hot molecular cores, photo-ionized regions and photo-evaporating disks.

# 1 Introduction

Despite high mass stars ($> 8$ M$_\odot$) are more scarce than their lower-mass counterparts, they play an important role in many respects. Massive stars control the dynamics and chemical evolution of the Galaxy, and their powerful radiation heats, ionizes, photodissociates and evaporates the cloud where they are born and their surroundings ([80]). Therefore, they have a strong influence on the formation of the next generation of stars. Two decades ago, the main signposts of the early stages in the life of massive stars were the photoionized regions that these stars themselves produce in their surroundings (the so-called HII regions). Since HII regions are in expansion, the size of these ionized regions is taken as indicative of their youth, with the smaller ones (compact HII regions) being younger. More recently, the study with millimeter interferometers focussed on even earlier stages, the so-called Hot Molecular Cores (HMCs), considered as forerunners of the HII regions, and on the Infrared Dark Clouds (IRDCs), which are so dense and cold that they appear in absorption even at mid-infrared wavelengths.

The very high sensitivity and angular resolution offered by SKA will be essential to study the earliest phases of massive stars. Although ALMA is already doing many contributions on this topic through dust and molecular tracers, SKA will be the only instrument able to detect and map the ionized emission coming from high-mass protostars located at remote distances. That will provide the first surveys of HII regions associated with high-mass stars of a wide range of spectral types. Furthermore, SKA will be able to detect radio-jets driven by low luminosity objects (including intermediate- and low-mass protostars) born near massive stars. This will allow us to identify and record the number of companions in crowded regions, in order to find differences between isolated and clustered modes of star formation. But SKA will not only provide large surveys for statistical studies, but also it will permit detailed analysis of individual sources, which will be extremely useful, for example, to distinguish radio-jets from HII regions. On the other hand, SKA will allow us to test theoretical predictions by carrying out a sort of observations that currently are difficult to perform. This is the case of the weak ionized winds arising from photo-evaporating disks around massive protostars that may be detected by means of radio recombination lines (RRLs). In addition to these lines of research, SKA will gather complementary information through molecular lines at centimeter wavelengths to study the kinematics and evolution of the filamentary structures whose fragmentation originates the formation of dense cores where stars are formed. These molecular tracers will also allow us to study accretion and ejection processes (infall, outflow, accretion disks, etc) that will take place in cores in order to build up a star.

In this Chapter, we discuss several topics on the early phases of star formation with special emphasis in those aspects of massive star formation where SKA can make a decisive contribution.

# 2 Initial Conditions

It is well accepted that molecular clouds, which are the cradles of star formation, present a complex filamentary morphology. Despite filaments were identified more than 30 years ago ([71]), the omnipresence of such structures in star-forming complexes as revealed by recent Herschel observations ([2, 52, 7]) triggered again special attention on their formation mechanism and their role in the star formation process. It has been suggested that massive stars form at the intersection of massive filaments, the so-called hubs (e.g., [53, 29]), but still many questions remain open regarding this topic. How do filaments form and evolve?, how do filamentary structures fragment to form dense cores, and hence form stars? Several studies performed toward massive star-forming regions reveal supersonic non-thermal motions and suggest the formation of filaments by the convergence of flows or by filament-filament collisions (e.g. [72, 17, 39, 32]). However, other agents like large-scale turbulence and magnetic fields could also play a role in their formation (e.g. [58, 54]). In this sense, the SKA, specially in its spectroscopic mode, will play a major role in the study of the formation of large-scale filamentary structures and their subsequent fragmentation. The sensitivity reached by the SKA in line mode will be much higher than what other facilities can offer currently. By observing some key atomic/molecular lines we can obtain the physical conditions of the gas. The HI line at 21 cm as well as carbon radio recombination lines can trace the cold and warm neutral gas of the interstellar medium around filamentary structures probing the low-density material. One of the important molecules that have been extensively used to trace the dense gas material is ammonia, whose inversion transitions are at frequencies that will be accessible during the second phase of the SKA (SKA2). With this molecule, we can trace the dense filament itself where star formation takes place and derive the initial conditions of the gas that leads to the formation of massive stars (e.g. [63, 14]). Clearly, the velocity information provided by line observations is a crucial ingredient to study the initial conditions of star formation. It offers the possibility to investigate the level of turbulence at several scales, from the large-scale and low-density material to the dense filaments that subsequently fragment to form dense cores and hence stars, and to study the core-to-core velocity and the velocity dispersion. Not only the kinematics can reveal the presence of colliding flows but also observations of shock tracers such as $CH_3OH$ are expected to be present where flows converge. Therefore, it is clear that the SKA will open a new window for spectroscopic line studies that will shed light on the initial conditions of massive star formation.

# 3 Clustered vs Isolated Star Formation

It has been recognized for decades that the star formation process within filamentary clouds takes place in different modes. The most common is the "clustered" mode, where stars are formed in gravitationally bound pc-scale groups with stellar densities larger than $\sim$100 pc$^{-3}$, so that interactions between different stars may occur during their formation. On the contrary, stars can also form in a more distributed or "isolated" mode, where interactions between stars are very scarce. [12] studied the formation of stars in both modes through a numerical simulation of a turbulent molecular cloud and found that the star formation mode is determined by the local gravitational binding of the cloud and ultimately the density in the cloud, but other ingredients such as magnetic field, and radiative feedback [33], could play a role as well, and the question of what determines the mode of star formation remains open. This is in part because there is a clear lack in the literature of observational works characterizing a sample of clusters in their first evolutionary stages of formation, which should constrain both theory and simulations. Such a lack of observational works is due to the fact that protoclusters are deeply embedded in their molecular clouds, being obscured in the optical/infrared and mostly emitting at radio wavelengths. Thus, achieving the high sensitivity ($\sim$0.2 $M_\odot$) and spatial resolution ($\sim$1000 au) of optical/infrared studies at longer wavelengths (millimeter, centimeter), to properly study those protoclusters, is very challenging, given the current instrumentation capabilities, and the typical large distances of massive star-forming regions. Some approaches ([13, 60, 61]) are being conducted in the millimeter range by compiling sensitive interferometric observations with subarcsecond angular resolutions towards massive and intermediate-mass star forming regions located at distances <3 kpc. In these works, the authors find regions with high fragmentation levels and regions showing no signs of fragmentation (see Fig. 1), and suggest that the density of the cloud might play a role in determining the fragmentation level ([61]), in line with simulations of [12]. [60] explore the relation between fragmentation degree and other cloud parameters and suggest that strong magnetic fields could produce low fragmentation levels as well. The measurement of OH Zeeman splitting using the SKA (see Chapter by Girart et al. of this book) will allow to estimate the intensity of the local magnetic field and hence, its role in the fragmentation process. Since most of the known massive star forming regions are located further than 3 kpc, the more numerous intermediate-mass star forming regions have been used to test theoretical models ([60, 61]). However the results are not directly extrapolable to the most massive stars in which some effects, such as the radiative feedback, are expected to be more important. The SKA will allow, for the first time, to extend these works to a statistically significant sample of massive star forming regions and compare them with theoretical models.

Furthermore, because the aforementioned works focus on the millimeter range, it would be extremely useful to assess the richness of the protoclusters using an independent tracer of lower luminosity objects (i.e. of intermediate- and low-mass protostars) in massive star-forming regions. Low mass protostars are extremely difficult to detect in the millimeter range because their emission represent only a small fraction (<0.1%) of the total emission coming from the parental molecular cloud. Since low-mass protostars emit significantly in the centimeter range through the ionization of shocks generated by the propagation of their protostellar jets (also known as thermal radio jets), the centimeter range is an excellent window to study the number of low-mass protostars in a massive star-forming region. The highest frequency band available with the SKA1 (SKA-MID band 5: 4.6-13.8 GHz) will play a major role in the study of the continuum emission in protoclusters thanks to its excellent sensitivity and high angular resolution. Protostellar radio jets associated with low-mass young stellar objects are typically weak (e.g. [4, 5]). Figure. 3 of Chapter by Anglada et al. (this book) shows a correlation between the radio luminosity and the bolometric luminosity. Using the $L_{\rm bol}$ range of 10-100 $L_\odot$, the expected radio luminosity is $S_\nu d^2$ = 0.03-0.13 mJy kpc$^2$. Given that the aim is to detect emission arising from radio jets associated with the low-mass members of the protocluster at the typical distances of massive star-forming regions ($\sim$4 kpc), the expected flux density is then $\sim$2-8 $\mu$Jy. Then, the sensitivity required is $\sim$0.5 $\mu$Jy, which will be reached with the SKA in just one hour. Observations at different frequencies will allow us to determine the spectral index of the continuum emission in the centimeter range. By combining with ALMA observations in the millimeter/submillimeter range it will be possible to separate the emission coming from ionized material from the thermal dust continuum for each member of the protocluster, measure the level of clustering at centimeter wavelengths, and compare it with that found in the millimeter/submillimeter range. The SKA will open a new window to investigate the star formation process in clustered mode, by allowing to assess the number of low-mass protostars driving radio jets in distant massive star-forming regions.

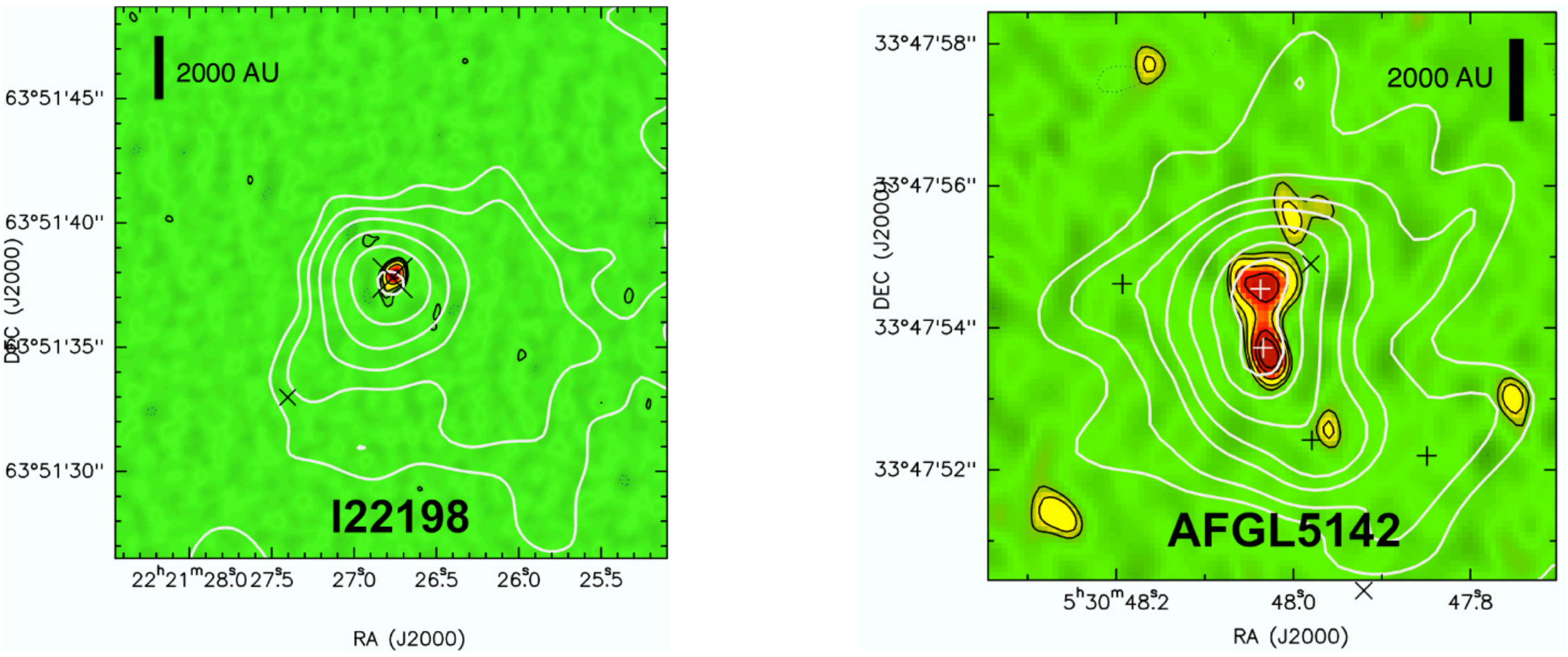

Figure 1: *Left:* IRAS 22198+6336 (I22198) star-forming region of $\sim$ 500 $L_\odot$. *Right:* AFGL5142 star-forming region of $\sim$ 3000 $L_\odot$. In both panels, the color scale corresponds to the PdB 1.3 mm emission at $0\overset{''}{.}4$ angular resolution ([60]) and the white contours correspond to the SMA 1.3 mm emission at $1''$-$2''$ angular resolution ([79, 67]). The plus signs indicate the sources identified by [79] in AFGL5142, and the tilted crosses correspond to mid-infrared sources. Note that for I22198 the large-scale envelope (white contours) does not split up in different subcondensations, while for the AFGL5142 case the emission splits up into around 6-7 subcondensations. The field of view corresponds to the same spatial scale for both regions (see [60]) for further details).

## 4 Observing the Signatures of Gravitational Collapse

Although gravitational collapse should play an essential role in the star formation process, infall motions have been always elusive to a detailed study. So far, only a few observational signatures, based on the shape of molecular line profiles (1D) have been commonly used (e.g., [48]). More robust signatures, based on images that spatially resolve the infalling gas (3D) can be obtained with sensitive high angular resolution observations ([3]). These images can provide additional spatially-resolved information that allows us to further investigate the kinematics and physical parameters of the molecular core around the protostar.

Low-mass protostars provide some of the best targets to study gravitational collapse due to their proximity and often-isolated nature. Even in these objects, observations with angular resolution better than one arcsecond are required to explore the gravitational acceleration region and to separate the infalling motions from the bipolar outflow ejection. The SKA will naturally satisfy this need of high angular resolution, especially when working at the highest achievable frequencies, where sub-arcsecond observations should be routinely possible. High frequency observations are also needed to detect the lines of ammonia ($NH_3$), which lie in the vicinity of 24 GHz and represent the most reliable tracer of infalling dense gas. Ammonia seems immune to the freeze out problem that affects most other molecules, and is therefore the ideal species to determine the complex gas kinematics expected near the protostar, where infall, outflow, and rotation motions coexist in a small volume. It is therefore critical that the SKA receivers reach the high observable frequencies already planned for the full SKA design.

High mass protostars can also provide high-quality targets for the study of gravitational infall. This is illustrated by the recent VLA observations of ammonia inversion transitions that reveal, for the first time, the expected 3D signatures of protostellar infall in the very massive hot molecular core (HMC) near G31.41+0.31 ([47]; see Fig. 2). The intensity of the ammonia emission is compact and sharply increases towards the center in the blue-shifted velocity channel maps, while it shows a more flattened distribution in the red-shifted velocity channels. Additionally, the emission becomes more compact with increasing (relative) velocity for both red and blue-shifted channels (Fig. 2, central panel). A new infall signature, the "central blue spot", easily identifiable in the first-order moment maps is introduced (Fig. 2, left panel). Also, it is shown that rotation produces an additional, independent signature, making the distribution of the emission in the channel maps asymmetric with respect to the central position, but without masking the infall signatures (Fig. 2, right panel). All these imaging (3D) signatures, which are identified in G31 HMC for the first time, can be used to study other protostars, provided a high enough sensitivity and angular resolution are reached. The SKA appears as an ideal instrument to carry out a deep survey of these new 3D kinematic signatures in star-forming regions.

## 5 Hot Molecular Cores

Hot molecular cores are compact (diameters $\leq 0.1\,\mathrm{pc}$), dense ($n \geq 10^7\,\mathrm{cm}^{-3}$), hot ($T \geq 100\,\mathrm{K}$) and dark ($A_v \geq 100\,\mathrm{mag}$) molecular clumps of gas and dust in or near sites of recent massive star formation (e.g. see [43]). Nevertheless, the nature of these objects and the complex chemical and physical processes occurring in them are not fully understood. In some cases, hot cores are believed to be the formation sites of massive stars [56], more specifically the precursors of ultracompact HII regions [19]. In others, the central energizing source has not been identified and seem to be only externally heated dense cores (see, e.g., the controversial case of the Orion-KL hot core; [77]). In all cases, hot molecular cores are associated to luminous IR sources in which massive bipolar outflows, accretion disks and inflow motions are observed ([16, 59, 47]).

A common characteristic in all these objects is their extremely rich chemistry. Due to the phenomena associated with massive star formation, complex molecules are released from the grain mantles and subsequent gas-phase reactions give rise to high abundances of a large variety of large molecules such as $CH_3CH_2CN$ or $HCOOCH_3$ ([55, 75, 25]). However there is some controversy about the mechanism(s) that released these molecules from the grain mantles into the gas phase. Hot cores are supposed to trace the innermost parts of the condensation where the massive star is being formed. In this region, dust grains are radiatively heated by the newly formed star, increasing their temperature to ∼100 K and producing the evaporation of their mantles ([57]). However, the radiative heating mechanism is questioned, as some observations suggest that the complex molecules could be associated with shocks as well ([45, 18, 28, 27, 24, 77]). Recent high sensitivity and high spatial resolution millimeter interferometric observations in Orion-KL and nearby (∼1 kpc) intermediate-mass star forming regions revealed that for some targets, the region where the emission of the complex molecules arises is actually the rotating disk around the young star ([59]), while in others, is associated to shocked regions along the bipolar outflow ([24, 27]). However, this kind of work is difficult to repeat for other massive star forming regions because most of them are located at a distance >3 kpc. In this sense, the SKA instrument will allow to observe the low frequency transitions of these complex molecules in large samples of massive hot cores.

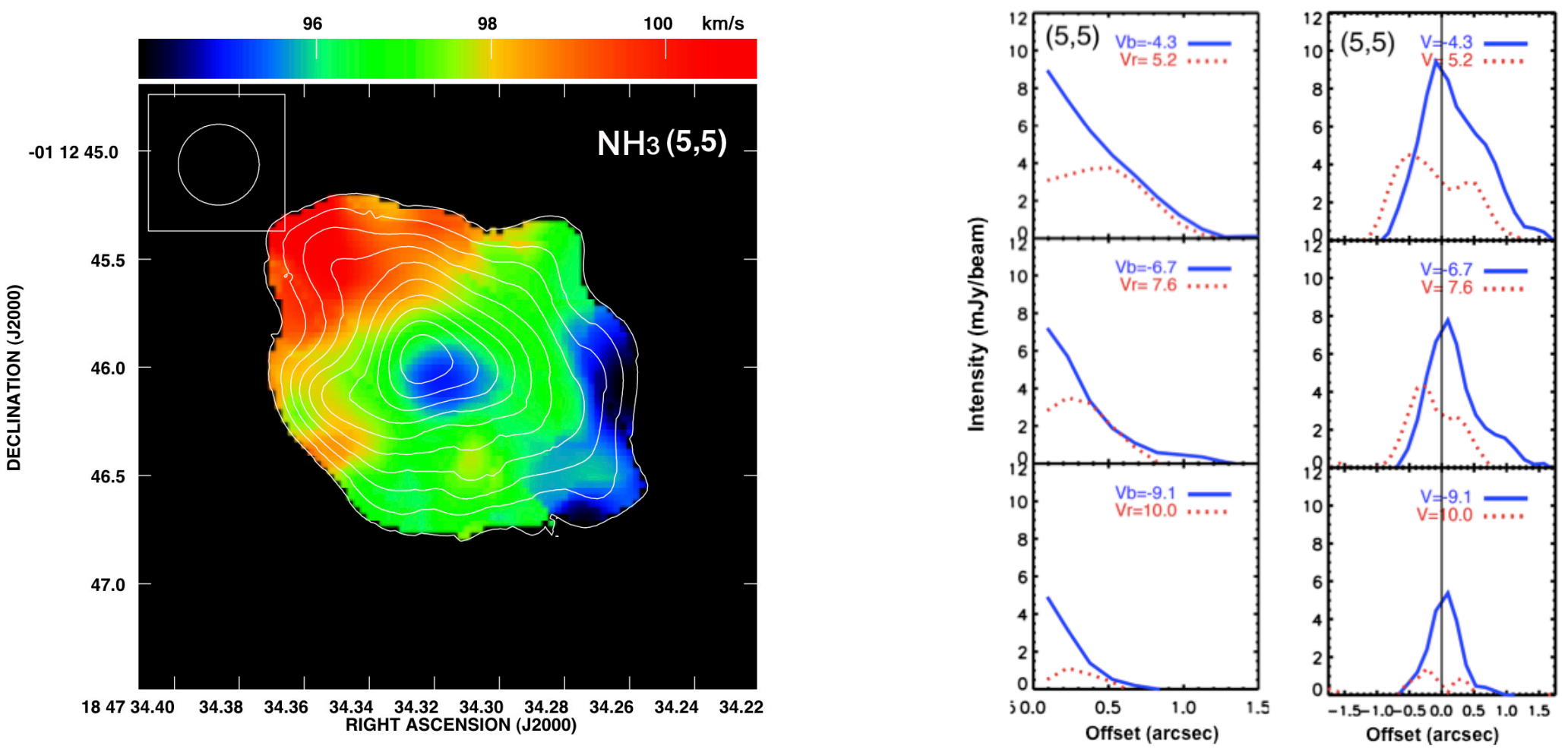

Figure 2: VLA observations of the ammonia (5,5) inversion transition in G31 HMC (from [47]). *Left:* Overlay of the integrated intensity (zero-order moment; contours) and the intensity weighted mean velocity (first-order moment; color scale) maps. The synthesized beam is shown in the upper left corner. The image shows a northeast-southwest velocity gradient and a central blue spot, indicative of infall, towards the center. *Center:* Azimuthally averaged observed intensity as a function of radius, for different pairs of blueshifted (solid blue line) and redshifted (dotted red line) channel maps. Labels indicate the channel velocity (in km $s^{-1}$) relative to the assumed systemic velocity of the cloud. *Right:* Observed intensity, averaged over half-annuli, as a function of radius. Negative offsets correspond to the NE half (redshifted first-order moment) of the source, and positive offsets correspond to the SW half.

Spatial distribution of molecular species such as $HCOOCH_3$, $CH_2CHCN$, $CH_3CH_2CN$, $HC_3N$, $CH_3CN$ and their vibrationally excited states have been studied in the millimeter and submillimeter domain (see, e.g., [46, 22, 20, 8]) but never at the SKA wavelength coverage due to the limited sensitivity of current instruments. The high sensitivity of SKA will provide the detection of these species, even for high energy vibrationally excited states. Frequencies of low $J$ rotational transitions in the vibrationally excited states of these complex molecules are inside the SKA coverage, most of them between 4 and 20 GHz (see, e.g., [20, 42]; JPL catalog).

At the final phase of SKA, full SKA will cover frequencies up to 24 GHz. At this point, we will have a unique opportunity for studying ammonia emission in hot cores with a high sensitivity. The Orion-KL hot core was discovered by [34], who identified it as a compact and hot source of ammonia emission, an ubiquitous molecule in hot cores. Because of its high abundance and its spectroscopic characteristics, ammonia will probably be the most valuable tool to determine the physical conditions and study the kinematics of the warm ($T_k$>100 K) gas associated with hot cores. Recent high spatial resolution observations of the ammonia inversion transitions using the upgraded VLA showed that ammonia molecules in Orion-KL have been released into the gas phase through the passage of shocks and not by stellar radiation ([27]). Thus far, this kind of study can only be done in the closest massive hot core, Orion-KL, that might not be representative of most hot cores. The full SKA will allow to observe large samples of hot cores in ammonia emission and provide the first statistically significant sample to investigate the nature of these objects and their relation with the mid/high mass star formation process.

# 6 Ultracompact (UC) HII regions

High-mass stars with high temperatures ($> 10^4$ K) emit large amounts of energetic photons ($h\nu > 13.6$ eV) that ionize the surrounding gas (mainly hydrogen), which results in the development of a region of ionized gas, known as HII region. This is one of the clear signposts and main differences of high-mass star formation with respect to low-mass star formation. In the last stages of the formation of a high-mass star, HII regions appear as gigantic structures (size $\sim$ 10–100 pc, electron density $\sim$ 10–100 cm$^{-3}$) that have already dispersed the material of the natal cloud. However, at the onset of the process, HII regions are much more compact ($<$ 0.1 pc) and have higher densities ($> 10^4$ cm$^{-3}$; [44]). This kind of objects — known as hypercompact (HC) and ultracompact (UC) HII regions — are deeply embedded in dense gas, and possibly still accreting in the form of ionized accretion flows ([41]). Their lifetimes have been measured to be longer than expected when assuming a simple model of expansion, which implies the existence of certain mechanisms that prevent the expansion and confine the HII regions [66]. Co-existing with these "young" HII regions, or even prior to their development, the first manifestation of ionized gas is the presence of outflowing ionized material in the form of collimated jets or winds (with sizes $\sim$ 100–1000 au, and velocities $\sim$ 500 km s$^{-1}$), that are likely directly related to the large-scale molecular outflows seen in many star forming regions ([6]). The study of the ionized gas is therefore, crucial to understand the accretion and feedback processes at the onset of high-mass star formation. The observations are, however, challenging due to the high sensitivities and angular resolutions that are required. HC HII regions have sizes $\sim 0.01$ pc, which at the typical distances of high-mass star forming sites ($> 3$ kpc) correspond to $< 0\farcs5$ [68]. It is also necessary to reach sensitivities $\sim 2\ \mu$Jy at radio wavelengths (1–10 GHz) in order to detect the free-free emission of HII regions associated with all high-mass stars (from B3 to O4 spectral types). Finally, a high dynamic range and good image fidelity are necessary to properly map the compact and weak HC and UC HII regions that are likely to be found in the vicinity of large (1 pc), bright (1 Jy) and more evolved HII regions.

The SKA, with its high sensitivity, angular resolution, and good image fidelity, becomes the ideal instrument to study the photo-ionized gas in the first stages of high-mass stars formation by means of continuum and spectral line observations.

In the continuum, the high sensitivity that SKA offers will permit to carry out large surveys of galactic HII regions over the whole Galaxy, providing for first time a statistically significant sample of HII regions associated with high-mass stars of all spectral types (and not only detections of the more massive ones as is possible nowadays). Observations at different frequencies (from 1 to 24 GHz) will permit to construct the spectral energy distribution of ionized gas sources to help to determine the nature of the emission (optically thick/thin photoionized HII region, wind, radio jet) and the spatial structure (density gradients) of the ionized gas. We expect optically thick free-free emission (spectral indices close to +2) for deeply embedded, compact HII regions, and flatter spectral indices for thermal radio jets or winds (see Chapter by Anglada et al. in this book). High-angular resolution, multi-frequency continuum observations will resolve the structure of HC HII regions and permit to study the morphology, size, and flux dependence with frequency. These properties, together with the luminosity of the sources, will allow us to characterize the emission and structure of the sources. Given the positive spectral indices, observations in the high frequency bands of SKA (10–20 GHz) are fundamental for their study. Finally, recent works ([65], [66]) have proposed that HII regions should undergo flickering (i.e., changes in their brightness and morphology) on short periods of time due to changes in the accretion rate. SKA will easily test this scenario via multi-epoch, high-angular, and sensitive continuum observations.

Complementary to the continuum, spectral line observations of HI and recombination lines will give us information on the kinematics of the ionized and atomic gas at scales of $\sim$100 au (see Fig. 3). This high spatial resolution is required to probe the thin gas layers where the energy interchange between the parent molecular cloud and the nascent massive star occurs and eventually determine the evolution of the HII region ([69, 70]). The intensity of these lines is estimated to be only 1–10% the intensity of the continuum in the radio regime. SKA will observe hundreds of recombination lines in the different bands, that when stacked will improve the sensitivity and provide clear detections.

## 7 Disks Photoevaporation

There is increasing evidence that massive stars form by gas accretion through neutral, molecular circumstellar disks ([62, 15, 38]). These neutral massive disks are expected to suffer strong photo-evaporation as a consequence of the intense UV radiation field arising from the central star. Theoretical models predict that the photo-evaporation of circumstellar disks may lead to the formation of ionized disk winds at the late stages of massive star formation ([36, 30]).

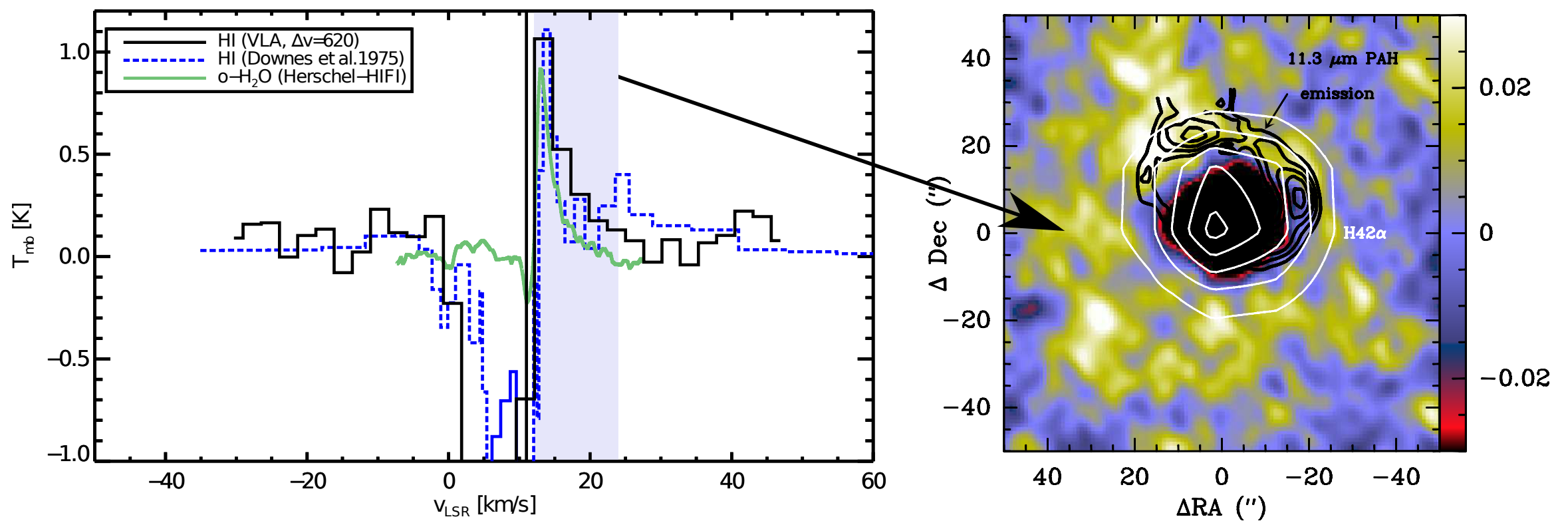

Figure 3: *Left:* Comparison of the HI spectra observed with the VLA, the single-dish observations of [21] (blue line), and the ortho-$H_2O$ ($1_{1,0}$→$1_{0,1}$) observations observed with Herschel ([64]; green line). *Right:* Image of the integrated intensity of the HI emission at red velocities (12−24 km $s^{-1}$) obtained from the HI VLA map. The HI emission arises in correspondence with the emission at 11.3 $\mu$m of polycyclic aromatic hydrocarbon (PAH) molecules ([11]; black contours) tracing the photodissociation region. The HI absorption is observed toward the UC HII region (in white contours), which is traced by the integrated intensity of the H42$\alpha$ recombination line observed at the IRAM 30m telescope (S. Treviño-Morales & A. Sánchez-Monge, private communication).

The detection of ionized winds in photo-evaporating disks has been attempted mainly by performing observations of the continuum emission at centimeter wavelengths where the contribution from thermal dust emission is much smaller. Some examples are the emission line star MWC349A ([49, 74]), S140 IRS1 ([35], [51]), Cepheus A HW2 ([38]) or LkH$\alpha$ 101 ([76]). For most of these sources, the radio continuum emission from their ionized photo-evaporating component is very faint (ranging from 0.1 to 1 mJy/beam; [35, 38]), which gives an idea of how challenging the detection of these winds is with current instrumentation. The detected continuum emission follows the disk surface (as in e.g. Cepheus A HW2 or MWC349A; see Fig. 4), which supports the idea that ionized winds are produced by disk photo-evaporation. We note however that only for the exceptional case of MWC349A, which shows RRLs with non-LTE (maser) emission, we know in detail the launching radius and physical structure of the photo-evaporating ionized wind ([50, 10, 9]).

The SKA1, which will operate at wavelengths between 50 MHz and 13.8 GHz (from 6 m to 2 cm), will represent a break-through in the study of photo-evaporating disk winds in massive star forming regions thanks to its unprecedented sensitivity and high-angular resolution imaging capabilities at centimeter wavelengths. The continuum emission associated with photo-evaporating disks (of the order of 0.1-1 mJy) will be readily detected with the SKA1 in Band 5 since the expected rms is $\sim$1 $\mu$Jy hr$^{-0.5}$ for a 0.″1-beam and $\sim$ 4 $\mu$Jy hr$^{-0.5}$ for a 0.″03-beam. In addition, based on our model of MWC349A ([10, 9]), the predicted peak intensity of the H80$\alpha$ line emission at 12.6 GHz arising from the wind is 5.6 mJy for a 10 km s$^{-1}$ velocity resolution (note that the RRL linewidths are, at least, a few tens of km s$^{-1}$; [37]). This emission will be detected with the SKA in Band 5 with a signal-to-noise ratio $\geq$35 for a 0.″1-beam (rms $\sim$160 $\mu$Jy) in just 1 hour of integration time. The high signal-to-noise ratio of the RRLs detected toward MWC349A will allow to image with SKA2 the rotation and angular momentum transport of its ionized wind with extraordinary detail down to spatial scales of $\sim$ 100 au. We would like to stress that the higher the frequency of the observed RRL, the brighter the predicted intensity of these lines (their flux increases as $\nu^{1.1}$), making the Band 5 receivers of the SKA1 (from 4.6 GHz to 13.8 GHz) an essential component in the study of photo-evaporating disk winds around massive protostars.

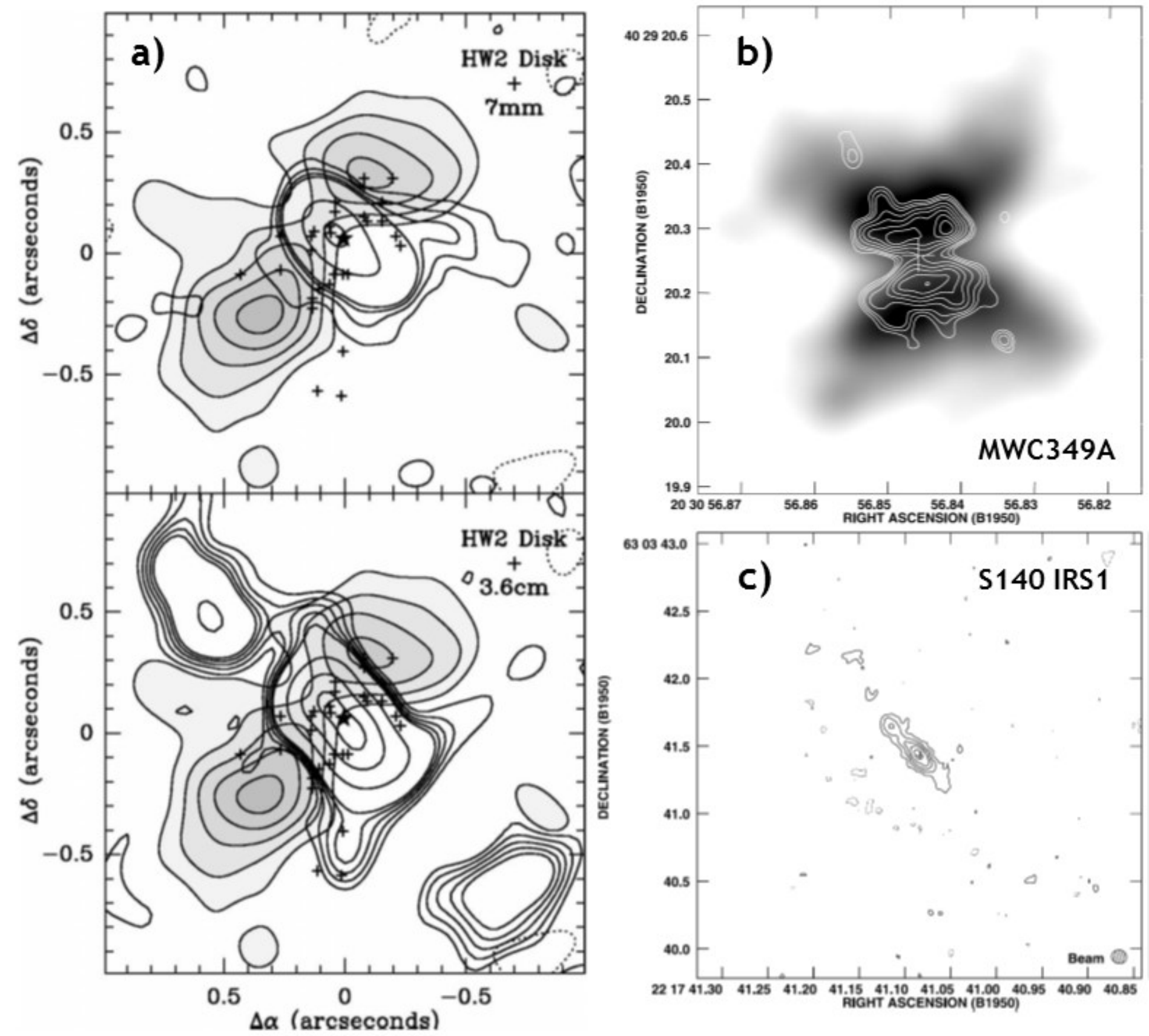

Figure 4: *(a)* Interferometric images of the Cepheus A HW2 massive star forming region (from [38]). The neutral molecular disk is detected in $SO_2$ with the VLA (gray scale and think contours). In addition to the thermal radio jet (in the direction perpendicular to the disk), the radio continuum emission at 3.6 cm and 7 mm shows an extra, fainter component that fills the surface of the disk. This component is possibly associated with the photo-evaporation of the neutral disk. *(b)* Radio continuum images obtained toward the emission line star MWC349A at 1.3 cm (in gray scale) and 7 mm (in contours; [74]). The photo-evaporating material fills the cavity left by the neutral disk (in the east-west direction). *(c)* Radio continuum emission reported by [35]) at 6 cm toward the S140 IRS1 massive protostar. The continuum emission arises from an equatorial wind produced by the disk photo-evaporation.

For an object at a distance of 1 kpc with a mass loss rate a factor of 50 lower than that of MWC349A (i.e. similar to those of Cepheus A HW2 or S140 IRS1), the detection of the H80$\alpha$ RRL emission from the photo-evaporating disk would require 50 hours of integration time to achieve a signal-to-noise ratio $\geq$4 (intensity of 31 $\mu$Jy) in a 1″-beam and a velocity resolution of 30 km s$^{-1}$ (rms$\sim$7.5 $\mu$Jy). We note however that this integration time is expected to be reduced by a factor of 100 during the second phase of the SKA (SKA2). This will open the possibility not only to detect many more of these photo-evaporating disks but also to understand the physical processes involved in the formation of ionized disk winds around massive protostars.

## 8 SKA Synergy with Herschel and ALMA

Infrared surveys, in combination with complementary multiwavelength observations, have produced an extensive sample of well studied protostars, covering a wide range of luminosities, evolutionary stages and a broad range of initial environmental conditions (e.g., the Herschel Orion Protostar Survey (HOPS) in the Orion Molecular Cloud at a distance of 420 pc; [1, 26]). These observations yield protostellar Spectral Energy Distributions (SEDs) that are well-sampled over several orders of magnitude in wavelength. Combined with detailed radiative transfer modeling, these powerful data-sets have made it possible to infer physical properties (total luminosity, mass, density, temperature, mass infall rate, geometry parameters, etc.) of the dusty envelopes and disks around protostars. The excellent sensitivity, high-angular resolution and large field of view of SKA make it an ideal instrument to perform complete, deep surveys of the protostellar population in large fields. We expect that these observations will detect radio jets essentially in all the previously known sources, including very low luminosity objects, and extremely young protostars (the so-called PBRs; [73]) providing a tracer of outflow (and thereby accretion) unbiased by extinction and the capability to identify previously unknown multiple systems. It is well known that there is a dependence on outflow momentum rate and source luminosity; the next question we want to answer is whether there is a dependence on evolution. The SKA data will complement the observations at shorter wavelengths which primarily trace the evolution of the infalling envelopes, and thereby provide an important component in constructing a comprehensive audit of infall, accretion and outflow in these sources.

On the other hand, ALMA will characterize the physical conditions and chemistry of the molecular gas in hundreds of individual protostars through the rotational lines of a wealth of molecular species from CO and its isotopologues to light complex organic molecules [40]. In contrast with low mass stars, the gas in the vicinity of a massive star is ionized by the stellar UV radiation and forms a compact HII region while the young star is still accreting matter from the molecular cloud. Therefore, we need to probe the ionized, atomic and molecular gas to understand the complexity of the accretion and outflow processes in massive protostars. The SKA will provide access to different tracers of the ionized and atomic gas at comparable spatial resolution and sensitivity as provided by ALMA in molecular lines. For individual protostars, the SKA will resolve and characterize the youngest compact HII regions and their interfaces with the molecular cloud through observations of the hydrogen and carbon RRLs and of the HI 21cm line. Together with the continuum emission, these lines will also inform on the kinematics and physical conditions of the gas in the ionized jets and of the atomic gas in the innermost layers of the PDRs formed on the walls of the cavity excavated by the outflow. The high spatial resolution and sensitivity of the SKA will be enough to resolve the UV illuminated surfaces of circumstellar disks and detect the photoevaporating flows emanating from them. In conclusion, the tandem SKA and ALMA will allow to study at high spatial resolution the morphology, dynamics and chemistry of all the gas (ionized, atomic, molecular) involved in the formation and early evolution of a massive star, and surely will lead to an unprecedented advance in the comprehension of this complex process.

## Acknowledgments

A.F. thanks the Spanish MINECO for funding support from grants CSD2009-00038, FIS2012-32096 and AYA2012-32032. A.P. acknowledges the financial support from UNAM, and CONACyT, México. Á.S.-M. acknowledges support by the collaborative research project SFB 956, funded by the Deutsche Forschungsgemeinschaft (DFG). G.A., G.B., J.M.M.-G. and M.O. acknowledge support from MICINN (Spain) grant AYA2011-30228-C03 (co-funded with FEDER funds) and from Junta de Andalucía (TIC-126). I.J.-S. acknowledges funding from the People Programme (Marie Curie Actions) of the European Unions Seventh Framework Programme (FP7/2007-2013) under REA grant agreement number PIIF-GA-2011-301538, and from the STFC through an Ernest Rutherford Fellowship (proposal number ST/L004801/1). J.R.G. and B.T. thank the Spanish MINECO for funding support under grants CSD2009-00038, AYA2009-07304 and AYA2012-32032 and also the ERC for funding support under grant ERC-2013-Syg-610256-NANOCOSMOS

# Protoplanetary disks, jets, and the birth of the stars

**G. Anglada[1], M. Tafalla[2], C. Carrasco-González[3], I. de Gregorio-Monsalvo[4], R. Estalella[5], N. Huélamo[6], Ó. Morata[7], M. Osorio[1], A. Palau[3], and J.M. Torrelles[8]**

[1] Instituto de Astrofísica de Andalucía, CSIC, E-18008 Granada, Spain
[2] Observatorio Astronómico Nacional-IGN, E-28014 Madrid, Spain
[3] Centro de Radioastronomía y Astrofísica, UNAM, 58090 Morelia, Michoacán, Mexico
[4] Joint ALMA Observatory, Santiago de Chile, Chile
[5] Departament d'Astronomia i Meteorologia, Univ. de Barcelona, E-08028 Barcelona, Spain
[6] Centro de Astrobiología (INTA-CSIC), E-28691 Villanueva de la Cañada, Spain
[7] Institute of Astronomy and Astrophysics, Academia Sinica, Taipei 106, Taiwan
[8] Instituto de Ciencias del Espacio (CSIC)-UB/IEEC, E-08028 Barcelona, Spain

## Abstract

Stars form as a result of the collapse of dense cores of molecular gas and dust, which arise from the fragmentation of interstellar (parsec-scale) molecular clouds. Because of the initial rotation of the core, matter does not fall directly onto the central (proto)star but through a circumstellar accretion disk. Eventually, as a result of the evolution of this disk, a planetary system will be formed. A fraction of the infalling matter is ejected in the polar direction as a collimated jet that removes the excess of mass and angular momentum, and allows the star to reach its final mass. Thus, the star formation process is intimately related to the development of disks and jets. Radio emission from jets is useful to trace accurately the position of the most embedded objects, from massive protostars to proto-brown dwarfs. One of the key problems in the study of jets is to determine how they are accelerated and collimated. High angular resolution observations at centimeter wavelengths are very useful to trace the base (at 100 AU scales) of the ionized component of the jets, close to the young central star and its accretion disk, where optical or near-infrared imaging is hampered by the high extinction. This kind of observations are also useful to trace the emission of dust grains that have grown up to large (centimeter-sized) scales in the process of planetary formation. In this chapter we review the main properties and recent results on jets and protoplanetary disks associated with young stars. We discuss the main expected contributions to this topic from the very sensitive, high-angular resolution centimeter wavelength observations that will be feasible with SKA.

# 1 Protoplanetary disks and planet formation

The birth of a star is accompanied by the formation of a surrounding disk of gas and dust containing the high-angular-momentum material of the collapsing envelope that could not fall directly onto the central (proto)star. This disk constitutes a reservoir of mass that slowly accretes onto the central star via the redistribution of its orbital angular momentum. The disk, in addition, provides the raw material to form planets. To this end, the dust component must undergo a complex process of size growth that starts with micron-sized particles, continues with cm-sized "pebbles" and meter-sized objects, and ends with km-sized planetesimals and fully-formed planets. The ubiquitous detection of IR-excess emission towards the youngest stars ([21]) and the increasingly stronger evidence for planetary companions in a majority of field stars ([59]) suggest that the above sequence of events represents an almost universal aspect of the star-formation process.

At the distance of the nearmost star-forming regions (120-140 pc), a protoplanetary disk of several hundred AU in diameter spans only a few arcseconds in size. Any detailed study of its structure therefore requires observations with sub-arcsecond resolution. The newly-operating ALMA interferometer will routinely reach this resolution, and will thus provide high-quality images of the millimeter/submillimeter emission from the protoplanetary disks in the most nearby regions. The dust grains, however, emit more efficiently at wavelengths close to their diameter, so any ALMA continuum observation at millimeter/submillimeter wavelengths will trace preferably the small dust aggregates present in a protoplanetary disk.

Exploring the complex path of dust growth toward planet formation requires observations at longer, cm-sized wavelengths, which lie outside the reach of ALMA but can be detected with the SKA. Also, moving to longer wavelengths will make the dust emission optically thinner and, therefore, better suited to trace the fine details of the dense central regions where the millimeter emission becomes optically thick. Indeed, the SKA, with its unique combination of high sensitivity and high resolution promises to be a key instrument in the exploration of the path from dust to planets, since the cm regime observable with the SKA presents a challenge to our understanding of grain growth. Up to millimeter sizes, grains are expected to stick to each other when colliding at the low velocities typical of disks, and this sticking mechanism provides a reliable starting point in the path of grain growth. Reaching the cm pebble size, however, requires an additional and not well-understood mechanism. At these larger sizes, sticking becomes inefficient, and grain-grain collisions lead to fragmentation or to bouncing, which can potentially stall the growth process ([12, 52, 18]). SKA observations of well-resolved disks will allow to explore potential solutions to the cm-barrier by mapping the distribution of grain growth in protoplanetary disks.

One of the best places to test planet formation theories are the so-called "transitional disks". Transitional disks are accretion disks that have developed large central cavities devoid of dust that might have been created by tidal interaction with protoplanets or substellar objects ([19]). These central cavities, initially inferred only from the SED modeling, have been recently imaged through mm/submm interferometry ([2, 26]) and polarized light observations (e.g., [41]). In several of these sources, the disk radial and azimuthal structure can be spatially resolved, revealing additional signatures of the gravitational interaction with proto-planets or young brown dwarfs, such as gaps, spiral arms, rings, and shadows.

Up to now, several substellar candidates have been detected within the gaps of transitional disks (e.g. [24, 30]). According to [40], a proto-planet can influence the growth and radial distribution of dust grains within a disk, due to the generation of pressure bumps outside the planetary orbit. The mm-sized grains are expected to accumulate in these pressure maxima (dust trapping), where they can grow to larger sizes and overcome the cm-barrier. The position and width of the pressure bumps depend on the orbital separation and mass of the proto-planet (among other parameters), so we expect to find different dust size populations at different disk radii. SKA will definitely help to confirm the presence of pressure bumps by imaging the cm dust emission in transitional disks. In combination with ALMA, it will be possible to spatially resolve the different dust populations as a function of the disk radius, a necessary step to understand if dust trapping is working and can explain grain growth in protoplanetary disks.

So far, 7 mm is the longest wavelength at which dust emission from transitional disks has been imaged. An interesting example is the transitional disk around HD169142, where recent sensitive high angular resolution VLA observations reveal the presence of two annular gaps and signs of circumplanetary material ([36]; see Fig. 1). A detailed modeling of this source suggests that large dust grains should accumulate near the walls of these gaps. It is expected that SKA will be able to image this and other similar disks even at longer wavelengths in order to trace cm-sized "pebbles". In this way, maps of the spectral index and, therefore, of the dust grain-size distribution across the disk would be obtained. This would make it possible to study the grain growth and migration processes in protoplanetary disks.

A unique feature of the SKA observations of protoplanetary disks (compared to, e.g., ALMA) is the possibility of observing simultaneously large numbers of nearby targets. The SKA field of view at band 5 (5-14 GHz), the best-suited band for thermal dust emission observations, will exceed 5 arcminutes in diameter, and in a cluster-forming cloud like L1688 in Ophiuchus, it will contain dozens of young stars at different stages of disk evolution. A single SKA observation of such a field will thus provide a statistically-significant sample of the disk population in the cloud, and will allow to combine the study of planet formation with the analysis of disk evolution.

While the emphasis of the SKA observations of protoplanetary disks will undoubtedly lie on the continuum emission and the study of the evolution of dust, line observations will provide additional information on different disk processes, in addition to revealing kinematical information not obtainable in the continuum. [28], for example, have shown that SKA observations should have enough sensitivity to detect the 21-cm line from neutral hydrogen (HI) arising from the UV-exposed outer layer of a disk. These HI observations will provide a unique view of the feedback of the star on the disk, and the role of UV irradiation and photoevaporation in the evolution of protoplanetary disks. Finally, SKA will also play an important role in the detection of Complex Organic Molecules (CMOs) in protoplanetary disks. They can be considered the building blocks for life and the key elements to understand the origin of life on Earth.

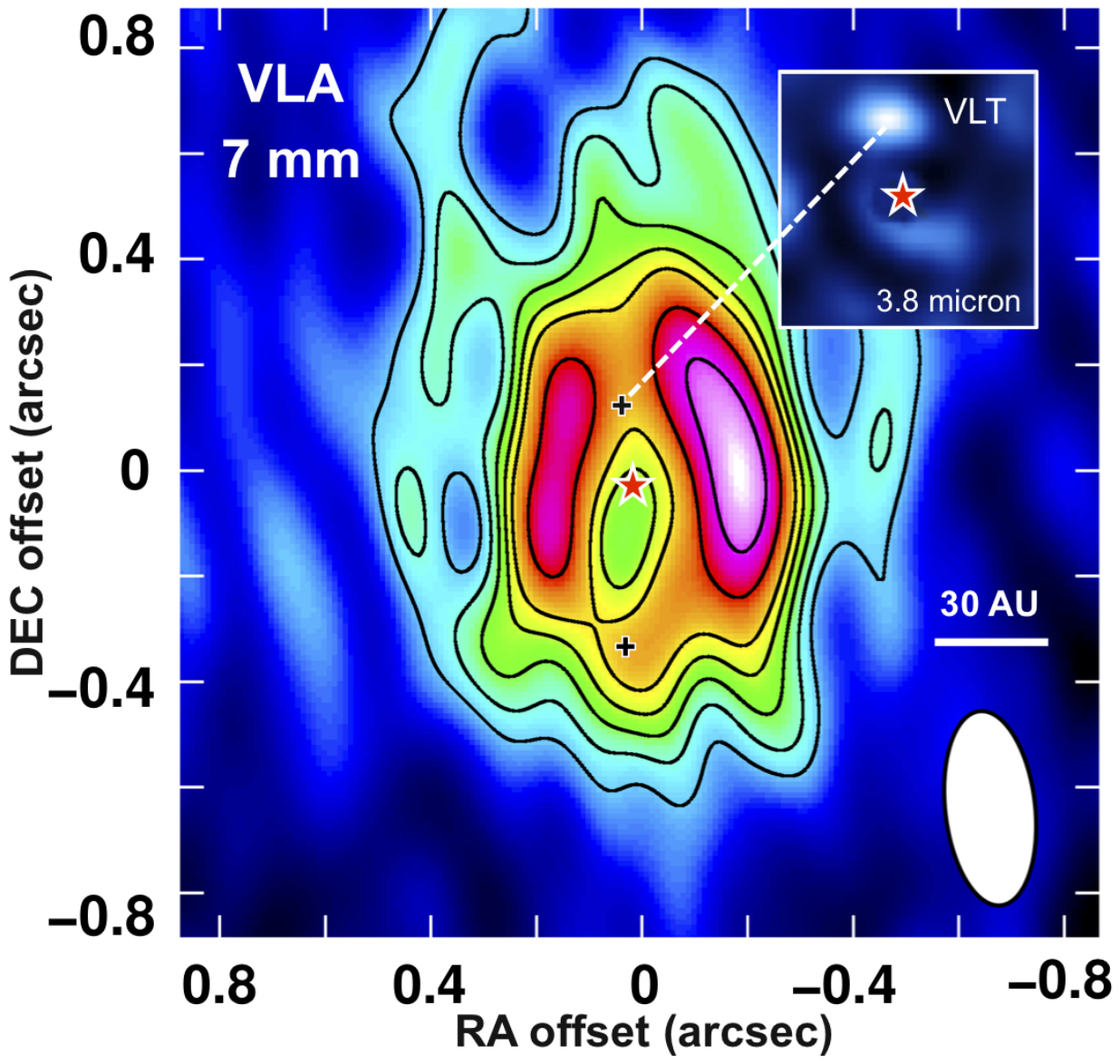

Figure 1: Image at 7 mm wavelength of the dusty disk around the star HD 169142 obtained with the VLA ([36]). The inset in the upper right corner shows, at the same scale, the bright infrared source in the inner disk cavity, as observed with the VLT at 3.8 micron wavelength ([42]). Plus signs (+) mark the positions of 7 mm and 3.8 micron protoplanet candidates.

## 2 Jets from young stars

As part of the star formation process, a fraction of the infalling matter is ejected along the polar direction, perpendicular to the disk equatorial plane. Thus, collimated bipolar outflows or jets are intrinsically associated with accretion. Outflows associated with young stars frequently exhibit a central weak centimeter emission source ([6, 3]). In the best studied cases, these sources are resolved angularly at the sub-arcsec scale and found to be elongated in the direction of the large-scale tracers of the outflow, indicating that they trace the region, very close to the exciting star, where the outflow phenomenon originates (see [4] for a review). Radio jets are present in young stars across all the stellar mass spectrum, from O-type protostars ([46]) to brown dwarfs (see Section 3), supporting the hypothesis that the disk-jet scenario might be valid to describe the formation of stars of all masses. Also, radio jets have been found in all phases of the stellar formation process, from very young protostars to more evolved stars, associated with transitional disks, where accretion (and radio emission) is very low ([48]; see Fig. 2).

The study of jets associated with young stars at radio wavelengths is important in several aspects. Given the large obscuration present towards the very young stars, the detection of the radio jet provides so far the best way to obtain their accurate positions. For the typical sizes (∼100-1000 AU) and velocities (∼200-1000 km $s^{-1}$) observed in radio jets, dynamical timescales <10 years are inferred ([4]). Thus, for angularly resolved sources radio observations also provide information on the direction and collimation of the gas ejected by the young system in the last few years, that can be compared with the gas in the molecular outflows and optical/infrared HH jets, which traces the ejection over timescales several orders of magnitude larger (e.g., [13, 33, 32]). This comparison allows us to detect the changes in the ejection direction, possibly resulting from precession or orbital motions in binary systems ([5, 20]). Additionally, we note that a good knowledge of the jet properties is indispensable for grain growth studies in protoplanetary disks to separate the dust emission of the disk from the free-free emission of the jet, as even transitional disks present free-free emission in their central regions ([48]).

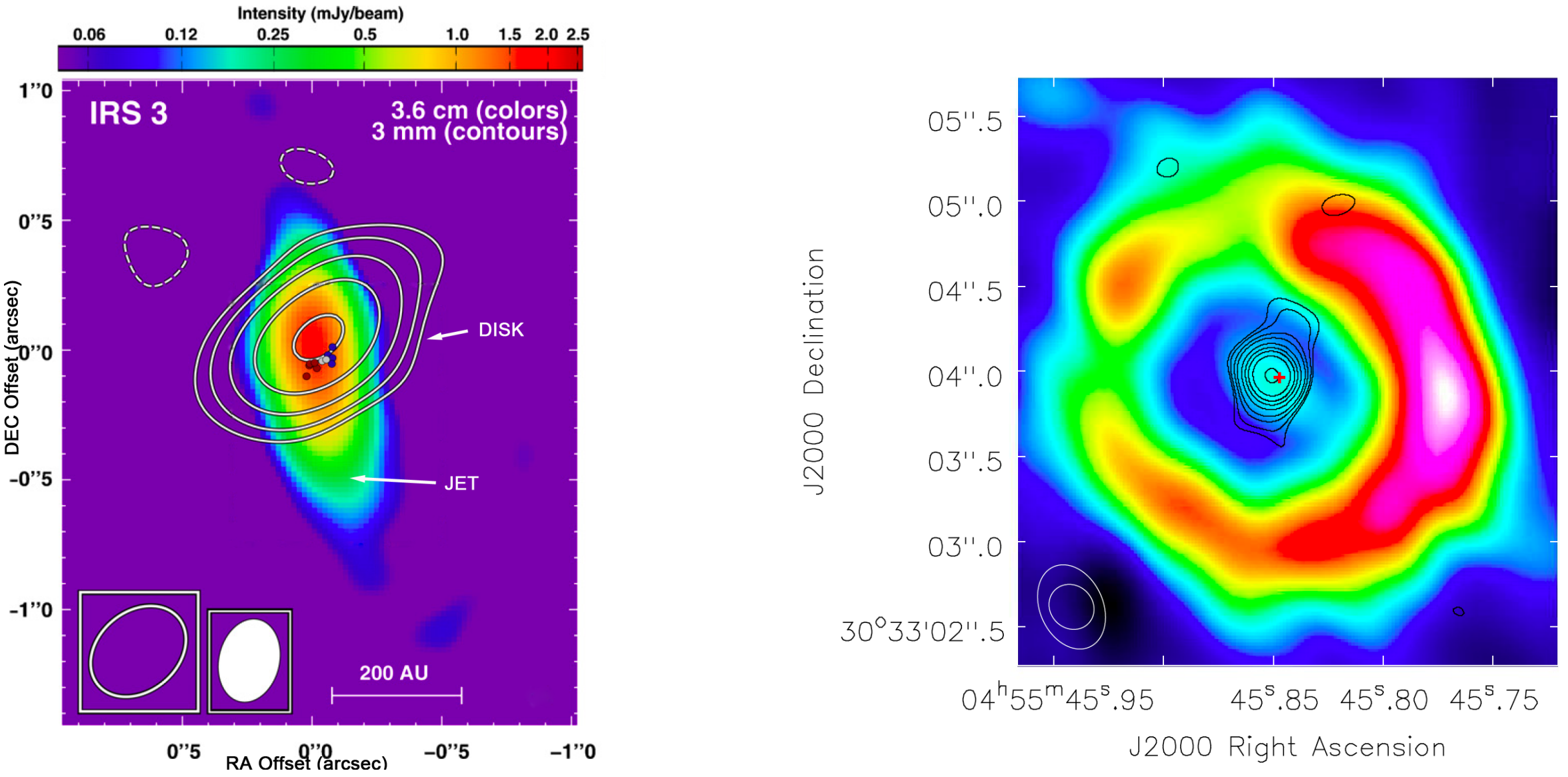

Figure 2: *Left:* Radio jet (color image) at 3.6 cm wavelength and disk (contours) at 3 mm wavelength in the young protostar NGC1333-IRS3 ([14]). *Right:* Radio jet at 3.6 cm wavelength (contours) and disk at 3 mm wavelength (color image) in the AB Aur star, which is in a more evolved phase ([48]). Synthesized beams are shown in the lower-left corner of each panel.

Physical parameters of jets, such as the ionized mass loss rate and the radius at which ionization starts can be derived from radio observations ([44, 8, 11]). The jet velocity, necessary to determine the ionized mass loss rate, can be estimated from proper motion measurements obtained from multi-epoch observations (proper motions of $\sim$0.2″-1″, depending on the stellar mass, are expected in a $\sim$5-year timespan for a source located at 1 kpc). The best studied jets have centimeter flux densities of a few mJy. However, most of the jets associated with low mass stars or even brown dwarfs present centimeter flux densities in the order of tens of $\mu$Jy, and their detection and study will require of the high sensitivity of SKA.

One of the key problems in the study of jets is to determine how they are accelerated and collimated. Sensitive, very high angular resolution observations at centimeter wavelengths can trace the base of the jets down to the injection radius, at scales of a few AUs, where the ionized jet is expected to begin. Exploration of this inner region will eventually shed new light on the jet acceleration and collimation mechanisms, helping to distinguish between different theoretical approaches such as the X-wind ([49]) and disk-wind ([29]) models. A comparison of detailed modeling results and high quality observations can provide an accurate description of the jet physical parameters.

Photoionization does not appear to be the ionizing mechanism of radio jets, since in the sources associated with low-luminosity objects, the number of UV photons from the star is clearly insufficient to produce the ionization required to explain the observed radio continuum emission (e.g., [47, 3]). The radio luminosity of radio jets is correlated with the bolometric luminosity of the source and with the momentum rate in the molecular outflow ([3, 4, 7]; see Fig. 3). These correlations include young stellar objects (YSOs) with luminosities spanning from 1 to $10^5$ $L_\odot$. In particular, these diagrams can be used to discriminate between the thermal radio jets (that should follow these correlations) and HII regions (that should fall close to the Lyman continuum line). Also, these diagrams can be used to estimate the expected radio emission of a given source from the bolometric luminosity and the outflow parameters.

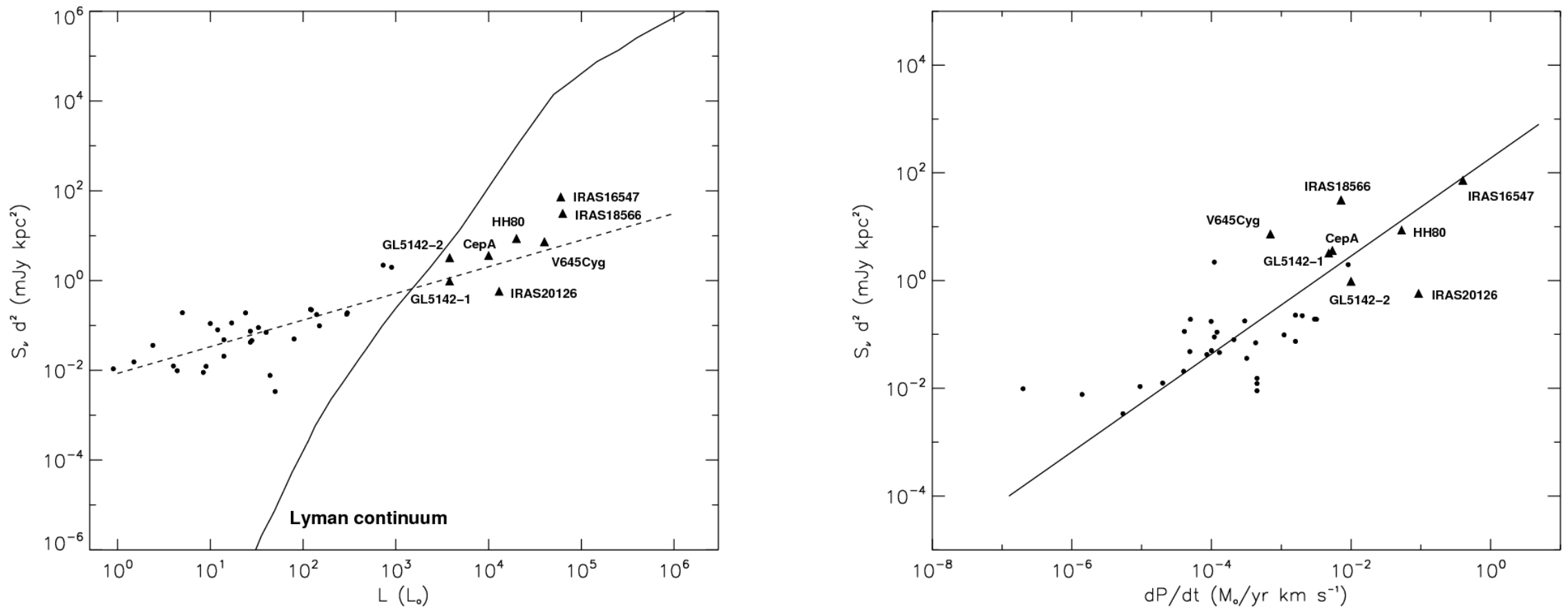

Figure 3: *Left:* Radio luminosity versus bolometric luminosity correlation for YSOs (dashed line). The solid line (Lyman continuum) represents the radio luminosity expected from photoionization. *Right:* Radio luminosity versus outflow momentum rate (solid line). The high luminosity objects are labeled in the figures. (from [7]).

The study of radio jets will greatly benefit from the higher frequency bands of SKA. In particular, given the rising spectrum and compact size of the radio jets, we anticipate that most observations of these sources will be carried out in band 5 (5-15 GHz). SKA will allow us to survey the southern hemisphere for radio jets associated with young stars across the mass spectrum, including proto-brown dwarfs (see Section 3). With SKA1-MID detection of the jets from the most massive protostars, with expected radio luminosities of $S_\nu d^2 \simeq$ 10-100 mJy kpc$^2$ (see Fig. 3) will be feasible across the whole Galaxy with a signal-to-noise ratio better than 10.

Observations at different frequencies across band 5 (providing angular resolutions in the 30 to 80 mas range) will determine the variations of the physical parameters along the jet axis and allow to image the region around the injection radius of the ionized gas in the jet. However, the high angular resolution study of radio jets with SKA1-MID only will be possible for jets of relatively high flux density (a few tenths of mJy). For weaker jets the full SKA (SKA2) will be required, since the uniform weighting SKA1 sensitivity is expected to be several times worse than for natural weighting (unless the antenna configuration is changed to favor longer baselines).

Radio jets are expected to also show radio recombination lines (RRLs) as part of the emission processes of the plasma. RRLs in jets (in combination with proper motions) will provide a 3D view of the kinematics at very small scales (near their origin). Also, detection of radio recombination lines in jets could be useful to distinguish between jets and HII regions, since lines are expected to be broader in jets, as noted by [23] (see their Fig. 6).

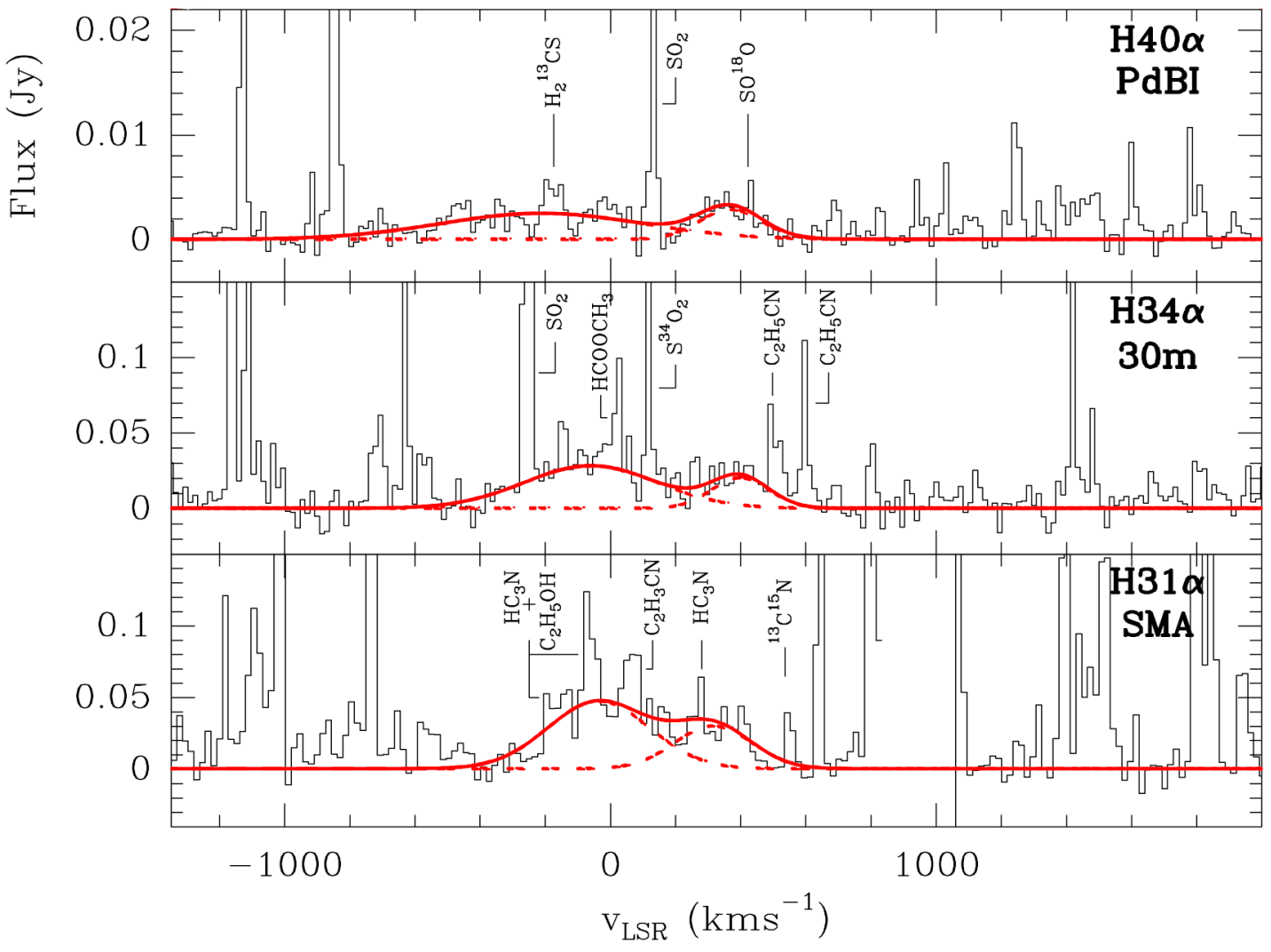

Figure 4: Extremely broad radio recombination maser lines toward the Cepheus A HW2 radio jet observed with the PdBI (H40$\alpha$), 30 m radio telescope (H34$\alpha$), and SMA (H31$\alpha$). Red-dashed lines are individual Gaussian fits to the blueshifted and redshifted velocity components of the RRL emission and the red-solid line is the total Gaussian fit. Some molecular lines in the RRL spectra are identified. (from [27]).

So far, there are no detections of RRLs in jets at the expected LTE level. [27] report the detection, at millimeter wavelengths, of broad recombination maser lines toward the jet in Cepheus A HW2 with flux densities about 5 times larger than those expected for LTE (see Fig. 4). Very sensitive observations, such as those that can be carried out with SKA, are needed to understand the nature of RRLs from jets and start using them as tools to study the outflow kinematics. For a radio jet with a continuum flux density of order 3 mJy, a peak RRL flux density of 75 $\mu$Jy is expected (assuming a line width of about 100 km s$^{-1}$; see formula 4.1 in [7]). Adopting a velocity resolution comparable to the line width, a 6-$\sigma$ detection will be achieved with an on-source integration time of 1 hour with SKA1-MID (4 hours in the "early science" phase of SKA1). On the other hand, many recombination lines will be available in the SKA bands and it will be possible to stack them and improve the detectability. This type of observations will reach its full potential with the full SKA (SKA2), when it will be possible to map the radial velocity distribution of the jet and in combination with proper motions obtain the 3D kinematics.

As discussed above, jets from YSOs have been long studied at radio wavelengths through their thermal free-free emission, which traces the base of the ionized jet and shows a characteristic positive spectral index. However, in the last two decades, negative spectral indices have been found in some regions of YSO jets (e.g., [17, 34, 46]; see [16] and references therein). This negative spectral-index emission is usually found in pairs of strong radio knots moving away from the central protostar at velocities of several hundreds of km s$^{-1}$. Because of these characteristics, it has been proposed that the knots trace strong shocks of the jet against dense material in the surrounding molecular cloud. Their negative spectral indices have been interpreted as indicating non-thermal synchrotron emission from a small population of relativistic particles that would be accelerated in the ensuing strong shocks.

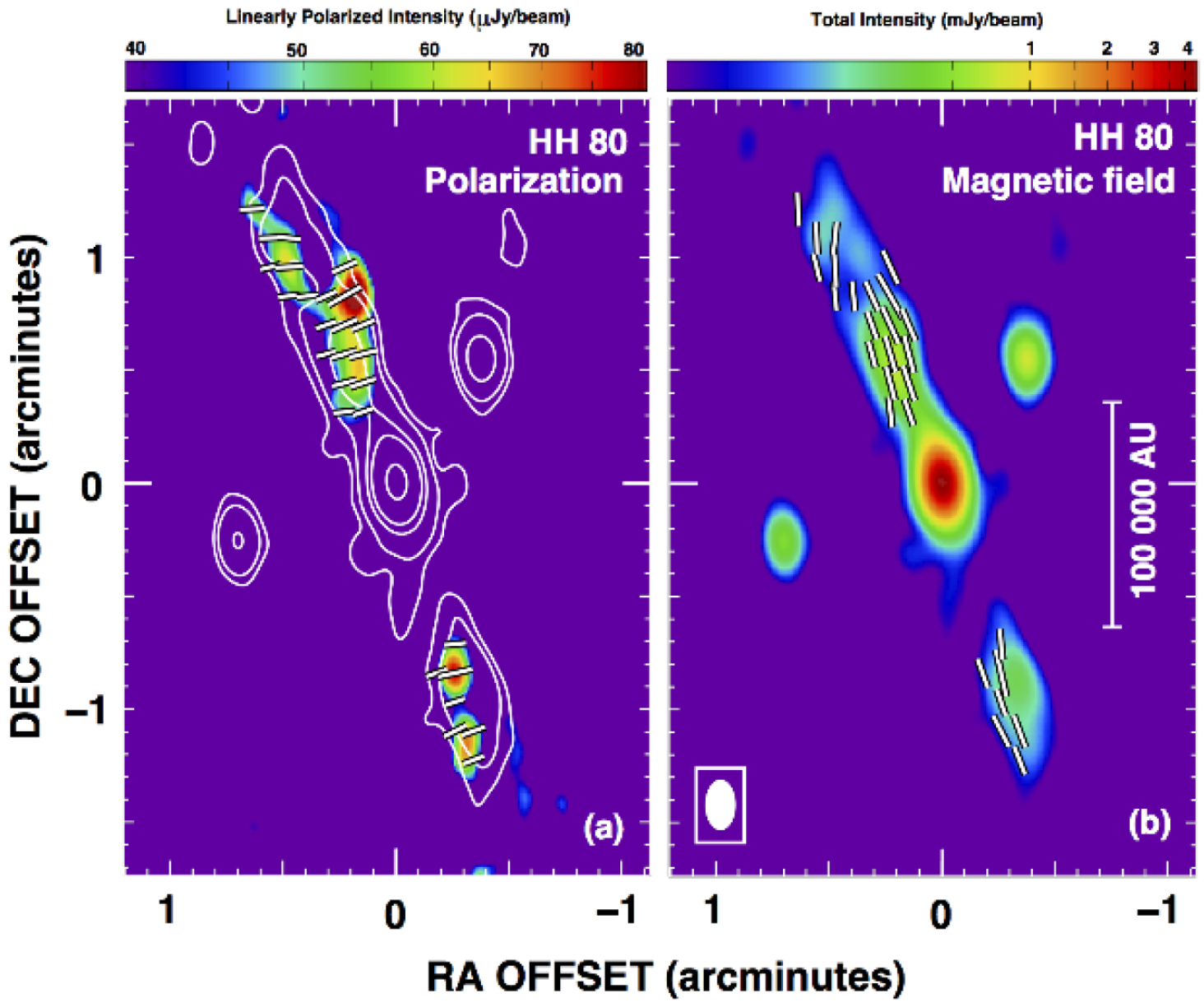

Figure 5: VLA images at 6 cm wavelength of the HH 80-81 jet showing the first detection of linearly polarized synchrotron emission in a jet from a YSO. Panel (a) shows the total continuum intensity (contours), the polarized emission (color scale), and the polarization direction (white bars). Panel (b) shows the direction of the magnetic field (white bars) and the total continuum intensity (color scale). (from [15]).

This scenario has been confirmed only very recently, with the detection of linearly polarized emission from the HH 80-81 jet ([15]; see Fig. 5). This result provided for the first time conclusive evidence for the presence of synchrotron emission in a YSO jet making possible the direct measure and study of the properties of the magnetic field strength and morphology. Measuring linear polarization in YSO jets is difficult because only a small fraction of the total emission is polarized. Ultrasensitive radio interferometers, such as the SKA, will allow us to detect and image the magnetic field in a large sample of YSO jets. In combination with the physical parameters (density, temperature, velocity) derived from observations of the thermal component, the measurement of the magnetic field from the non-thermal component will help in understanding YSO jet acceleration and collimation mechanisms, that appear to be similar for all kinds of astrophysical objects.

A search for linear polarization using SKA1-MID will be feasible in the jets with relatively bright non-thermal knots. These knots have characteristic flux densities of about 100 $\mu$Jy. Assuming a linear polarization degree of 10%, the detection of the Stokes parameters at a signal-to-noise ratio of 10, will require on-source integration times of order 30 minutes (about 2 hours in the “early science” phase of SKA1). With the full SKA (SKA2) it will be possible to search for linear polarization, and even to perform a high angular resolution mapping, in a much more extended sample.

## 3 On the formation of brown dwarfs

Brown dwarfs (BDs) are substellar objects not massive enough ($M_* \sim$ 13-80 $\mathrm{M_{Jup}}$) to sustain hydrogen fusion reactions in their cores. Although they are a natural product of the star formation process, their formation mechanism is not yet understood. One of the most accepted scenario is the turbulent fragmentation of molecular clouds with the eventual formation of BDs as a scaled-down version of low-mass stars ([37, 22]). However, BDs could also result from the ejection of a substellar object from a fragmented protostellar core, or from the fragmentation of a circumstellar disk ([43, 45, 50]). Finally, the photoevaporation of cores due to ionizing radiation from massive stars ([58]) can also produce BDs.

To shed light on the BD formation process, several works have been focused in the study of two type of objects: (i) sources in a possible very early pre-BD stage (e.g. [1, 38]), which are molecular cloud cores prior to collapse whose properties point to the formation of a future BD; and (ii) objects in the young BD stage, when the BD is almost free of surrounding gas and dust, but still presents some infrared excess due to circumstellar material (Class II stellar analogs; e.g., [10]). The missing gap between these two stages corresponds to the so-called proto-BD stage, and it would be equivalent to the Class 0/I stage of low-mass YSOs. If BDs form as a scaled-down version of low-mass stars, we should find substellar objects embedded in extended envelopes, and characterized by strong accretion and outflow activity, as observed in low mass protostars.

Several projects have been focused on identifying bona-fide proto-BDs in nearby molecular clouds (e.g., [9, 38, 31, 39]). One example are the 11 proto-BD candidates identified by [38] in the B213-L1495 cloud in Taurus. These sources were recently observed with the VLA at 3.6 and 1.3 cm to look for outflow/ejection phenomena ([35]). As a result, they detected extended 1.3 cm emission with positive spectral indexes in four proto-BDs for the first time, strongly suggesting the presence of thermal radio jets (see Fig. 6). Although this result supports the "turbulent fragmentation" scenario as the most plausible formation mechanism of proto-BDs, the number of identified sources is still very small.

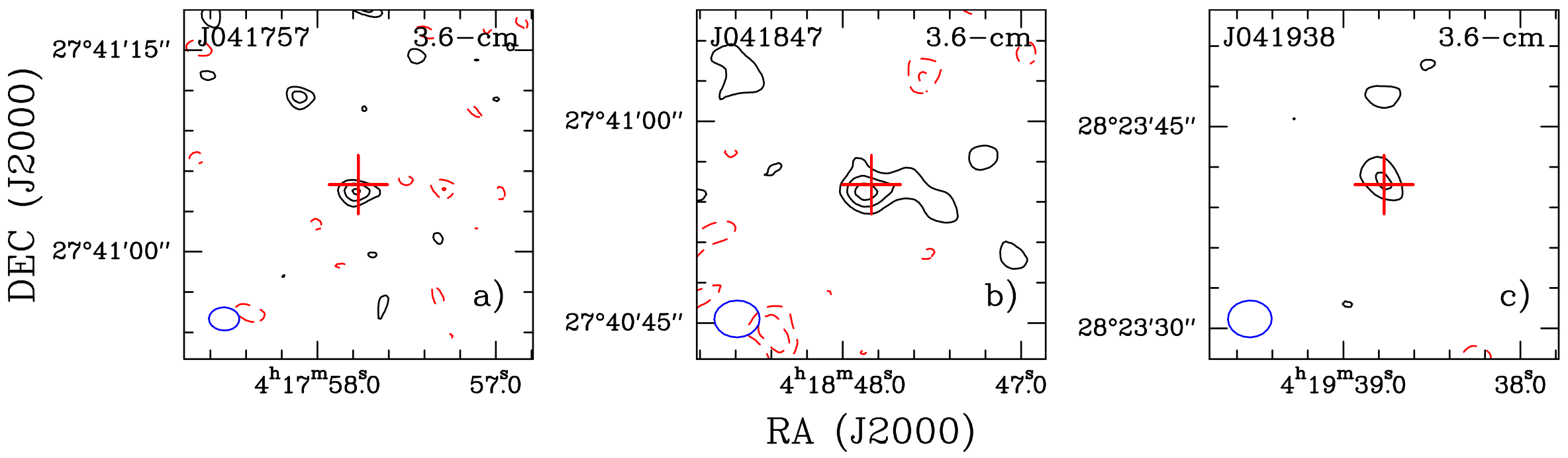

Figure 6: VLA maps of the 3.6 cm emission of three proto-BDs in the B213-L1495 cloud in Taurus ([35]). The red crosses mark the position of the sources in Spitzer mid-IR observations ([9]). The blue ellipse at the lower left corner of each panel indicates the beam size.

SKA, with a sensitivity about five times better that that of the VLA, will be able to identify large samples of proto-BDs in a very efficient way. From the results of [35] we estimate a radio luminosity $S_\nu d^2 \simeq 3$ $\mu$Jy kpc$^2$ at 10 GHz, for a proto-BD radio jet, in agreement with the extrapolation of the Fig. 3 (left) diagram. Thus, a flux density of $\sim$12 $\mu$Jy is expected for a source located at a distance of 500 pc. The SKA1-MID will detect this weak emission at the 10-$\sigma$ level in only 20 minutes of on-source integration time. Therefore, SKA will be able to survey all the Southern nearby clouds and search for radio jets in proto-BDs in a reasonable amount of time. Moreover, SKA opens the possibility to spatially resolve radio jets around substellar objects, providing important spatial and kinematical information. Mapping the jet of a proto-brown dwarf at 500 pc is expected to take less than one hour with SKA2, although it would take about one hundred hours with SKA1. As a result, SKA will provide crucial observations to understand the formation mechanism of proto-BDs.

## 4 Masers as tracers of disks and jets

Molecular maser emission at cm wavelengths (e.g., $H_2O$, $CH_3OH$, OH) is commonly found at the early stages of the evolution of the protostar-disk-jet systems. Such maser emission is usually very compact and strong, with brightness temperatures exceeding in some cases $10^{10}$ K, allowing us to observe protostar-disk-jet systems at very small scales. Through observations of the spatio-kinematical distribution of the maser emission and of its polarization, it has been possible to study the main properties of the densest and hence most obscured portions of star-forming regions located at distances up to ∼1 kpc, allowing (for example): to identify new protoplanetary disks in YSOs at scales of 50 AU ([54]; see top-left panel in Fig. 7); to discover very young (tens of years), compact (∼100 AU) outflows associated with deeply embedded protostars ([53]; see top-right panel in Fig. 7); to reveal new protostars not seen at any other wavelengths; to discover new phenomena such as the short-lived, episodic non-collimated outflow events ([55]); and to determine the distribution and strength of the magnetic field very close to protostars allowing to know its role in the star formation processes ([51, 56]; see bottom panel in Fig. 7). However, due essentially to a lack of enough sensitivity of the existing radio interferometers, only in a very few objects has it been possible to carry out a full study of the main physical ingredients of the formation and evolution of protostars.

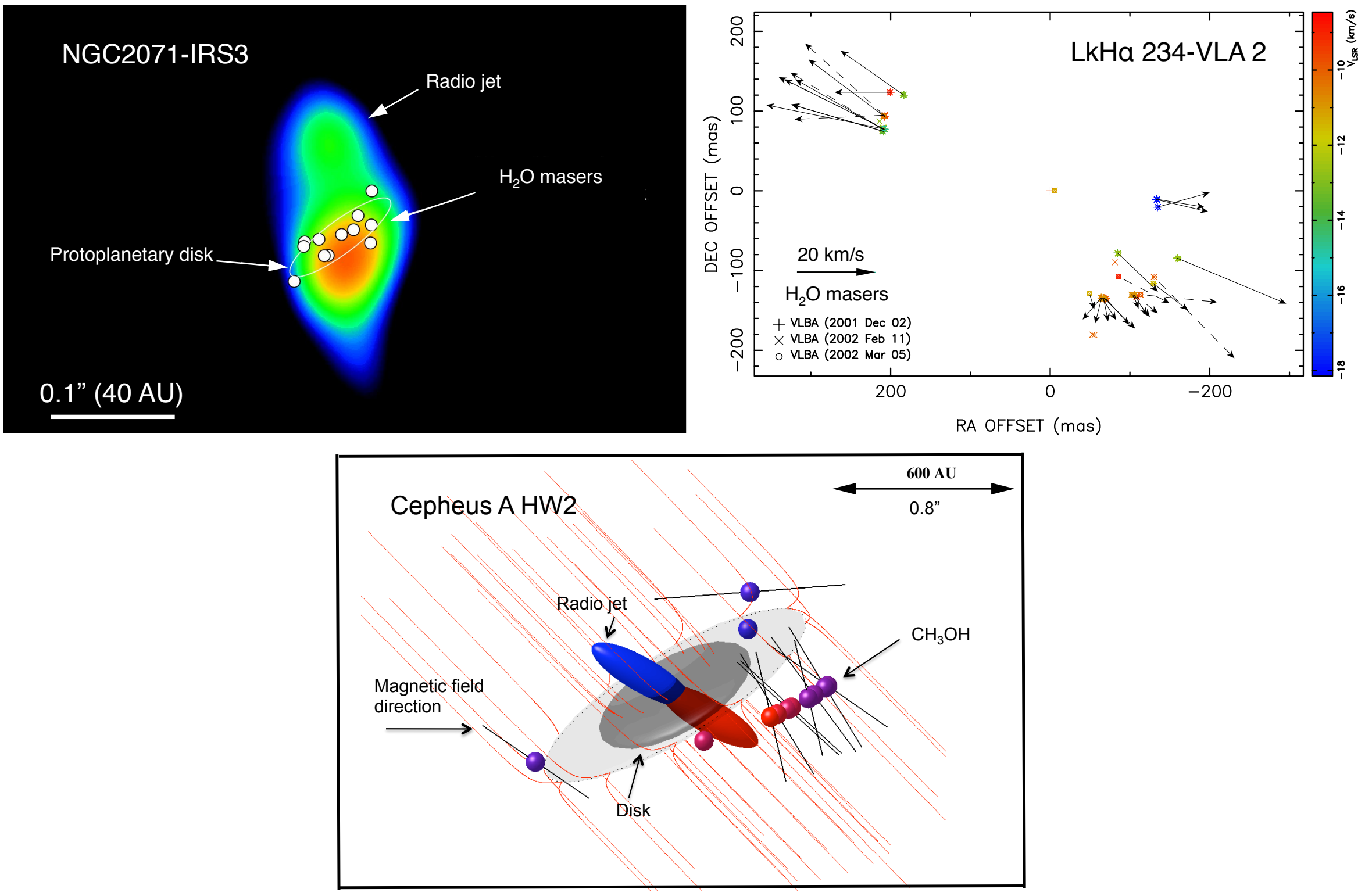

Figure 7: *Top-left panel*: $H_2O$ masers (white circles) tracing a protoplanetary disk of 50 AU size oriented perpendicular to the thermal radio jet associated with the protostar NGC2071-IRS3 ([54]). *Top-right panel*: The distribution and proper motion of the $H_2O$ masers show a very compact (∼180 AU), short-lived (∼40 yr), bipolar outflow from a very embedded protostar of unknown nature in the LkH$\alpha$234 star-forming region ([53]). *Bottom panel*: Magnetic field structure around the protostar-disk-jet system of Cepheus A HW2. Spheres indicate the $CH_3OH$ masers and black vectors the magnetic field direction ([56]). This kind of research can be carried out with the SKA with very high sensitivity and high angular resolution.

SKA spectral line observations of different species of masers, together with simultaneous very sensitive multi-frequency radio continuum observations (see above), will constitute a very powerful tool to study and significantly advance in our knowledge on how stars form. SKA-1 will provide angular resolutions of ∼30-80 mas (∼15-40 AU at 500 pc) at bands 5, where there is the 6.7 GHz methanol ($CH_3OH$) maser line. By observing the polarized emission and Zeeman-splitting of the methanol lines, it will be possible to probe magnetic fields and gas kinematics along outflows and on the disk surface around protostars. This is necessary to integrate them with the parameters obtained through radio continuum (e.g., gas density and velocity of the jet, mass loss rate, magnetic field within the jet; see above) for a full modeling of the protostar-disk-jet systems. In addition, observation of different species of masers in full polarization mode, such as OH ($\lambda = 18$ cm), $CH_3OH$ ($\lambda = 4.5$ cm), and $H_2O$ ($\lambda = 1.3$ cm; available with SKA-2), which are sensitive to different gas densities and temperatures, will tell us about what magnetic field ($B$) configuration we have in the region from scales of tens to thousands of AU (e.g., solar-type, $B \propto 1/r^2$, or toroidal-type, $B \propto 1/r$; [57]).

## Acknowledgments

GA, RE, MO, and JMT acknowledge support from the Spanish Ministry of Economy and Competitiveness (MINECO) through grant AYA2011-30228-C03 (co-funded with FEDER funds). MT acknowledges support from projects FIS2012-32096 and AYA2012-32032 of Spanish MINECO and from the MICINN program CONSOLIDER INGENIO 2010 CSD2009-00038. C.C-G. acknowledges support by UNAM-DGAPA-PAPIIT grant number IA101214. NH has been funded by Spanish grant AYA2012-38897-C02-01. AP acknowledges the financial support from UNAM, and CONACyT, México.

# Chemical Complexity in the Universe: Pre-biotic Chemistry with the SKA

**Izaskun Jiménez-Serra**[1] **and Jesus Martín-Pintado**[2]

[1] University College London, Department of Physics and Astronomy, 132 Hampstead Road, London NW1 2PS, UK
[2] Centro de Astrobiología (INTA-CSIC), Ctra. de Torrejón a Ajalvir, km 4, 28850 Torrejón de Ardoz, Madrid, Spain

## Abstract

The advent of the SKA will open up the possibility to detect amino acids and other pre-biotic species in the ISM and to constrain their formation processes. It will have the potential to link their chemistry with their subsequent delivery onto protoplanetary systems and thus, with the origin of life on Earth. Furthermore, SKA will open up the possibility to study the evolution of the chemical complexity in the Universe over cosmic time.

## 1 Introduction

In the late 1930s, the first molecular species were found in Space with the discovery at optical wavelengths of the diatomic radicals CH and CN and the molecular ion $CH^+$. Thanks to the advent of radioastronomy, in the 1960s other molecules such as OH [31] and $H_2O$ and $NH_3$ [8, 9] were discovered at centimeter wavelengths. Forty-five years later, more than 180 molecular species have been found in Space, which reveals an extremely high level of chemical complexity in both the interstellar medium (ISM) and the circumstellar envelopes around AGB stars (see the Cologne Database for Molecular Spectroscopy, or CDMS, for an updated list of the detected molecular species)[1].

Current studies of the chemistry in the ISM show the presence of large and heavy molecular species, some of them with more than 12 atoms in their molecular structure. These molecules are typically called Complex Organic Molecules (or COMs), and they are defined as carbon-based molecular species with more than 6 atoms in their structure [14]. It is now accepted that COMs mostly form on the surface of dust grains by atomic hydrogen addition and radical-radical reactions [13]. Most of the detections of COMs in the ISM have been reported toward either the central region of our Galaxy, the Galactic Center [15, 22, 26, 25]; hot molecular cores, representative of the early stages of massive star formation [6, 3, 4, 5]; or toward hot corinos, the low-mass counterparts of massive hot cores [18]. This is due to the active chemistry that these objects present as a result of the evaporation of the ices from dust grains.

[1]See https://www.astro.uni-koeln.de/cdms/molecules.

This chemical complexity is not unique of regions within our own Galaxy. In the past decade, a series of works focusing on the chemical inventory toward galaxies in the vicinity of the Milky Way, have reported the detection of over 56 molecular species in nearby extragalactic sources such as the starburst galaxies NGC253 and M82, the Ultraluminous Infrared Galaxy Arp220 and the Active Galactic Nucleus NGC1068 ([20], [27], [21], [1]). In fact, it has been shown that this chemical complexity is not exclusive of our nearby Universe but it is likely present at higher red-shifts, as recently found toward the quasar PKS 1830-211 at z=0.89 [23].

In this chapter, we will present a summary of the major challenges that we will be facing in the future decades in the area of Astrochemistry and Astrobiology in both Galactic and extragalactic sources, and how the SKA will be a key instrument in our understanding of not only the chemical complexity at all scales in the Universe, but also of the chemical pathways to life.

## 2 Detecting amino acids in the ISM. Molecular spectroscopy at cm wavelengths

Pre-biotic species such as amino acids have attracted significant attention in the past decades due to their important role in biological processes such as the synthesis of proteins. Over 70 amino acids have been found in meteorites, which supports the idea of their extraterrestrial origin [12]. However, although several attempts have been made to detect the simplest amino acid, glycine ($NH_2CH_2COOH$), in the ISM toward massive hot cores [19], its firm detection is to be reported [28, 11, 17].

The discovery of amino acids in hot sources (with temperatures of a few 100 K) is indeed challenging for several reasons. Hot cores, hot corinos and the Galactic Center present very rich spectra in molecular lines whose linewidths are broad (from some to tens of km $s^{-1}$). In addition, previous searches mostly targeted transitions in the millimeter/sub-millimeter wavelength range where the COM spectra peak, and which gets very crowded due to the high temperatures. All this yields high levels of line blending and line confusion, which complicates the identification of molecular lines, especially of those from low-abundance species such as amino acids.

## 3 The potential of the SKA

One way to circumvent this problem is by observing rotational lines of amino acids at lower frequencies – and in particular at centimeter wavelengths, where the frequency span between transitions gets larger – significantly reducing line blending and line confusion. However, we note that *cold sources* (with temperatures below 10 K) should be targeted instead of *hot sources* at low frequencies in order to maximize the probability of detection. This is due to the fact that the peak of the amino acid (and other COMs) spectra shifts to low frequencies at cold temperatures. The SKA, therefore, represents a unique instrument for the detection of pre-biotic species in the ISM, because it covers exactly the frequency range at which the emission peak of cold amino acids is expected to be found.

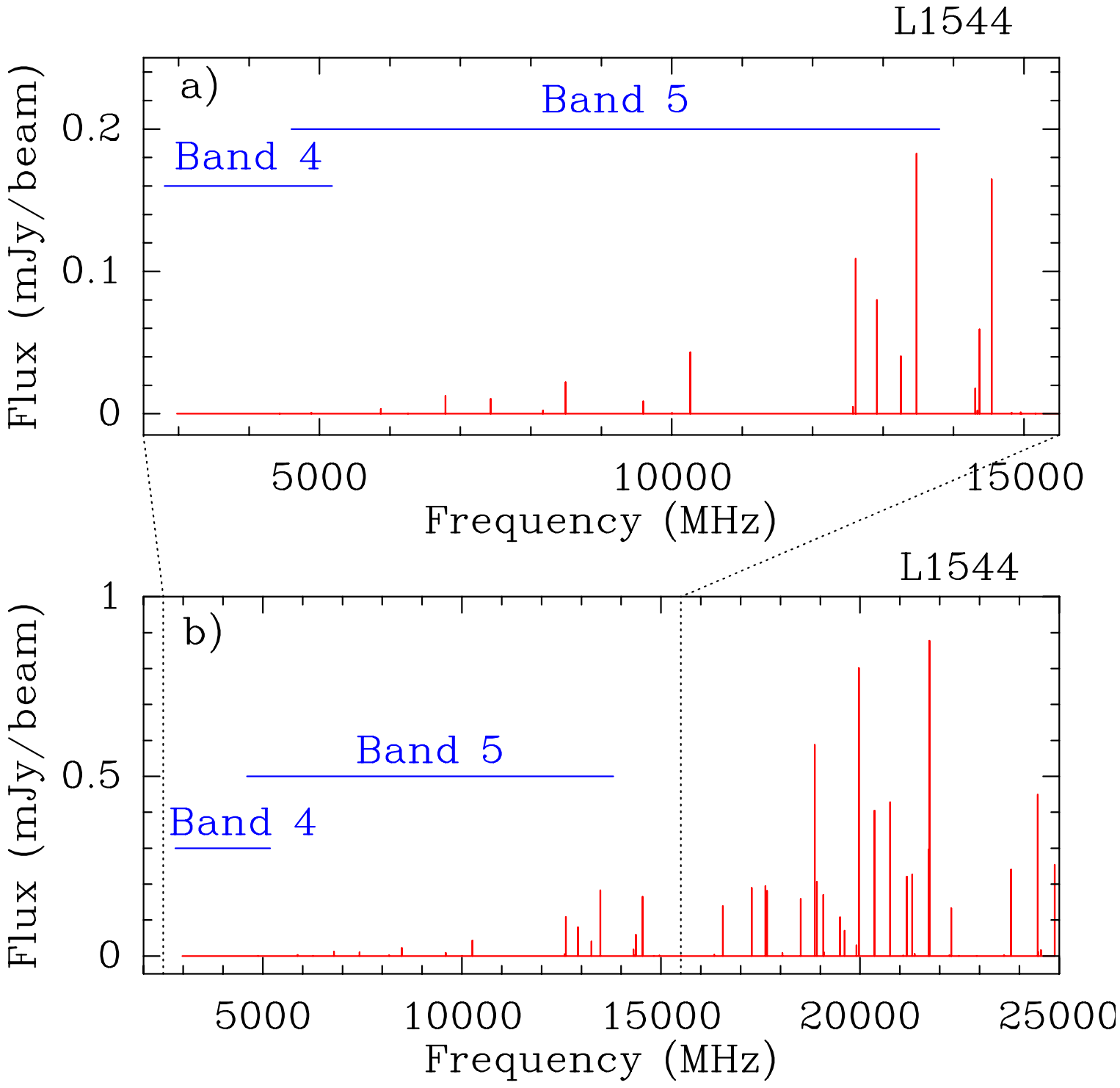

Figure 1: *Top panel:* Spectrum of glycine predicted toward the cold, pre-stellar core L1544 for the frequency range between 2.5 GHz and 15 GHz. The physical structure and gas-phase distribution of glycine is taken from [16]. The horizontal blue lines show the coverage of the SKA-MID Band 4 and Band 5 receivers. *Bottom panel:* Predicted spectrum of glycine toward L1544 for frequencies between 2.5 GHz and 25 GHz. The extension of the upper end of the frequency band of the Band 5 receivers during SKA2 will allow the detection of glycine in just few tens of hours of observing time.

This has been illustrated recently by [16], who have provided predictions for the spectrum of glycine toward the cold, quiescent pre-stellar core L1544 (a precursor of Solar-type Systems). Under some reasonable assumptions of the abundance of solid glycine in the ices (~0.01% with respect to water ice; see [24, 7]), these predictions show that the glycine lines in pre-stellar cores reach *detectable* levels for frequencies below 80 GHz. If we extend these results to the frequency bands of SKA, we find that glycine could be detected in just a few tens of hours of observing time with the Band 5 receivers of SKA-MID (expected to cover frequencies up to 24 GHz) in its second phase (SKA2; see Codella et al. 2014).

We note that the detection of amino acids could be attempted already with SKA-MID in its first phase of operations (SKA1) by observing cold glycine and alanine i) toward protoplanetary disks in emission; and ii) toward the hot gas in the Galatic Center (background temperature of 100 K) in absorption. For the protoplanetary disk case (i), the predicted glycine emission lines arising from the cold disk mid-plane in e.g. TW Hya are expected to be ~0.17-1.36 mJy at 15 GHz within a 3"-beam and a linewidth of $3\,\mathrm{km\,s^{-1}}$ [29]. These lines will be detected with SKA1 with a signal-to-noise ratio larger than 3 in just ~20 hours of observing time in Band 5. For the case of the Galactic Center clouds (ii), COMs have indeed been detected toward the ISM in the Galactic Center with excitation temperatures as low as ~5 K (see e.g. [26]). By assuming such low temperatures, the predicted intensities of the glycine and alanine lines seen in absorption against the Galactic Center are 0.7 mJy for linewidths of $20\text{-}30\,\mathrm{km\,s^{-1}}$ and within a beam of 12". These lines will thus be detected with signal-to-noise ratios larger than 5 in just 5 hours of integration time and for a velocity resolution of $5\,\mathrm{km\,s^{-1}}$.

The advent of the SKA will therefore open up the possibility not only to constrain the formation processes of amino acids and other pre-biotic species in the ISM, but also to link their chemistry with their subsequent delivery onto protoplanetary systems and thus, the origin of life on Earth.

## 4 COM chemistry in the early Universe with the SKA

COM species such as methanimine, formamide and acetaldehyde have been already detected toward nearby galaxies like Arp220 [27] and the quasar PKS 1830-211 at red-shifts of z=0.89 [23]. The detection of gas-phase water at even higher red-shifts (up to z=6; [30]) indicates that grain chemistry becomes active very early-on in the Universe, suggesting that the formation of complex organics in the ices of dust grains can also occur at these cosmological time-scales.

The detection of COMs and of the precursors of pre-biotic species at high red-shifts however cannot be tackled in the millimeter and sub-millimeter wavelength regimes. This is due to the following. The molecular emission of COMs in galaxies peaks at millimeter/sub-millimeter wavelengths *in the rest frame*. The molecular gas will be characterized by temperatures ∼40 K since $T_{bg}$= 2.7(1+z) K; see Fig.2). However, as the light travels from high red-shifts to us, the peak of the COM spectra gets shifted to lower frequencies. As an example, for a galaxy at a red-shift z=7, the COM emission peak is expected to be shifted to frequencies between 15-25 GHz, which will be covered by the SKA in its phase 2 of operations.

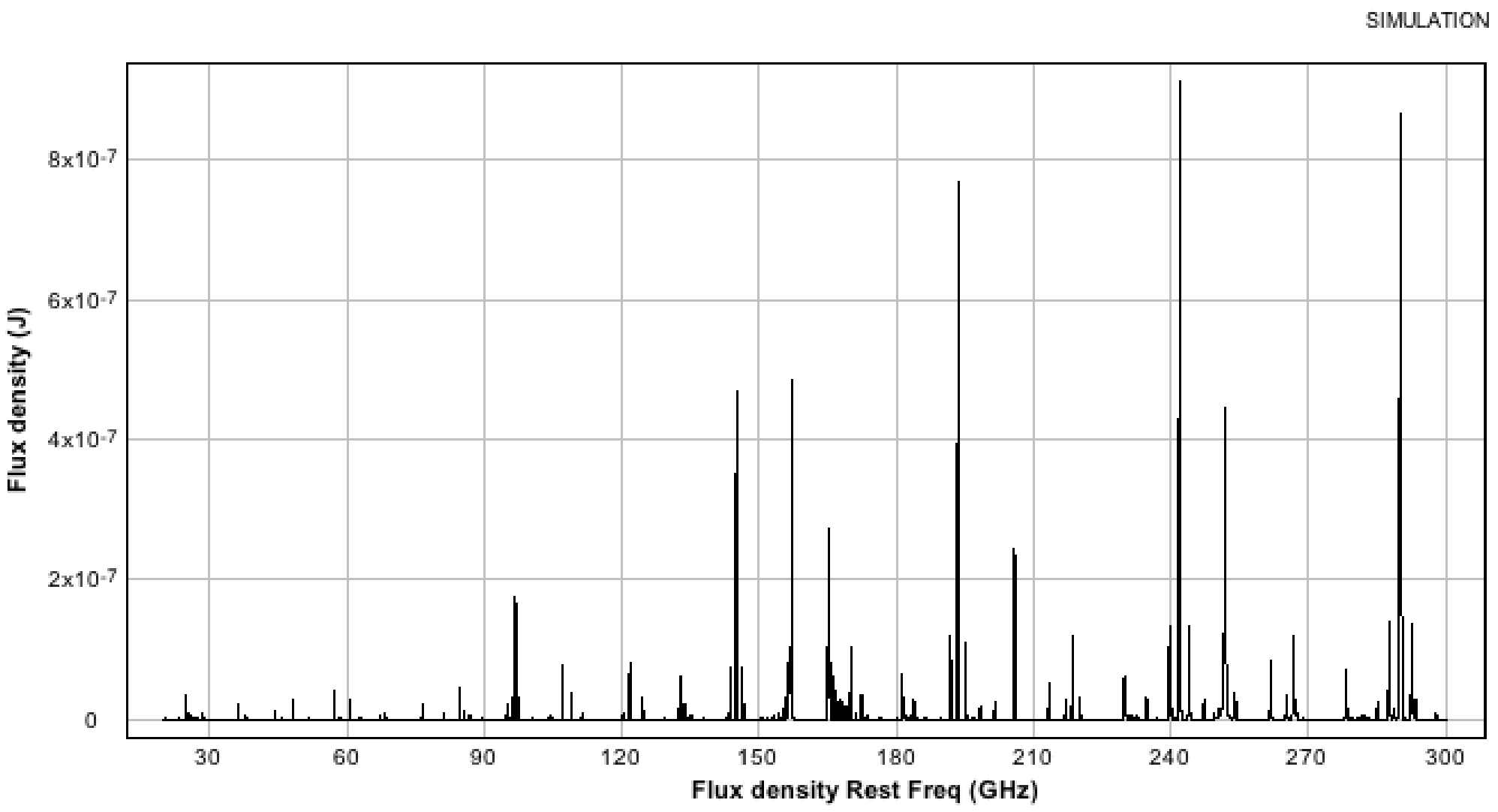

Figure 2: Simulated spectrum of the observed $CH_3OH$ emission in the rest frame of a galaxy. The intensity scale corresponds to the observed flux density, in Jy, for a galaxy at a redshift of 7. The emission is considered to be emitted under LTE conditions with the excitation temperature corresponding to that of the Cosmic Microwave Background of $T_{bg}$= 2.7 (1+z) K, $\sim$ 40 K. The assumed abundance of $CH_3OH$ is few $10^{-7}$, similar to that found toward Arp220 ([21]). The $CH_3OH$ spectrum peaks at frequencies between 150-200 GHz at the rest frame, while the observer will measure it shifted to cm wavelength, i.e. 15-25 GHz for a red-shift of z=7 (150-200 GHz/(1+z)). This frequency range is covered by the Band 5 of the SKA. For lower values of the background temperature, the peak of the $CH_3OH$ spectrum is expected to shift to even lower frequencies covered by other SKA Bands.

The expected sensitivity for Band 5 at 10 GHz during SKA1 is 63 $\mu$Jy $h^{-1/2}$ for a bandwidth of 100 kHz. This corresponds to 0.77 K $h^{-1/2}$ for a velocity resolution of 3 km $s^{-1}$ in a 1"-beam. The detection, at a 3–sigma level, of the strongest and most abundant COM in galaxies, methanol ($CH_3OH$), assuming an abundance of few $10^{-7}$ (similar to that found in the 1 Kpc disk of Arp220; see [21]), would require several hundreds of hours of on-source integration time for a velocity resolution of 300 km $s^{-1}$. Phase 2 of the SKA will improve the sensitivity of the phase 1 by a factor of 10 which will allow to observe the methanol lines in galaxies at high red-shifts in just a few hours.

Therefore, the SKA will open up the possibility to study the evolution of the chemical complexity over cosmic time.

## Acknowledgments

"I.J.-S. acknowledges funding from the People Programme (Marie Curie Actions) of the European Union's Seventh Framework Programme (FP7/2007-2013) under REA grant agreement number PIIF-GA-2011-301538, and from the STFC through an Ernest Rutherford Fellowship (proposal number ST/L004801/1). JM-P acknowledges partial support by the Spanish MINECO under grant numbers AYA2010-21697-C05-01, FIS2012-39162-C06-01 and ESP2013-47809-C3-1-R."

# Gamma-ray bursts with SKA

**Alberto J. Castro-Tirado[1,2], Javier Gorosabel[1,3,4,†] Bin-Bin Zhang[1], Samantha R. Oates[1], Soomin Jeong[1], You-Dong Hu[1], Rubén Sánchez-Ramírez[1,3,4] and Juan Carlos Tello[1]**

[1] Instituto de Astrofísica de Andalucía (IAA-CSIC), P.O. Box 03004, 18080 Granada
[2] Unidad Asociada Ingeniería de Sistemas y Automática (ISA-UMA), Universidad de Málaga, C/. Doctor Ortiz Ramos s/n (Campus de Teatinos) 29071 Málaga
[3] Unidad Asociada Grupo Ciencias Planetarias (UPV/EHU, IAA-CSIC), Departamento de Física Aplicada I, E.T.S. Ingeniería, Universidad del País Vasco (UPV/EHU), Alameda de Urquijo s/n, E-48013 Bilbao
[4] Ikerbasque, Basque Foundation for Science, Alameda de Urquijo 36-5, E-48008 Bilbao
[†] Deceased on April 21, 2015

## Abstract

The unique capabilities of SKA will allow us to study the different phases of the GRB afterglow emission with unprecedent sensitivity to determine the afterglow properties and the radiation mechanisms.

## 1 Introduction

Gamma-ray Bursts (GRBs) are transient events with the bulk of the emission arriving as high energy photons lasting 0.1-100 s that originate at cosmological distances with energy releases of $10^{49}$–$10^{51}$ ergs considering that the emission is collimated.

The multi-wavelength emission (from radio to X-ray wavelengths; i.e., the "afterglow"), that follows the gamma-ray emission satisfies the predictions of the "standard" relativistic fireball model [3], which generally decays rapidly as $t^{-2\sim-1}$. These detections have recently been possible for more afterglows due to the rapid and precise localization capabilities of the *Swift* satellite [1].

There is a bimodality in the burst duration distribution [2], with short bursts (lasting $<$ 2s) and long bursts (lasting $>$ 2s). As a matter of fact, long-duration GRBs, those usually lasting more than 2s, have been found at redshifts in the range of z = 0.1-9. The central engines that power most of these extraordinary events are linked to the explosion of massive stars [4] related to highly energetic type Ibc supernovae and can be used as tracers of star formation and as beacons to point to the location of these high-z galaxies, thanks to their extreme luminosities (with energy releases of $10^{51}$–$10^{53}$ ergs). On the contrary, short-duration GRBs have been found in galaxies with different morphology and with no underlying supernova. The merger of two compact objects is the preferred model.

The aforementioned "standard" relativistic fireball model considers a black hole as the end product of either the collapse of a massive star or the merger of two compact objects.

The collimated ejecta, which are launched by the black hole central engine, expand outward relativistically with Lorentz factors $\Gamma$ of several hundred initially. Internally, the ejecta release their energy through internal shocks [7, 6, 5], magnetic dissipation (e.g, ICMART model; [8]) or photospheric dissipation (e.g., [9, 10, 11, 15, 12, 13, 14]) and produce the prompt $\gamma$-ray emission of GRBs. Externally, the ejecta are further decelerated by an ambient medium (e.g., a constant density interstellar medium, ISM; or a stellar wind environment with density inversely proportional to distance squared) and produce long term broadband afterglows through external shocks (see e.g., [16], for a review). The accelerated non-thermal electrons give rise to the typical synchrotron spectrum [20, 19], which is theoretically expected to be accompanied by an SSC (synchrotron self-Compton, [18, 17]) component.

Early observations of the afterglows allow to provide redshifts and additional spectral information about the host galaxies. This is most essential as some of these galaxies (the ones a $z > 6$) are responsible of a significant proportion of ionizing radiation during the reionization era.

## 2 GRB observations at radio wavelengths

Since the detection of the first afterglow at radio wavelengths in 1997 [21], more than 100 afterglows have been detected so far at these frequencies. Most of them have been achieved for long-duration GRBs (with a success detection rate of 1/3) whereas only very few radio afterglows for short GRBs have been detected.

Observations are normally performed within 0.1-100 days after the trigger with a predominant frequency of 8.4 GHz, but other frequencies (4.8 GHz, 1.4 GHz) have been also used to constrain the spectrum. Radio observations do not need to be performed as fast as those at shorter wavelengths because the peak of the synchrotron spectrum moves to the radio wavelengths in timescales of few days. The low detection rate (compared to optical/nIR afterglows) is due to the lack of sensitivity of most facilities as $\mu$Jy is required to detect most radio afterglows.

Amongst the most important results obtained so far, we highlight the following: i) radio scintillation (GRB 970508, [21]) which showed that the outflow is relativistic; ii) the late time ($\sim$100 days) flattening of the light curve (GRB 980703, [22]) has been interpreted as the jet becoming nonrelativistic; iii) late time (100 – 450 days after the burst) radio calorimetry for several events (GRB 030329, [23] amongst others) sets some constraints on the total kinetic power of the jet; iv) the observation of a ultra-high redshift event (GRB 090423A, [24]) was interpreted as the reverse shock emission; v) radio monitoring of local SN Ibc has put some constraints on the GRB/SN association ([25]); and vi) the detection of a few GRB hosts [26] in the radio band has provided an estimate of the host unobscured star formation rate.

# 3 Expected results from the SKA study of GRBs

## 3.1 Early time (1-30 days) observations

- The reverse shock. Reverse shock and self-absorption effects are important in the early phase of GRB afterglow in the radio and sub-mm regime. For ultra-high redshift (z $> 5$) events, observations within $\sim$ 1 day might reveal the rapidly decaying reverse shock emission, as was attributed to GRB 090423A. This will help to constrain some of the unknown parameters such as the fraction of the shock energy in electrons and post-shock magnetic field ($\epsilon_B/\epsilon_e$).

- The forward shock. At early times (few days post burst), the radio afterglow emission is suppressed by synchrotron selfabsorption until (due to expansion) the emitting region becomes optically thin and the flux peaks (typically around 10 days with a flux density of $\sim$100 $\mu$Jy). After the peak, the flux decreases as $t^{-2}$ until the outflow becomes non-relativistic (at $\sim$ 100 days with a flux density - e.g. at 8.5GHz - of few $\mu$Jy). According to the expected MeerKAT and full SKA (SKA phase 2) continuum sensitivities, $\sim$ 70% and $\sim$95% of the long-duration burst population can be detected by MeerKAT and SKA2. In the case of short-duration events, the number of radio afterglows to be detected by MeerKAT and SKA2 should significantly increase on the basis of the improved continuum sensitivities, with respect to the low detection number currently.

## 3.2 Late time (30-300 days) observations

According to the standard afterglow model, the outflow should become nonrelativistic (NR) at $t_{NR} = 275(1 + z){E_{53}}^{1/3}n^{1/3}$ and a flattening of the radio emission is predicted following a temporal decay index variation $\Delta\alpha = (21 - 5p)/10$ where $p$ is the slope of the electron energy distribution at the shock front. This can vary if there is a steep electron energy distribution differing than the canonical value $p = 2.2$ or if the surrounding medium is not a constant density one such as a wind-profile medium (resulting of the late stages evolution of Wolf-Rayet massive star progenitor). According to the expected sensitivities, MeerKAT will be able to monitor $\sim$10-15% of the events whereas SKA2 will be able to monitor the trans-relativistic transition up to $\sim$50% of the events detected at earlier times.

## 3.3 Very late time ($>$300 days) observations

Several hundred days after the initial explosion, the host galaxy flux will dominate the overall *radio emission* and the properties of the GRB host population and of the burst environment will allow us to use GRBs to trace the cosmic star formation history up to very high redshifts. Assuming a typical SFR and a typical spectrum we expect that the host galaxies should have typical fluxes between 0.01 and 0.5 $\mu$Jy. Whereas MeerKAT sensitivity is likely too low, we estimate that SKA will be able to detect $\sim$ 50% of the host galaxies.

Thus, in conjunction with other multi-wavelength observations, MeerKAT and SKA observations of GRB will allow us to address fundamental physics questions such as:

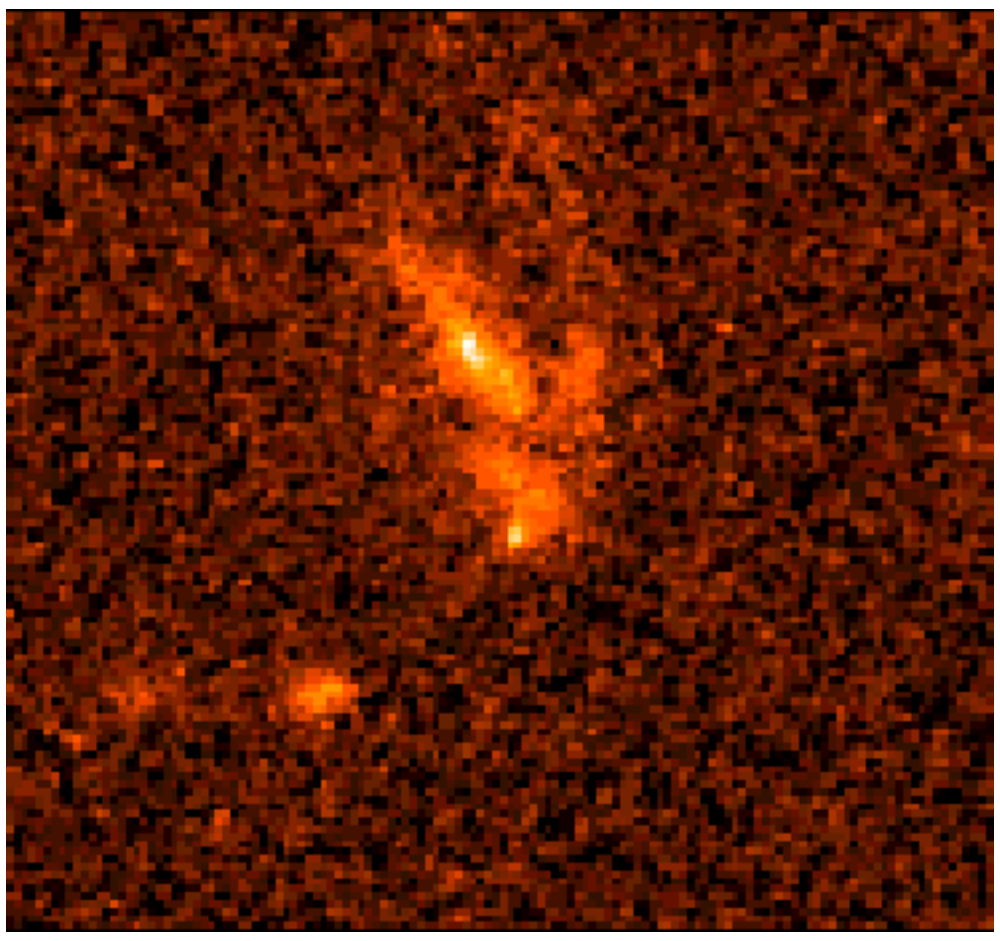

Figure 1: The host galaxy of GRB 990123 as imaged by *HST* on 23 March 1999. The optical afterglow (the faint point source close to the centre of the image), which reached V = 8.9 simultaneously to the burst, faded to V=27.7 by that time [30]. A break observed in the light curve $\sim$ 1.5 days after the high energy event suggested for the first time the presence of a beamed outflow in a GRB [29, 30, 31].

- What is the range of GRB explosion energies? For every GRB, the following six observables can be measured: the synchrotron peak, break and self-absorption frequencies, the maximum flux and the power-law decay exponent (all from the multi-wavelength spectrum) and $z$ (from optical or X-ray spectroscopy). The properties of the blast wave can be derived from the classical synchrotron spectrum produced by a population of electrons with the addition of self absorption and a cooling break. This will allow us to obtain the total energy per solid angle, the fraction of the shock energy in electrons and post-shock magnetic field, and the density of the ambient medium (e.g., [28, 27]).
- What is the environment of the circumburst medium? This can be addressed by performing a detailed study of the time evolution of the multi-wavelength afterglow emission over the first 2-4 weeks after the event. This will enable us to trace the evolution of the characteristic synchrotron self-absorption frequency, the peak frequency and the peak flux density. Taken together, this can constrain the different theoretical models (eg. homogeneous or wind-generated ambient media with a spherically symmetric outflow; a relativistic collimated outflow, etc).
- What is the nature of "dark" GRBs? These are a fraction of long-duration events ($\sim$20-30%) that remain undetected at optical wavelengths. A radio detection will determine the position of the optically obscured GRBs with better accuracy than high-energy (X-ray telescopes and gamma-ray detectors) can provide, thus pinpointing the host galaxies of a fraction of dark events even if no afterglow is found at optical wavelengths. In principle one should expect that bright X-ray afterglows to be accompanied by strong radio afterglows, according to the canonical model, but the recent claim of significant absorption in the X-ray spectra of GRB afterglows makes this statement inconclusive.

- Are high-z events a different population? The nature of the GRB explosions at z > 5 may trace the first generation of massive stars and their host galaxies, may show distinct properties.

## 4 Conclusions

The fundamental topics mentioned above will be addressed by MeerKAT and SKA thanks to their broad-band coverage and unprecedented sensitivity. This will be complemented with all other available data regarding both the prompt gamma-ray emission and and the afterglow emission from X-rays to optical/nIR and mm wavelengths. The Spanish community should should take the advantage of having access to the largest optical telescope in the world (GTC) as well as additional Spanish ground-based facilities (optical/nIR: Calar Alto, mm: 30 m PV) and other resources worldwide (optical: BOOTES robotic telescope network; optical/nIR/mm: ESO; mm: PdBI). The expertise we have gathered over the last 20 yr will allow to better understand the most energetic phenomena in the Universe (after the Big Bang), taking advantage of the unique capabilities of the SKA.

## Acknowledgments

We acknowledge the support of the Spanish Ministry grant AYA2012-39727-C03-01.

# Ultraluminous X-ray sources with the SKA

**Mezcua, M.**[1], **Caballero-Garcia, M. D.**[2], **Casares, J.**[3,4], **Gonzalez-Martin, O.**[5], **Hernández-García, L.**[6], **Negueruela, I.**[7], **and Torres, M. A. P.**[8,9]

[1] Harvard-Smithsonian Center for Astrophysics (CfA), 60 Garden Street, Cambridge, MA 02138, USA
[2] Czech Technical University in Prague, Faculty of Electrical Engineering, Technická 2, 166 27 Praha 6 (Prague), Czech Republic
[3] Instituto de Astrofísica de Canarias, E-38200 La Laguna, Tenerife, Spain
[4] Departamento de Astrofísica, Universidad de La Laguna, Avda. Astrofísico Francisco Sánchez s/n, E-38271 La Laguna, Tenerife, Spain
[5] Centro de Radioastronomía y Astrofísica (CRyA-UNAM), 3-72 (Xangari), 8701, Morelia, Mexico
[6] Instituto de Astrofísica de Andalucía, CSIC, Glorieta de la Astronomía, s/n, 18008 Granada, Spain
[7] Departamento de Física, Ingeniería de Sistemas y Teoría de la Señal, Universidad de Alicante, Apdo. 99, 03080, Alicante, Spain
[8] European Southern Observatory, 3107 Alonso de Córdova, Vitacura, Santiago, Chile
[9] SRON Netherlands Institute for Space Research, Sorbonnelaan 2, 3584 CA Utrecht, the Netherlands

## Abstract

The discovery almost three decades ago of non-nuclear, point-like X-ray sources with X-ray luminosities $L_X \geq 3 \times 10^{39}$ erg s$^{-1}$ revolutionized the physics of black hole accretion. If of stellar origin, such Ultraluminous X-ray sources (ULXs) would have to accrete at super-Eddington rates in order to reach the observed high X-ray luminosities. Alternatively, ULXs could host sub-Eddington accreting intermediate-mass black holes, which are the long-time sought missing link between stellar and supermassive black holes and the possible seeds of the supermassive black holes that formed in the early Universe. The nature of ULXs can be better investigated in those cases for which a radio counterpart is detected. Radio observations of ULXs have revealed a wide variety of morphologies and source types, from compact and extended jets to radio nebulae and transient behaviours, providing the best observational evidence for the presence of an intermediate-mass black hole in some of them. The high sensitivity of the SKA will allow us to study the faintest ULX radio counterparts in the Local Universe as well as to detect new sources at much larger distances. It will thus perform a leap step in understanding ULXs, their accretion physics, and their possible role as seed black holes in supermassive black hole and galaxy growth.

# 1 Introduction

Ultraluminous X-ray sources (ULXs) are identified as extragalactic, non-nuclear X-ray objects with X-ray luminosities exceeding the Eddington limit of a $\leq$20 $M_\odot$ stellar-mass black hole (BH). Several types of objects are able to produce these high luminosities (e.g. BHs of different masses and accretion rates, highly accreting neutron stars, ...), hence the debate about the nature of ULXs is still open.

Stellar-mass BHs of up to 100 $M_\odot$ can produce X-ray luminosities up to 5 $\times$ $10^{40}$ erg $s^{-1}$ if they accrete at rates around or above the Eddington limit. This can cause the launch of radiatively-driven outflows from the inner part of the disk, producing a mild collimation or beaming (e.g., [32]; [2]; [33]; [22]). This, together with the finding of good correlations between the number of ULXs in spiral galaxies and the star-formation rate ([64];[65]) plus studies of the X-ray luminosity function of nearby galaxies (e.g., [66]; [50]; [46]), suggests that the majority of ULXs with X-ray luminosities $\leq$2 $\times$ $10^{40}$ erg $s^{-1}$ are high-mass X-ray binaries with near to or super-Eddington accreting stellar-mass BHs. This is supported by X-ray timing analysis of some ULXs, which show that they have properties similar to those observed in BH binaries, such as variations in timescales of $\sim$100 s, spectral breaks and quasi periodic oscillations (QPOs; e.g. [29]; [3]), and that their timing and spectral characteristics are consistent with super-Eddington emission models ([63]). This is the case of the ULX M82 X-2, for which X-ray timing analysis revealed the presence of a 1.37 s pulse and a 2.5 d sinusoidal modulation in its emission caused by, respectively, the pulsation and orbital period of a ($\leq$1.4 $M_\odot$) neutron star in orbit around a stellar companion ([1]).

In the optical, some ULXs are found surrounded by bubbles of ionised gas that emit nebular lines ([52]; [31]). When detected, these emission lines allow us to determine the redshift and thereby to confirm if a ULX candidate is associated to the galaxy. Some of the donor stars in ULXs are expected to be blue supergiants on the basis of their location in or near young star clusters (e.g. [23]) and on the blue colours of the optical counterparts. However, a blue colour is also consistent with optical emission from an irradiated accretion disc. While a recent work advocates for the dominance of an irradiated accretion disc in the UV/optical ([67]), the detection of bright near infrared counterparts of 11 ULXs in a sample of 67 sources within 10 Mpc suggests that these may contain red supergiant donor stars ([28]). This implies that it may be possible to measure dynamical masses in some ULXs using infrared photospheric lines. Measuring dynamical masses of ULXs in the optical has proved extremely difficult given the need of high-resolution images to resolve the optical counterpart from unrelated objects in the host galaxy, the usual faintness of the optical counterparts (V > 23), the weakness of the absorption features from the donor star and the strong contamination from superimposed nebular lines. The most reliable mass determination has been obtained for the ULX P13 in NGC 7793, a binary system with a B9Ia donor star and an orbital period of $\sim$64 days, for which the BH mass was constrained to less than 15 $M_\odot$ ([51]). P13, together with M82 X-2, constitute the most compelling evidence of a high-mass X-ray binary in a ULX, indicating that some ULXs are powered by supercritical accretion on to stellar-mass BHs and even neutron stars.

However, some ULXs, in particular those whose X-ray luminosities of $\sim 5\times10^{41}$ erg $s^{-1}$

cannot easily be explained by the stellar-mass scenario (see e.g., [19]), could be intermediate-mass BHs (IMBHs) of $10^2 - 10^5$ $M_\odot$ accreting at sub-Eddington rates ([6]) or, in a reduced number, recoiling supermassive BHs (see e.g. [30]). IMBHs may play a key role in the formation of supermassive BHs ([18]) and in the formation and evolution of galaxies. They could come from very young and massive stars ([39]), from stellar mergers in dense stellar clusters ([57]), from the direct collapse of pre-galactic gas disc ([37]), or from the merging of compact objects in the discs of active galactic nuclei (AGN; [40]). Despite their importance, observational evidence of IMBHs is scarce.

Some IMBH candidates were suggested in globular clusters (e.g. [21]; [69]; [38]), for which the non-detection of a radio counterpart, in some cases down to 1.5 $\mu$Jy beam$^{-1}$, yielded stringent BH mass upper limits (down to 360 $M_\odot$, e.g., [7]; [71]; [62]). The most compelling evidence for IMBHs has been found in the nucleus of low-mass and dwarf galaxies (e.g. [56]; [25, 26]; [68]; [16, 17]; [58]) and in ULXs showing X-ray variability and/or radio emission ([19]; [12]; [42]; [43, 45, 47]; [54]). [24] showed that the amplitude of the X-ray variations in some ULXs was fully consistent with them being accretion-dominated objects with IMBHs of $10^4$ $M_\odot$. Based on a mass-scaling relation between the soft X-ray time lag (estimated from variability studies) and the BH mass obtained for AGN ([13]), a BH mass of $\sim 10^4$ $M_\odot$ was suggested for the ULX NGC 5408 X-1 ([14]), in favour of the IMBH scenario ([15]; [4], although see [48] supporting the stellar super-Eddington accretion scenario, instead). This is also the case for the ULX M 82 X-1, in which the presence of an IMBH was also argued by identifying the QPO frequency and applying mass-frequency scaling relationships ([20]; [5]; [54]).

ULXs offer thus a unique testbed for studies of high accretion rate physics (i.e. super-Eddington accretion on to stellar-mass BHs or neutron stars) as well as open a new window in where to look for the elusive IMBHs. However, despite the breadth of studies carried out in the optical, near-infrared, and X-ray regimes, even for the most studied ULXs a consensus on the mechanism producing the high X-ray luminosities has not yet been reached [e.g. the spectral and timing properties of NGC 5408-X-1 suggest either a stellar mass BH ([48]) or an IMBH ([15]; [4]; [14])].

Determining the ULX BH masses in the optical and X-rays has been proven to be very difficult and time-consuming. An alternative method to estimate the BH mass has however been provided by radio observations, which have in the past few years shed light on understanding the nature of ULXs.

## 2 Radio investigation of ULXs

The detection of radio emission allows us to measure the brightness temperature, spectral index (from which the physical mechanism responsible for the radio emission can be assessed, e.g., synchrotron radiation or brehmstrahlung emission), and study possible flux variability. Furthermore, very long baseline interferometry (VLBI) radio observations can possibly resolve the radio emission and indicate the presence of a relativistic jet. Combined with *Chandra* X-ray observations, radio observations can be also used to derive accurate positions to search for optical and infrared counterparts.

Studies of ULX radio counterparts have allowed us to reveal the nature of some ULXs as powerful nebulae, which discards relativistic beaming as the origin of the high X-ray luminosities (e.g., [49]; [35]; [8]; [9]; [61]), while interferometric and VLBI radio observations of ULXs have yielded the resolved structure of one of the youngest known supernova remnants (SNR 4449-1; [44]) and the detection of jet radio emission (compact and extended) in a few ULXs (N4088-X1, N4861-X2, [42], [46]; N5457-X9, [43]; Holmberg II X-1; [10]; HLX-1, [70], [11]; NGC2276-3c, [45]; [47]). In the case of compact radio emission, the location of a ULX in the fundamental plane of accreting BHs (e.g. [41]; [34]) can be used to estimate its BH mass. The fundamental plane is a correlation between radio core luminosity, X-ray luminosity, and BH mass, valid from stellar to supermassive BHs ([27]) accreting at sub-Eddington rates. The combination of radio observations with the use of the fundamental plane revealed the existence a non-nuclear IMBH of 0.3–30 $\times$ $10^4$ $M_\odot$ in the galaxy ESO 243-49 (HLX-1, e.g., [19]; [12]; [70]; [11]), the discovery of a parsec-scale radio jet from an off-nuclear IMBH of 5 $\times$ $10^4$ $M_\odot$ in the spiral arm of NGC 2276 (NGC2276-3c; [47]), and the possible presence of an IMBH in an HII region in the spiral arm of NGC 5457 (ULX N5457-X9; [43]). In all these three cases, the IMBH is thought to be the nucleus of a stripped satellite galaxy. The study of ULX radio counterparts is thus crucial not only for clarifying the ULX nature and their environment but also for revealing their possible role in galaxy evolution in the case of hosting IMBHs.

Unfortunately, no radio emission is detected from most ULXs due to the sensitivity limits of the current radio observing facilities. The cross-match of the VLA FIRST survey with the ULX catalogs of [64] and [36] yielded the detection of only 11 and 7 ULX radio counterparts, respectively ([59]; [55]; which corresponds to 27% and 19%, respectively, of the total number of ULXs in the survey area covered by FIRST), while only 1 out of 7 extreme ULXs was detected with the VLA at 5 GHz down to an rms of 0.01 mJy/beam ([45]). The faintest radio detections of ULXs correspond to radio luminosities of $10^{34}$ erg $s^{-1}$ and distances up to 15 Mpc (with the exception of HLX-1 and NGC2276-3c, which are located at 95 Mpc and 33.3 Mpc respectively). An effort to increase the number of ULX radio counterparts is thus required and will be possible thanks to the advent of the SKA.

# 3 A new era: studying ULXs with the SKA

The sensitivity and large-scale area of the SKA will allow us to monitor known ULX radio counterparts, determine their spectral and timing properties, detect new radio counterparts beyond distances $>$ 100 Mpc, investigate the properties of jets in ULXs and their feedback on the environment, and study ULX powered radio nebulae.

## 3.1 The ULX environment

Some of the ULXs at distances $<$ 5 Mpc are observed to power radio nebulae of a few tens of pc across and radio luminosities $L_{5\mathrm{GHz}} \sim 10^{34}$–$10^{35}$ erg $s^{-1}$ (e.g. [9]; [10]; [43]; [60]; [61]).

ULXs radio bubbles of $10^{33}$ erg s$^{-1}$ could be detected up to distances of 10 Mpc with only one hour of integration time per field with the SKA1-MID (rms $\sim$0.6 $\mu$Jy beam$^{-1}$), to up to 30 Mpc with one hour of integration time per field with the SKA-MID, and to up to 40 Mpc when considering the 1.4 GHz all-sky SKA Ultra Deep survey of 1 deg$^2$ and rms 0.05 $\mu$Jy beam$^{-1}$ (see [72], fig. 1). Phase 1 of the SKA will thus already allow us to extend the study of radio nebulae to distances two times larger than the ones permitted by the current facilities, reaching distances far beyond the Local Universe for SKA Phase 2.

### 3.2 Core radio emission

The compact object powering a ULX can present either steady extended jet emission or radio outbursts denoting possible state transitions as that observed in XRBs. Examples of these have been observed with the currently available facilities up to distances of $\sim$100 Mpc and radio luminosities of $\sim 10^{37}$ erg s$^{-1}$ (NGC2276-3c, [45]; [47]; HLX-1, [70], [11]), corresponding to the brightest ULX radio counterparts detected so far. Identifying more of these bright ULX radio counterparts will be possible with the wide-field SKA surveys, though a combination of SKA and VLBI observations (SKA-VLBI; [53]) will be required to resolve the ULX radio emission from the host galaxy nuclear emission for distances $>$100 Mpc. The accurate positions, on milliarcsec scales, provided by the SKA-VLBI observations will in addition allow us to properly search for the optical and infrared ULX counterparts in crowded environments.

Moreover, the high sensitivity of the SKA and its rapid survey speed will permit detecting new flaring events and monitoring transient emission from ULXs with much shorter integration times than current facilities. For example, bright transients with peak radio luminosities like the microquasars MQ1 ([61]) or S26 ([60]) will be detected up to distances of 40 Mpc with only 1 hour on-source integration time with the SKA1-SUR and at distances of $\sim$100 Mpc with the same integration time with the SKA1-MID, while transients with peak luminosities like the Galactic GRS1915+105 ($\sim 10^{32}$ erg s$^{-1}$) will be detected up to distances of $\sim$10 Mpc with 1-hour on-source integration time with the SKA-MID (see [72], fig. 2).

The all-sky surveys at 1.4 GHz that will be performed with the SKA already in its phase 1 will reach resolutions of 0.5 arcsec and sensitivities of 2 $\mu$Jy beam$^{-1}$. This will permit detecting and distinguishing from galactic nuclear emission faint ULXs (down to $10^{33}$ erg s$^{-1}$) up to distances of 5 Mpc and bright ULXs ($> 10^{35}$ erg s$^{-1}$) up to 150 Mpc. The SKA, with a sensitivity for the SKA-MID already 10 times better than the current most sensitive array, will thus not only increase the number of ULX radio counterparts and allow us to distinguish between nebular, steady or transient radio emission, but also to detect new radio sources beyond the Local Universe. For those ULXs where timing/X-ray spectroscopy or dynamical mass measurements are not possible, the detection of a radio counterpart with the SKA will permit estimating the BH mass using the fundamental plane of accreting BHs, a correlation that will be further tested and understood with SKA observations of BH XRBs and AGN (see Chapters by J.M. Paredes & J. Martí and I. Agudo et al., this book).

The SKA will be thus a key tool for revealing the nature of ULXs either as IMBHs or stellar-mass sources that, combined with observations from state-of-the-art instruments in the optical/infrared and X-ray regimes, will provide a breadth understanding of the accretion mechanism governing ULXs.

## Acknowledgments

To the memory of María Dolores Pérez Ramírez (Mariló), for her large dedication and contribution to the study of ULXs. This publication was supported by the European social fund within the framework of realizing the project "Support of inter-sectoral mobility and quality enhancement of research teams at Czech Technical University in Prague, CZ.1.07/2.3.00/30.0034. LHG acknowledges financial support from the Ministerio de Economía y Competitividad through the Spanish grant FPI BES-2011-043319.

# Core-collapse and Type Ia supernova studies with the SKA

**M.A. Pérez-Torres[1,2], A. Alberdi[1], R. Herrero-Illana[1], E. Ros[3,4,5], J.C. Guirado[4,5], I. Martí-Vidal[6], and J.M. Marcaide[4]**

[1]Instituto de Astrofísica de Andalucía (IAA-CSIC), 18008 Granada, Spain
[2]Centro de Estudios de la Física del Cosmos de Aragón, 44001 Teruel, Spain
[3]Max-Planck-Institut für Radioastronomie, Auf dem Hügel 69, 53121 Bonn, Germany
[4]Dept. Astronomia i Astrofisica, Universitat de València, 46100 Burjassot, Spain
[5]Observatori Astronòmic, Universitat de València, 46980 Paterna, València, Spain
[6]Onsala Space Observatory, Chalmers University of Technology, 43992 Onsala, Sweden

## Abstract

We discuss in this chapter some of the possibilities opened by the future availability of the SKA in the study of both core-collapse and thermonuclear runaway supernovae.
**Core-collapse SNe (CCSNe):** Optical searches of CCSNe miss a significant fraction of them due to dust obscuration; CCSN radio searches are thus more promising for yielding the complete, unobscured star-formation rates in the local universe. The SKA yields the possibility to piggyback for free in this area of research by carrying out commensal, wide-field, blind transient survey observations. SKA1 should be able to (blindly) discover several hundreds of CCSNe in just one year. SKA, with an expected sensitivity ten times that of SKA1, is expected to detect CCSNe in the local Universe by the thousands. Therefore, commensal SKA observations could easily result in an essentially complete census of all CCSNe in the local universe, thus yielding an accurate determination of the volumetric CCSN rate. **Type Ia SNe:** We encourage the use of the SKA as the most sensitive trigger machine to unveil the putative prompt ( $\lesssim$ first few days after the explosion) radio emission of any nearby type Ia SN, via target-of-opportunity observations. The huge improvement in sensitivity of the SKA with respect to its predecessors will permit to unambiguously discern which progenitor scenario (single-degenerate vs. double-degenerate) applies to them.

## 1 CCSN searches with the SKA

The limited sensitivity of pre-eMERLIN/VLA interferometric arrays has biased past radio observations of CCSNe towards the study and monitoring of only the brightest events, thus preventing systematic radio follow-ups of CCSNe. This makes the currently existing radio observations of CCSNe of rather limited use.

In this section, we discuss the benefits that can be obtained from simply making use of commensal, wide-field, transient surveys with the SKA. In particular, such surveys could potentially allow us to obtain a complete census of CCSNe in the local universe, and therefore permit us to determine the true CCSN rate and the star-formation rate of the population of massive stars in the local universe. They could also yield important information on the Initial Mass Function (IMF) of the host galaxies. In addition, some specific, relevant questions that will be tackled by those observations include the following:

- *Unveiling the hidden CCSN population and the true volumetric CCSN rate in the local universe*, $\Re$. [10] found out an apparent mismatch between the measured CCSN rate (mostly from optical observations) and the cosmic massive star formation rate, which was twice as large as the measured one. However, [13] and [17] have shown that a significant fraction of the exploding CCSNe in the local universe are hidden behind dust, and that this fraction seems to increase significantly as one goes back in the history of the universe (see Fig. 1, left panel), i.e., there seems to be no mismatch when the *hidden CCSNe* are taken into account. Radio observations have the advantage over both optical and near-IR that the emission from CCSNe is not hampered by dust, and thus offers an excellent opportunity to determine the true core-collapse supernova rate in the local universe. Wide-field SKA observations covering a significant area of the sky will discover many CCSNe in the nearby universe, and therefore will allow us to accurately determine this relevant parameter.

- *Probing the SN-CSM interaction for all CCSN types*, from the relatively faint Type IIP to the extremely radio bright Type IIn SNe, thanks to the superb sensitivity of SKA. Probing the SN-CSM interaction for all CCSN types will allow us to obtain basic, crucial information to characterize their progenitors, including mass-loss rates and, for synchrotron-self-absorbed SNe, the shock radius and the magnetic field–directly from the light curves (see, e.g., [3]).

- *Bridging the gap between Type Ibc SNe and (long) $\gamma$-ray bursts.* Type Ibc are arguably the CCSNe that show the highest blastwave speeds, yet most of them are energetically much less powerful than GRBs. Recently, however, cases like SN 2009bb, with $\beta \sim 0.9$ and energy $\sim 10^{49}$ erg seem to be intermediate cases. These "engine-driven" CCSNe could be detected with the high-sensitivity offered by the SKA, thus filling this gap in the energy-blastwave velocity parameter space of SNe-GRBs ([8] and references therein).

- *Typing CCSNe from their radio behaviour.* A systematic monitoring could permit us to type CCSNe from their radio light curves. This is crucial for the study of the hidden SN population in (Ultra)Luminous Infra-Red Galaxies, where a spectroscopical classification, or even an optical discovery, is essentially impossible (see [17] and Fig. 1).

- *Correlating optical and radio properties.* The combined use of optical information for both SN and host galaxy in dust-free environments, together with the obtained peak (radio) luminosities will allow us to check whether there is a correlation between the optical and radio properties of CCSNe, as well as with their host galaxies. This will be possible by, e.g., making a combined, commensal use of wide-field surveys programmed at radio wavelengths with SKA, and at optical wavelengths with, e.g., the LSST or similar telescopes. Obviously, the most interesting cases will likely be subject of targeted, monitoring observations with these and other facilities. For example, [8] clearly showed the impact of carrying radio and optical follow-up observations of all possible radio transients discovered in surveys covering a significant fraction of the sky area, in terms of GRBs and SNe studies.

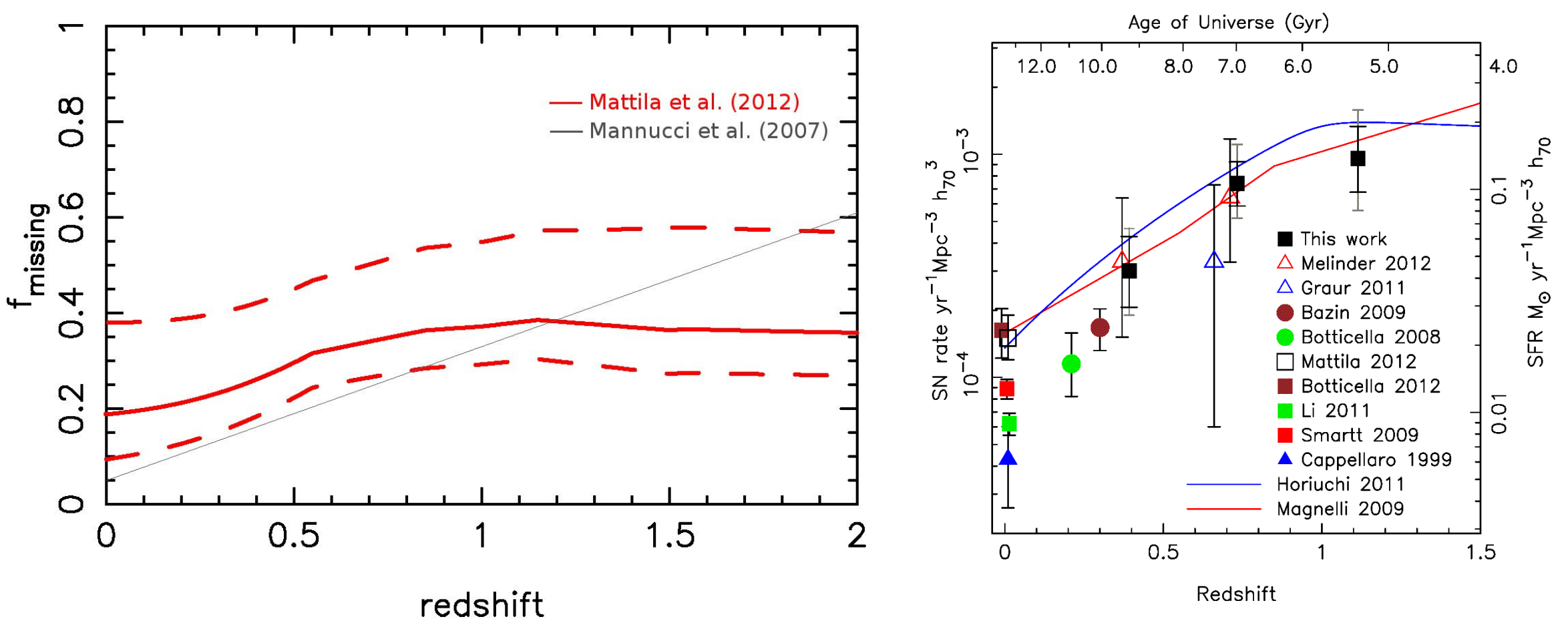

Figure 1: *Left:* Fraction of CCSNe missed by rest-frame optical searches as a function of redshift. The solid red line shows the best estimate, with the upper and lower bounds of the missing fraction shown as dashed lines. The solid grey line corresponds to the missing fraction from [13]. (Figure from [17].) *Right:* Core-Collapse Supernova rate as a function of redshift. (Figure from [5].) The use of the SKA as a CCSN discovery machine willl reduce the relatively large uncertainty in the missing fraction of CCSNe in the local universe ($z \leq$0.3), as well as in the volumetric CCSN rate (right figure).

## 2 CCSN searches with the SKA

Several wide-field sky surveys will be carried out with the SKA pathfinders MeerKAT and ASKAP, as well as with the upgraded Very Large Array. Those surveys can be used commensally for transient studies, by profitting from programmed wide-field observations. For example, the planned Very Large Array Sky Survey (VLASS) is contemplating the possibility of observing wide field areas (a few hundreds to about one thousand of square degrees) with nominal sensitivities of $1\sigma \simeq 70\mu$Jy/beam per epoch, aiming at reaching r.m.s. values of $\simeq 30\mu$Jy/beam after stacking multi-epoch observations. However, those sensitivities are just too shallow to be of any real use for CCSN studies. Indeed, a $5\sigma$ figure of merit corresponds to 500 $\mu$Jy/beam per epoch, so that the maximum distance to detect a type IIP event would be ∼9 Mpc (Table 1). Unless sky areas close to the full celestial sphere are surveyed–which is very unlikely– VLASS will only pick up a handful of CCSNe after one year of observations (Table 1). Those few CCSNe will be discovered first by optical searches, and some of them will be subject of targeted radio observations anyway, given their vicinity. Thus, deeper sensitivies are needed to make a substantial contribution to the field.

### 2.1 SKA survey strategy for commensal CCSN searches

The best strategy for transient studies with the SKA is one that combines good angular resolution ( $\lesssim$ 1.5 arc sec) and a large field of view (FoV) at frequencies around and above ∼1.7 GHz. Those parameters warrant essentially unambiguous source identification and large numbers of potential SNe in the field of view. After the SKA rebaselining in March 2015, SKA1 will consist of two instruments, SKA1-LOW and SKA1-MID.

SKA1-LOW is not suitable for searching young CCSNe. First, disentangling the radio emission of a CCSN from that of the host galaxy will be essentially impossible with the angular resolution provided by SKA1-LOW. Second, at the low frequencies of SKA1-LOW ( $\lesssim$ 350 MHz), there will be significant absorption due to H II regions, which would prevent the discovery of many CCSNe. Finally, the radio light curve evolution at low frequencies is very slow, which severely impacts on the cadence time needed to ascertain the variability of a CCSN candidate. In fact, given the radio spectral evolution of supernovae, the cadence time should be roughly inversely proportional to the observing frequency. Therefore, visiting the same field at 110 MHz implies a cadence time about 16 times larger than at 1.7 GHz. Since at the latter frequency the cadence time is $\sim$ 90 days (thus taking about one year to complete five visits of the same field), this would take almost 16 yr at 110 MHz, which is unrealistic for any sensible blind search of CCSNe.

The other instrument approved after the rebaselining, SKA1-MID, will observe at frequencies above 350 MHz, in principle up to 14 GHz, and will have a maximum baseline of 150 km. At the nominal frequency of 1.7 GHz, SKA1-MID will thus yield an angular resolution of ∼0.25 arcsec. It will have a nominal continuum sensitivity of 1.14 $\mu$Jy/b after one hour, for an assumed effective area of 33000 $m^2$ and a bandwidth of 770 MHz. Since SKA1-MID will be made of ∼15 m dishes, the FoV will be of $\sim$ 0.50 $deg^2$ at its nominal frequency of 1.7 GHz.

We assume here that SKA1-MID will observe the sky for $\gtrsim$ 2000 hr in its first year of operations. For the sake of simplicity, we also assume that the observing band will be centered at 1.7 GHz, and that one hour of on-source time is devoted to each field, which means that the area will be of ∼1000 $deg^2$ after one year, with a typical r.m.s. of 1.14 $\mu$Jy/b per pointing. To be of use for CCSNe searches, each field of view should be visited five times (each time for an on-source time of 12 minutes), at a cadence of one visit every $\leq$ 90d. This scheme is also handy to obtain an homogeneous rms across the whole FoV, since each pointing can be done at slightly different positions. Since we aim at a final, stacked r.m.s. of $\sim 1.14\,\mu$Jy/beam, the r.m.s. attained in each individual visit will be $\sim 1.14 \times \sqrt{5} = 2.55\,\mu$Jy/beam.

Table 1: Expectations for CCSN detections from commensal radio surveys for the VLASS, SKA1-MID, and SKA. $D_{max}$, in Mpc; $L_{\nu,26} = L_{\nu,peak}/10^{26}$ erg/s/Hz; $\nu_5^{-1} = \nu/5$ GHz.

| SN Type | $\Delta t_{peak}\, \nu_5^{-1}$ [days] | $L_{\nu,26}$ | VLASS | | SKA1-MID | | SKA | |
|---|---|---|---|---|---|---|---|---|
| | | | $D_{max}$ | $N_{det}$ | $D_{max}$ | $N_{det}$ | $D_{max}$ | $N_{det}$ |
| Ib/c | 30 | 20 | 69 | 8 | 362 | 126 | 1145 | 3976 |
| IIb, IIL | ∼150 | 10 | 49 | 1 | 256 | 21 | 422 | 654 |
| IIP | 40 | 0.5 | 11 | 0 | 57 | 1.1 | 94 | 33 |
| IIn | 1000 | 100 | 154 | 11 | 810 | 162 | 1334 | 5129 |
| 87A | 2 | 0.04 | 3 | 0 | 16 | 0 | 27 | 0 |
| Total | | | | ∼20 | | ∼310 | | ∼ 9790 |

## 2.2 Expectations for CCSN discoveries with wide surveys

Table 1 summarizes the expectations for detecting CCSNe using the VLASS, SKA1-MID, and SKA. We show the expected number of detected CCSNe, $N_{det}$, up to the maximum distance, $D_{max}$ to which CCSNe of a given type are expected to be detected above $5\sigma$, for the nominal r.m.s. values of VLASS ($\sim 70\mu$Jy/b/pointing) and SKA1-MID ($\sim 2.55\,\mu$Jy/b/visit), as well as for SKA (ten times more sensitive than SKA1). We assumed that each survey observes at a nominal frequency of 1.7 GHz. We also assumed that the area covered by the observations after one year is of 1,000 $deg^2$ (SKA1-MID and SKA) and ten times larger (10,000 $deg^2$) for the VLASS.

We refer the interested reader to [22] for details on how the numbers in Table 1 are obtained, and highlight here that the expected number of detected CCSNe from commensal surveys will be much larger than expected from currently envisioned surveys with state-of-the-art arrays.

The main limitation is due to the relatively small value of the maximum distance that will allow a detection of a type IIP SN, which is why so few detections of them are expected. The relatively high luminosity of spirals at those frequencies ($\sim 7 \times 10^{27}$ erg $s^{-1}$ $Hz^{-1}$) may also prevent the unambiguous detection of even Type IIn at those distances with SKA1, as the synthesized beam of $1.0''$corresponds to 2.1 kpc, which could pick up a significant amount of the background galaxy luminosity, thus the need for high angular resolution. This limitation will be overcome once SKA is completed, as is foreseen to have a twentyfold better angular resolution. In addition, one has to take into account that around 10% of the massive star-formation already at $z \sim 0.1$ will come from Luminous Infrared Galaxies [12]. While those galaxies are prolific CCSN factories, their detection with SKA1, or even with SKA, will not be possible in general, as the compact starburst (size $\lesssim$ 500 pc) will have a typical 1.4 GHz brightness well in excess of our 5-$\sigma$ sensitivity limit. For particular targets of interest, SKA-VLBI (VLBI observations incorporating the phased-array of the inner core of SKA; see chapter by Ros et al. in this book) will be able to detect individual radio supernovae and supernova remnants in the circumnuclear starbursts.

# 3 Type Ia supernovae with the SKA

Type Ia SNe are the end-products of white dwarfs (WDs) with a mass approaching, or equal to, the Chandrasekhar limit, which results in a thermonuclear explosion of the star. While it is well acknowledged that the exploding WD dies in close binary systems, it is still unclear whether the progenitor system is composed of a C+O white dwarf and a non-degenerate star (single-degenerate scenario), or both stars are WDs (double-degenerate scenario). In the single-degenerate scenario, a WD accretes mass from a hydrogen-rich companion star before reaching a mass close to the Chandrasekhar mass and going off as supernova, while in the double-degenerate scenario, two WDs merge, with the more-massive WD being thought to tidally disrupt and accrete the lower-mass WD [14]. This lack of knowledge makes difficult to gain a physical understanding of the explosions, and to model their evolution, thus also compromising their use as distance indicators.

Radio and X-ray observations can potentially discriminate between the progenitor models of SNe Ia. For example, in all scenarios with mass transfer from a companion, a significant amount of circumstellar gas is expected [1], and therefore a shock is bound to form when the supernova ejecta are expelled. The situation would then be very similar to circumstellar interaction in core-collapse SNe (see above), where the interaction of the blast wave from the supernova with its circumstellar medium results in strong radio and X-ray emission [2]. On the other hand, the double-degenerate scenario will not give rise to any circumstellar medium close to the progenitor system, and hence essentially no prompt radio emission is expected. Nonetheless, we note that the radio emission increases with time in the double-degenerate scenario, contrary to the single-degenerate scenario. This also opens the possibility for confirming the double-degenerate channel in Type Ia SNe via sensitive SKA observations of decades-old Type Ia SNe.

Radio [18, 9] and X-ray [11, 23] observations of SNe Ia resulted in upper limits on the wind density around SN Ia progenitors of the order of $\dot{M} = 1.2 \times 10^{-7}$ $M_\odot\,yr^{-1}$, assuming a wind velocity of 10 km $s^{-1}$. At the moment, the deepest radio limits on circumstellar gas come from SNe 2011fe and 2014J. The limits on mass-loss rate from the progenitor system of SN 2011fe are $\dot{M} = 6 \times 10^{-10}$ $M_\odot\,yr^{-1}$ and $\dot{M} = 2 \times 10^{-9}$ $M_\odot\,yr^{-1}$ from radio [4] and X-rays [15], respectively, assuming a wind velocity of 100 km $s^{-1}$. Similarly, the mass-loss rate limits for the progenitor system of SN 2014J are $\dot{M} = 7 \times 10^{-10}$ $M_\odot\,yr^{-1}$ and $\dot{M} = 1.2 \times 10^{-9}$ $M_\odot\,yr^{-1}$ from radio [21] and X-rays [16], respectively, for a wind velocity of 100 km $s^{-1}$. The above limits permit to rule out all symbiotic systems and the majority of the parameter space associated with stable nuclear burning WDs, as viable progenitor systems for either SN 2011fe or SN 2014J. Recurrent novae with main sequence or subgiant donors cannot be ruled out completely, yet most of their parameter space is also excluded by those radio observations (see Fig. 2.)

## 3.1 Unveiling the progenitor scenarios of type Ia SNe with the SKA

With the advent of the SKA, we will be able to obtain significantly deeper radio limits: for a SNe Ia exploding at the distance of M 82, we should eventually detect it; for more distant supernovae, we will be able to obtain similar or even more constraining limits to those obtained for SNe 2011fe and 2014J, which will allow us to build a global picture thanks to the availability of a larger statistical sample of observed SNe Ia.

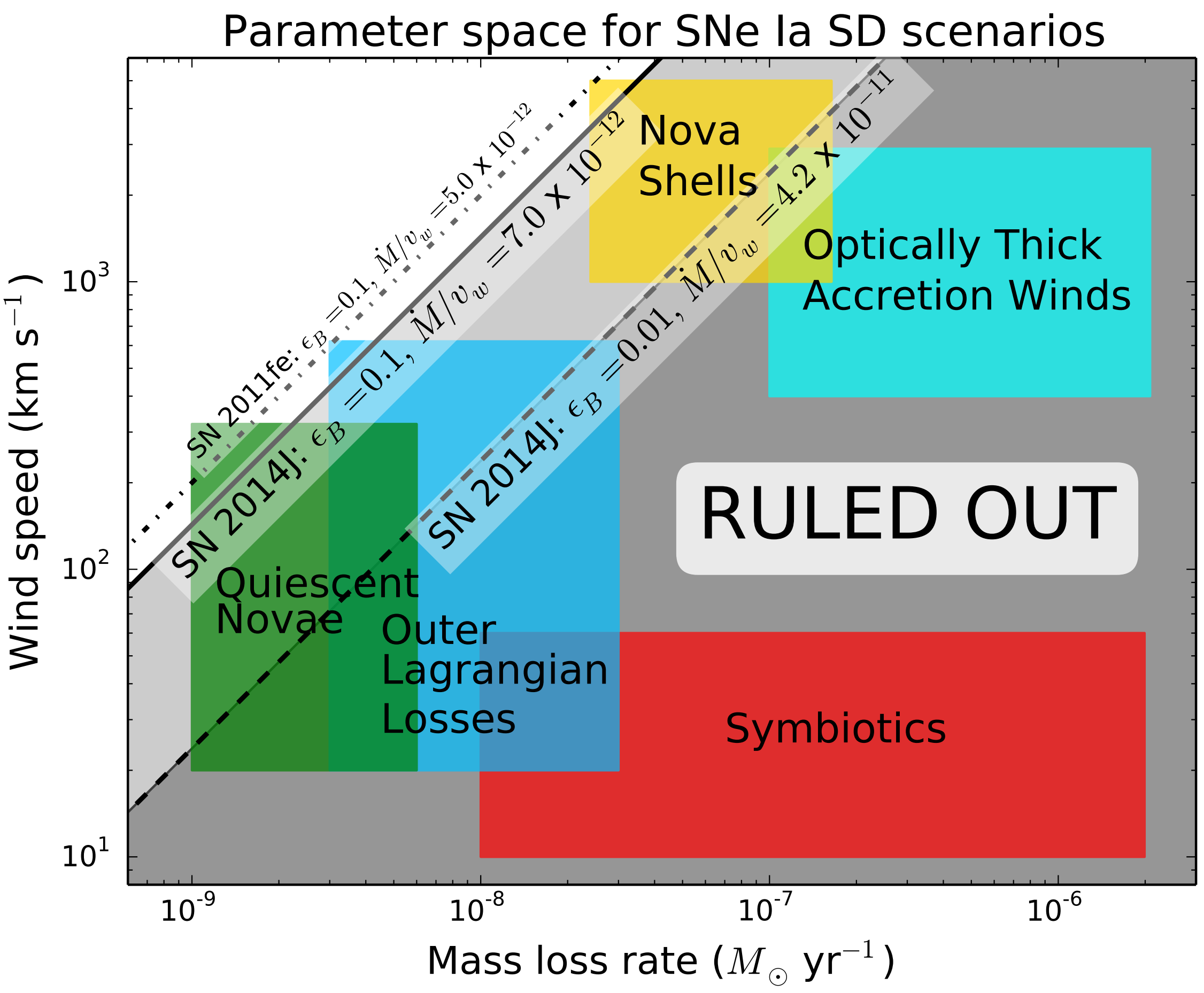

Figure 2: *Constraints on the parameter space (wind speed vs. mass-loss rate) for single degenerate scenarios for SN 2014J. Progenitor scenarios are plotted as schematic zones, following [4]. We indicate our 3$\sigma$ limits on $\dot{M}/v_w$, assuming $\epsilon_B$ = 0.1 (solid line) and the conservative case of $\epsilon_B$ = 0.01 (dashed line). Mass loss scenarios falling into the gray-shaded areas should have been detected by the deep radio observations, and therefore are ruled out for SN 2014J. For a comparison, we have included also the limit on SN 2011fe (dash-dotted line) for the same choice of parameters as the solid line for SN 2014J, which essentially leaves only room for quiescent nova emission as a viable alternative among the single-degenerate scenarios for SN 2011fe (see [21] for details).*

SKA1-MID promises to yield 1$\sigma$ sensitivities of $\sim 0.7\mu$Jy/beam in one hour at a fiducial frequency of 1.7 GHz. This figure is five times better than currently provided by the most sensitive array, the VLA. SKA1-MID will be able to either detect the putative radio emission of SN 2014J-like objects up to distances $\lesssim 8$ Mpc in less than one hour, or put significantly better constraints on the parameter space of single-degenerate scenarios for the next Type Ia SN that explodes at a distance no larger than that to M 82. However, the expected number of SNe Ia per year in such a volume of the local universe is small. Since the volumetric rate of Type Ia SNe is $\sim 3 \times 10^{-5}$ SN yr$^{-1}$ Mpc$^{-3}$ [6], we should expect on average one Type Ia SN every $\sim$15 yr within a distance of $\lesssim$ 8 Mpc (more than twice the distance to M 82), which is a value too small to allow for any statistical improvement in a reasonable amount of time. [24] found 26 Type Ia SNe out of 132 SNe from a 10.5 year-long survey within 28 Mpc. This figure corresponds to about 1 SN Ia every 13 yr within a distance of 8 Mpc, in agreement with the value found by [6].

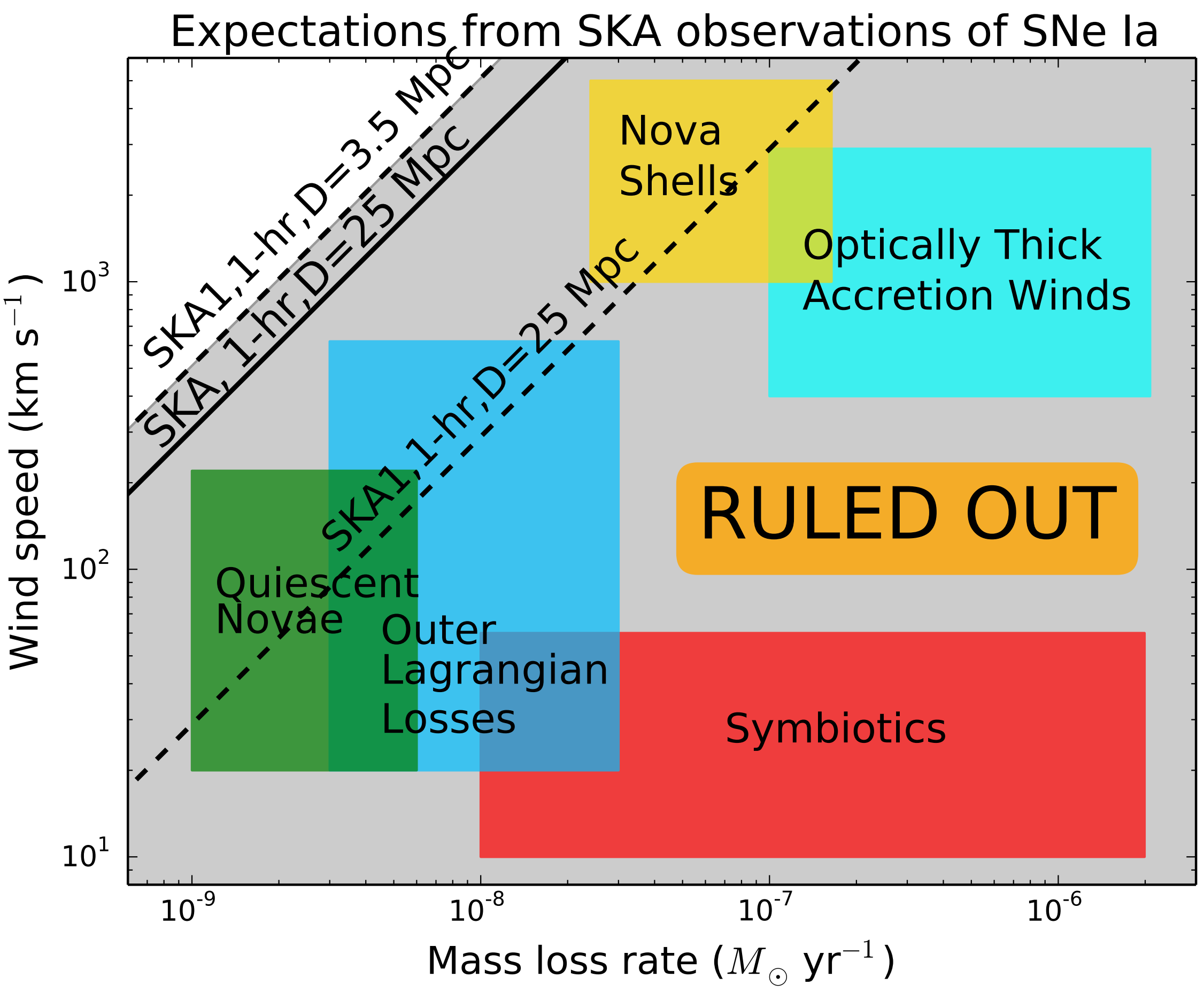

Figure 3: *Constraints on the parameter space (wind speed vs. mass-loss rate) for the same single-degenerate scenarios as in Fig. 2, as expected with SKA1 (dashed lines) and the full SKA (solid line). SKA1 will be able to unambiguously probe all single degenerate scenarios for SNe Ia exploding at distances similar to that of M 82 (3.5 Mpc), and will be more sensitive than current state of the art, deep radio observations of SN 2014J in M 82, up to a distance of 25 Mpc, or even larger. When SKA is completed, we will be able to unambiguously probe the prompt radio emission within the single degenerate scenario up to distances of* $\geq 25$ *Mpc. All lines correspond to 3-$\sigma$. (Figure taken from [22].)*

To obtain a statistically significant sample of SNe Ia observed in radio and with similar upper limits to those obtained for SNe 2011fe and 2014J, we need to sample significantly larger volumes and need much more sensitive radio observations. For example, by sampling out to a distance of 25 Mpc, we can expect $\sim$2 SNe Ia per year within this sampled volume, which in 10 years would result in a total of $\sim$20 SNe Ia, enough to extract statistical results. At this maximum distance, we need a sensitivity $\sim$50 times better than obtained by the deep radio observations discussed in [4] and [21], i.e., 80 nJy/beam (1-$\sigma$), to be as constraining. The fiducial 1-$\sigma$ sensitivity of SKA should be 10 times better than that of SKA1-MID, or about $\sim$ 70 nJy/beam in one hour, which will allow to obtain deep radio limits (or eventually the detection) of type Ia SNe in a short amount of time and out to 25 Mpc, or even further. To get a more clear idea of what can be reached with SKA1 and SKA, we plot in Fig. 3 the constraints on the mass-loss rate parameter for an upper limit of 3-$\sigma$ ($3\times0.7\,\mu$Jy/beam for SKA1, or $3\times70$ nJy/beam for SKA) for a Type Ia SN exploding at the distance of M 82 and at 25 Mpc. It is evident that, at this level of sensitivity, a non-detection would be essentially as meaningful as a detection, since the former would imply that only the double-degenerate scenario is viable, while the latter would tell us which of the single-degenerate channels results in Type Ia SNe. The overall time needed to carry out such a target-of-opportunity programme will require no more than about 12$-$24 hr/year, overheads included, for an average of two targets/year within a radius of 25 Mpc. Such modest time requests can be easily accommodated within a sensible period of time.

# 4 Summary

We have presented the prospects for advancing our understanding of the physics of supernovae via their study at radio wavelengths with the SKA. Our suggested approach for core-collapse supernovae is a commensal one, taking advantage of the deep, sensitive surveys that are planned with SKA. We have discussed the expectations for CCSN studies under the specific case of SKA1-MID ($\sim$2000 hr in one year, covering an area of $\sim$1000 deg$^2$ in one year, rms=1.14 $\mu$Jy/beam in 1-hr, bandwidth=770 MHz, central observing frequency = 1.7 GHz). The expected number of new CCSNe discovered after one year would be $\sim$310. For SKA, the expected number of expected CCSN discoveries is close to 10,000 in one year. Therefore, the number of detections is expected to approach the number of explosions of all CCSNe in the local universe, thus allowing us to obtain a dust-free, complete census of CCSNe. The only request from such a programme is a multi-epoch approach, observing a cadence-time of $\lesssim 90\nu_{1.7}^{-1}$ days, where $\nu_{1.7} = \nu/1.7$ GHz.

We have also discussed the prospects for probing Type Ia SNe progenitor scenarios with the SKA via ToO observations. The SKA can be used at very low time-cost for searching the putative prompt radio emission arising, in the single-degenerate scenario, from the circum-stellar medium around Type Ia SN progenitors in the nearby universe. Complementarily, in the double-degenerate scenario, the radio emission of Type Ia SNe is expected to increase with time and, therefore, the SKA should observe decades-old, nearby Type Ia SNe. In conclusion, the huge improvement in sensitivity of the SKA with respect to their predecessors should allow us to unambiguously discern which progenitor scenario (single-degenerate vs. double-degenerate) applies to them, thus solving this long-standing issue.

## Acknowledgments

MAPT, AA, and RHI acknowledge support from the Spanish Ministerio de Economía y Competitividad (MINECO) through grant AYA2012-38491-C02-02. ER, JCG, and JMM acknowledge support from the MINECO through grant AYA2012-38491-C02-01, as well as from Generalitat Valenciana through grants PROMETEO/2009/104 and PROMETEOII/2014/057.

# Pulsar science with the SKA

**Nanda Rea[1,2] and José A. Pons[3]**

[1] Anton Pannekoek Institute for Astronomy, University of Amsterdam, Science Park 904, Postbus 94249, Amsterdam, The Netherlands
[2] Institute of Space Sciences (ICE-CSIC, IEEC), Campus UAB, Carrer de Can Magrans, s/n 08193 Barcelona, Spain
[3] Departament de Fisica Aplicada, Universitat d'Alacant, Ap. Correus 99, E-03080 Alacant, Spain

## Abstract

The unprecedented sensitivity and large field of view of SKA will be of paramount importance for pulsar science, and for many related research fields. In particular, beside the obvious discovery of many more pulsars (even those with very low luminosity), and the extremely accurate timing analysis of the current pulsar population, SKA will allow to use pulsars to measure or put strong constraints on gravitational waves, Galactic magnetism, planet masses, general relativity and nuclear physics.

## 1 Introduction

Pulsars are the most common observational manifestation of neutron stars, which are considered *ideal laboratories* to study matter under extreme conditions of gravity, density, and magnetic fields, not reachable in terrestrial facilities. They are very compact objects with the mass of the Sun in a $\sim$10 km radius. Their rotational periods can span a very wide range between about 1.5 ms–1000 s, and their magnetic fields are in the $10^6 - 10^{15}$ G range [5]. They are possibly the only environment in the Universe where such extreme conditions can be reached simultaneously, and tested. The recent past has seen substantial advances in our understanding of the astrophysics of compact stars, thanks to the availability of new telescopes and space observatories. The launch of a new generation of satellites (e.g., *Chandra, XMM-Newton, Swift, Integral, Agile, Fermi*) has allowed us to collect unprecedented data on the high-energy emission (X-ray and gamma-ray) of compact stars. This data, together with ground-based astronomical observations at different wavelengths (mainly radio and optical) has significantly advanced our understanding of neutron stars, yielding more accurate measurements of their physical properties, as well as adding new classes of objects to the ones already known. Among the many examples, it is worth highlighting the accurate mass measurements and probe of general relativity in binary systems consisting of two pulsars and the increasing numbers of magnetars [7], relatively young neutron stars with extremely high magnetic fields which power their emission. The accurate measurements via radio observations of neutron stars with masses around two solar masses has set important constraints on the nuclear equation of state, ruling out some theoretical models.

In addition, new ambitious projects, such as the gravitational-wave detectors *Advanced LIGO* and *Advanced Virgo* are being completed, and even more ambitious projects, such as the X-ray mission *Athena* are planned in the near future, and others as *LOFT* (Large Observatory for X-ray Timing) are in an advanced design study. In the radio domain, LOFAR (the LOw Frequency ARray) is already taking exciting new data, while SKA (the Square Kilometer Array) will provide us with an unprecedented wealth of information about pulsars. The combined input of all these instruments will be crucial for our understanding of neutron stars and some aspects of fundamental physics, and SKA will play a leading role.

# 2 Pulsar science with the SKA

## 2.1 Pulsar surveys

The SKA sensitivity will allow to increase the number of currently known pulsars by about an order of magnitude [9]. In particular, simulations predict that $\sim$ 30000 new pulsars will be discovered in the first few years of operations, comprising also pulsars hosted in other Galaxies. Furthermore, about 3000 new recycled millisecond pulsars will be discovered, which are key systems for many of the studies mentioned in the following sections (as 2.2, 2.5 and 2.6). Using SKA for monitoring of pulsars will allow a precision in the pulse arrival time determination of the brightest pulsars a factor of 100 better than currently achievable, that will set pulsars as the best tool for the absolute *Time* calibrations.

## 2.2 Pulsars as a test of General Relativity

Millisecond pulsars in binary systems are important tests for General Relativity. In fact, when a pulsar is orbiting in a close system with another compact object with a very short orbital period, can lead to a strong space-time deformation that results in several relativistic effects. Timing the orbital parameters in double neutron star systems can provide stringent constraints on the theory of gravity in the strong field limit. In the point-mass approximation, the masses of the objects are the only two free parameters. Thus, measuring three or more relativistic corrections to the orbit (the so-called post-Keplerian parameters, e.g. periastron precession, Shapiro delay ...) overdetermines the systems and therefore provides a test of the theory of gravity. As an example, current monitoring of the Double Pulsar [1] has constrained the prediction of the General Relativity with an accuracy of 0.05 % [4]. SKA will greatly enlarge the number of relativistic binaries, as well as improving the timing analysis of known relativistic systems enabling the measurement of second order relativistic effects.

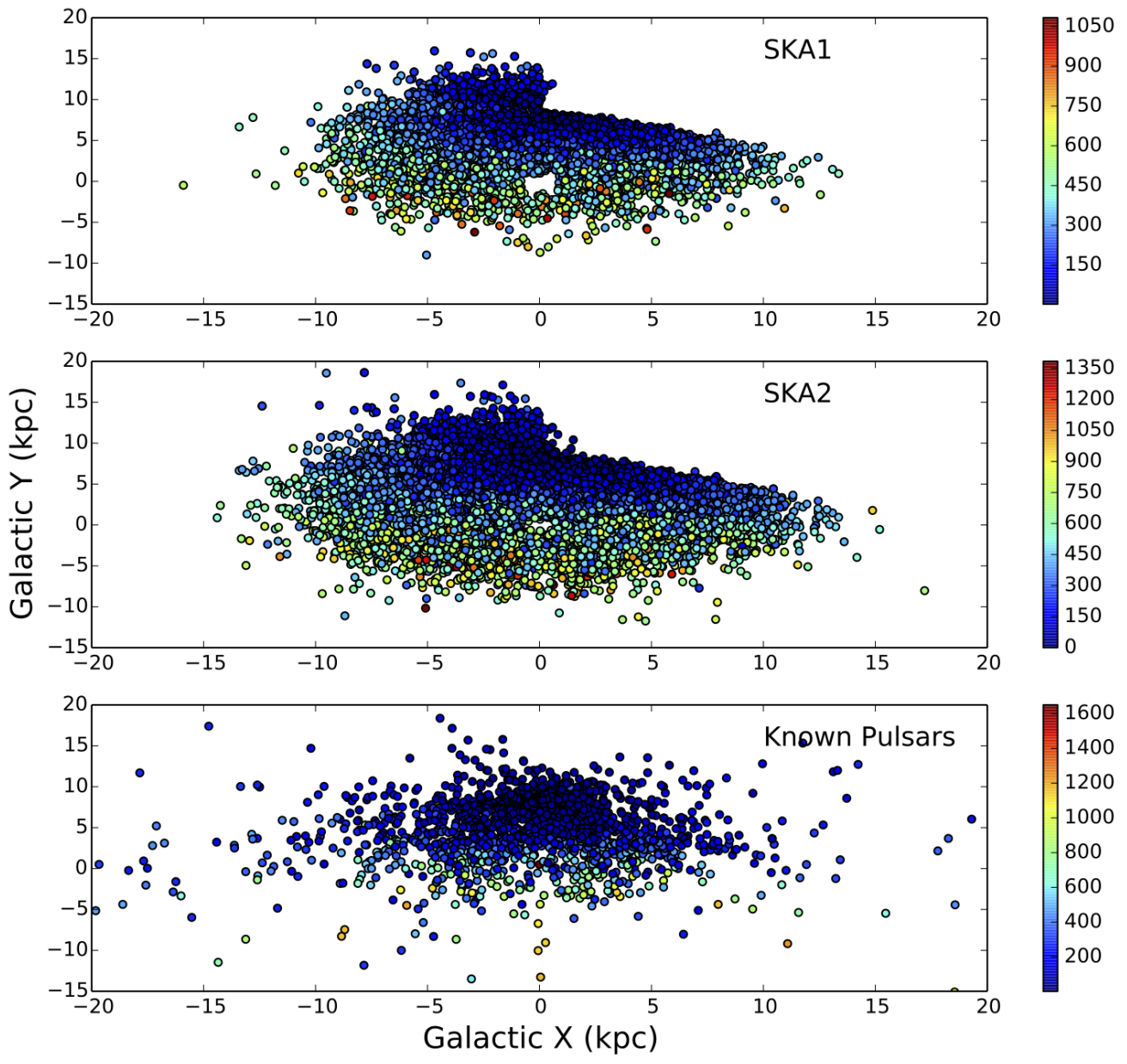

Figure 1: Distribution in Galactic coordinates of pulsars expected to be found with the SKA, compared to the known pulsar distribution. The color coding indicates the approximate range of dispersion measures of the simulated pulsars [3].

## 2.3 Pulsars as probe of ISM and Galactic magnetism

Pulsars are efficiently used to measure the warm ionized medium in the Milky Way, using Faraday rotation measures, dispersion and scattering by the interstellar medium (ISM) of the radio pulse arrival. SKA observations will enable to construct, via pulsar observations, the most detailed map of the Galactic ionized interstellar medium. This detailed map will shed light on the turbulent interstellar medium, but will also improve considerably our knowledge on the pulsar distances, and of the Galactic magnetic field.

## 2.4 Pulsars as a tool to measure Planet Masses

The unprecedented accuracy that pulsar timing will reach with SKA will enable an extremely accurate measure of our Solar System planets, taking advantage of pulsars as one of the most precise clocks. In particular, we can measure the delay in pulsar arrival times due to the motion of all planets with respect to the Earth. Modeling in the pulsar arrival times the residuals due to the planet motions will enable to measure the planet masses very accurately, with uncertainties comparable to dedicated space missions as Cassini or Galileo [2].

### 2.5 Pulsars to detect Gravitation Waves

The passage of a gravitational wave can alter the space-time metric. This alteration is expected to produce a very subtle signature in the arrival time of pulsars. The Pulsar Timing Array (PTA) project is currently monitoring several pulsars ($\sim$ bi-weekly since years) to reach an extremely accurate timing solution to be sensible to these tiny signatures. A gravitational wave due to a coalescing supermassive black-hole binary is expected to be detected by SKA thanks to the very high sensitivity and accuracy of the SKA pulsar monitoring [8].

### 2.6 Pulsars to constrain Nuclear Physics

Beating the present records of maximum observed mass and spin frequency of a neutron star will further constrain the equation of state of dense matter. Because of their history of accretion, recycled millisecond pulsars in binary systems are the most promising systems to search for very massive or fast rotating pulsars. With SKA, about 3000 of such systems will be discovered, increasing by a factor of ten the current sample. On the other hand, the improvement in the timing solutions for the known systems will allow to measure at least two relativistic effects for many systems, hence allowing for more mass measurements, and improving the precision of the existing measurements. Even in the case that no period shorter than 1.5 ms is found, this would require a theoretical explanation, probably connected to hydrodynamical instabilities and associated to the emission of gravitational waves.

In the other end of the spectrum, increasing the number of observed isolated radio-pulsars with long periods (several seconds, or longer), will also help discriminating between different models of the long-term magnetic field evolution and constraining the neutron star crust composition [6].

## 3 Conclusions

SKA will certainly lead a revolution in the field of pulsar astrophysics. It will increase significantly the number of pulsars observed in our galaxy, and allow the detection of many pulsars in the closest galaxies. It also allows to detect giant pulses from pulsars as distant as the Virgo cluster. Among the about 30 thousands new pulsars that will be discovered, we expect about 3 thousand millisecond pulsars, among which the most rapidly rotating and most massive neutron stars are expected to appear, placing new constraints on fundamental physics (the nuclear interaction and the equation of state of dense matter). We may also be witnessing the discovery of the, up to now elusive, first neutron star / black hole binary system, as well as detected gravitational waves via pulsar timing. In addition, the structure of the Milky Way and its magnetic field will also be studied in far more detail than at present, all thanks to the extraordinary sensitivity and large field of view of the SKA.

## Acknowledgments

NR acknowledges support from an NWO Vdi Grant, the MINECO grant AYA2012-39303, and the grant SGR2009-811 from the Generalitat de Catalunya. JP acknowledges support from the MINECO grant AYA2013-42184-P, and the Generalitat Valenciana grant PrometeoII-2014-069. Partial support comes from the NewCompstar COST action MP1304.

# Microquasar Perspectives with the Square Kilometre Array

**Josep M. Paredes**[1], **and Josep Martí**[2],

[1] Departament d'Astronomia i Meteorologia, Institut de Ciències del Cosmos, Universitat de Barcelona, IEEC-UB, Martí i Franquès 1, E-08028 Barcelona, Spain
[2] Departamento de Física, Escuela Politécnica Superior de Jaén, Universidad de Jaén, Campus Las Lagunillas s/n, Edif. A3, 23071, Jaén, Spain

## Abstract

The first half of the twenty-one century Astrophysics will be marked by different ground-based observatories that provide a major leap in photon collecting area across the whole electromagnetic spectrum, coupled with state-of-the-art instrumentation. Among them, the low-energy end of the spectrum in the domain of radio wavelengths will be covered by the Square Kilometer Array (SKA). Here we address the potential offered by this new observational tool when applied to the study of microquasars. The superb SKA capabilities will enable us to explore in depth many issues, such as revealing the Galactic and extragalactic census of these enigmatic systems, opening the window of high-time resolution radio photometry, how do they behave in quiescent states, polarization of the jets, and their weak signatures of interaction with the ambient medium.

## 1 Introduction

The concept of microquasars refers to a special kind of X-ray binary stars (XB) in our Galaxy, one with the ability to generate collimated beams, or jets, of relativistic plasma. The ejection takes place in a bipolar way perpendicular to the accretion disk associated with the compact star, a black hole or a neutron star. The word 'microquasar' was chosen due to the extraordinary analogy between these astronomical objects and quasars and other active galactic nuclei (AGN) at cosmological distances [16].

The relativistic jets of plasma are probably the most reliable fingerprints of microquasar sources and are responsible for the non-thermal emission, of synchrotron origin, that is often detected from them. The upcoming construction of the SKA will provide an unprecedented tool to study many aspects of this emission at radio wavelengths. This papers addresses different paths of progress that we naturally expect in the microquasar field thanks to SKA.

The predicted SKA-mid continuum sensitivity of 0.72 $\mu$Jy h$^{-1/2}$, in the 0.35-14 GHz frequency range[1], will be used hereafter for the performance estimates discussed here. Nevertheless and as it often occurs in science, some of the most important advances will likely result from findings not being possible to anticipate now.

## 2 The microquasar population

Microquasars, as XB with relativistic radio jets, represent a subset (15 numbers) of the XB population in the Galaxy [19]. XB in general are binary systems containing a compact object (a neutron star (NS) or a stellar mass black hole (BH)) accreting matter from the companion star. Depending on the spectral type, the optical companions are classified into High Mass X-ray binaries (HMXB) or Low Mass X-ray Binaries (LMXB). The most recent catalogue of HMXB contains 114 sources [11], while the catalogue of LMXB amounts to 187 objects [12]. A significant part of them, about 22% of all XB, have been detected at radio wavelengths with flux densities $\geq$ 0.1–1 mJy, being 9 of them HMXB and 56 LMXB. The 9 radio emitting HMXB include 6 persistent and 3 transient sources, while among the 56 radio emitting LMXB we find 18 persistent and 38 transient sources. Some of these radio sources have been revealed to be microquasars after VLBI observations resolved the radio structure showing the presence of relativistic jets. So far, whenever radio emission has been resolved, it appears with a clear elongated shape, as would be expected from a collimated jet flow. In this sense, it is very likely that the simple detection of radio emission from an XB in a particular X-ray spectral state has to be considered as a tell-tale sign of jets[8].

It has been estimated that the total number of XB in the Galaxy brighter than $2 \times 10^{34}$ erg s$^{-1}$ is about 705, these being distributed as $\sim$ 325 LMXB and $\sim$ 380 HMXB [9]. This suggests an upper limit on the population of microquasars in the Galaxy of about one hundred and fifty systems. This number could increase if we consider that the fraction of radio emitting XB observed today (22%) is an underestimate.

The non-detection of radio emission from most XB could be due to a low level of emission undetectable for the current radio telescopes and/or variability. Confirming this suspicion will require a substantial improvement in interferometers, both in terms of their sensitivity and angular resolution. SKA is one of the projects currently underway with the potential to achieve this. SKA will contribute to increase significantly the population of microquasars, allowing a more general study of this class of sources that now is still very limited by the small number of them.

The discovery of new microquasars in nearby galaxies could become easily feasible with SKA. A typical microquasar, within 10 kpc from us with a flux density of $\sim$ 10 mJy at 6 cm wavelength, would emit at the $\sim$ 2 $\mu$Jy level if placed in the Andromeda galaxy (0.8 Mpc away). This object could be detected using for instance SKA-mid with signal-to-noise ratio of $\sim$ 10 after about 12 hour of integration. Therefore, most microquasars known in our Galaxy would be detected by SKA in a reasonable observing time period if they were located in Andromeda. The above estimates are based on the radio emission level of an average microquasar when in quiescence. During strong flaring events, where the radio luminosity may increase up to two or three orders of magnitude, microquasars belonging to different galaxies in the Local Group could also be within SKA reach.

[1] http://www.skatelescope.org/wp-content/uploads/2012/07/SKA-TEL-SKO-DD-001-1_BaselineDesign1.pdf

## 3 High-time resolution radio photometry

The expected SKA sensitivity will allow the capability of routinely monitoring the radio variability of microquasars and gamma-ray binaries[2] with second, and possibly sub-second, detail too. This will open the high-time resolution window to radioastronomy in a way similar to X-ray astronomy today. In all these objects, the genesis of jets and outflows occur in a relatively small volume ($\ll 1$ AU) within the binary system. Consequently, the associated time scales can be very short. Previous radio photometric studies, such as those of the highly variably microquasars Cygnus X-1 and Cygnus X-3, have revealed a richness of radio flaring phenomena but they remain still limited by moderate time resolution of a few minutes at most [13, 18]. The predicted SKA-mid continuum sensitivity implies that an rms noise of a few mJy could be attained in just one second. For instance, this will improve our sampling capability of targets such as Cygnus X-1 and Cygnus X-3 by at least two orders of magnitude in time, and still achieve a signal-to-noise ratio well above 10.

Another variability issue that SKA will satisfactory address is related to the fast flux density and structural changes that both microquasars and gamma-ray binaries experience at radio wavelengths in time scales of hours and days. This morphological variability has created strong difficulties when mapping these systems with present day interferometers, that require an on-source time of a few hours to produce reliable images of the ejecta [21]. This is because most deconvolution algorithms implicitly assume a constant source in the sky, and this is clearly violated in this case throughout the duration of the observations. Indeed, these sources can be simultaneously variable both in brightness level and structure in contrast to other less changing targets. The SKA possibility of performing quick snapshots in a matter of minutes with sub-mJy sensitivity will overcome this problem. Our physical understanding of these systems will strongly benefit from deep radio maps free from variability artifacts.

## 4 Testing the X-ray vs radio correlation at very low luminosities

The black hole X-ray binaries go through transitions between different spectral states that are defined according to X-ray spectral and timing properties [14]. Continuous steady jets are observed in radio during the low-hard state whereas the ejection of blobs is produced during the transition between spectral states. In the high-soft state the radio emission is strongly suppressed, indicating the disapearance of the radio jet. There exists a correlation between the radio and X-ray fluxes in these systems in the hard and quiescent states, demonstrating a coupling between the jet and the accretion disk [4, 6]. However, the quiescent part of this correlation has been poorly explored because the limited sensitivity of the available instruments. With the good sensitivity that SKA will provide it will be possible to determine the coupling between ejection and accretion at low radio luminosities.

[2]Similar Galactic systems also thought to host relativistic outflows.

## 5 Polarization properties of radio emission from microquasars

The observational study of radio polarization is a mature tool in the domain of AGN physics as evidenced in recent reviews such as [23] and references therein. Being able to produce not only intensity maps, but also maps in all the four $IQUV$ Stokes parameters provides the strongest physical constrains. The three-dimensional components of the jet magnetic field (e.g. toroidal or helical), its degree of order, and the jet particle content and distribution of Faraday rotating material can be inferred or, at least, significantly restricted. In the domain of microquasars and gamma-ray binaries, the radio polarization studies are much less developed because we are usually dealing with very faint radio jets and compact cores sometimes with typical centimetric flux densities as low as 0.1 mJy or fainter. Therefore, progress has been achieved only in a reduced sample of bright cases. These include for instance systems such as LS I +61 303[20], GRS 1915+105[5], GRO J1655−40[15], SS 433[22] or XTE J1748−288[3]. Thanks to the exquisite SKA sensitivity, this situation will radically change and polarization studies of microquasars and gamma-ray binaries will become routine in the future on an equal basis with today's AGN situation.

## 6 Unveiling the weak signatures of interaction between microquasar jets and the Interstellar Medium

The microquasar jets are known to inject a substantial amount of energy into their nearby surrounding ISM whose effects have not been seriously considered until recent times [2, 1, 17]. The signatures of jet-ISM have the potential to be used as calorimeters to assess the true power of relativistic outflows in a stellar system. In a few cases, the observational effects of this interaction have been detected at a radio wavelengths in the form plumes or jet-driven bubbles [7, 21]. However, for the majority of systems the expected effects are too weak and elusive to be imaged by the previous generation of interferometers. This is mainly due to the fact that most microquasars are located in regions less dense than the canonical ISM density of $\sim 1$ cm$^{-3}$ [10]. In this context, deep SKA imaging of a representative sample of microquasars is very likely to reveal a wide range of faint extended structures created as a result of the interaction between relativistic particles and the cold ISM, from which independent constrains on the time-averaged jet power could be inferred.

## Acknowledgments

We acknowledge support by DGI of the Spanish Ministerio de Economía y Competitividad (MINECO) under grant AYA2013-47447-C3-1-P and AYA2013-47447-C3-3-P. J.M.P. acknowledges financial sup-port from ICREA Academia. J.M. also acknowledges support from Consejería de Economía, Innovación y Ciencia of Junta de Andalucía as research group FQM-322 and FEDER funds.

# Circumstellar envelopes and their descendents

**Javier Alcolea[1], Valentín Bujarrabal[1], Jean-François Desmurs[1], José Francisco Gómez[2], and Carmen Sánchez Contreras[3]**

[1]Observatorio Astronómico Nacional, IGN. [2]Instituto de Astrofísica de Andalucía, CSIC. [3]Centro de Astrobiología, INTA-CSIC.

## Abstract

Copious mass-loss in thermally pulsating evolved stars is responsible for the formation of circumstellar envelopes and planetary nebulae, which in turn constitute the major source for the galactic enrichment in dust and light elements. The detailed study of these envelopes and nebulae requires normally both high angular resolution and sensitivity, as these sources are typically at distances of several hundreds of pc. These needs will be fully met at cm-wavelengths by the SKA. This instrument will significantly contribute to the better knowledge of these sources in many fundamental aspects [14]. Such as the compact ionized cores of nascent planetary nebulae (observations of radio continuum and radio-recombination lines); the cool outer parts of their envelopes (observations of HI, CH and OH); or the onset of the bipolarity in the pre-planetary nebulae phase, and its possible relationship with magnetic dynamos (observations of OH masers and their Zeeman splitting).

## 1 Introduction

The thermally-pulsating asymptotic giant branch (TP-AGB) is one the latest stages of evolution of stars with masses $\sim 1-8\,\mathrm{M}_\odot$. At this point, these asymptotic red giant stars (aRGs) experience a huge mass-loss ($\sim 10^{-8}-10^{-4}\,\mathrm{M}_\odot\,\mathrm{yr}^{-1}$) that determines their ulterior evolution. These mass ejections lead to the formation of dense circumstellar envelopes (CSEs), with total masses up to some $\mathrm{M}_\odot$, that gentle expand into the interstellar medium (ISM). These CSEs will later transform into planetary nebulae (PNe), when the host star, denuded from its mantle as a result of the mass-loss, gets hot enough ($\gtrsim 25{,}000\,\mathrm{K}$) in its way to the white dwarf phase and ionizes its surroundings. Very similar mass-loss processes also occur in the heavier counterparts of TP-AGB stars, the red super-giants (RSGs) and yellow hyper-giants (YHGs), diminishing the final mass of the core-collapse supernova and constituting the circumstellar environment the supernova blast will run into. In addition to this, CSEs of both aRGs and RSGs/YHGs constitute the most important source of dust and light metals enrichment in the galaxy. Yet, we are still far from understanding how these mass-loss processes occur. For example, we still do not have a reliable analytical mass-loss rate formula to use in population synthesis models, and we are far from understanding the extreme latest mass-loss event which results in the birth and shaping of PNe.

### 1.1 The neutral atomic gas: observations of the HI 21-cm line

CSEs are formed by gas and dust particles, with typical gas-to-dust mass ratios of a few hundreds. The most abundant constituent of the gas phase is obviously hydrogen, that is usually found in molecular ($H_2$) or neutral atomic (HI) form depending on whether the temperature of the central star is less or more than 2500 K respectively [8]. $H_2$ rotational and ro-vibrational lines and strongly forbidden, and can only be observed in the IR from hot gas or ionized regions close to the star, or in the presence of strong shocks, and therefore it can not be used to trace the cool outer parts of the CSEs [11]. The study of these circumstellar regions is usually done by means of the observation of low-excitation rotational lines of other molecules, mainly CO. However, CO (and other molecules too) can only trace the mass-loss history of the stars up to its photo-dissociation radius, 0.1 pc at most, which means only 10 000 yr back in time, about 10% of the whole TP-AGB phase duration. Indeed, we know that CSEs can be much larger, $\sim 0.1-$to 1.0 pc, as shown by the IR emission of the circumstellar cold dust [13]. Unfortunately, dust emission does not provide any information on the kinematics (i.e., mass-loss rates can not be computed), and total mass estimations are model dependent (grain composition and size, gas-to-dust ratio, etc.). These outermost layers are very interesting, as they are where both circumstellar and interstellar media interact, sometimes resulting in the formation of very extended bows and/or cometary tails of shocked excited gas [16].

These very outer parts of CSEs can in principle be probed by observing the 21-cm (1420 MHz) hyperfine structure transition of HI. However, this is a very difficult task for the current instrumentation, because of the weakness of this emission in CSEs and the strong contamination due to the ISM. Both problems can largely be circumvent by the SKA expected capabilities: the much higher sensitivity and the ability of performing high-fidelity mapping of structures over a wide range of spatial scales. Due to the present observational limitations, the existing observations of HI in CSEs are meagre, and sometimes difficult to interpret. Using the Nançay radio-telescope, the VLA, the GMRT, and the GBT, about a dozen sources have been detected in the 21-cm HI line [21, 7]. In some cases (e.g. IRC $+10^{\circ}216$ and R Cas) there are still doubts that the detected signal is indeed of circumstellar origin and not due to ISM gas along the line of sight. However, in the cases of Ep Apr, $o$ Cet, RC Cnc, Y Cvn, and RX Lep, the detections are clear and the estimated HI mass-loss rates are sometimes comparable to those inferred from CO. In particular, the case of $o$ Cet is very interesting, as HI is detected in a extremely long ($\sim 3$ pc) cometary tail also seen in X-rays [20]. This tail would consist of circumstellar gas stripped off as result of the motion of the star with respect to the local ISM. This could also be the case of RS Cnc and RX Lep.

In view of these few but promising results, it is expected that the revolutionary HI 21-cm line observations that SKA should provide, will open a new window for studying CSEs at large scales. This, in combination with other data, will give us a much complete view of the mass-loss history in evolved stars and its impact on the chemistry and dynamics of the ISM.

### 1.2 Thermal emission of light molecules: OH and CH Λ-doubling lines

Molecules in CSEs and young-PNe (yPNe) are usually observed at mm- and submm-wavelengths. However, some key species can also be observed at lower frequencies. In particular, the CH and OH radicals have Λ-doublet transitions at cm-wavelengths. These molecules, which show widespread emission in both interstellar and circumstellar environments, are important carriers of abundant elements and thus constitute basic chemical ingredients.

OH is well known to display strong maser emission in several Λ-doubling lines at 18-cm, from the $\Pi$=3/2 $J$=3/2 rotational ground level (see later). These lines must also show thermal (i.e. non maser) emission under certain conditions. The advantage of observing the thermal emission is that the interpretation of the results is more straightforward than for masers, and that it traces the overall structure and dynamics of the emitting nebula more directly. Of course, the disadvantage is that the lines are weaker by orders of magnitude. To date, such thermal lines have been only observed with single-dish telescopes, from very extended nebulae, and with very poor spatial resolution. The very high sensitivity of SKA should allow systematic observations of thermal emission of the four OH 18-cm lines. We expect these lines to be optically thick, and therefore, brightness temperatures of the order of the kinetic temperature in the OH emitting regions: $\sim 100$ K or higher in both CSEs and in shocks propagating in young PNe [10, 3]. This results in line intensities detectable for the SKA, even at $1''$ resolution, in reasonably short integration times.

CH, the counterpart of OH in C-rich environments [24], very often shows weak maser emission at 3.3 GHz, from the three Λ-doubling components of the ground rotational levels [2]. Because of the weak amplification, for CH we can just expect brightness levels over 50 K, even if population inversion is practically universal in this case. For this reason, high spatial resolution observations of CH are very rare, and most data have been obtained using single-dish radio-telescopes. However, the SKA sensitivity at 3.3 GHz, in units of brightness temperature, must be significantly higher than for the OH lines (about 4 times better). Therefore, the CH cm-wave lines should be detectable with SKA, with enough angular resolution for detailed mapping, in a large number of sources.

### 1.3 Ionized cores of pPNe: radio continuum and RRLs observations

PNe are ionized remnants of CSEs, but while the latter are roughly spherically symmetric, the former are mostly bipolar and often show most bizarre structures. The physical mechanisms responsible for this transformation is yet unknown, but it must be active in the very early stages of the evolution beyond the TP-AGB, the pre-planetary nebula (pPN) phase. Therefore, pPNe, and yPNe too, hold the key for understanding this complex and fast ($\sim 1000$ yr) nebular evolution. Studies of pPNe support the idea that the multiple lobes and high-velocities observed in pPNe and PNe are produced by the interaction of fast collimated winds (jets) with the spherical and slowly expanding CSE. However, this jet+CSE two-wind interaction scenario remains unconfirmed by the direct observation of the jets themselves, and of the central nebular regions from where these jets should be launched (a few 100 AU). Studying these central regions is difficult due to their small angular extent (sub-arcsec.)

and because they are heavily obscured by dense central circumstellar dust layers. Progress requires sensitive high-angular resolution observations at long wavelengths, which could be swimmingly provided by SKA in the future.

Typically, central stars of PNe start ionizing their envelopes when they reach a B-type spectral classification. Optical spectroscopic observations of pPNe and yPNe have revealed the presence of widespread broad ($\sim$100–1000 km s$^{-1}$) H$\alpha$ emission, often with blue-shifted absorption features (P Cyg profiles) produced at their nuclei [27]. Surprisingly, H$\alpha$ emission is also found in some pPNe with much cooler central stars, suggesting the presence shocks in the stellar vicinity, which could be linked to on-going jets sculpting the inner regions of the old CSE, or to mass-accretion by a compact companion. Due to the large dust opacity of these regions in the visible, in most cases these nuclei can only be indirectly observed via their dust reflected light. This situation not only makes the observations more difficult, but also complicates the interpretation of the observed profiles [26]. These central ionized regions of pPNe and yPNe are visible through radio-continuum and radio-recombination line (RRL) emission at long wavelengths, which have the key advantage of being dust-extinction free.

In constrast to more evolved PNe, continuum flux measurements at wavelengths beyond $\sim$1 cm are lacking for the vast majority of pPNe and yPNe. This is because at these long-wavelengths their compact, just emerging ionized cores are very weak emitters. Typically, reliable detections in pPNe/yPNe would require sensitivities better than $\sim$0.01 mJy, which are difficult to achieve with the current facilities in reasonable observing times. However, it is at centimetre (and longer) wavelengths where the thermal bremsstrahlung (free-free) radiation, produced by these nascent ionized cores, becomes observable and much stronger than any possible contamination from thermal cold dust emission. The unprecedented capabilities of the SKA, which will routinely reach sub-$\mu$Jy sensitivities at sub-arcsecond resolution, should allow to carry out sensitive radio continuum imaging observations of pPNe/yPNe. Resolving the morphology of the central ionized regions, down to $\approx$1–100 AU spatial scales, holds the promise for unveiling some of the pivotal structures postulated by most PNe shaping theories. The SKA could perform time-series imaging of the expansion and development of these central HII regions, as well as of the changes in the shock activity. The SKA could also provide maps of the free-free emission turnover frequency, and thus maps of the emission measure. This can be used to deduce the spatial distribution of the electron density, and to estimate the present-day (post-AGB) mass-loss rate, a long sought key parameter in the post-AGB evolution. Imaging the continuum at long wavelengths where the free-free emission becomes optically thick, will provide a measure of the electron temperature throughout the nebulae [18]. Radio continuum maps can also be compared with H$\alpha$ images at optical wavelengths, to obtain dust extinction maps of these objects [30]. Mapping the spectral index variation across the nebulae will show whether other emission mechanisms, such as synchrotron radiation from magnetically collimated jets [23], whose contribution to the radio continuum flux cannot be excluded, are present at these early post-AGB phases.

Observations of RRLs are challenging, especially in the cm- to m-wavelength range, where these lines are extremely weak in comparison to the continuum level (typically $\sim$1%). To date, mm- and cm-wavelength RRLs studies have been carried out for some of the best-known and most luminous evolved PNe [1, 30], but this field remains largely unexplored for pPNe and yPNe. RRLs have only been sought and detected in the mm-wavelength range towards the C-rich pPN CRL 618 [17]. In the frequency range covered by SKA there are hundreds of hydrogen recombination lines (H$n\alpha$, with $n > 90$) that are ideal tools to image the structure, physical conditions (electron temperature and density), and kinematics of the central ionized regions of pPNe/yPNe. These low-frequency transitions, although much weaker than those in the mn-wave range, are particularly well suited to study the current, fast but tenuous ($< 100\,\mathrm{cm}^{-3}$) post-AGB winds that fill in the central cavities inside CSEs that result from the abrupt decrease of the mass-loss rate (by several orders of magnitude), after the central star departs from the TP-AGB. RRLs studies with SKA will fully complement analogous observations at higher frequencies (e.g. with ALMA), which are better probes of the inner and denser ($\sim 10^5$–$10^7\mathrm{cm}^{-3}$) environments. RRLs arising from other elements, such as He and C, can also be observed simultaneously to the hydrogen RRLs, providing a direct measurement of the chemical composition of the material ejected in the late-AGB and early post-AGB phases. We are confident that multi-frequency RRLs observations, with the exceptional high-sensitivity and high-angular resolution imaging capabilities of SKA, will certainly produce significant new insights into the structure, dynamics, and physicochemical conditions of the central regions of pPNe/yPNe, and will help in addressing the problem of the formation and shaping of of post-AGB nebulae.

## 1.4 Maser emission from CSEs and beyond

Maser emission is often observed in the CSEs of aRGs and RSGs. In particular, sources in which oxygen is more abundant than carbon, O-rich CSEs, can show strong maser emission in several lines of silicon monoxide (SiO), water vapour ($H_2O$) and hydroxyl (OH). These emissions have been widely used to characterize the geometry and velocity field of these envelopes at high angular resolution.

OH masers are typically observed in the $\Lambda$-doubling hyperfine transitions at 1.6–1.7 GHz, from the $\Pi$=3/2 $J$=3/2 ground-rotational states. These lines have been observed towards hundreds of CSEs and some PNe. Maser emission from OH hyperfine transitions in vibrationally excited states is also very common in star-forming regions. On the contrary, the lines from the first vibrationally excited state at 6 GHz are either undetected or extremely rare and weak in evolved stars: just a few possible detections have been reported both in CSEs, NML Cyg and AU Gem, and in yPNe, Vy 2_2 and K 3-35 [4]. All these studies will greatly benefit of the advent of the SKA, as its frequency coverage will embrace not only the ground-rotational lines at 1.6–1.7 GHz, but also excited lines at 4.7, 6, and 13 GHz.

One of the possible key programs for the SKA, due to its high sensitivity, will be to perform deep maser surveys. In particular, OH surveys promise to be very fruitful, not only to detected new masers much farther away (beyond galactic center, or towards near galaxies such the Magellanic Clouds, M 31, M 33, etc.), but also to extend the searches towards much weaker luminosities, and thus making possible the detection of possible new excited masers. The increase in number of known OH masers will be especially interesting for dynamical studies. Using phase referencing techniques, it is possible to determine distances with a high accuracy, not only for stars in our local galactic arm, but also for the Galactic Center, and even nearby galaxies. This will result in a much better sampling of the rotation curve of our galaxy, as well as a new trigonometric parallax distances to Local Group galaxies.

The widespread maser emission characteristic of CSEs of TP-AGB stars disappears shortly after the end of this phase, when the mass-loss rate drops dramatically, from $\simeq 10^{-4}$ to $10^{-7}$ $M_\odot$ yr$^{-1}$ [29]. After the end of the TP-AGB, the survival time-scales are roughly $\simeq$ 10, 100, and 1000 years for SiO, $H_2O$, and OH masers, respectively [9, 15]. Therefore, maser emission is less likely to be found as the star further evolves into the post-AGB phase, and it is extremely rare in PNe. However, despite their lower detection rates, maser emission can provide key insights on the collimated mass-loss processes during these post-AGB phases, which determine how PNe are shaped.

The spectra of OH maser emission in CSEs typically shows two well-defined peaks, with separations $\simeq$ 30 km s$^{-1}$ [28], corresponding to the approaching and receding sides of the circumstellar envelope, which is expanding with a roughly spherical symmetry at a constant velocity. However, the post-AGB and early PN phases are characterized by mass-loss in the form of fast collimated jets, and maser emission from these objects shows a clear departure from the double-peaked pattern: an easily recognizable hallmark of non-spherical mass-loss [5]. Important maser-emitting objects in this phase are the so called *water fountains* (WFs) [12, 6]. These are evolved stars with $H_2O$ maser emission tracing high-velocity ($>$ 80 km s$^{-1}$) jets of very short dynamical ages ($<$ 100 yr). They could represent one of the first manifestations or collimated mass-loss in evolved objects and thus, are key targets to understand the subsequent shaping of PNe. While studies of WFs normally focus on their $H_2O$ maser emission, they also show OH emission tracing jets [25]. SKA observations of OH masers in WFs and other post-AGB stars will allow us to study their jet morphology with unprecedented detail. Using SKA1-mid, in one hour of observing time and a velocity resolution of 0.2 km s$^{-1}$, the relative positional accuracy between two different OH maser spots at the level of 0.5 Jy will be $\simeq$ 150 micro-arcsec. This means that we can study maser distributions of $\simeq$ 1 AU for sources as distant as the Galactic center. This level of precision, which is impossible to obtain with current radio interferometers, will be key to understand where and how mass-loss is produced in these evolved stars. Moreover, observations in different epochs would give us the opportunity to measure proper motions. Therefore, it will be possible to obtain, both the spatial distribution and the 3-dimensional velocity structure of masers in these objects. With this, we could study the launching location and the mass-loss history of jets (e.g., whether they are continuous or explosive events). This can be related to the physical origin of the jets, in particular whether they are powered by accretion disks [19] or via energy release in magnetic dynamos [22].

In addition, because of its non zero electronic angular momentum, the OH radical is a natural magnetometer. By measuring the intensity of the Zeeman splitting effect, we can determine the value of the magnetic field along the line of sight, $B_{||}$, in the regions where the maser emission arises. The role of magnetic fields in evolved stars has not been explored in detail, though it must influence the dust formation process, the launching of circumstellar winds, and obviously provide information on the magnetic properties of the star. Measuring the magnetic field in these OH-emitting regions in a large number of sources will provide invaluable information on these issues, but it can only be done with SKA capabilities.

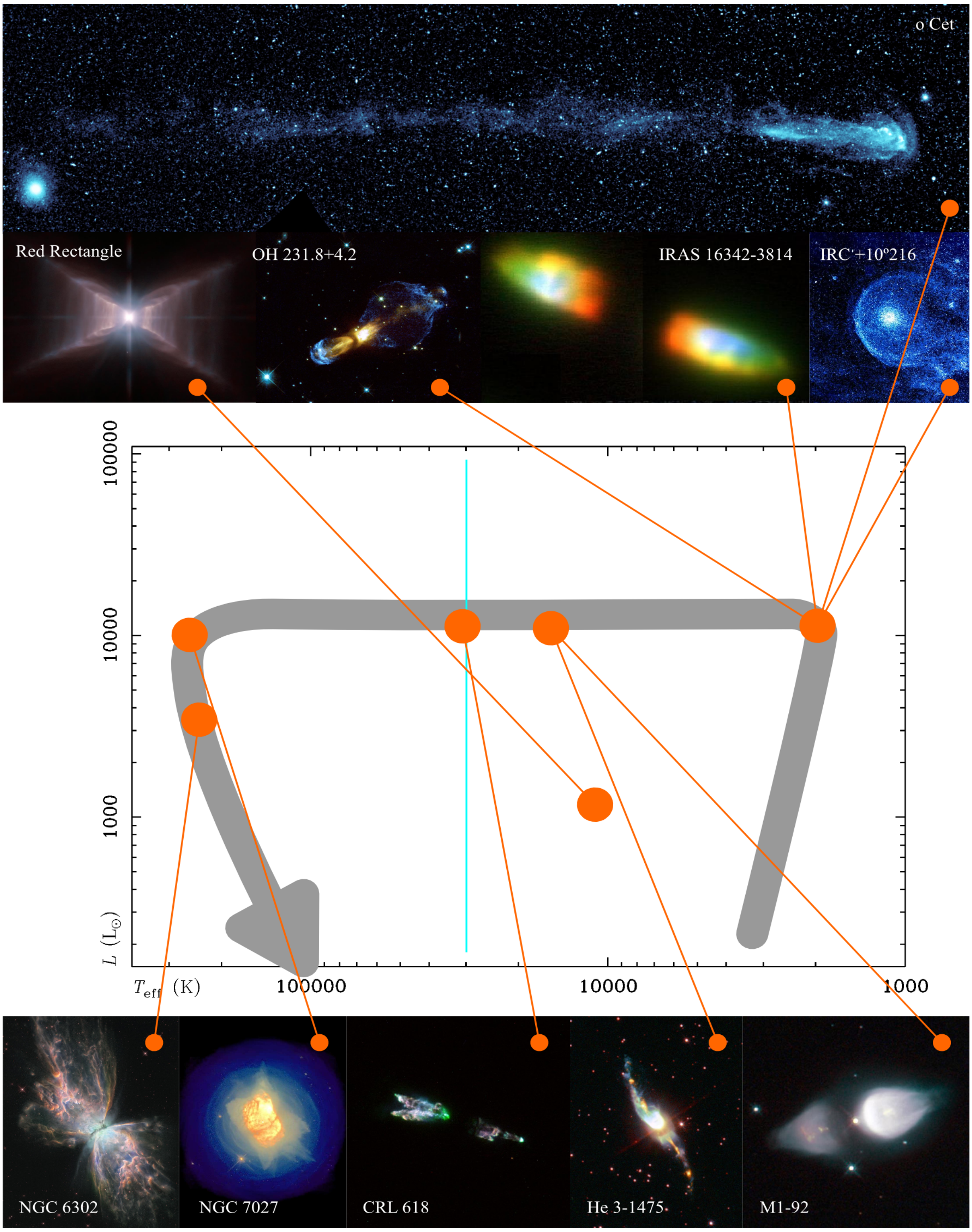

Figure 1: HR-diagram of the late evolution of intermediate mass stars ($\sim 1-8\,M_\odot$), including the location of some of the sources mentioned in the text, as well as images of their envelopes and nebulae.

# Acknowledgments

This work has been financially supported by MINECO (Spain) grants CSD 2009-00038, AYA 2009-07304, AYA 2011-30228-C03-01, AYA 2012-32032, FIS 2012-32096, and AYA 2014-57369-C3-3-P.

# Spanish Involvement in Precursors and Pathfinders for the SKA.

**The Editorial Board of the SKA Spanish White Book.**

## Abstract

In this contribution, we describe the main SKA precursors and pathfinders, focusing on those facilities in which Spanish astronomers are involved. SKA will benefit -scientifically, technically and operationally- from all previous experiences with these facilities. For the Spanish community, it will be very convenient a significant participation in these projects and their planned scientific key-projects to optimize the scientific exploitation of the SKA in the future.

## 1 Introduction

In 2008, and in order to protect the SKA project, the SKA Science and Engineering Committee (SSEC) established a clear definition of what constitutes a SKA Precursor and a SKA Pathfinder facility. The following designations were established:

- Precursor facility: A telescope on one of the two candidate sites. The Precursor telescopes are the South African MeerKAT, the Murchison Widefield Array (MWA) in Australia and the CSIROs Australian SKA Pathfinder (ASKAP).
- Pathfinder: SKA-related technology, science and operations activity. The Pathfinder telescopes and systems are spared around the world and are engaged in SKA related technology and science studies. The list of pathfinders in Europe includes, among others, the electronic European VLBI Network (eEVN), e-MERLIN, the Aperture Tile in Focus (APERTIF), the Electronic MultiBeam Radio Astronomy Concept (EMBRACE), NenuFar and the Low Frecuency Array (LOFAR).

To be officially designed as an SKA Contribution, the facility must contribute to the development of technologies, science, software and analytical techniques and/or operation modes, of interest for SKA, according to the following criteria: i) technology: it should contain new technical elements that are considered as part of the SKA Baseline design and that have not previously been tested on a large telescope; ii) science: observational tests will be carried out, both simulated and real, in order to explore new capabilities at flux density and dynamic range levels similar to or scalable to the full SKA; iii) operations: it provides robust tests of methods of scheduling and time allocation similar, or scalable to, what is needed for the SKA.

In the following, we briefly present the SKA precursors and the main european SKA pathfinders, making emphasis on those scientific projects in which the spanish community is participating.

# 2 SKA Precursors

## 2.1 ASKAP

ASKAP is the Australian SKA precursor. It comprises 36 antennas of 12-m diameter each. It is located in a radio quiet zone in Western Australia, where part of the SKA will be installed in the future. The ASKAP array is using Phased-Array Feeds (PAFs), which allow a 30-square-degree field-of-view. These PAFs are very efficient for the performance of surveys and searches for unidentified transients. However, they suffer from a loss of sensitivity (narrower total bandwidth and higher antenna temperature), compared with the new, very broadband feeds. In that sense, the use of both technologies -single-pixel feeds or phased-array feeds- is being pursued for cutting-edge observations and its use depends on the sicientifc application.

ASKAP will operate in the 0.7 - 1.8 GHz frecuency range and will provide a maximum angular resolution of $\sim$8 arcseconds. A 6-antenna test array (BETA) was completed in 2013 and is being used to test the phased-array feed (PAF) technology. ASKAP is primarily a survey instrument. During the first five years of observations, around 75% of its observing time will be dedicated to large survey science projects. For that purpose, ten projects[1] (needing more than 1500 hours each) have been selected based on their scientific merit. None of these projects are lead by a European PI. However, these programs comprise 363 investigators from 131 institutions with 28% of the Co-Is from Europe (33% from Australia, 30% from North America and 9% from the rest of the world).

Spanish Astronomers are involved in some of the legacy projects (VAST, "Variable and Slow Transients"; EMU, "Evolutionary Map of the Universe); GASKAP, "The Galactic ASKAP Spectral Line Survey").

## 2.2 MeerKAT

The MeerKAT interferometric array consists of 64 dishes of 13.5-m diameter each, and is currently under construction in South Africa (with a completion planned for 2016). It is located on the radio-quiet, future SKA South-Africa site. MeerKAT will have a dense inner core, with 70% of the dishes and with baselines ranging from 29 m to 1 km; the other 30% of the dishes, will provide baselines ranging from 2.5 km to 8 km (with a possibility of an extension up to 20 km with 7 additional antennas). The antennas have been designed to achieve high sensitivity -a large 4-GHz bandwidth- and high imaging dynamic range in several observing windows ranging from 0.6 to 15 GHz. The construction of the KAT-7 science prototype array was successfully completed in 2012, and several refereed scientific publications have already been published.

[1] http://www.atnf.csiro.au/projects/askap/ssps.html

The first years of activity of MeerKAT will mainly be reserved for key science projects[2] (70% of the available observing time) following an international call for proposals in 2009. Ten large proposals have been selected1, each with a total observing time ranging from 1900 to 8000 hours. European PIs lead (or co-lead) a significant fraction (55%) of these large programs, whereas 27% of the proposals are led by South African PIs (18% are lead by the rest of the world).

Spanish Scientists are members of the initiative to incorporate MeerKAT to global VLBI operations with all major radio astronomy observatories around the world, improving the sensitivity and enhancing the southern VLBI arrays.

# 3 SKA Pathfinders

## 3.1 EVN

The European VLBI Network (EVN) is a collaboration of radio telescopes in Europe, with joint observing campaigns with telescopes in China, Russia, South Africa, and other single dishes in the world. Up to 20 telescopes participate in the array. With a frequency range from 327 MHz to 43 GHz, the EVN provides a unique resolving power: from a few milliarcseconds to sub-milliarcsecond scales. Due to its high angular resolution, high sensitivity and superb astrometric precision, the EVN has become a unique astronomical facility. The EVN covers many different research fields: i) the study of the inner regions of AGNe, blazars and radio galaxies; ii) the study of young radio supernovae, GRBs, pulsars and transients in general; iii) the study of circumstellar rings and the process of high-mass star formation, considering the gas dynamics, molecular excitation and the magnetic fields very close to the forming star; iv) the study of distant star-forming systems and the central regions of nearby galaxies; v) the astrometric capabilities of the VLBI arrays, providing very accurate determinations of the proper motions, parallaxes and distances to the star-forming regions within the Milky Way; vi) the extremely precise determination of state vectors of planetary probes.

Within the framework of the RadioNet program NEXPReS, the EVN has supported and developped real-time electronic VLBI operations. In fact, the EVN have routinely carried real-time science observations since 2006 for baselines up to 12000 km. In that sense, the eEVN is an SKA pathfinder in terms of data collection and transportation (with data rates > 1 Gbps), distribution of the clock signal and operations (quick response to external triggers).

Spanish Astronomers are frequent users of the (e-)EVN. Spain is one of the members

[2]http://public.ska.ac.za/meerkat/meerkat-large-survey-projects

of the EVN, contributing to it with the Yebes antenna, and also member of the recently approved JIVE-ERIC (European Research Infrastructure Consortium).

### 3.2 LOFAR

The Low-Frequency Array (LOFAR) is a next generation telescope based on phased-array technology, in which thousands of low-cost, stationary collecting elements are digitally combined in a central processor. The radio maps for a given region of the sky are reconstructed with innovative software. LOFAR offers two observing bands: low band, 10 - 90 MHz, and high band, 110 - 240 MHz. LOFAR was dessigned to address the Key Science Projects: i) Epoch of Reionization: understanding how the first stars and black holes ionized the Universe; ii) Extragalactic Surveys: probing the history of star formation and black hole growth with time; iii) Transients and Pulsars: exploring extreme and explosive astrophysical events; iv) Cosmic Rays: identifying the origin and energy distribution of the most energetic particles in the Universe; v) Solar Physics: mapping the solar wind, solar bursts, and their interaction with the Earth; vi) Cosmic Magnetism: mapping the large-scale magnetic field in the Universe.

LOFAR is the largest pathfinder on the road to SKA-Low. From a technical point of view, LOFAR provides a pathfinder scenario in terms of distributed operation, high bandwidth data links, massive correlation and post-processing, among others.

Furthermore, the LOFAR core and the LOFAR international telescopes are being used for a number of experiments intended to test challenging concepts in radio astronomy. The NenuFAR experiment at the Nanay station extends a station to the lowest frequencies at the ionospheric limit, with much greater collecting area. The ARTEMIS (Advanced Radio Transient Event Monitor and Identification System) experiment at Chilbolton does real-time beam forming and searches for very fast pulses such as Fast Radio Bursts, pulsar giant pulses, and possibly unknown phenomena. The AARTFAAC (Amsterdam-ASTRON Radio Transients Facility And Analysis Centre) and DRAGNET (Dynamic Radio Astronomy of Galactic Neutron Stars and Extragalactic Transients) experiments on the LOFAR core correlate the innermost sets of dipoles one-by-one, and search for fast transients in all-sky imaging and in beam formed signals, respectively.

Spain is not a member of the LOFAR collaboration, yet a number of Spanish astronomers participate at a personal level in some LOFAR projects.

### 3.3 eMERLIN

e-MERLIN is operated by the University of Manchester, as a national facility, through a contract with the Science and Technology Facilities Council (STFC). e-MERLIN consists of 7 telescopes distributed across the UK, with a maximum baseline of 217 km. eMERLIN provides sub-arcsecond imaging with very high (microJy) sensitivity. The telescopes are all connected to the central correlator via optical fibre. e-MERLIN is a uniquely sensitive instrument for imaging on milliarcsecond to arcsecond scales at centimetre wavelength. A number of approved legacy projects have been approved[3]. e-MERLIN achieves 50 to 150 milliarcsecond resolution, with sub-microJy sensitivity in deep observations, across relatively large fields-of-view. With this angular resolution, many astrophysical scenarios can be probed in a unique way such as the pebble-sized material in planet formation, the disc-jet zone (linear sizes of tens of AUs) in newly forming stars, the powerful jets of AGNe and X-ray transients and the star-forming regions in distant galaxies.

e-MERLIN has been considered as an SKA pathfinder in terms of high resolution science observations, long-distance data transport and phase transfer over optical fibre links.

Spanish scientists are involved in the commissioning and observations of some legacy projects (LIRGI, "Luminous Infrared Galaxy Inventory", with Miguel Ángel Pérez-Torres as one of the PIs; LEMMINGS, "Legacy e-MERLIN Multi-band Imaging of Nearby Galaxies Survey").

### 3.4 APERTIF

APERTIF (Aperture Tile in Focus) constitutes an operational demonstrator for the focal plane array technology, with the associated developments of algorithms and software to calibrate and process APERTIF data both in real time and off-line. With APERTIF, WSRT science will be driven full-time by large all-sky survey.

Several spanish astronomers are involved in some of the Apertif survey key-projects.

### 3.5 Other Pathfinders

We should mention other pathfinders outside Europe, in which the spanish astronomers regularly submit proposals and obtain observig time. This is the case of the:

- The Jansky Very Large Array (JVLA), located in New Mexico (USA), is composed of 27 antennas of 25-m diameter along a reconfigurable array providing a wide range of angular resolutions. It operates at frequencies from 1 to 50 GHz. The JVLA has just completed a very major upgrade with a large, 8-GHz broadband backend, making it the most sensitive radio telescope in the world for sources accessible from the Northern Hemisphere.

- The Giant Metre wave Radio Telescope (GMRT) consists of thirty antennas with a diameter of 45 m spread over distances of up to 25 km. It is located near Pune (India). It is a versatile telescope, with unique capabilities (high sensitivity and high resolution) in the metre wavelength range.

[3]http://www.e-merlin.ac.uk/legacy/projects/

## 4 Synergies with other Future Facilities

SKA surveys will be synergetic with current complementary multi-spectral-range facilities (e.g. ALMA, SDSS, Chandra) and other future projects like the LSST, Euclid, or eROSITA. Using synoptic observations from SKA, LSST, and Euclid, it will be possible to perform large scale structure and very high redshift cosmology (up to z 10 and beyond) studies. In the field of Galaxy Evolution, SKA will accurately determine the Star Formation history over cosmic time while LSST will provide redshifts and stellar masses. Moreover, it will be possible to trace the pathway from neutral gas (HI with SKA), to molecular gas (with ALMA) to star formation (with SKA radio continuum, probing galaxies with Star Formation Rates of up to 10 solar masses per year up to redshifts of around 3-4). Such synergies are discussed in the different chapters of this White Paper.

# Synergies between SKA and J-PAS.

**A. Fernández-Soto[1,2], M.Á. Pérez-Torres[3,4], on behalf of the J-PAS collaboration**

[1] Instituto de Física de Cantabria (CSIC-UC), Santander
[2] Unidad Asociada Observatori Astronòmic (IFCA-UV), Valencia
[3] Instituto de Astrofísica de Andalucía (CSIC), Granada
[4] Centro de la Física del Cosmos de Aragón (CEFCA), Teruel

## Abstract

In this contribution, we present the basic aspects of the Javalambre-Physics of the Accelerating Universe Survey, and the possible synergies that will be generated with the arrival of the Square Kilometer Array.

## 1 Introduction

The Javalambre-Physics of the Accelerating Universe Survey (J-PAS ) is a Spanish-Brazilian 8500□ cosmological survey that will be carried out from the Javalambre Astrophysical Observatory (Teruel, Spain; see Figure 1) using a purpose-built, dedicated 2.5m telescope and a 4.7□, 1.2Gpix camera. Starting in 2015, J-PAS will use 59 filters to measure high precision [$\sigma_z \sim 0.003(1+z)$] photometric redshifts for $\gtrsim$ 90 million galaxies plus several million QSOs, i.e., about 50 times more than the largest current spectroscopic survey, and sampling an effective volume of $\sim$ 14 Gpc$^3$ up to z = 1.3. J-PAS will not only be the first radial baryonic acoustic oscillation (BAO) experiment to reach Stage IV; it will also detect and measure the mass of $\sim 7 \times 10^5$ galaxy clusters and groups, setting constraints on Dark Energy that rival those obtained with state-of-the-art BAO measurements.

The combination of a set of 145Å medium-band filters, placed 100Å apart from each other, and a multi-degree field of view makes the J-PAS camera an extremely powerful "redshift machine", equivalent to a 4000 multiplexing spectrograph, yet many times cheaper to build. The J-PAS camera is also equivalent to a very large, 4.7□ "Integral Field Unit", which will produce a time-resolved, 3D image of the Northern Sky with a very wide range of scientific applications in Galaxy Evolution, Stellar Physics and the Solar System.

J-PAS will be, by construction, a Northern Hemisphere survey. This renders possible synergies with SKA apparently less important or interesting. However, both practical and scientific considerations have guided the selection of the J-PAS footprint to include a significant area in the Southern Galactic Hemisphere, close to the celestial equator. This area coincides, for the very same practical reasons, with the region observed by the Sloan Digital Sky Survey (SDSS) that includes, for example, the well-known SDSS Stripe82 and DEEP2 fields. This area will be part of the main J-PAS survey, and has also been selected for a possible extension to deeper magnitudes. We note that JPAS will be among the deepest and widest optical surveys (see Fig. 2). Moreover, SKA will certainly be able–from both its proposed sites–to routinely observe targets with $\delta > 0$.

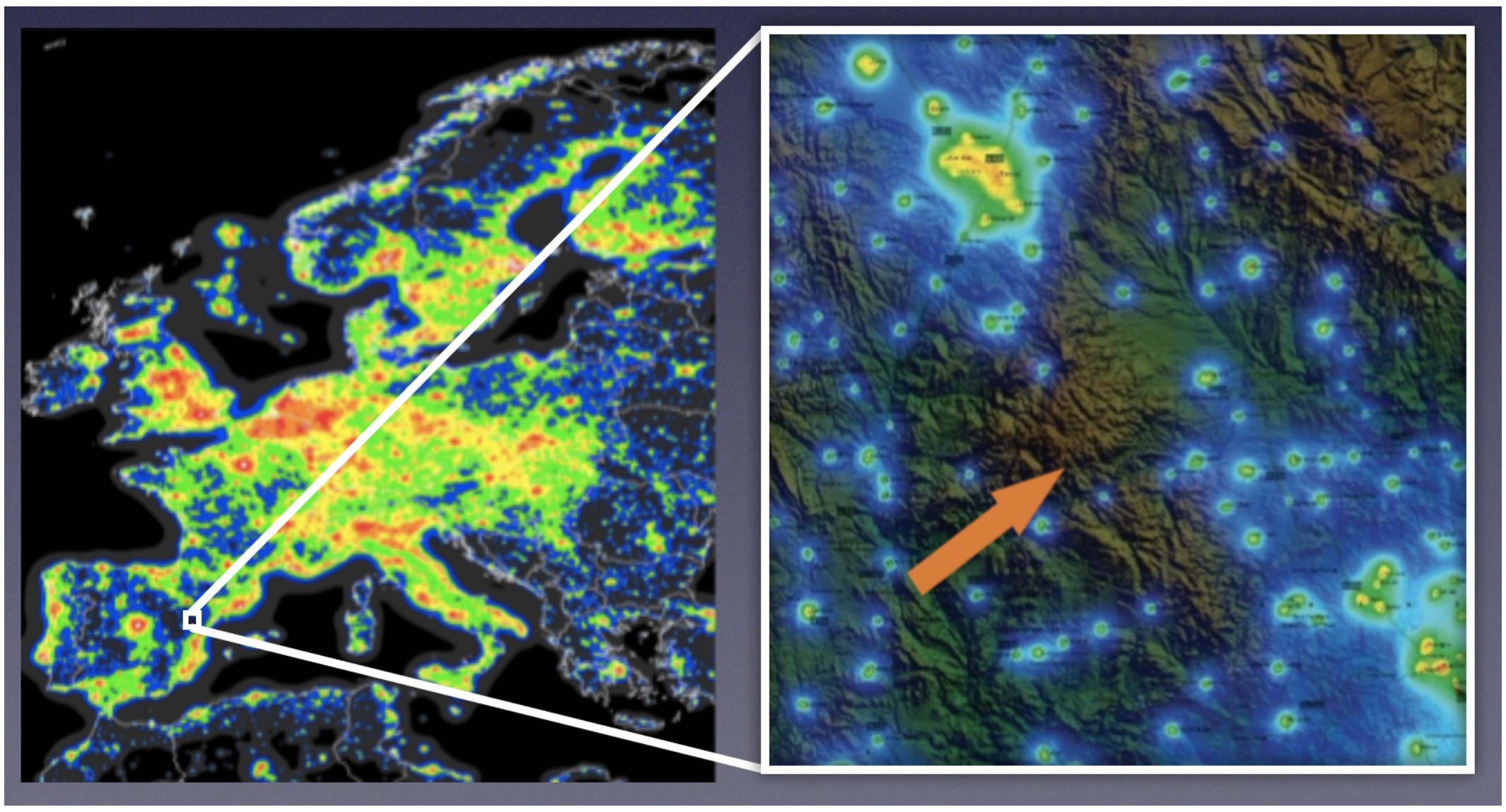

Figure 1: Light pollution map of Europe showing the position of Javalambre Observatory in Teruel (Spain), one of the very few remaining dark areas in continental western Europe.

## 2 Survey design

The scientific objective driving the design of J-PAS, as presented in [1], is to measure the properties of Dark Energy through the observation of BAOs in the space distribution of Luminous Red Galaxies and other cosmological objects. The determination of the redshifts of these sources will be based on photometric methods [1], and the necessary accuracy $[\sigma_z \sim 0.003(1 + z)]$ demands the production of a purpose-built set of 56+3 filters.

J-PAS will survey two separate areas, a larger one in the Northern Galactic Hemisphere (NGH) and a smaller one in the Southern Galactic Hemisphere (SGH, see Figure 3). These areas have been defined by combining three different motivations. First, we only selected areas whose observability from Javalambre at cenit distances $< 50°$ is larger than a minimum, significant number of hours ($> 200$) per year. Second, areas at low galactic latitudes are avoided based on the dust maps by [3]. Finally, we artificially add an extra weight to areas in the SGH to avoid possible conflicts in the transition epochs around December and June,

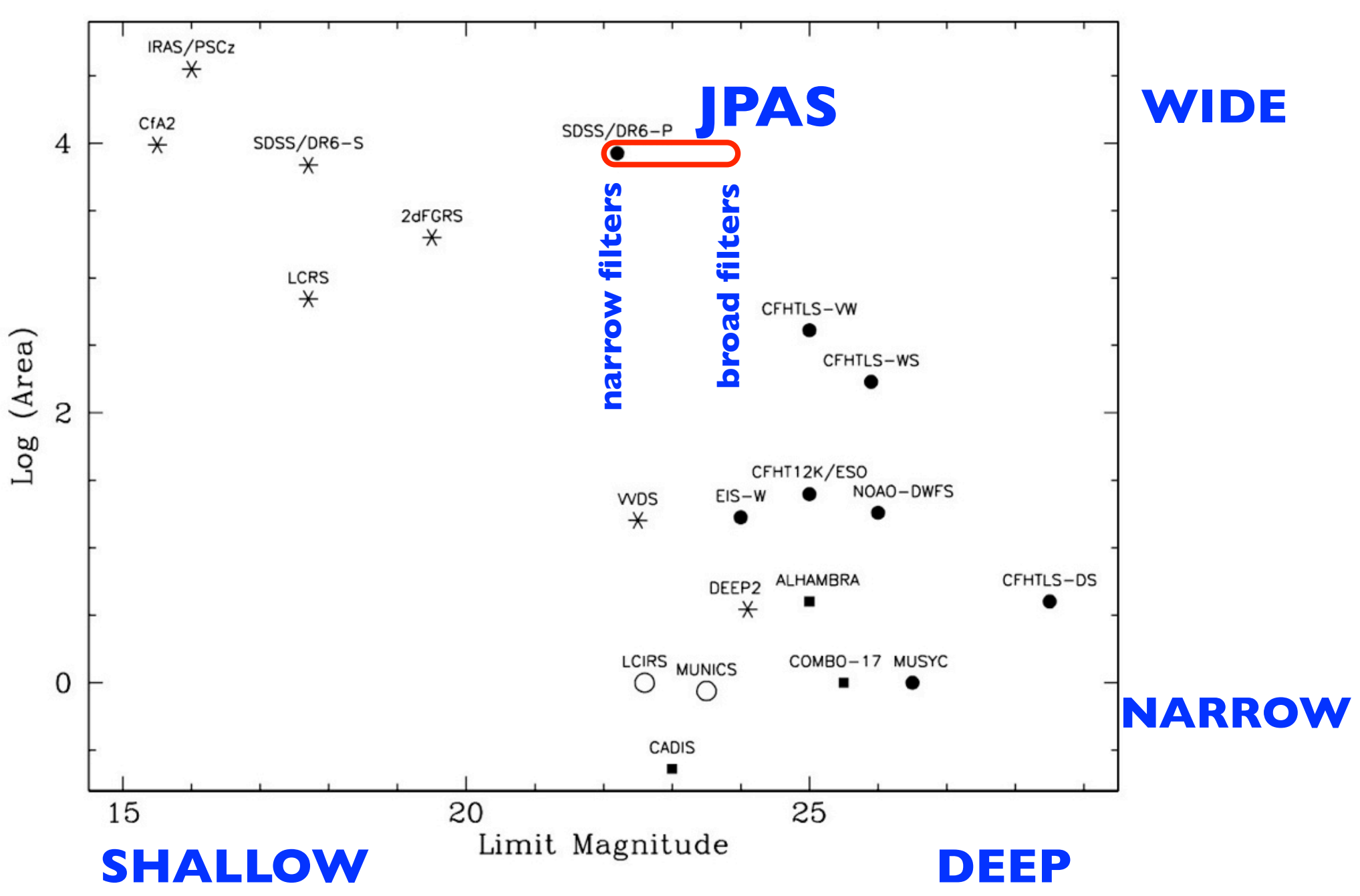

Figure 2: Surveyed angular area (in squared degrees) vs. apparent limiting magnitude for a number of well-known optical (both spectroscopic and photometric) surveys. JPAS will be among the deepest and widest optical surveys, reaching a limiting magnitude significantly deeper than the SLOAN DSS with their broadband filters.

when the (excluded) Milky Way is high in the sky for large fractions of the night time.

With these premises we have defined the two separate areas, which cover approximately 6500 and 2300 square degrees in the sky. These regions will be imaged through a set of 56+3 filters that encompass the whole visible range, from the UV to the NIR atmospheric cuts. The bulk of the filters (53 of them) are top-hat filters of width $\simeq$ 145 Å, with central wavelengths that cover from 3900 to 9100 Å in steps of 100 Å. Two other filters cover the blue and red ends of the visible spectrum, and the final three are the regular *ugr* from the Sloan set. Figure 4 shows the J-PAS filters together with a $z = 1$ elliptical galaxy template.

There will be at least four exposures for every point of the surveyed area with every filter, taken following a 2+1+1 strategy, with a cadence time of one month between each subset. This strategy has been chosen to maximize the detection rate of supernovae and other variable/transient effects. The total real exposure time for every pixel will be ~5.0 hours.

The focal plane of the Javalambre 2.5m telescope covers $4.7^{\square}$, which means that the effective survey speed is $1^{\square}$/h, and that $\simeq$ 9000 hours of real observing time will be needed to complete the survey. Taking into account technical and instrumental overheads, visibility issues, the Moon cycle, and the local meteorology, we estimate that seven years will be necessary to complete the project, with incremental releases becoming public after year 2.

# 3 Data products

The main deliverable of J-PAS will be a photometric catalogue of all galactic and extragalactic sources in its $8500^{\square}$ footprint, down to AB$\simeq$ 22.3 in the narrow filters from 3500 to 7000 Å, and AB$\simeq$ 21.5 in those from 7000 to 9000 Å, reaching AB$\simeq$ 24 in the broad-band *gr* filters. This shall amount to over 15 million red galaxies and 70 million emission line galaxies having redshift quality $dz/(1+z) < 0.003$, the *rms* accuracy target of our survey. The catalog will also include photometric information for approximately 300 million galaxies with worse photometric redshifts. It will also include and identify approximately two million quasars, reaching $g \simeq 24$ and $z \simeq 5$.

The data releases will be incremental both in terms of area and spectral coverage. The first two years of the survey (currently expected to begin in 2016) will mostly be devoted to observations with the reddest filters. This is driven by practical considerations—these filters will be the first to arrive at the telescope—as well as by scientific reasons: luminous red galaxies at moderate redshift will be targeted in a particularly efficient way with these filters, yielding some of the most exact results on dark energy parameters from the beginning of the survey.

# 4 Synergies with SKA

Both sites that have been accepted for the SKA telescope lie at a latitude of approximately 25° South. Depending on how relaxed the observing conditions can be acceptable, as well as on the particular instruments under consideration, this translates to a maximum reachable declination of $\approx 35°$ North. This means that most of the SGH portion of the J-PAS footprint, and a fraction of the NGH portion, will be available to SKA.

In order to feed the SKA community and the large surveys that are planned with information on J-PAS and the catalogs it will produce, these are some of the most important pieces of information:

- For the community at large, perhaps the best way to understand J-PAS will be to think of a *deeper*, *finer-grained* and *close to spectroscopic* version of the photometric SDSS catalogue.

- *All* galaxies with $L > L^*$ and $z < 1$ in the $\simeq 8500^{\square}$ covered by the survey will be fully characterized, with their redshift and spectral energy distributions accurately measured. SKA will be able to measure the neutral gas content and dynamics for large samples of this kind of objects.

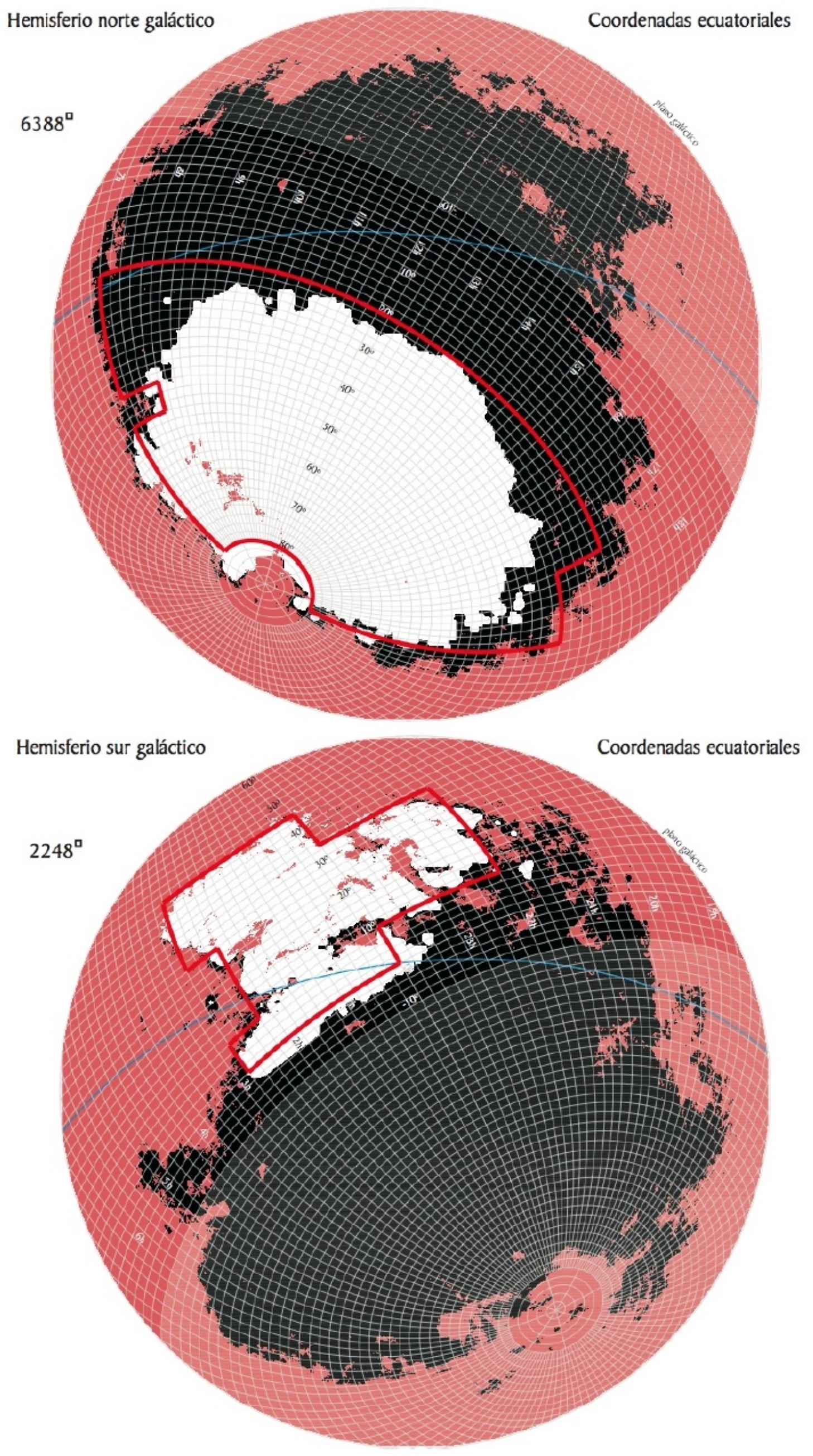

Figure 3: Representation in Lambert Projection of the Northern and Southern Galactic Hemispheres. Each plot shows in pink the area with relatively high galactic extinction (as given by E(B-V)>0.1 in the Schlegel et al. 1998 maps), and in white the area that is chosen when taking the best $(6000 + 2000)^{\square}$ selected separately in both hemispheres. The blue line is the ecliptic. We take the areas marked in red as compactified versions of the white ones, that define the J-PAS North and South areas. They cover approximately $(6400 + 2250)^{\square}$.

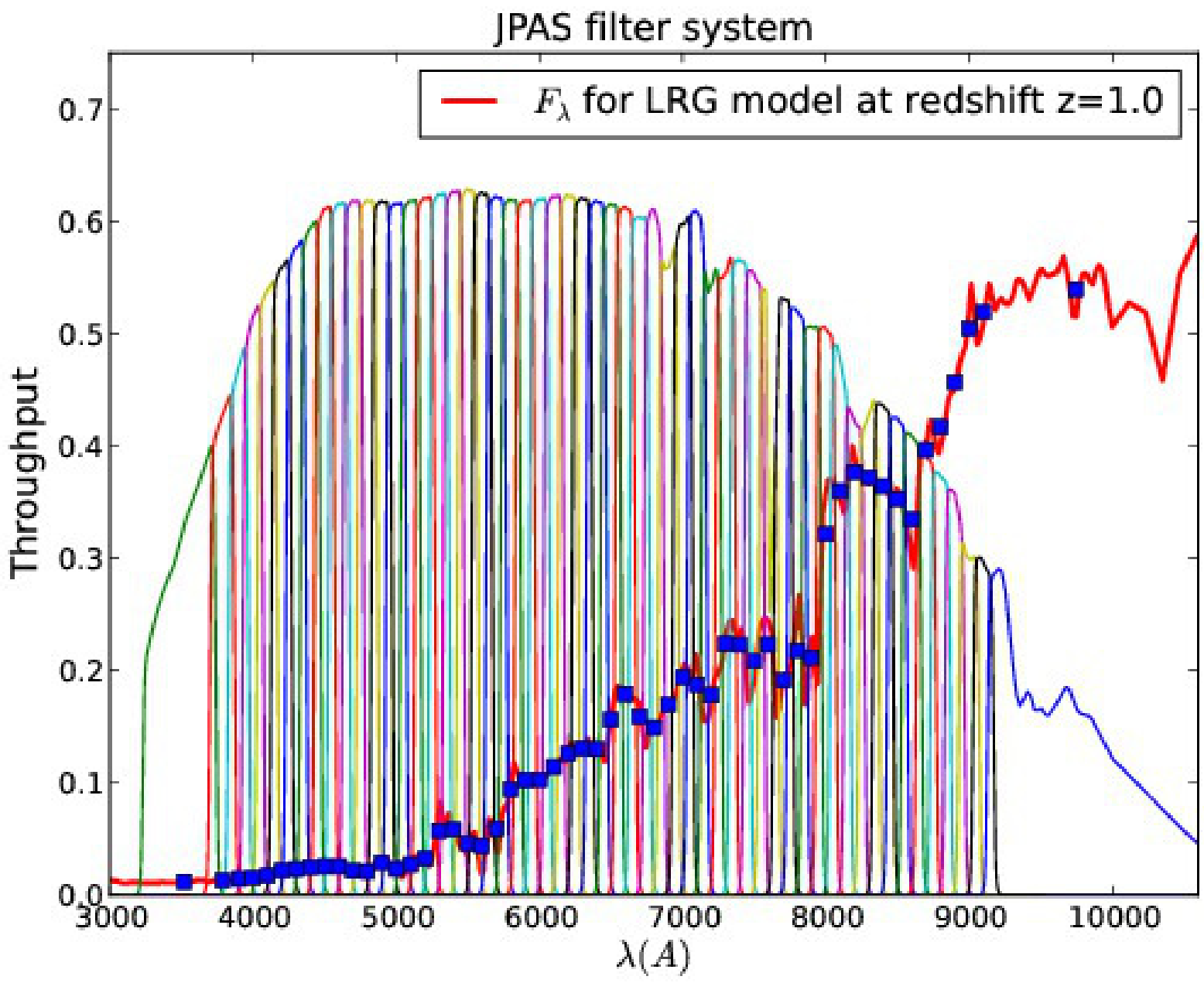

Figure 4: The J-PAS filter system. We have included the redshifted spectrum of an early type galaxy at z=1.0 from [2]. The filters are spaced by about 100 Å but have FWHM of 145 Å, what produces a significant overlap among them. The blue squares represent the flux which would be observed through the filters. Note that many spectral features apart from the 4000 Å break are resolved, which is why the precision in redshift is much larger than that which would be produced by a single break, $\Delta z/(1+z) \simeq \Delta\lambda/\lambda \simeq 0.02$.

- For galaxies in the local Universe J-PAS will offer detailed resolved colour/morphology data with resolution of $\simeq 1$ arcsecond, that can be complemented with detailed measurements of the neutral gas and molecules as measured by SKA and other large facilities in the future.
- As was mentioned before, J-PAS will identify approximately 2 million quasars up to high redshifts $z \approx 5$. It will be possible to use these objects as markers for large mass overdensities at high redshift, that will be further explored with SKA. The properties of the neutral/ionised gas surrounding these targets will be of particular interest.
- The ability to cross-identify sources combining data from J-PAS and the SKA surveys will be a basic ingredient in both cases, as they will cover widely different ranges in wavelength/energy.

## 5 More Information

We advice those interested in the latest status of the survey to check the following sites:

- `http://arxiv.org/abs/1403.5237`, the permanently updated J-PAS "Red Book",
- `http://www.cefca.es/es/proyecto-oaj`, status of the Javalambre Observatory,
- `http://j-pas.org` with the most up-to-date project information.

## Acknowledgments

AFS acknowledges support from the Spanish Ministry for Economy and Competitiveness (MINECO) and FEDER funds through grants AYA2010-22111-C03-02 and AYA2013-48623-C2-2, and Generalitat Valenciana project PROMETEOII/2014/060. MAPT acknowledges support from the Spanish MINECO through grant AYA2012-38491-C02-02.

# Synergies of high-energy facilities with the SKA

**Javier Moldón[1,2], and Marc Ribó[1,3]**

[1] Departament d'Astronomia i Meteorologia, Institut de Ciències del Cosmos, Universitat de Barcelona, IEEC-UB
[2] ASTRON, the Netherlands Institute for Radio Astronomy
[3] Serra Húnter Fellow

## Abstract

The Square Kilometer Array (SKA) promises to revolutionise radio astronomy thanks to a significantly better sensitivity over the currently existing radio facilities. In parallel, the high-energy facilities that will be available during the early science and operations of SKA will also bring significant insight into the physics of non-thermal sources in the Universe. The combination of SKA with such high-energy facilities will allow a better understanding of the ongoing physical processes in the known sources and, hopefully, allow the discovery of new types of sources among a large fraction of still unidentified sources at high and very-high energy gamma-rays. Here we mention some of the high-energy facilities that will be available in the SKA era and comment on some physical cases where synergy between SKA and high-energy facilities could provide a significant impact.

## 1 Introduction

During the last two decades we have seen a strong development of X-ray astronomy thanks to X-ray satellites such as *RXTE*, *XMM-Newton* or *Chandra*, which provided timing, spectral and angular resolution capabilities, respectively. In the gamma-ray domain, old satellites like *CGRO* provided around two hundred of unidentified EGRET sources at High Energy (HE, $E > 100$ MeV) gamma-ray energies [20]. The orbiting *Fermi* satellite has already provided a few thousands of new sources and yielded the discovery of many new types of GeV sources [29]. At Very High Energy (VHE, $E > 100$ GeV) gamma-ray energies, Imaging Atmospheric Cherenkov Telescopes (IACTs) like H.E.S.S, MAGIC or VERITAS have already discovered more than one hundred TeV sources of different types [21][1].

Non-thermal sources in the Universe emit across the whole electromagnetic spectrum. Therefore, observations of these sources combining radio and high-energy facilities can help to better understand the ongoing physical processes. The superb sensitivity of the Square Kilometer Array (SKA), combined with existing and planned high-energy facilities, promises to revolutionise the study of non-thermal sources both in the Galaxy and beyond. This includes pulsars, binary systems of different types, supernova remnants (SNRs), active galactic nuclei (AGN) of different types, unidentified sources, etc.

[1]See `http://tevcat.uchicago.edu/` for updates.

In this chapter we mention high-energy facilities that could work together with SKA in Sect. 2, we comment on a few physical cases where the synergy between SKA and HE facilities could provide a significant impact in Sect. 3, and we make some concluding remarks in Sect. 4.

## 2 High-energy facilities in the SKA era

The early science phase of SKA should start in 2020. By that time, several X-ray and gamma-ray facilities covering the 0.1 keV to 100 TeV energy range should be available. Here we provide a non-exhaustive list of these high-energy facilities and their characteristics:

- X-ray facilities (see for example [27] and references therein). While some of the existing missions might be decommissioned before 2020, it is likely that *XMM-Newton* and *Chandra* will overlap the early science phase of SKA. These missions cover the ∼0.2–12 keV energy range and operate in pointing mode with a relatively small field of view (FoV) of $< 1^{\circ}$. The MAXI instrument onboard the *International Space Station* (*ISS*) provides an all sky monitor in the 1–20 keV band, and thus it is ideal for detecting new transient sources. The *INTEGRAL* and *NuSTAR* satellites operating above 10 keV and in the 10–80 keV energy range, respectively, are also likely to extend until 2020 and beyond, providing hard X-ray energy coverage for pointed observations. In 2015 the *Astro-H* mission, operating in the 0.3–600 keV range, should be launched, while *eROSITA*, to survey the whole sky in the 0.1–30 keV range, should be launched in 2016. Other future missions are *ASTROSAT*, with an all sky monitor similar to the one that was provided by *RXTE* (in 2015) or the *ISS* instruments NICER (in 2016), and the proposed LOBSTER with a very good all sky monitor. The chinese *HXMT* should operate between 2016 and 2020 with a huge FoV and sensitive between 20 and 200 keV. The *SVOM* mission, to follow GRBs but much more sensitive than *SWIFT*, is scheduled for launch in 2021. Planned missions include the X-ray timing *LOFT* mission in the 2–30 keV range (with a hard X-ray monitoring) recently proposed as an ESA mission for the M4 call and to be launched in 2025 if selected, and finally *Athena+*, due for launch in 2028 and with a superb sensitivity and spectral resolution in the 0.1–12 keV energy range. As can be seen, several X-ray satellites with different capabilities will allow detailed observations in the whole X-ray energy range during the early science of SKA and beyond.

- MeV facilities. The energy range between 1 and 100 MeV has been relatively poorly explored up to now, with old contributions by *CGRO*/COMPTEL and more recent ones by *INTEGRAL*. A proposal for an M4 mission to operate in the 0.3–100 MeV energy range, namely *ASTROGAM*, has recently been submitted to ESA. The goal is to improve by at least one order of magnitude the sensitivity and by a factor of several the angular resolution of previous facilities. With a large 2.5 sr FoV, *ASTROGAM* promises a revolution at these low gamma-ray energies. If selected, it should be launched in 2025 and thus would provide a plethora of new sources to be followed at low energies with SKA during its operation.

- GeV facilities (see [13] and references therein). The *AGILE* and *Fermi* satellites are currently continuously scanning the sky in the $\sim$100 MeV to 300 GeV energy range. While *AGILE* will probably stop operations before the early science phase of SKA, *Fermi* will likely continue oparting through 2020 and potentially beyond, with both survey and pointed observations. In the future the *GAMMA-400* mission, planned for launch in 2021, will explore the sky in the 100 MeV up to 3000 GeV energy range in pointing mode, but reaching about an order of magnitude improvement in angular and energy resolution over *Fermi* at energies of 100 GeV. Therefore, detailed HE gamma-ray observations of known and newly identified sources should lead to a significant progress in our understanding of particle acceleration processes, which could be complemented with SKA ones already in the early science phase.

- TeV facilities. The IACTs H.E.S.S., MAGIC and VERITAS are currently operating above 50 GeV and up to several tens of TeV. With a limited energy resolution of $\sim$10% and angular resolution of $\sim 0.1°$, they provide a FoV between 3 and 5°. More than a hundred sources of different classes have been discovered by these facilities, both in pointed and survey mode. The future in this energy range will be provided by the Cherenkov Telescope Array (CTA, [3]). CTA will have two arrays of Cherenkov telescopes, one in the north (either in Spain, Mexico or the USA) and a larger one in the south (either in Chile or Namibia) with a better access to the Galactic plane. CTA should operate from a few tens of GeV to above 100 TeV, improving the sensitivity by about an order of magnitude, and with enhanced angular and energy resolutions and larger FoV, over existing VHE gamma-ray observatories. The High Altitude Water Cherenkov (HAWC) will start full operations in 2015-2016, and will be used to perform a high-sensitivity synoptic survey of the sky at energies between 100 GeV and 100 TeV. At similar energies the LHAASO proposal is currently under consideration in China. All these facilities should discover more than a thousand new VHE sources during the early science phase of SKA, and should allow unprecedented studies of cosmic accelerators. Detailed studies of such sources with SKA will greatly help in modelling their multi-wavelength emission, thus allowing us to improve our physical understanding of them.

## 3 Possible major contributions of SKA to HE astrophysics

### Pulsars

Pulsars have been traditionally discovered and studied at radio wavelengths. However, a number of pulsars have been discovered at X-rays and gamma-rays. In particular, the *Fermi* satellite has detected more than 117 pulsars at GeV energies [1]. Some of them do not have a radio counterpart. The Crab pulsar has also been detected with IACTs at sub-TeV energies [6, 30, 5], and it is expected that CTA will discover new gamma-ray pulsars [15]. Detecting the pulsed radio counterpart is important to understand the pulsar properties, but also because the dispersion measure of pulsars inferred with radio observations yields their (approximate) distance [12]. The distance to the pulsar is fundamental to scale its energy output at higher energies. Deep radio observations of pulsars discovered in X-ray or gamma-ray blind searches are also key for understanding the neutron star (NS) luminosity distribution.

Some pulsars display giant radio pulses (GRPs), with a sudden luminosity increase that can last microseconds, although they can show profile variability down to nanoseconds at radio wavelengths [19]. In the Crab pulsar, giant pulses have been also observed in the optical band [11], while searches have provided null results up to now at X-ray, GeV and TeV energies [9, 8, 7]. But do giant pulses occur at higher energies? If so, how does the energy output relate to the structure of the pulsar and its magnetic field? While SKA will be fundamental to understand the origin of GRPs, the possible high-energy counterpart will be searched and explored by fast and sensitive telescopes such as LOFT in X-rays and CTA at TeV energies.

**Galactic binary systems**

Accreting X-ray binaries have revealed a radio/X-ray correlation while in the so-called low-hard state, which is a fundamental tool to understand the behaviour and the different states of these sources. In the case of low-mass X-ray binary systems (LMXBs) the correlation has been densely explored for black hole (BH) systems even down to low luminosities [17], but this has not been the case for NS systems at low accretion rates ($L_{\rm X} \sim 10^{34} - 10^{36}$ erg s$^{-1}$). The slope of the NS-LMXBs radio/X-ray correlation seems to be different than the correlation for BH-LMXBs [23], but since the radio emission from NS systems is about two orders of magnitude fainter than for BH systems, that possibility has not yet been tested. Joint observations between SKA and the new generation of X-ray satellites are required to explore the correlation at very low accretion rates.

In the case of accreting high-mass X-ray binaries (HMXBs), Cygnus X-1 lies in the radio/X-ray correlation, but it always displays a very high X-ray luminosity. The recent discovery of the first HMXB containing a Be star and an accreting BH [10] has led to the X-ray detection of the first BH-HMXB in quiescence [25]. The source, namely MWC 656, remains undetected in radio [24], but its position in the radio/X-ray luminosity diagram may be consistent with the radio/X-ray correlation observed in BH-LMXBs. This suggests that this correlation might also be valid for BH-HMXBs with X-ray luminosities down to $10^{-8} L_{\rm Edd}$ [25]. Other quiescent BH-HMXB akin to MWC 656 could be discovered in the following years after performing extensive searches. Their detailed study to understand accretion at low luminosities will also benefit from combined observations between the SKA and sensitive X-ray satellites such as *Astro-H* or *Athena+*.

Interestingly, a different correlation has been recently proposed for the low-mass X-ray binary/MSP transition objects (tMSP) [14]. These systems, which have been observed to switch between radio millisecond pulsars and NS-LMXB, are a fundamental link between these two types of sources. They are also gamma-ray emitters that can show different states. Wide and deep surveys at radio wavelengths combined with high energy data will increase the number of such systems, which would help to understand the details of pulsar recycling models [28]. For instance, in X-rays the Wide Field Monitor (WFM) onboard LOFT, which will cover at least 50% of the sky simultaneously in the 2–50 keV energy band, will be a perfect discovery machine of X-ray transients.

On the other hand, gamma-ray binaries are high-mass systems that also display gamma-ray emission, in this case clearly detected up to TeV energies [16]. Their radio emission has been resolved at milliarc-sec scales [24], revealing cometary-tail like morpholgies that rotate with the orbital period. These systems are excellent but complex physical laboratories that need to be studied through the whole electromagnetic spectrum from radio to TeV energies. Since they are highly variable, joint simultaneous multiwavelength campaigns are required to understand them. Therefore, this type of systems will exploit the high-sensitivity, the fast response, and the flexibility of the next generation of instruments, from radio with SKA to TeV energies with CTA [26].

### SNR and the origin of high energy cosmic rays

The work of identifying two SNRs as Cosmic Rays (CRs)' accelerators was selected as the Breakthough of the Year 2013 by the magazine *Science*. SNR constitute an important population of Galactic sources that generate VHE gamma-rays, and have been confirmed to accelerate protons, the main component of cosmic rays, to very high energy by their shocks [2]. Multi-wavelength observations at radio wavelengths and HE/VHE gamma-rays are a powerful tool to probe the particle acceleration mechanism in SNRs.

Magnetic field amplification is needed to accelerate cosmic rays to VHEs in SNRs. This could happen in small scale density perturbations and filaments in SNRs. The study of this filamentary emission requires radio observations at high resolution to reduce the depolarization caused by small scale distortion of the magnetic field. The low-frequency synchrotron emission ($< 2$ GHz) tracks the MeV-GeV part of electrons spectrum. The SKA1-LOW, with a frequency range between 50 MHz to 350 MHz, provides an excellent opportunity to distinguish the spectrum features produced from protons or secondary electrons [31]. This low frequency radio emission is linked to the GeV energy particles, whereas the high energy TeV particles can be studied through the X-ray synchrotron spectrum features. In the coming golden era of multi-wavelength astronomy, sensitive gamma-ray, X-rays and low frequency radio observations will simultaneously trace the emission produced by GeV and TeV particles from SNRs, and thus contribute significantly to solve the CR's origin issue.

### Particle acceleration in extragalactic jets (AGN)

Relativistic jets in AGNs are among the most powerful astrophysical objects. However, despite their ubiquity, we still lack a comprehensive understanding of their internal physics and energy budget, their composition, and how they are formed and collimated. Some of the main open questions will be addressed with the deep, all-sky surveys (both total and polarimetric flux) from SKA1. These questions (from [4]) include: a) why are jets produced efficiently only in some systems, and what is the relation between jet power and the properties of the black hole and accretion system? b) what influence do magnetic fields have on jet formation, collimation, and maintenance up to distances above 100 kpc? c) what is the actual plasma composition at different jet scales and how does it evolve down the jet? d) how does the particle acceleration mechanism make the jet an efficient emitter on scales exceeding the size of the host galaxy? and e) how does the feedback between the jet and the (inter-)galactic medium influence the evolution of galaxies and clusters, and how do AGN jets and their central black holes evolve on cosmological time scales up to $z \sim 10$?

In particular, SKA will provide insightful information related to the high energy processes in AGN jets that are also producing the high-energy emission, and therefore coordinated multiwavelength studies will be mandatory. For instance, observations with SKA1-MID in South Africa should be coordinated with VHE gamma-ray observations conducted with IACTs like H.E.S.S. or CTA-South, optical telescopes like SALT, and supported by X-ray and HE gamma-ray observations from satellites.

To understand how jet propagation depends on the jet power and the environment, we need to observe large samples of weak jets at subarcsecond resolution down to noise levels of 1$\mu$Jy/beam with SKA1-MID [22]. In more powerful jets, we want to test if the X-ray emission along the jet can be explained by inverse Compton with the CMB, or by a separate synchrotron component. These questions will be definitely solved by deep observations with SKA2, because we need a resolution better than 0.05 arcsec, ideally in the 1–10 GHz frequency range, with rms noise levels of roughly 10 nJy/beam and extremely high dynamic range, imaging fidelity and polarization purity. In summary, high-quality radio observations combined with X-ray observations will reveal the underlying non-thermal particle distributions, which can be used to constrain the acceleration processes at work.

**Unidentified gamma-ray sources**

Overall, about 30% of *Fermi* sources lack a high-confidence low-frequency counterpart [29]. Also, about 20% of the sources detected by the new generation of IACTs remain unidentified or with no firm low-energy counterpart[2]. The main reason is the large uncertainty of the gamma-ray positions, and also the fact that faint gamma-ray sources are usually associated with low flux density radio sources, whose space density is larger and which often lack a radio or optical spectrum. Deep and multifrequency radio surveys are fundamental to identify gamma-ray sources. By the time SKA will be operational, *Fermi* shall have completed at least a 10-year survey. At the same time, CTA will be starting its operation, and a moderately shallow wide area survey should reveal a large number ($\sim$ 100) of week and/or transient sources to be identified through carefull studies conducted at other wavelengths such as radio. A similar situation should probably happen with the unidentified multi-TeV sources to be discovered by HAWC during the $\sim$ 5 years prior to SKA early science.

Observations of these gamma-ray sources in radio with SKA1-MID will provide detailed properties of possible radio counterparts thanks to its sensitivity, flexibility, and sub-arraying capability [18]. When faint gamma-ray sources are considered, pure positional coincidences are not significant enough for selecting counterparts and we need additional physical criteria to pinpoint the right object. The criteria can be the radio spectral index, variability, polarization, or compactness, requiring the high angular resolution of SKA1-MID. Thus, detailed studies with SKA can help to identify the counterparts of the large population of unidentified gamma-ray sources.

[2] See `http://tevcat.uchicago.edu/`

## 4 Concluding remarks

The superb sensitiviy of SKA, in combination with the use of high-energy facilities, should allow us to better study and constrain different physical processes in a veriety of interesting objects, from pulsars to SNR and black holes of all masses. In addition, the follow-up by SKA of unidentified HE and VHE gamma-ray sources discovered by *Fermi*, CTA or HAWC should allow us to pinpoint the low-energy counterparts and potentially unveil new types of high-energy non-thermal sources in the Universe.

## Acknowledgments

J. M. acknowledges support from the Spanish Ministry of Economy and Competitiveness (MINECO) through grant AYA2013-47447-C3-1-P. M. R. acknowledges support from the Spanish MINECO through grant FPA2013-48381-C6-6-P.

# SKA and VLBI synergies

**Eduardo Ros[1,2,3], Antxon Alberdi[4], Iván Agudo[4], Francisco Colomer[5], José L. Gómez[4], José C. Guirado[2,3], Iván Martí-Vidal[6], Mar Mezcua[7], Javier Moldón[8], Miguel Á. Pérez-Torres[4,9,10], Manel Perucho[3,2], Marc Ribó[11], and María Rioja[12,13,6]**

[1] Max-Planck-Institut für Radioastronomie, Bonn, Germany
[2] Observatori Astronòmic, Universitat de València, Spain
[3] Departament d'Astronomia i Astrofísica, Universitat de València, Spain
[4] Instituto de Astrofísica de Andalucía-CSIC, Granada, Spain
[5] Observatorio Astronómico Nacional, Alcalá de Henares, Spain
[6] Onsala Space Observatory, Chalmers Univ. of Technology, Onsala, Sweden
[7] Harvard-Smithsonian Center for Astrophysics, Cambridge, MA, USA
[8] Netherlands Foundation for Radio Astronomy, Dwingeloo, The Netherlands
[9] Centro de Estudios de la Física del Cosmos de Aragón, Teruel, Spain
[10] Dep. Física Teórica, Universidad de Zaragoza, Spain
[11] Dep. d'Astronomia i Meteorologia, Inst. Ciències del Cosmos, Univ. de Barcelona, Spain
[12] Korea Astronomy and Space Science Institute, Daejeon, Korea
[13] International Centre for Radio Astronomy Research, Univ. Western Australia, Australia

## Abstract

VLBI is a unique tool in astronomy to study the compact radio emission of celestial bodies in extreme detail. This technique has been applied in fundamental astronomy for establishing celestial reference frames and in astrophysics to study the non-thermal continuum and maser emission of galactic and extragalactic objects. The main targets of VLBI studies are the black-hole powered active galactic nuclei and their relativistic outflows, the surfaces of stars, the atomic and molecular gas outflows, disks, and shells in star forming regions, young stars, supernovae and their products (neutron stars and stellar black holes with a companion star). Combining the SKA with VLBI arrays will provide improvement of structural sensitivity on larger scales (covered by SKA baselines alone) and help reach unparalleled flux density sensitivity on compact angular scales probed by VLBI. Here we describe some of the scientific targets of the combined SKA-VLBI and we discuss some technical issues such as the enhancement in sensitivity and the joint observations with existing VLBI elements at present.

# 1 Introduction

The technique of very-long-baseline interferometry (VLBI) probes compact objects in the microwave range, at angular scales from 100 millarcseconds (mas) down to fractions of mas. This corresponds to scales of several astronomical units at the distance of few kpc (e.g., microquasars in our Galaxy or the Galactic Centre), and parsec scales at distances of Mpc to Gpc (active galaxies and quasars). VLBI arrays can be combined with connected interferometers such as the VLA or eMERLIN, with the latter being the only instrument which covers properly the intermediate (mas-to-arcsecond) scale resolution. The addition of the Square Kilometre Array will provide new capabilities for the full range of resolutions between the milliarcsecond and the arcsecond. It will, as well, enhance dramatically the sensitivity of VLBI arrays, if combined as a single element. In a first approach, SKA-VLBI will be possible by adding phased-array outputs for SKA1-mid to the VLBI arrays, using these sensitive stations with other radiotelescopes. In a later stage, SKA2 would add VLBI antennas to the SKA correlator, providing a VLBI instrument with baselines of thousands of km.

The high-resolution capability of the SKA has been discussed in several works (see [13] and references therein). Recent works have described in detail the options and the science of VLBI in the SKA era, such as [35, 46]. Here we will show some of the highlights, in extragalactic (Sect. 2), intermediate black holes (Sect. 2.2), and stellar astronomy (Sect. 3), respectively, and discuss some technical considerations in Sect. 4.

# 2 Extragalactic objects

## 2.1 AGN Physics & Galaxies

the study of relativistic jets in active galactic nuclei (AGN) [44] is one of the key targets for VLBI since its pioneering times. A fraction of AGN features powerful collimated outflows of plasma, which can reach distances of hundreds of kpc from the central engine. These objects have high brightness temperatures, reaching up to $10^{12}$ K and even higher in exceptional cases due to Doppler boosting and other effects [37]. These jets contain enhanced regions of emission which appear to move at speeds that imply high Lorentz factors, and the time evolution of these features has been studied in detail in many individual objects (e.g., [9, 10] for CTA 102, [38] for NGC 1052, [1] for OJ 287, [11] for 3C 120, or [18] for 3C 111), and extensively in imaging surveys (e.g., [45, 23]).

The inclusion of the SKA in new arrays, with its enhanced capabilities in image fidelity and detection of low brightness temperature features, will have an impact on the understanding of the jet physics, the mechanisms which govern the structure and jet kinematics. This is complementary to the present developments in the studies of the jet base at higher frequencies, especially with the inclusion of ALMA in VLBI networks in a near future [8, 39].

Magnetic fields play a major role in extracting energy from the central region and feeding the jet, which is confined by strong, collimating magnetic forces [43]. The study of the polarised emission of relativistic jets probes the nature and structure of the magnetic fields. It is often assumed that the magnetic field structure is helical from the launching in the central black hole (e.g., [4]), but it is unclear if this helical structure persists after iteration with shocks or the external medium. The field orientation can reveal poloidal or toroidal structure, and the compression of the field by shocks perpendicularly to the jet direction can create enhanced polarised emission [21, 16]. Exhaustive studies have been performed by combining large data sets in surveys (see [22]). The polarised emission is affected by Faraday rotation when propagating through plasma, and it is being studied in detail, to see which is its effect in depolarisation, if it is caused by an external screen or if it is intrinsic, etc. Recent studies of large samples of sources reveal few cases of internal rotation, but most of the measurements are compatible with external screens ([12, 15]). The high fidelity and the increase of sensitivity provided by the SKA will help much in solving the open questions about the strength and structure of the magnetic field and its influence in the jet formation and dynamics.

The energy released by an AGN affects strongly the surrounding interstellar medium [2, 3]. The feedback regulates the growth of the central massive black hole and the star formation rate in the galaxy [26, 6]. HI spectral-line imaging of the AGN 4C +12.50 showed how the jet contributes to AGN feedback [33]. SKA-VLBI will image ultraluminous infrared galaxies (for which the infrared excess is caused by re-processed UV light from hot young stars) at unprecedented sensitivity at 18-cm wavelength. Exquisite observations of feedback with the enhanced sensitivity of the SKA-mid and combined with VLBI at 18-cm will be possible. The complex interaction between the relativistic outflows and the interstellar regions will shed light on the question of why giant elliptical galaxies have switched off star formation. The synergy of SKA-VLBI observations with future X-ray missions should provide the final answer.

Stimulated spectral line (maser) emission has been observed in extragalactic sources [24], with large luminosities, over 8 orders of magnitude larger than in galactic masers. Successful detections were performed for hydroxyl (OH) at 1.6–1.7 GHz, water ($H_2O$) at 22.2 MHz, formaldeyhde ($H_2CO$) at 4.83 GHz. Recent ALMA detections of megamasers in NGC 1068 were reported for mm-transitions of methanol ($CH_3OH$) at 84.52 GHz and silicon monoxide (SiO) at 85.04 GHz [42]. More detections at millimetre-wavelengths are expected with ALMA, and the SKA with its sensitivity should access to further, weaker targets and potentially be able to detect other transitions such as the 23.87 GHz line in ammonia ($NH_3$), if this frequency is reached at some point. Redshifted megamasers will then reach the detection threshold to be observed and imaged. In this sense, the effort of the Megamaser Cosmology Project (see the white paper here and recent results at [20, 19]) with water masers could be extended to other species and with a far deeper scope.

Due to lack of sensitivity, the population of quasi-stellar objects with low radio emission has not been studied in detail. Different explanations have been proposed to address the radio-quiet nature of these objects, being a fraction of them associated to broad absorption lines (for which the wind from the accretion disk would switch off the jet), or being just thermal radiation [5]. The combination of the high sensitivity and high resolution of SKA-VLBI observations will address in a much better detail the morphology and the spectrum of the radio cores (if detected) of these objects.

### 2.2 Intermediate-mass black holes

Intermediate-mass black holes are the missing link between super-massive black holes and stellar-mass black holes. They should be present in the nuclei of low-mass galaxies and could be also observed in the halos of large galaxies and in the arms of spiral galaxies, where they manifest themselves as ultraluminous X-ray sources [7]. Stellar-mass black holes accreting at Eddington or super-Eddington rates could explain the X-ray emission of ultraluminous X-ray sources up to $L_{\rm X} = 5 \times 10^{40}\,{\rm erg\,s^{-1}}$, but X-ray luminosities above this value need larger black hole masses (see [41, 32]). Recent results based on combined X-ray and VLBI observations show that some of these ultraluminous X-ray objects are intermediate-mass black holes accreting at sub-Eddington rates [28, 29, 30] and powering strong radio jets [31]. The highest sensitivity provided by SKA-VLBI at cm-wavelengths will allow for further detections and deeper studies in the radio regime of these ultraluminous X-ray sources.

## 3 Stellar physics

VLBI observations have remarkably strong impact in the Stellar physics studies. We can mention the study of Gamma Ray Bursts (GRBs), Radio Supernovae (RSNe) and Supernova Remnants (SNRs), Microquasars or Pulsars, among others. Moreover, VLBI astrometry is a unique tool providing stellar parallax measurements and proper motion determinations as well as the search for planets via the reflex motion of the star.

The addition of SKA to the Global VLBI Arrays will not increase the angular resolution of the arrays. However, it will provide an excellent sensitivity on the long baselines assuring that the scientific targets will be imaged both in total intensity and polarisation with better detail than is possible today. Additionally, the accuracy of the source sizes obtained via model fitting will also benefit from the increased sensitivity [27]. Assuming observations at 5 GHz and observations with a SNR 100-1000, the minimum source size determination will be of the order of a few tenths of milliarcsecond.

One of the immediate applications of SKA-VLBI will be the studies of explosive events in stellar objects in the Milky Way and nearby galaxies. Thanks to the sensitivity of SKA, the number of detected and monitored core-collapse supernovae will increase significantly. It will be possible to resolve any of them, studying their angular expansion for much longer periods of time, yielding information about both the circumstellar and interstellar environments and the progenitor star. For nearby galaxies, a good estimate for their star-formation rate can be obtained [36] through the detection of individual radio supernovae and supernova remnants in the circumnuclear starbursts. The extra sensitivy provided by the SKA1 phased-arrays will permit the detection of Type I-a radio supernova, giving unique insights on the nature of the progenitor systems.

A particular case of study are the Gamma Ray Bursts (GRBs). According to the collapsar model, the long duration GRBs are related to the collapse of a fast rotating star in a low metallicity environment. In the case when the star is rotating quickly enough, then the fallback to the black hole will produce relativistic jets (the GRB); on the other hand, if such a star is rotating slowly, then it will produce a faint supernova. The type Ic supernovae are associated with GRB explosions; however, this connection has been shown only for a few cases [34]. SKA-VLBI will have the potential to study the launch of the relativistic jets and the expansion of the GRB relativistic outflows during a period of 1-2 weeks after the explosion.

Due to the superb-sensitivity provided by SKA-VLBI, the number of detected stars will be very large (hundreds to thousands). For many of them, it will be possible to measure parallaxes with precisions better than 10 microarcsec, permitting a direct comparison with *GAIA* and favouring a cross-calibration among both reference frames (the International Cellestial Reference Frame (ICRF) and *GAIA*). SKA-VLBI will deliver a remarkably accurate astrometry, tracing very precisely the trajectories of stars on the plane of the sky. They will be modeled as the superposition of their trigonometric parallax and a uniform proper motion, permitting the accurate determination of distances to the parental molecular clouds [25].

For the case of stellar mass black holes, it will be possible to monitor the ejection of new plasma components following radio bursts and to determine proper motion of these components, giving insights on the launching, collimation and acceleration of the relativistic jets. Moreover, the accurate determination of the luminosities will permit to test the accretion physics in a wide range of accretion rates well below the Eddington limit.

Special mentioning should be made about studies of the proper motion of pulsars. One of the main key science drivers of SKA1 will be the study of binary pulsars in order to test gravity in the strong field case. With SKA-VLBI, it will be possible to study both pulsar proper motions and parallaxes, with an accuracy of the parallax determination becoming better than 3 microarcsec. This information will be one of the keys for exploiting the complementary pulsar timing measurements. In order to achieve the required time precision, the effects of the intervening ISM (dispersion and scintillation) should be accurately determined.

Finally, it should be mentioned that the sensitivity and accuracy of SKA-VLBI brings the possibility to detect planets directly via astrometric observations of the small periodic motions resulting from planets orbiting around active stars.

# 4 Technical considerations

There are two options for the potential synergy betwen the SKA and VLBI: i) adding VLBI dishes to the SKA as an interferometer, for which the SKA will have to correlate the external stations together with the rest of the interferometer, and ii) VLBI networks get the SKA as an additional element, either completely phased or in parts. Option i) will need optical fibre transmission of data (e-VLBI), and option ii) is in principle possible with "classical" disk-recording-plus-shipping operations, but world-wide e-VLBI capabilities should be enhanced and favoured as well.

The SKA1-low version (mostly compact) is only suitable for the option ii), and for a limited set of available antennas. Only a few VLBI antennas have receivers at 327 MHz (90 cm), namely, the VLBA, GBT, Jb-1, Ar, Wb, Nt, Ur, Sm, and the GMRT. Those are non-cooled and do not offer a high performance, which results in a limited applicability to science.

Table 1: SEFD system noise in Jy for selected antennas

| $\lambda$ [cm] | 90 | 50 | 30 | 21 | 18 | 13 | 6 | 4 | 2 | 1 |
|---|---|---|---|---|---|---|---|---|---|---|
| SKA1-low | 2.8 | – | – | – | – | – | – | – | – | – |
| SKA1-mid | 4.4 | 4.4 | 4.4 | 2.1 | 2.1 | 2.1 | 2.3 | 2.8 | 2.8 | 2.8 |
| VLBA[a] | 2742 | 2744 | – | 289 | 314 | 347 | 210 | 327 | 543 | 640 |
| Y27 | 119 | – | – | 17 | 17 | – | 17 | 15 | 55 | 47 |
| Gb[b] | 35 | 28 | 24 | 10 | 10 | 12 | 14 | 15 | 18 | 22 |
| Jb-1 | 132 | 83 | – | 36 | 65 | – | 80 | – | – | – |
| Jb-2 | – | – | – | 350 | 320 | – | 320 | – | – | 910 |
| Wb | 150 | – | 120 | 30 | 40 | 60 | 120 | 120 | – | – |
| Eb | 600 | 600 | 65 | 20 | 19 | 300 | 20 | 20 | 44[c] | 90 |
| Ar | 12 | – | 3 | 3.5 | 3 | 3 | 5 | 6 | – | – |
| Ro70 | – | – | – | 35 | 20 | – | – | 18 | – | 83 |
| KVN | – | – | – | – | – | – | – | – | – | 1288 |
| Tm65 | – | – | – | 39 | 39 | 46 | 26 | 48 | – | – |
| Hh | – | – | – | – | 430 | 410 | 650 | 630 | – | – |
| Pk[d] | 40 | 40 | 42 | 30 | 30 | 110 | 110 | 43 | 370 | 81 |
| ATCA | 68 | 68 | 68 | 106 | 106 | 106 | 70 | 86 | – | 106 |
| Tb70 | – | – | 23 | 16 | 16 | – | – | 25 | – | 60 |
| Mp | 340 | 340 | 340 | 530 | 530 | 530 | 350 | 430 | 1300 | 530 |
| Hb | 470 | 420 | 650 | 650 | 640 | 640 | 1240 | 560 | 1200 | 1800 |
| Cd | 900 | 900 | 800 | 400 | 400 | 450 | 550 | 600 | 750 | 2500 |
| Ww | – | – | – | 8000 | 8000 | – | – | 8000 | – | – |
| Tg | – | – | – | – | – | – | – | 7700 | – | – |

Antenna keys: VLBA–Very Long Baseline Array, Ar–Arecibo, Eb–Effelsberg, Wb–Westerbork, Jb–Jodrell Bank (1: Lovell, 76-m; 2: Mk2, 25-m), Rob70–DSN Robledo, Hh–Hartebeesthoek, Tm65–Tianma 65-m (Shanghai), KVN–Korean VLBI Network dishes (Yonsei, Ulsan, or Tamna), Tb70–DSN Tidbinbilla, Mp–Mopra, Pk–Parkes, ATCA–Australia Telescope Compact Array (5×22-m tied array), Hb–Hobart, Cd–Ceduna, Ww-Warksworth, Tg–Tigo (earlier in Chile, planned to move to Argentina). [a]: VLBA Observational Status Summary 2015B.; [b]: GBT performance at NRAO web; [c]: Effelsberg Wiki Page [d]: For the LBA antennas, see their nominal SEFD values

The SKA-mid will include dishes observing at frequencies of 350 MHz to desirably 14 GHz. The antennas will be arranged with a central core of up to 3 km with arms extending to a radius of 100 km from the centre. This makes both options i) and ii) feasible. The frequency setup is compatible with the standard VLBI wavelengths of 49, 30, 21, 18, 13, 6, and 5 cm (2 cm in the future). The typical system equivalent flux density[1] will be of 1.7 Jy.

[1]SEFD, defined as the flux density of a radio source that doubles the system temperature. The baseline sensitivity for two antennas with $\mathrm{SEFD}_1$ and $\mathrm{SEFD}_1$ is $\Delta S = \sqrt{\mathrm{SEFD}_1 \times \mathrm{SEFD}_2}/[\eta_s \cdot \sqrt{2 \cdot \Delta\nu \cdot \tau_{\mathrm{ff}}}]$ Jy, where $\eta_s \leq 1$ accounts for the VLBI system inefficiency, $\Delta\nu$ is the bandwidth in Hz, and $\tau_{\mathrm{ff}}$ is the fringe-fit interval

Eventually, the core of the SKA in the Phase 2 would reach values of 0.27 Jy. In Table 1 we show the sensitivity for selected antennas to be part of such an array. Following the EVN status table, most atennas have receivers at 18/21 cm, and at 6 cm.

Option i) will only be feasible in the full SKA2 realisation, at a later point, with antennas connected by optical fibre (e-VLBI), and with an enhanced correlator.

Option ii) with SKA1-mid will have a VLBI network with a much enhanced image sensitivity at long baselines. SKA-mid will participate as a very sensitive element in conventional VLBI arrays. SKA-mid beats in sensitivity of most telescopes, and is comparable to Arecibo (altough this antenna is only available for a limited region of the sky in a short period of time due to the elevation constrains). In Table 2 we show the common range of time for different elevations for a realisation of the SKA in South Africa and Western Australia.

When SKA-mid is phased to participate as the most sensitive dish of the SKA-VLBI array, the beams of the tied-array are narrower than the field of view of the individual telescopes that form the VLBI array. They can be even smaller than the typical separation between the target and the phase calibrator source. Thus, an additional SKA-phased array beam will be required per phase calibrator. Additionally, it has been shown that the use of multi-view approaches (observations of a number of phase calibrators ($\geq$3) to map the spatial atmospheric distortions around a certain VLBI taget) improves the phase determination, image quality and astrometric accuracy [17]. In this case, some additional tied array beams (as many as phase calibrator sources) will be needed.

The option ii) for VLBI operations will be of extreme interest when combined with future space VLBI missions, for which new frontiers in sensitivity (with a ground array enhanced with the SKA) and resolution will be reached.

It should also be mentioned that the total intensity and polarisation calibration (both leakage terms and the determination of the absolute orientation of the electric vector) of the VLBI data will benefit from the availability of both the phased-array (for the SKA-VLBI array) and the local SKA interferometric data.

## 5 Concluding Remarks

SKA-VLBI will improve the sensitivity of the current VLBI arrays, providing unique contributions in different research fields like the study of transients or the accurate determination of distances and luminosities. It will probe the central engines of stellar, intermediate mass and massive black holes with unprecedented sensitivity. HI observations will permit to test the feedback mechanisms between the central AGN and the host galaxy. Thanks to its superb astrometric capabilities, SKA/VLBI will determine accurately pulsar and stellar parallaxes, providing for the case of pulsars in binary systems the most stringest tests of gravity. The amplitude and polarisation calibration of SKA-VLBI will be very accurate thanks to the combination of the SKA1 and the SKA-VLBI interferometric data. The developments in the

---

in seconds (less than or about equal to the coherence time) [40]. The theoretical image sensitivity for an homogeneous array with natural weighting would be $\Delta I_m = \mathrm{SEFD}/\eta_c\sqrt{N(N-1)t_{\mathrm{int}}\Delta\nu}$ Jy beam$^{-1}$, where $N$ is the number of stations and $t_{\mathrm{int}}$ is the total integration time on source in seconds.

SKA will complement presend and future efforts in e-VLBI, and in space VLBI. In order to be competitive in terms of angular resolution and optimize the scientific return, the high frequencies (Band 5) are needed and should be promoted within the project.

Table 2: Common visible time in hours for the SKA sites with other array elements

| South African SKA | | | | | | | | | |
|---|---|---|---|---|---|---|---|---|---|
| Dec | W. Aus. | JP | CN | NZ | CL | USA-HI | USA-NM | USA-PR | Europe |
| +45° | 0.0 | 2.2 | **5.4** | 0.0 | 0.0 | 0.0 | 1.5 | **3.6** | **5.4** |
| +30° | 0.0 | 2.7 | **7.6** | 0.0 | 0.4 | 0.0 | 1.8 | **4.4** | **8.0** |
| +15° | 0.0 | 2.7 | **7.3** | 0.0 | 2.3 | 0.0 | 1.8 | **4.8** | **9.4** |
| 0° | 0.8 | 2.5 | **6.8** | 0.6 | **3.7** | 0.0 | 1.7 | **5.0** | **9.1** |
| −15° | 2.0 | 2.1 | **6.3** | 1.9 | **4.9** | 0.0 | 1.4 | **5.1** | **7.8** |
| −30° | 3.0 | 1.2 | **5.4** | 3.1 | **6.2** | 0.0 | 0.0 | **5.2** | 3.0 |
| −45° | **4.0** | 0.0 | 0.0 | **5.2** | **7.9** | 0.0 | 0.0 | **5.2** | 0.0 |
| −60° | **5.5** | 0.0 | 0.0 | **11.8** | **11.6** | 0.0 | 0.0 | 0.0 | 0.0 |
| −75° | **12.9** | 0.0 | 0.0 | **24.0** | **24.0** | 0.0 | 0.0 | 0.0 | 0.0 |
| −90° | **24.0** | 0.0 | 0.0 | **24.0** | **24.0** | 0.0 | 0.0 | 0.0 | 0.0 |
| Western Australian SKA | | | | | | | | | |
| Dec | ZA | JP | CN | NZ | CL | USA-HI | USA-NM | USA-PR | Europe |
| +45° | 0.0 | **5.3** | **5.3** | 0.0 | 0.0 | **4.0** | 1.3 | 0.0 | **5.3** |
| +30° | 2.0 | **7.9** | **7.9** | 3.2 | 0.0 | **4.8** | 1.7 | 0.0 | **4.5** |
| +15° | 3.5 | **8.6** | **9.2** | **5.2** | 0.0 | **5.1** | 1.7 | 0.0 | 3.8 |
| 0° | **4.5** | **8.5** | **6.7** | **6.6** | 0.0 | **5.3** | 1.6 | 0.0 | 3.2 |
| −15° | **5.5** | **7.8** | **8.3** | **7.8** | 0.4 | **5.4** | 1.3 | 0.0 | 2.3 |
| −30° | **9.0** | **5.1** | **6.2** | **9.1** | 1.9 | **5.5** | 0.8 | 0.0 | 0.4 |
| −45° | **7.7** | 0.0 | 0.0 | **10.8** | **5.2** | **5.4** | 0.0 | 0.0 | 0.0 |
| −60° | **11.5** | 0.0 | 0.0 | **14.1** | **11.7** | **5.1** | 0.0 | 0.0 | 0.0 |
| −75° | **24.0** | 0.0 | 0.0 | **24.0** | **24.0** | 0.0 | 0.0 | 0.0 | 0.0 |
| −90° | **24.0** | 0.0 | 0.0 | **24.0** | **24.0** | 0.0 | 0.0 | 0.0 | 0.0 |

Key: W. Aus.: Western Australia; ZA: South Africa; JP: Japan, CN: Eastern China; NZ: New Zealand; CL: Chile (latitude of former TIGO site); USA-HI: Hawaii; USA-NM: New Mexico; USA-PR: Puerto Rico; Europe: Id.

# Acknowledgments

We acknowledge A.P. Lobanov for useful discussions and comments to the manuscript. MINECO grants AYA2012-38491-C02-01 (AA & MAPT), AYA2012-38491-C02-02 (ER & JCG), AYA2013-40979-P (MP), AYA2013-48226-C3-2-P (MP), and AYA2013-40825-P (IA & JLG) are acknowledged. Generalitat Valenciana grant PROMETEOII/2014/057 (ER & JCG) is acknoweledged.

# SKA Astrometry

**J.C. Guirado[1,2], I. Agudo[3], A. Alberdi[3], R. Azulay[2], F. Colomer[4], J.L. Gómez[3], I. Martí-Vidal[5], J. Moldón[6], M.A. Pérez-Torres[3,7,8], M. Perucho[1,2], M. Ribó[9], M.J. Rioja[10,4], and E. Ros[11,1,2]**

[1] Observatorio Astronómico, Universidad de Valencia, Spain
[2] Departamento de Astronomía y Astrofísica, Universidad de Valencia, Spain
[3] Instituto de Astrofísica de Andalucía – CSIC, Granada, Spain
[4] Observatorio Astronómico Nacional, Alcalá de Henares, Spain
[5] Onsala Space Observatory, Chalmers Univ. of Technology, Onsala, Sweden
[6] Netherland Foundation for Radio Astronomy, Dwingeloo, The Netherlands
[7] Centro de Estudios de la Física del Cosmos de Aragón, Teruel, Spain
[8] Departamento de Física Teórica, Universidad de Zaragoza, Spain
[9] Dep. d'Astronomia i Meteorologia, Ins. Ciències del Cosmos, Univ. Barcelona, Spain
[10] International Centre for Radio Astronomy Research, Univ. Western Australia, Australia
[11] Max-Planck-Institut für Radioastronomie, Bonn, Germany

## Abstract

We report on the astrometric capabilities of the different configurations of the SKA. For SKA1-MID, the large enhancement in sensitivity will allow astrometric studies at sub-mas level of a wide variety of objects well below the detection threshold of present VLB arrays. Microarcsecond astrometric accuracy will necessarily come from combination of SKA1-MID with existing VLBI networks or from the SKA2 realization using conventional in-beam phase-referencing or Multiview techniques. We describe some astrometry projects that will become accessible with the SKA.

## 1 Introduction

The measurement of the motion and size of the celestial bodies constitutes the foundations of our present understanding of the universe. The progressive improvement of the astrometric accuracy has permitted to attack and solve new questions in astrophysics. During the past decades, radio interferometric techniques have provided a giant improvement of the precision of astrometric accuracy, going from arcseconds to the microarcsecond level at cm-wavelengths.

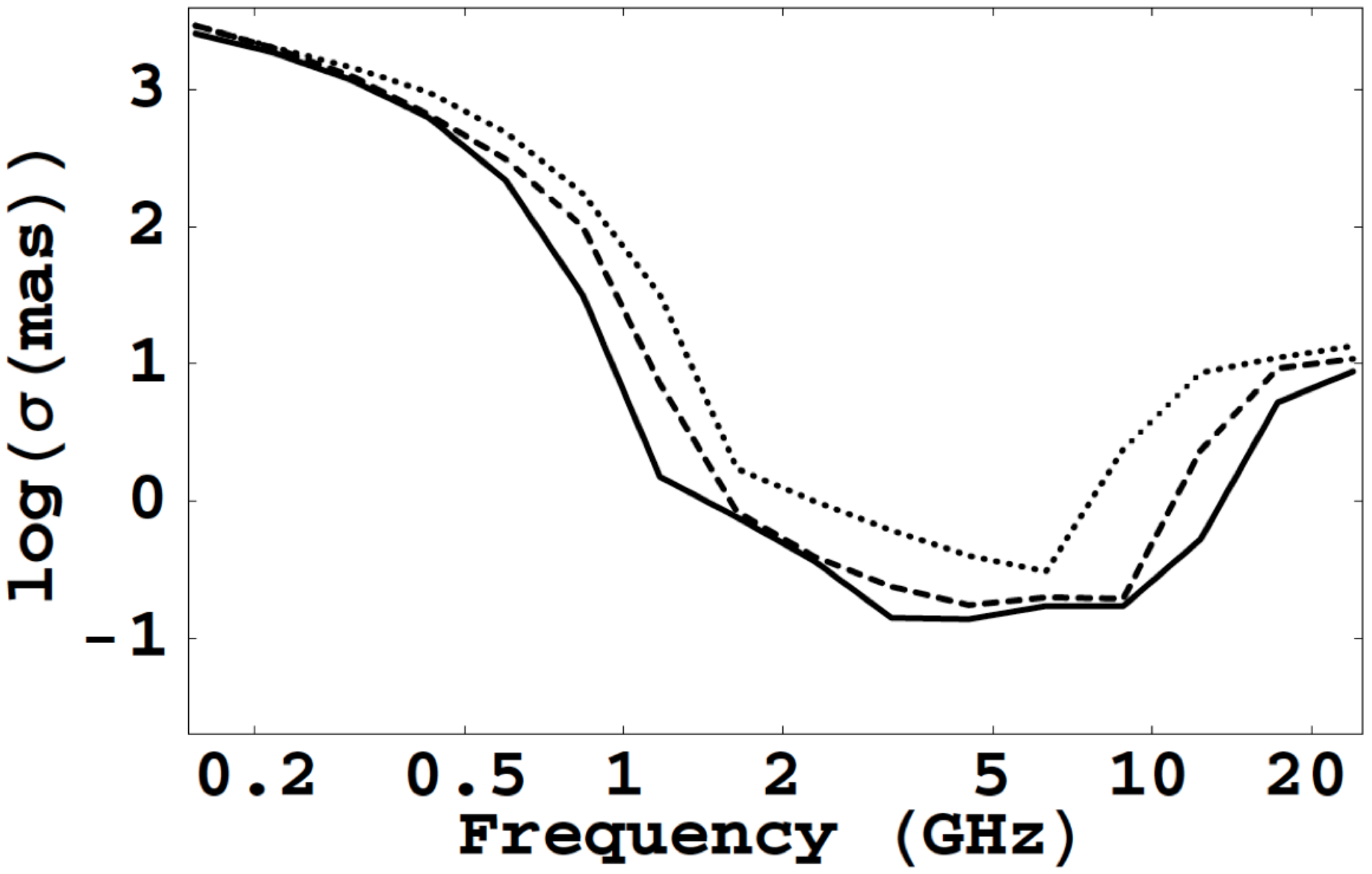

Figure 1: Astrometric accuracy as a function of frequency of an interferometric array with maximum baselines of 3000 km (after the simulations reported in [13]). Typical atmospheric turbulences and a target-calibrator separation of 5 deg are considered. Different flux densities of the target source are plotted, 10 $\mu$Jy (continuous line), 1 $\mu$Jy (dashed line), and 0.1 $\mu$Jy (dotted line).

Accordingly, a number of astronomical events have been either discovered or tested via radio astrometry [18] covering a wide range of fields from star formation until cosmology including tests of general relativity. The advent of the SKA will boost the sensitivity to the order of tens of nJy, and combined with dedicated calibration techniques will allow the access of radio astrometry to new astrophysical scenarios. The astrometric capabilities of SKA have been discussed in previous works [7, 17]. Here we explore the limits of the astrometric accuracy of SKA both working as a standalone array and in combination with existing radio telescopes.

## 2 Astrometry with SKA

In absence of atmosphere, the theoretical astrometric uncertainty of an interferometric array, $\sigma_{th}$, can be expressed as [21]:

$$\sigma_{th} \sim \left(\frac{\theta_0}{\mathrm{D}}\right) \tag{1}$$

where $\theta_0$ is the size of the beam, and D is the dynamic range of the image of the target source. Using present VLBI arrays at cm–wavelengths, typical values of $\sigma_{th}$ may reach the microarsecond ($\mu$as). However, systematic effects and atmospheric turbulences prevent VLBI astrometry achieving such a high accuracies in phase-referencing observations at other frequency regimes. These effects can be cancelled out, only to a certain degree, if using calibrator sources close to the target object. This is illustrated in Fig. 1: under the assumption of an interferometer array with a maximum baseline of 3000 km (actually, very similar to the planned SKA2 realization), Martí-Vidal et al. [13] reported a complete study based on Monte Carlo simulations of the behaviour of the astrometric accuracy and sensitivity of the array as a function of the frequency and calibrator source separation. These authors found that the astrometric performance of such an array is optimum between 1 GHz and 10 GHz; out of this range, the accuracy is degraded by the ionosphere (lower frequencies) or the wet troposphere (higher frequencies). We show in Fig. 1 the difficulty to reach $\mu$as-accurate astrometry with an array similar to the SKA in resolution and sensitivity. Therefore, the application of conventional in-beam phase-referencing and Multiview techniques [6, 19], which minimize the astrometric errors as those shown in Fig. 1, appear essential to achieve accuracies close to the values given by $\sigma_{th}$. According to the Multiview techniques, with multiple simultaneous observations from different lines of sight it is possible to solve for a full 2D correction to the atmospheric distortions around the target, thereby providing significantly improved calibrations and enabling astrometric measurements. This configuration allows the use of calibrator sources with separations beyond the ionospheric patch size. Therefore, the Multiview approach allows one to select as calibrators well-studied sources which have been shown to be steady and compact at higher frequencies, addressing the issue of whether the reference position is stable in weak sources.

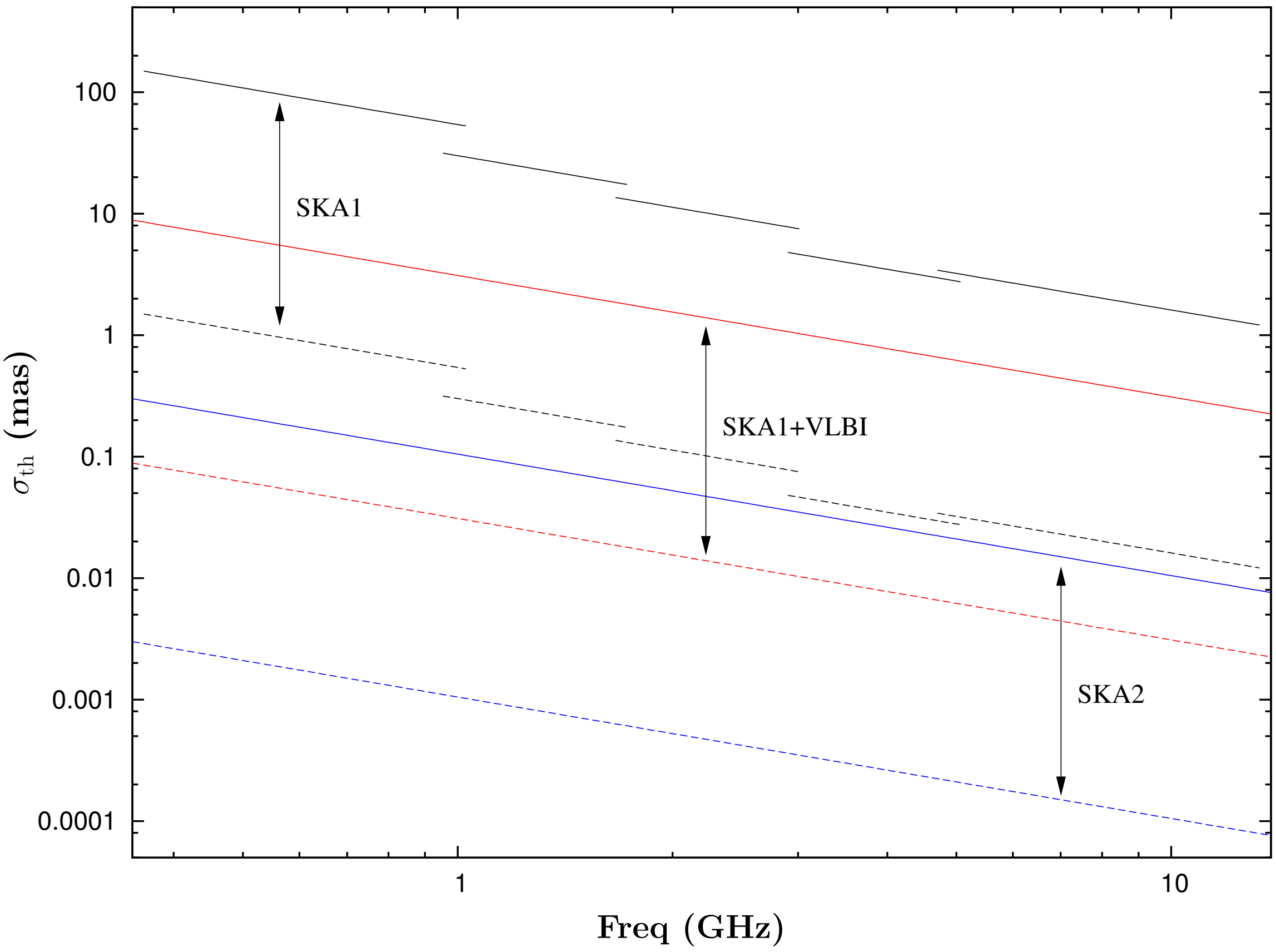

Figure 2: Theoretical astrometric accuracy of different SKA configurations (see text). For each configuration, two flux densities of the target source are considered, 10 $\mu$Jy (continuous lines) and 1 mJy (dashed line). Dynamic range values have been calculated using the sensitivities given in [20] for 1 hr integration time.

We show in Fig. 2 the behaviour of $\sigma_{th}$, i.e., the *maximum* accuracy attainable, for different configurations of SKA:

- **SKA1-mid**. Considering the present SKA Phase 1 design, the maximum baseline lengths will come from the SKA1-mid version, around 200 km, which are in principle certainly short if compared with the thousand-km size of present VLB arrays. Accordingly, even for substantial dynamic range and high frequencies, the accuracy would be limited to tenths of the milliarcsecond.
- **SKA1+VLBI**. To approach the $\mu$as accuracy, the resolution of SKA1 needs to be enhanced with the addition of VLBI stations to form longer baselines. In the case considered for Fig. 2, SKA1-mid is used as a hypersensitive phased-array within a VLBI network (in practice the array used to calculate the accuracy values extended the baseline lengths to 10000 km).
- **SKA2**. Alternatively to the previous item, the resolution can be improved by adding new antennas to the SKA1 configuration. This could be effectively done connecting electronically present VLBI stations to SKA1; however, the full development of this option will have to wait until completion of SKA2 realization, which will extend SKA1-mid spiral arms to baselines up to 3000 km. We have calculated the astrometric accuracy for this latter case.

SKA1 will have the capability for carrying out Multiview observations using the ability to form at least four independent pointed beams of phased-up VLBI outputs within the primary beam of a single antenna, which is ∼1 degree (i.e. multiple "in-beam" type calibrators). This type of observations are compatible with the capabilities of antennas from existing VLBI networks for joint observations with SKA1. Multiview with larger target-calibrator separations are possible with SKA1, using the capability for multiple tied sub-arrays (i.e. groups of antennas pointed simultaneously at different sources), although this is not widely available at other sites. On the other hand, with antennas equipped with phase array feeds (PAFs) spanning up to 3000 km baselines, planned for the SKA2 realization, any target-calibrator separation ("in-beam" type or not) for Multiview observations will be within reach.

In summary, for SKA in phase 1, the extraordinary continuum sensitivity, $\sim 1\,\mu$Jy for 1hr integration time, will allow astrometric studies at sub-mas level of a wide variety of ob-jects well below the detection threshold of present VLB arrays. Significant improvements in astrometric accuracy, even in the most favorable configuration of array/source geometry (allowing either in-beam phase-referencing or multibeaming), necessarily will come from the SKA2 realization or combination of SKA1-mid with existing VLBI networks.

## 3 Science from SKA Astrometry

### 3.1 Stellar astrometry

Although mostly based on targeted observations, our present understanding of stellar radio emission comprises from young stellar objects until supernova remnants, going through virtually every stage of the stellar evolution [8]. SKA, in any of its realizations, will provide a comprehensive study of radio stars from well-defined samples, and many astrometric projects will benefit from a precise determination of the proper motion, parallax, and possible further perturbations of their trajectories. Here we mention only some of the possible projects.

**Distances and the 3D structure of star forming regions** in the solar neighborhood can be definitively established. Present surveys on the *Gould's belt* [11] could be extended to hundreds of stars on those regions accesible by the SKA (i.e. star forming regions in Ophiucus, Taurus, Perseus, Aquila/Serpens, and Lupus). Kinematics within these regions could be measured, of relevance to the understanding of stellar formation activity. Similarly, **distance to stellar clusters** could be defined with high precision. The Hyades and the Pleaides clusters have been used as reference in astronomical distances, from which the cosmic distance ladder is defined. However, the "Pleaides distance controversy" put into question the Pleiades distance derived by HIPPARCOS, different from the distance derived by other techniques [23, 16]. VLBI observations have shown the possibility to use radio stars to solve this important ambiguity [15]. The contribution of SKA will extend the distance determinations to other clusters which, in turn, will be also observed by Gaia, surely producing new "Gaia/SKA controversy" episodes that could be studied at the microarcsecond level.

**Calibration of the stellar mass-luminosity relation**, in special for pre-main sequence (PMS) objects, is relevant as this relation is profusely used to derive masses of stellar, and substellar objects. Many of the stars belonging to nearby clusters or moving groups are known to be double or multiple systems. SKA astrometry would be directed to determine or refine measurements of the distance and/or orbital motion of hundreds of stars (where RV methods might be hampered by the activity inherent to the radio emission). Taking into account that the age of these objects is precisely determined given their membership to known clusters, the dynamical masses obtained by successive SKA observations would impose strong constraints to stellar evolutionary models [4, 5]. This will specially relevant to alleviate the deficient observational status of dynamical masses of PMS stars [9]. The $\mu$as accuracy provided by **SKA1+VLBI or SKA2 would reveal low-mass objects, brown dwarfs or planets**, around the sampled stars, whose mass could be measured with high pre-cision. The moving groups found in the vicinity of the Sun [22, 24] are specially well suited for this purpose with SKA: these associations are nearby (15–50 pc), young (5–100 Myr, favoring the possible radio emission of the host stars and –luckily– the low-mass companions [3]), and located well in the South ($\beta$ Pictoris, Tucana-Horologium, TW Hydrae, Columba, Carina, Argus, and AB Doradus moving groups). Possible microarcsecond astrometric signatures present in stars members of these groups could be accurately sampled by the SKA, with capability to detect from a Jupiter-like planet around a $1\,M_\odot$ star at 50 pc ($100\,\mu$as signature) to an Earth-like planet around a $0.2\,M_\odot$ star at 15 pc ($1\,\mu$as signature).

### 3.2 AGN astrometry

Present VLBI arrays may reach $\mu$as accuracy in astrometric observations of pair of AGN cores with the appropriate geometry on the sky. SKA1+VLBI (and SKA2) will provide a similar (or superior) accuracy extending the application of astrometric techniques to micro-Jy level sources. Opacity effects near the AGN core, that translate to shifts of the core position as measured at different frequencies [14, 10], could be studied with high precision for a significant number of sources.

Interestingly, Gaia will provide optical positions for hundreds of AGNs with accuracies up to the microarcsec level that can be compared with those measured in radio; this will lead to relevant testing of AGN theories about the different location of optical and radio emission (i.e., the corona or disk of the black hole for the former, and last scattering surface of the jet for the latter). For the case of blazars, the class of radio loud AGN with powerful jets better oriented to the line of sight, the relative location of the multi-spectral-range emission region with regard to that at radio wavelengths has drastic consequences on the feasible high-emission (gamma-ray) regions [12]. Currently, estimates of the relative locations can be made for a small number of sources, by cross correlation of monitoring light curves at different wavelengths with those from particular jet knots identified on ultra-high resolution VLBI monitoring projects [1, 2]. The combination of high precision astrometric programs with SKA on large samples of blazar jets with the Gaia positions will provide the needed volume of data to make independent and robust tests on the relative location of the radio, optical, and even gamma-ray emission regions in blazars.

Additionally, the International Celestial Reference Frame (ICRF), defined by the radio positions of non-variable radio sources, should be tied to the future Gaia reference frame. Since AGNs are expected to be used for the alignment of the radio and optical frames, the understanding of the previous effects is essential.

## Acknowledgments

Spansih MINECO grants AYA2012-38491-C02-01 (ER, JCG & RA), AYA2012-38491-C02-02 (AA & MAPT), AYA2013-4079-P (MP), AYA2013-48226-C3-2-P (MP), and AYA2013-40825-P (IA & JLG) are acknowledged. Generalitat Valenciana grant PROMETEOII/2014/057 (ER, JCG & RA) is ac-knowledged.